\documentclass[12pt]{article}
\pdfoutput=1
\usepackage{amsfonts}
\usepackage{amsmath, amssymb}
\usepackage{youngtab}
\usepackage{hyperref}
\usepackage{psfrag}
\usepackage{feynmp}
\usepackage[pdftex]{graphicx}
\usepackage{braket}
\usepackage{bm}
\usepackage{subfigure}
\usepackage[numbers,sort&compress]{natbib}

\usepackage[utf8]{inputenc}

\allowdisplaybreaks[4]

\Yboxdim{5pt}

\newcommand{\ba}{{\bm a}}
\newcommand{\be}{{\bm e}}
\newcommand{\bv}{{\bm v}}

\makeatletter
    
    \@addtoreset{equation}{section}
  \makeatother

\usepackage{color}

\renewcommand{\title}[1]{\vbox{\center\LARGE{#1}}\vspace{5mm}}
\renewcommand{\author}[1]{\vbox{\center\large#1}\vspace{5mm}}

\begin{document}
\bibliographystyle{utphys}


\begin{titlepage}
\begin{center}
\vspace{5mm}
\hfill {\tt 
UT-Komaba-20-3
}\\
\hfill {\tt 
IPMU20-0110
}\\
\vspace{20mm}

\title{
ABCD of 't Hooft operators
\\
\vspace{2mm}
}
\vspace{7mm}

Hirotaka Hayashi$^a$, Takuya Okuda$^b$, and Yutaka Yoshida$^c$

\vspace{6mm}

$^a$Department of Physics, School of Science, 
Tokai University\\
Hiratsuka-shi, 
Kanagawa 259-1292, Japan\\
\vskip 1mm
\href{mailto:h.hayashi@tokai.ac.jp}{\tt h.hayashi@tokai.ac.jp}
\vspace{3mm}

$^b$Graduate School of Arts and Sciences, University of Tokyo\\
Komaba, Meguro-ku, Tokyo 153-8902, Japan \\
\vskip 1mm
\href{mailto:takuya@hep1.c.u-tokyo.ac.jp}{\tt takuya@hep1.c.u-tokyo.ac.jp}

\vspace{3mm}
$^c$Kavli IPMU (WPI), UTIAS, University of Tokyo \\
 Kashiwa, Chiba 277-8583, Japan \\
\href{mailto:yutaka.yoshida@ipmu.jp}{\tt yutaka.yoshida@ipmu.jp}

\end{center}

\vspace{10mm}
\abstract{
\noindent
We compute by supersymmetric localization the expectation values of half-BPS 't~Hooft line operators in $\mathcal{N}=2$ $U(N)$, $SO(N)$ and $USp(N)$ gauge theories on $S^1 \times \mathbb{R}^3$ with an $\Omega$-deformation.
We evaluate the non-perturbative contributions due to monopole screening by calculating the supersymmetric indices of the corresponding supersymmetric quantum mechanics, which we obtain by realizing the gauge theories and the 't~Hooft operators using branes and orientifolds in type II string theories.
}
\vfill

\end{titlepage}

\tableofcontents

\section{Introduction}
\label{sec:introduction}

The 't~Hooft line operator, defined by a singular Dirac monopole boundary condition
\begin{equation}
\frac12 F_{\mu\nu} dx^\mu \wedge dx^\nu \sim \frac{\bm B}{2} {\rm vol}_{S^2} \qquad (\bm B \text{: magnetic charge}) 
\end{equation}
on a gauge field,
is an interesting disorder operator that universally exists in all four-dimensional (4d) gauge theories.
The 't~Hooft operator has played important roles in understanding the physics of gauge theories both in non-supersymmetric~\cite{tHooft:1977nqb} and supersymmetric (SUSY)~\cite{Kapustin:2005py,Kapustin:2006pk,Alday:2009aq,Alday:2009fs,Drukker:2009id,Gaiotto:2010be,Gomis:2011pf} settings.

In~\cite{Ito:2011ea} the supersymmetric localization method was developed for the computation of the expectation values of  half-BPS 't~Hooft operators in 4d $\mathcal{N}=2$ gauge theories on $S^1 \times \mathbb{R}^3$.
Recently, Brennan, Dey, and Moore modernized the localization technique for 't~Hooft operators, in the case of the~$SU(N)$ gauge group, and proposed to use the supersymmetric quantum mechanics (SQMs) that compute the non-perturbative monopole screening (often also called monopole bubbling) contributions~\cite{Brennan:2018yuj, Brennan:2018moe, Brennan:2018rcn}.
In our previous paper~\cite{Hayashi:2019rpw} that considered the $U(N)$ gauge theory with hypermultiplets in the fundamental representation (SQCD), we explored, by extending a result in~\cite{Assel:2019iae}, the relation between wall-crossing in the SQMs and the ordering of 't~Hooft operators along the line (in $\mathbb{R}^3$) on which the operators are inserted.

The main aim of this paper is to generalize the results of these works in two directions.
First, for the $U(N)$ gauge group, we compute the correlation functions involving 't~Hooft operators with non-minimal charges
and study the relation between wall-crossing and the ordering of such operators, whereas~\cite{Hayashi:2019rpw} treated only the cases with minimal charges.
Second, we consider gauge groups other than~$SU(N)$ or~$U(N)$ and perform localization calculations for $SO(N)$ and $USp(N)$ gauge groups.

Let us elaborate on the summary in the previous paragraph.
We consider the 't~Hooft operator $T_{\bm B}$ with a magnetic charge~${\bm B}$ (an element of the cocharacter lattice~$\Lambda_\text{cochar}$) of a gauge group $G$.
In general the operator~$T_{\bm B}$ is a linear combination of products of more fundamental operators, and we need to supply more information than the charge~$\bm{B}$ to fully specify the operator as we will do on a case-by-case basis.
 The vacuum expectation value (vev)~$\langle T_{\bm B}\rangle$ on $S^1\times\mathbb{R}^3$ with an $\Omega$-deformation along the $(x^1,x^2)$-plane is a function of a pair of parameters $(\bm{a},\bm{b})$ taking values in the complexified Cartan subalgebra, and takes the form
\begin{equation}\label{eq:TB-vev}
\Braket{T_{\bm B} } = \mathop{\sum_{\bm{v} \in \bm{B}+ \Lambda_{\text{coroot}}}}_{|\bm v|\leq |\bm B|} e^{\bm{v}\cdot \bm{b}} Z_\text{1-loop}(\bm{v}) Z_\text{mono}(\bm B,\bm v),
\end{equation}
where $\Lambda_{\text{coroot}}$ is the coroot lattice and $|\bullet|$ denotes the norm given by the Killing form~\cite{Ito:2011ea}. 
The vev also depends on the $\Omega$-deformation parameter, and the complexified masses, which we suppressed in~(\ref{eq:TB-vev}).
The one-loop contributions~$ Z_\text{1-loop}(\bm{v})$ were determined for general gauge groups and matter contents in~\cite{Ito:2011ea}.
In~\cite{Gomis:2011pf,Ito:2011ea} non-perturbative contributions $Z_\text{mono}$ due to monopole screening were evaluated for gauge group $U(N)$ based on the correspondence~\cite{Kronheimer:MTh} between monopoles with Dirac singularities and instantons on the Taub-NUT space.
The new method of~\cite{Brennan:2018yuj, Brennan:2018moe, Brennan:2018rcn} identifies the monopole screening contributions with the supersymmetric indices of the appropriate SQMs and evaluates them by localization~\cite{Hwang:2014uwa, Cordova:2014oxa, Hori:2014tda} using the Jeffrey-Kirwan (JK) residue prescription~\cite{MR1318878}%
\footnote{%
A summary of the prescription can be found in Appendix~A.2 of our previous paper~\cite{Hayashi:2019rpw}.}, possibly combined with extra approximate calculations to capture Coulomb branch contributions.
The SQMs can be found by realizing the 't~Hooft operators using branes in string theory.

In this paper we extend the analysis of~\cite{Hayashi:2019rpw}  to non-minimal 't~Hooft operators in the $U(N)$ gauge theory with $2N$ flavors, and to 't~Hooft operators in the $SO(N)$ gauge theory with $N-2$ flavors and the $USp(N)$ gauge theory with $N+2$ flavors.%
\footnote{%
Each number of flavors is such that the beta function for the (non-abelian) gauge coupling vanishes.
}
A flavor is a hypermultiplet in the fundamental (also called vector for $SO(N)$) representation of the gauge group.
The gauge theories whose matter hypermultiplets are in this representation will be referred to as SQCDs.
We extend the relation between wall-crossing and operator ordering to the cases with non-minimal higher charges.
We also study 't~Hooft operator correlators in the theories with an adjoint hypermultiplet ($\mathcal{N}=2^*$ theories) with gauge groups $U(N)$, $SO(N)$, and $USp(N)$.

To read off the SQMs we will realize the gauge theories, 't~Hooft operators, and monopole screening in D2-D4-NS5-D6-brane systems, together with an orientifold 4- or 6-plane for the $SO(N)$ or $USp(N)$ gauge group.
For gauge group $U(N)$ our brane systems are T-dual to the D1-D3-NS5-D7-brane systems considered in \cite{Brennan:2018yuj, Brennan:2018moe, Brennan:2018rcn}, and the 4d theory is realized on D4-branes as in \cite{Witten:1997sc}. 
Gauge groups can be modified from $U(N)$ to $SO(N)$ or $USp(N)$ by introducing an orientifold 4- or 6-plane to the brane systems~\cite{Landsteiner:1997vd, Brandhuber:1997cc,Landsteiner:1997ei,Uranga:1998uj}.
We will propose that the 't~Hooft operators engineered by our brane constructions have the magnetic charges that correspond to exterior powers $\wedge^k V$ of the fundamental (or vector) representation $V$ of the Langlands dual of the gauge group.
(For the unitary gauge group we will also realize the operators that correspond to $\wedge^k \overline V$.)

Throughout the paper, we do not distinguish different global structures (gauge group topologies and discrete theta angles~\cite{Aharony:2013hda}) associated with a given gauge algebra.
Our brane construction realizes\footnote{
There can be subtleties in the brane realization of 't Hooft operators. For example, in \cite{Brennan:2018yuj, Assel:2019iae, Hayashi:2019rpw}, it has been found that some SQMs for 't Hooft operators in the 4d $\mathcal{N}=2$ $SU(N)$ (or $U(N)$) gauge theory with an odd number of flavors 
require a Chern-Simons term to avoid an anomaly. However the necessity of the Chern-Simons term is not clear from the brane construction. 
} 
minimal 't~Hooft operators, and the global structure of the corresponding gauge theory must be one of those which admit the minimal operators. 

The organization of the paper is as follows. 
In Section~\ref{sec:brane}, we propose a brane construction of 't~Hooft operators in the $U(N)$, $SO(N)$ and $USp(N)$ gauge theories. 
Section~\ref{sec:UNhigher} studies non-minimal 't~Hooft operators in the $U(N)$ gauge theories.%
\footnote{%
Within each subsection 
one symbol denotes a single quantity, but in different subsections it may denote different quantities.
For example the symbol~$Z_{ij}$ denotes the same quantity~(\ref{UN-Zij}) in Sections~\ref{sec:UN-wedge2V-and-wedge2barV} and~\ref{sec:UN-wedge2V-wedge2barV} (both within a single subsection Section~\ref{sec:UNSQCD}), but it denotes different quantities in the other subsections.
}
In Sections~\ref{sec:SO} and~\ref{sec:USp} we study 't~Hooft operators in the $SO(N)$ and $USp(N)$ gauge theories, respectively. 
We summarize our results and discuss open problems in Section~\ref{sec:conclusion}.  
Appendix~\ref{sec:thooft-QFT} collects useful facts about the $U(N)$, $SO(N)$ and $USp(N)$ groups. 
Appendix~\ref{sec:formulas} contains formulas for the one-loop determinants in 4d and 1d gauge theories.
In Appendix~\ref{sec:subtle} we study the correlation functions that involve an operator with magnetic charge corresponding to $\wedge^2V$ in $SO(N)$ and $USp(N)$ SQCDs.
We discuss the subtleties exhibited by these correlators and discuss their possible interpretations.

\section{Brane construction of 't~Hooft operators}
\label{sec:brane}

In this section we propose a brane construction of 't~Hooft line operators in 4d $\mathcal{N}=2$ theories with a gauge group $U(N)$, $SO(N)$, or $USp(N)$. 
We will focus on 4d theories with a single gauge group.
The constructions of the 4d theories themselves have been well-known~\cite{Witten:1997sc,Landsteiner:1997vd,Brandhuber:1997cc,Uranga:1998uj}.
Our aim is to engineer 't~Hooft operators in a useful way by combining the results of these earlier works with the insights from the works of Brennan, Dey and Moore~\cite{Brennan:2018yuj, Brennan:2018moe, Brennan:2018rcn}.
In particular we explain how to read off the SQMs that capture the monopole screening contributions to the 't~Hooft operator correlation functions.
We also propose what we call the ``extra term prescription", motivated by earlier works on instanton counting.
This prescription will be used in the computations for $SO$ and $USp$ gauge theories in later sections.
Finally, we will generalize some conjectures we made in~\cite{Hayashi:2019rpw} regarding the SQMs, wall-crossing, and the operator ordering.

\subsection{$U(N)$ SQCD and $\mathcal{N}=2^{\ast}$ theory on D4-branes}
\label{sec:braneUNSQCD}

\begin{figure}[t]
\centering
\subfigure[]{\label{subfig:SQCD-branes}
\includegraphics[width=4cm]{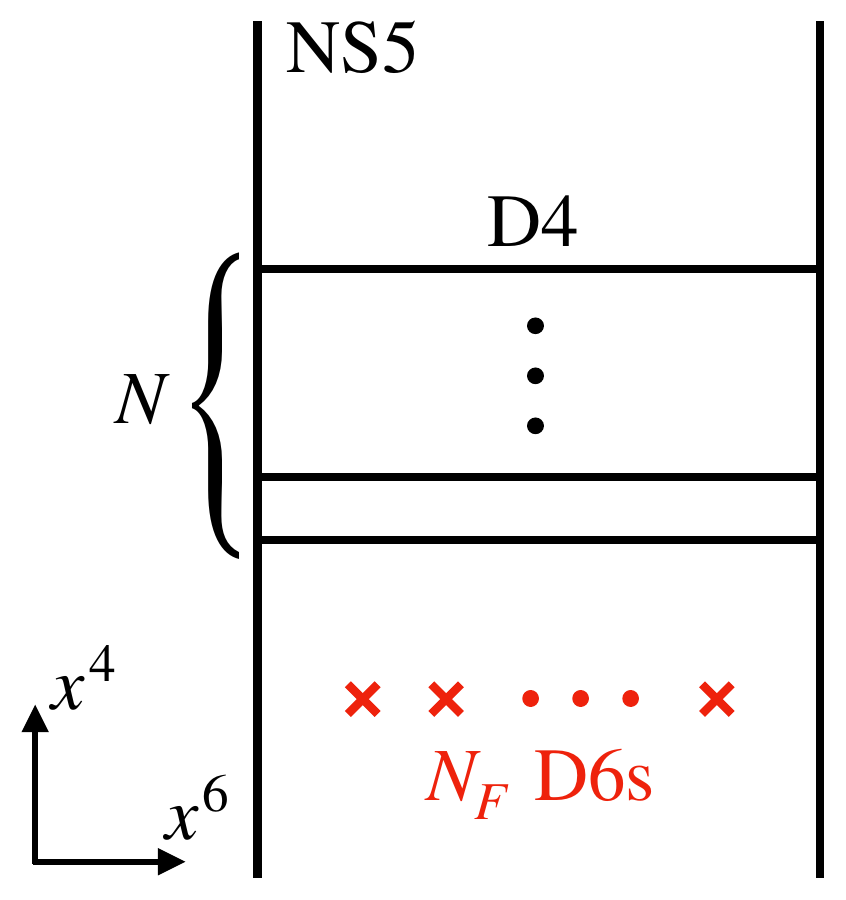}}
\hspace{0.5cm}
\subfigure[]{\label{subfig:SQCD-quiver}
\raisebox{.5cm}{\includegraphics[width=1cm]{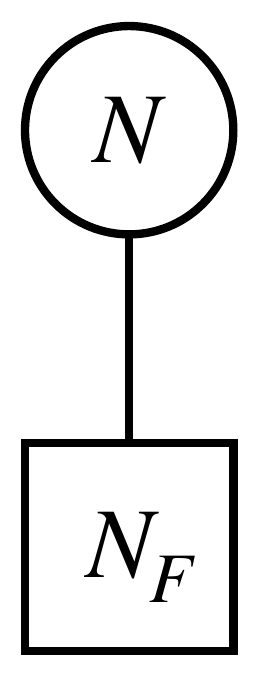}}
\hspace{.15cm}}
\hspace{0.35cm}
\subfigure[]{\label{subfig:SQCD-branes-4d}
\includegraphics[width=4.5cm]{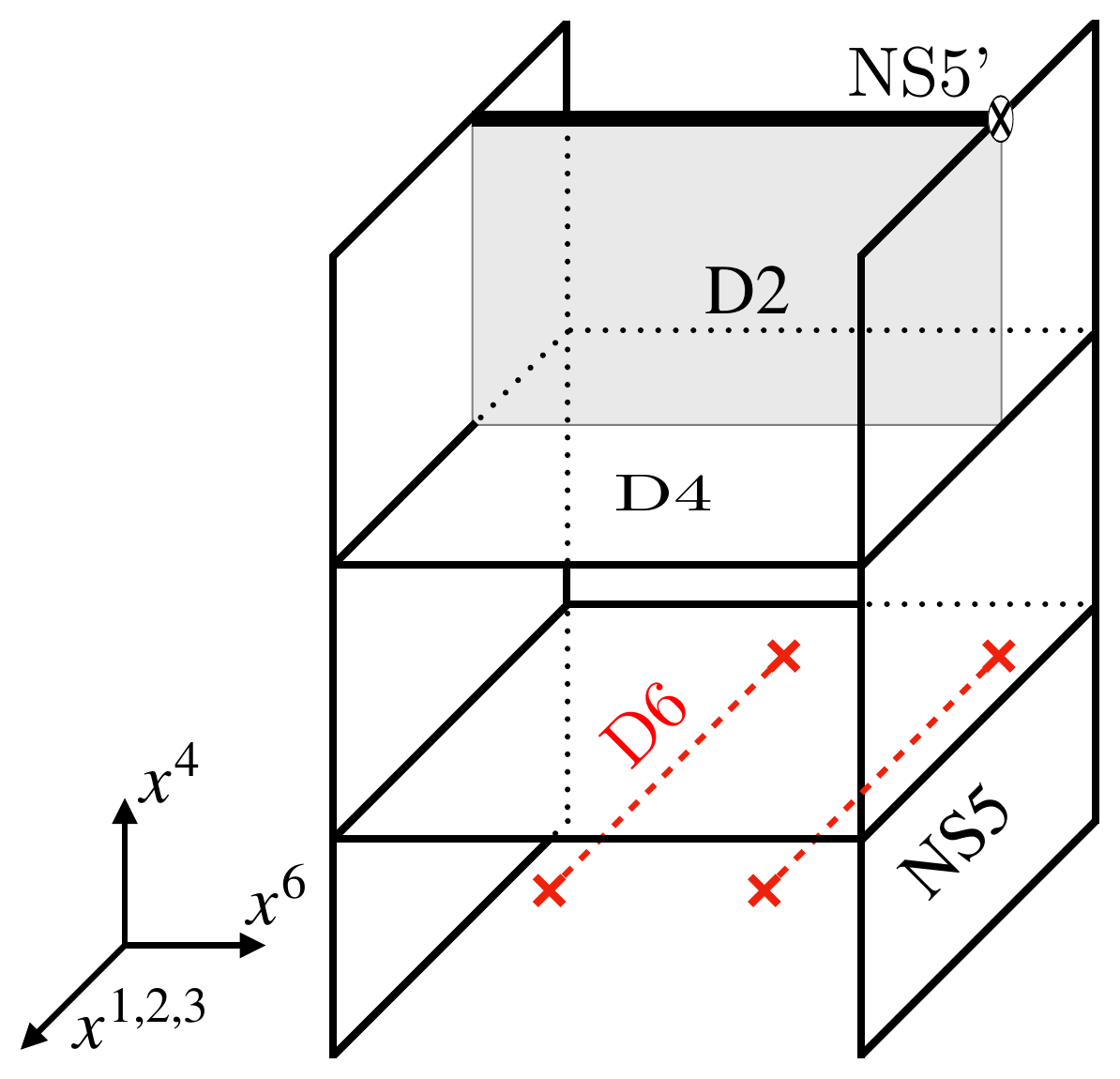}}
\hspace{.5cm}
\subfigure[]{\label{subfig:SQCD-branes-1234}
\raisebox{.5cm}{\includegraphics[width=4cm]{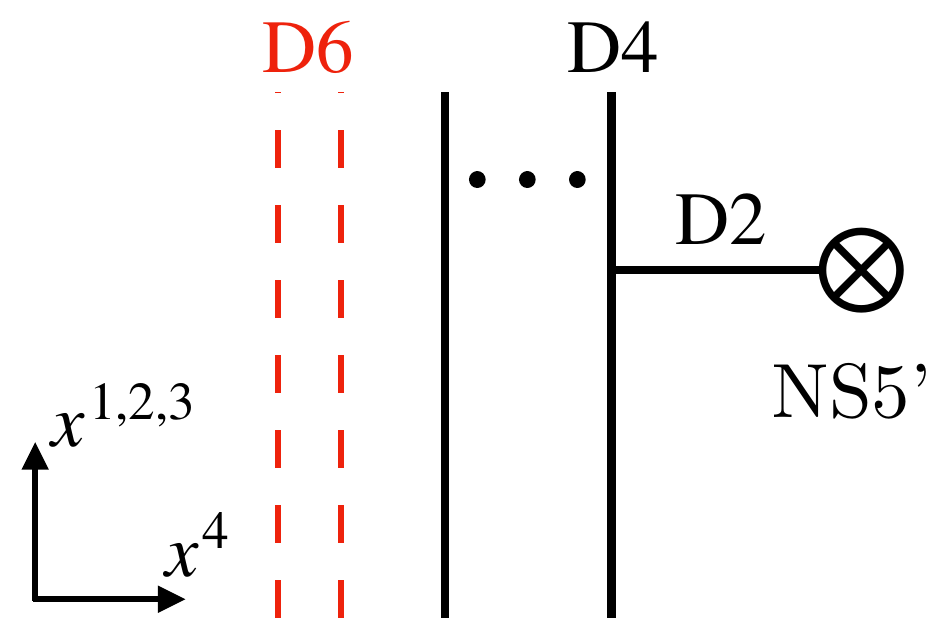}}}
\caption{(a): The brane configuration for the 4d $\mathcal{N}=2$ SQCD.
The $(x^4,x^6)$-directions are shown explicitly.
(b): The 4d $\mathcal{N}=2$ quiver diagram corresponding to~(a). A circle with $N$ means the $U(N)$ vector multiplets and a solid line represents a hypermultiplet in the fundamental representation. The number in a box implies the number of the fundamental hypermultiplets. We will use similar notation for 2d quivers later.
(c): The brane configuration for the SQCD with $N=N_F=2$, with a D2-brane and an NS5'-brane realizing an 't~Hooft operator.
In addition to the $(x^4,x^6)$-directions of (a), another direction that collectively represents $(x^1,x^2,x^3)$ is added to the figure.
While the D2-brane ends on the NS5-branes, the NS5'-brane (despite the $\otimes$ symbol in the figure) extends infinitely along the $x^6$-direction and intersects  the NS5-branes.
(d): The projection of the same brane system to the $x^4$-direction and another direction that collectively represents $(x^1,x^2,x^3)$.
The NS5'-brane and the D2-brane inserts an 't~Hooft operator with charge $\bm{B}=\bm{e}_N$.
}
\label{fig:SQCD}
\end{figure}

Let us begin with the brane construction of 't~Hooft operators in the 4d $\mathcal{N} = 2$ $U(N)$ SQCD with $N_F$ flavors.
We consider type IIA string theory in flat space~$\mathbb{R}^{1,9}$ with coordinates $x^\mu$ ($\mu=0,\ldots,9$) and introduce two NS5-branes placed at two points in the $(x^6,\ldots ,x^9)$-space separated only in the $x^6$-direction by a finite distance.
We introduce $N$ D4-branes that extend in the $(x^0,\ldots ,x^3,x^6)$-directions and are suspended between the two NS5-branes. 
Since the D4-brane is put on a finite segment in the $x^6$-direction, we obtain a 4d effective field theory at low energies living in the $(x^0,\ldots,x^3)$-spacetime.
To introduce hypermultiplets in the fundamental representation we include $N_F$ D6-branes between the two NS5-branes. The D6-branes are point-like in the $(x^4, x^5, x^6)$-space. 
This is the same construction as in~\cite{Witten:1997sc}, but we take the gauge group to be $U(N)$ rather than $SU(N)$;%
\footnote{%
One may first realize the $SU(N+1)$ theory by the NS5-D4-D6 system~\cite{Witten:1997sc} and then reduce to the $U(N)\subset SU(N+1)$ theory by giving an appropriate expectation value to the adjoint scalar.
} 
as we will see the $U(N)$ gauge group is more natural when considering 't~Hooft operators. The brane configuration realizing the 4d $U(N)$ gauge theory with $N_F$ flavors is depicted in Figure \ref{subfig:SQCD-branes}, where the $(x^4, x^6)$-space is shown explicitly. 

Magnetic charges can be introduced as the end points of D2-branes on the D4-branes~\cite{Diaconescu:1996rk}; the D2-branes are also bounded by the two NS5-branes.
To make the magnetic monopoles infinitely heavy, {\it i.e.}, to construct 't~Hooft operators rather than dynamical 't~Hooft-Polyakov monopoles, we let the D2-branes terminate at another NS5-brane at the opposite ends; we call it an NS5'-brane.%
\footnote{%
For the purpose of realizing an 't~Hooft operator, we can also make the D2-branes semi-infinite in the $x^6$-direction.
As shown in~\cite{Brennan:2018yuj} in a similar set-up, terminating the D2-branes by an NS5'-brane allows us to read off the SQMs that compute screening contributions.
} 
The brane configuration for an exmaple of an 't~Hooft operator in the 4d $\mathcal{N}=2$ $U(N)$ SQCD is depicted in Figures~\ref{subfig:SQCD-branes-4d} and~\ref{subfig:SQCD-branes-1234}. The magnetic charge of the 't~Hooft operator may be read off from which D4-brane the D2-brane ends on. 
We also summarize the directions in which these branes are extended in Table~\ref{table:brane-directions}.
\begin{table}[t]
\begin{center}
\begin{tabular}{c|c|ccc|cc|c|ccc}
&0&1&2&3&4&5&6&7&8&9
\\
\hline
D4/O4&$\times$&$\times$&$\times$&$\times$&&&$\times$ 
\\
D6/O6&$\times$&$\times$&$\times$&$\times$&&&&$\times$&$\times$&$\times$
\\
NS5 &$\times$&$\times$&$\times$&$\times$&$\times$&$\times$&&
\\
\hline
D2 &$\times$&&&&$\times$&&$\times$
\\
NS5'&$\times$&&&&&$\times$&$\times$&$\times$&$\times$&$\times$
\end{tabular}
\end{center}
\caption{The directions~($\times$) in which branes extend for a configuration engineering 't~Hooft operators in the 4d $U(N)$ gauge theories. For the 4d $\mathcal{N}=2^{\ast}$ $U(N)$ gauge theory, the $x^6$-direction is circle-compactified with a twist as in~(\ref{shift-for-two-star}) and no D6-branes are introduced. We also list the directions of the orientifolds which we will discuss in Sections~\ref{sec:SQCD-O4} and~\ref{sec:SQCD-2star-O6} to realize gauge groups different from $U(N)$. 
}
\label{table:brane-directions}
\end{table}

We can also construct the 4d $U(N)$  $\mathcal{N}=2^*$ theory and its 't~Hooft operators in a similar manner.
For this we compactify the $x^6$-direction on a circle and identify the two NS5-branes~\cite{Witten:1997sc}.
If we take the quotient by the simple shift $x^6 \rightarrow x^6 + 2\pi L$ with the other coordinates fixed, then the hypermultiplet in the adjoint representation in the 4d $\mathcal{N}=2^{\ast}$ $U(N)$ gauge theory is massless. 
To introduce the (complex) mass~$m$ for the adjoint hypermultiplet we include a shift in the $(x^4, x^5)$-directions when we go round the circle.  
Namely, we identify 
\begin{equation}\label{shift-for-two-star}
(x^4,x^5,x^6,x^\text{other}) \sim (x^4 + {\rm Re}\,m ,x^5 + {\rm Im}\,m,x^6+2\pi L, x^\text{other}) \,.
\end{equation}
We let $N$ D4-branes end on the single NS5-brane from the both sides.
The massless modes of the D4-D4 open strings that do not cross the NS5-brane form the $\mathcal{N}=2$ $U(N)$ vector multiplet.
The lightest modes of the D4-D4 open strings that cross the NS5-brane form the adjoint hypermultiplet whose mass $m$ is given by the shift.%
\footnote{\label{footnote:T-duality}%
Starting with this brane construction of the $\mathcal{N}=2^*$ theory, one can obtain the pure super Yang-Mills theory by taking the limit $m\rightarrow\infty$ and by integrating out the adjoint matter.
This procedure is related to a similar procedure in~\cite{Brennan:2018rcn} by the T-dualiy in the $x^6$-direction.
The single NS5-brane localized in the $(x^6,x^7,x^8,x^9)$-space turns into a 4d geometry with the topology of $\mathbb{R}^4$ in the $(x^6_\text{dual},x^7,x^8,x^9)$-directions, with background supergravity fields determined by~$m$ in~(\ref{shift-for-two-star}). 
(If $m$ were zero this geometry would be the single-center Taub-NUT space.)
The T-duality turns D4 into D3, and D2 into D1.
}

In our previous paper~\cite{Hayashi:2019rpw} we focused on minimal 't~Hooft operators, corresponding to the fundamental or anti-fundamental representation of the Langlands dual of the gauge group $G=U(N)$. The minimal 't~Hooft operator is realized by a single D2-brane ending on an NS5'-brane.  
In this paper we consider non-minimal 't~Hooft operators for the $U(N)$ gauge group.
We propose that

\vspace{.5cm}
\begin{minipage}[c]{15cm}
$k$ D2-branes stretched between a single NS5'-brane and the stack of D4-branes realize an 't~Hooft operator whose magnetic charge corresponds to the rank-$k$ anti-symmetric representation of the Langlands dual group $G^\vee=U(N)$.
\end{minipage}
\vspace{.5cm}

\begin{figure}[t]
\centering
\includegraphics[width=5.5cm]{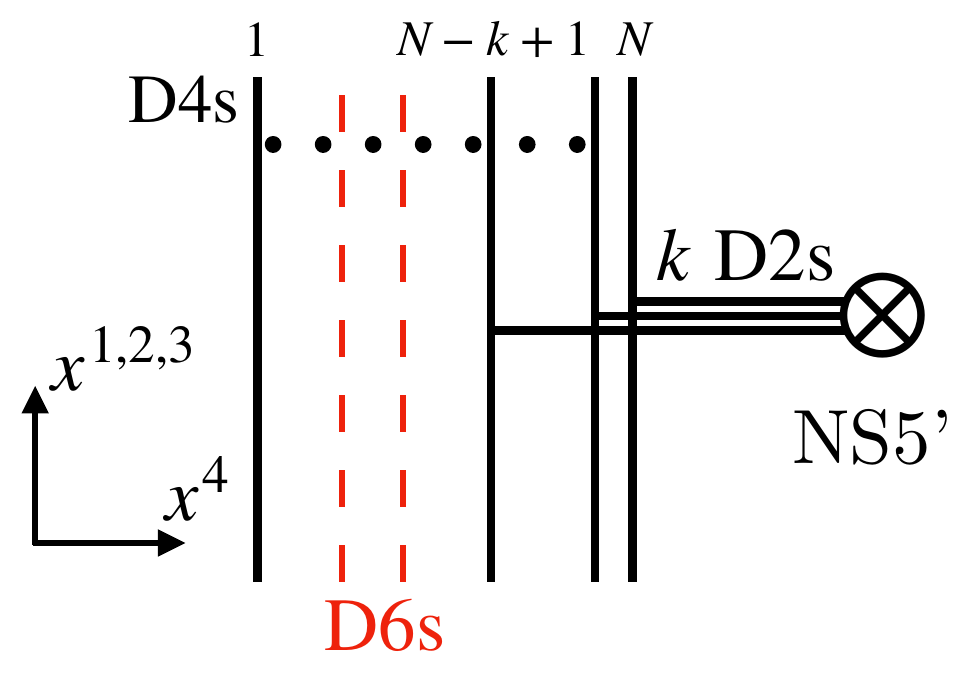}
\vspace{-3mm}
\caption{The $k$ D2-branes stretched between an NS5'-brane on the right and the stack of D4-branes realize an 't~Hooft operator with the charge corresponding to the anti-symmetric representation $\wedge^k V$ of $U(N)$.
An NS5'-brane on the left would give an operator corresponding to $\wedge^k \overline{V}$.
The locations of the D6-branes in the $x^6$-direction do not affect the vev.
}
\label{fig:SQCD-higher}
\end{figure}

\noindent
More precisely, for an NS5'-brane placed on the right (left) of the D4-branes, we propose that the $k$ D2-branes correspond to the $k$-th exterior power $\wedge^k V$ ($\wedge^k \overline{V}$) of the fundamental (anti-fundamental) representation of $G^\vee$.%
\footnote{%
We mentioned this possible correspondence in footnote~13 of~\cite{Hayashi:2019rpw}. 
See also~\cite{Brennan:2018moe} for a related discussion.
}
The brane configuration for~$\wedge^k V$
is depicted in Figure~\ref{fig:SQCD-higher}.
To realize BPS 't~Hooft operators, the s-rule~\cite{Hanany:1996ie} requires that for each pair of an NS5'-brane and a D4-brane there is at most a single D2-brane between them.

Since the D2-branes end on $k$ D4-branes, the corresponding 't~Hooft operator is charged under the Cartan generators of the Langlands dual group associated to the $k$ D4-branes. Hence it is natural to expect that the configuration in Figure~\ref{fig:SQCD-higher} realizes an 't~Hooft operator that corresponds to the rank-$k$ anti-symmetric representation of $U(N)$.
For the $\mathcal{N}=2^*$ $U(N)$ theory, our proposal above is S-dual to the fact~\cite{Gomis:2006sb}, well known in the context of AdS/CFT, that a single D5-brane with $k$ units of the fundamental string charge realizes a Wilson line operator for the $U(N)$ gauge theory realized on a stack of $N$ D3-branes.
In Section~\ref{sec:UNhigher} we will provide quantitative evidence for our proposal by generalizing our analysis for minimal operators in~\cite{Hayashi:2019rpw}.

\subsection{$SO/USp$ SQCD on D4-branes and an O4-plane}
\label{sec:SQCD-O4}

We can change the gauge group by introducing an orientifold. 
The inclusion of an O4-plane along the directions of the D4-branes in the brane configuration for the 4d $\mathcal{N}=2$ $U(N)$ SQCD described in Section \ref{sec:braneUNSQCD} engineers an SQCD with a gauge group $SO$ or $USp$~\cite{Landsteiner:1997vd,Brandhuber:1997cc}.%
\footnote{%
More generally one can realize a linear quiver theory with alternating $so$ and $usp$ gauge algebras by including more than two NS5-branes.
When an O4-plane crosses an NS5-brane its NS $\mathbb{Z}_2$ charge changes, while when an O4-plane crosses a D6-brane its RR $\mathbb{Z}_2$ charge changes~\cite{Hanany:2000fq}.
}
Both the O4-plane and the D4-branes extend in the $(x^0, x^1, x^2, x^3, x^6)$-directions. When the O4-plane is placed at $x^4 = x^5 = x^6 = x^7 = x^8 = x^9 = 0$, then the orientifold action on the spacetime coordinates is
\begin{equation}
(x^{0,1,2,3},x^{4,5},x^6,x^{7,8,9})
 \mapsto 
(x^{0,1,2,3},-x^{4,5},x^6,-x^{7,8,9}) \,.
\end{equation}
The brane configuration needs to be compatible with the orientifold action.

\begin{figure}[t]
\centering
\subfigure[]{\label{subfig:O4}
\includegraphics[scale=.25]{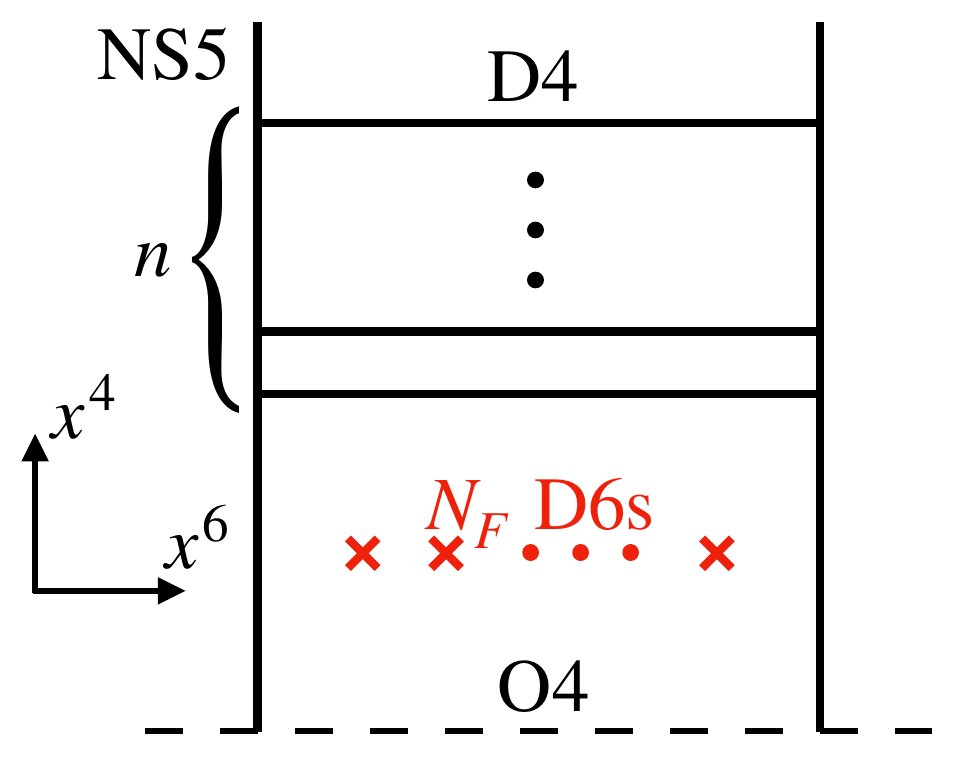}}
\hspace{1cm}
\subfigure[]{\label{subfig:O4-4d}
\raisebox{.0cm}{\includegraphics[scale=.25]{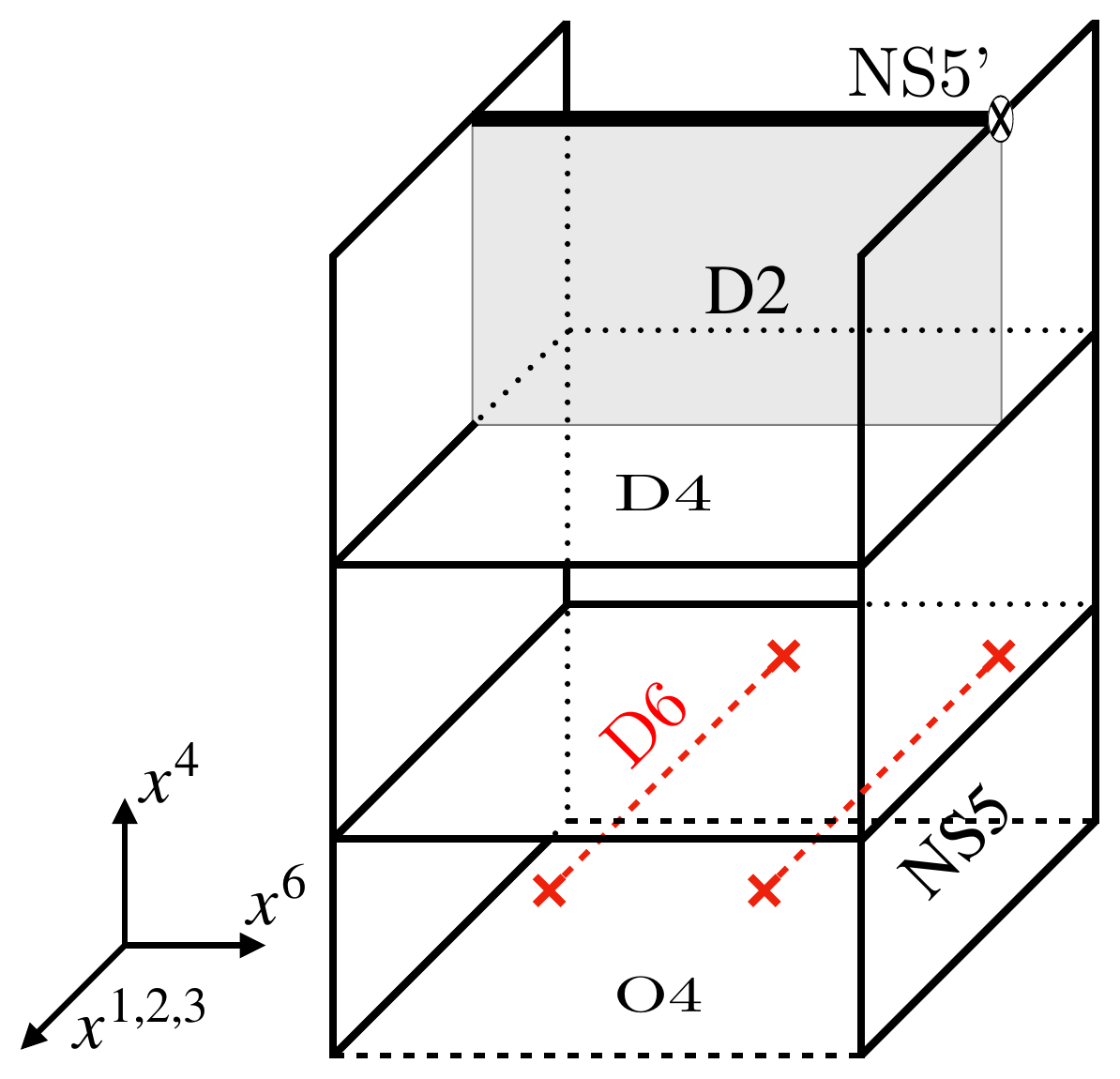}}}
\caption{(a): A brane construction of the $SO$/$USp$ SQCD with an O4-plane.
The $(x^4,x^6)$-directions are shown explicitly, and the mirror image is omitted.
(b): The brane configuration for the SQCD with $n=N_F=2$.
In addition to the $(x^4,x^6)$-directions of Figure \ref{subfig:O4}, another direction that collectively represents $(x^1,x^2,x^3)$ is added to the figure.
A D2-brane with an NS5'-brane realizing the minimal 't~Hooft operator is also shown.
}
\label{fig:O4}
\end{figure}

Let us look at the brane construction in more detail. We consider $N$~D4-branes, including physical as well as mirror image branes, stretched between two NS5-branes. 
Different types of the orientifold yield different gauge groups. 
A stack of~$n$ physical D4-branes on top of the O4$^-$-plane ($N=2n$) gives rises to the $SO(2n)$ gauge group.
On the other hand  a stack of~$n$ physical D4-branes on top of an $\widetilde{\text{O4}}{}^-$-plane  realizes the $SO(2n+1)$ gauge group. 
Because the D4-brane charge of an $\widetilde{\text{O4}}{}^-$-plane is the same as that of an O4$^-$-plane with a half D4-brane,\footnote{%
An $\widetilde{\text{O4}}{}^-$-plane may be referred to as an O4$^0$-plane \cite{Hori:1998iv} since the total D4-brane charge is zero. 
} 
the configuration effectively contains $N=2n+1$ D4-branes including the mirror images. 
Finally a stack of $n$ physical D4-branes on top of an O4$^+$-plane yields the gauge group $USp(2n)$. 
For the O4$^+$-plane a half D4-brane is not allowed to be stuck there, and $N=2n$ is the only possibility.\footnote{%
We do not consider an $\widetilde{\rm O4}{}^+$-plane
, which also leads to the $USp(N)$ gauge group.
}
Adding $N_F$ D6-branes between NS5-branes introduces~$N_F$ hypermultiplets in the fundamental (vector) representation. 
The gauge symmetry on D6-branes depends on the type of the O4-plane. 
We summarize the gauge symmetries on D4- and D6-branes in the following table.
\begin{center}
\begin{tabular}{c|cccc}
 & O4$^-$ &$\widetilde{\text{O4}}{}^-$ & ${\rm O4}{}^+$  \\
\hline
D4 & $SO(2n)$ & $SO(2n+1)$ & $USp(2n)$ \\
D6 & $USp(2N_F)$ & $USp(2N_F)$ & $SO(2N_F)$ 
\end{tabular}
\end{center}
This is consistent with the flavor symmetry group for $SO$ and $USp$ gauge theories.
Namely the flavor symmetry of an $SO$ gauge theory is the $USp$-type and that of an $USp$ gauge theory is the $SO(\text{even})$-type. The brane configuration for the $SO/USp$ SQCD is depicted in Figure~\ref{subfig:O4}.

\begin{figure}[t]
\centering
\subfigure[]{\label{subfig:O4-tHooft-V}
\raisebox{.0cm}{\includegraphics[scale=.25]{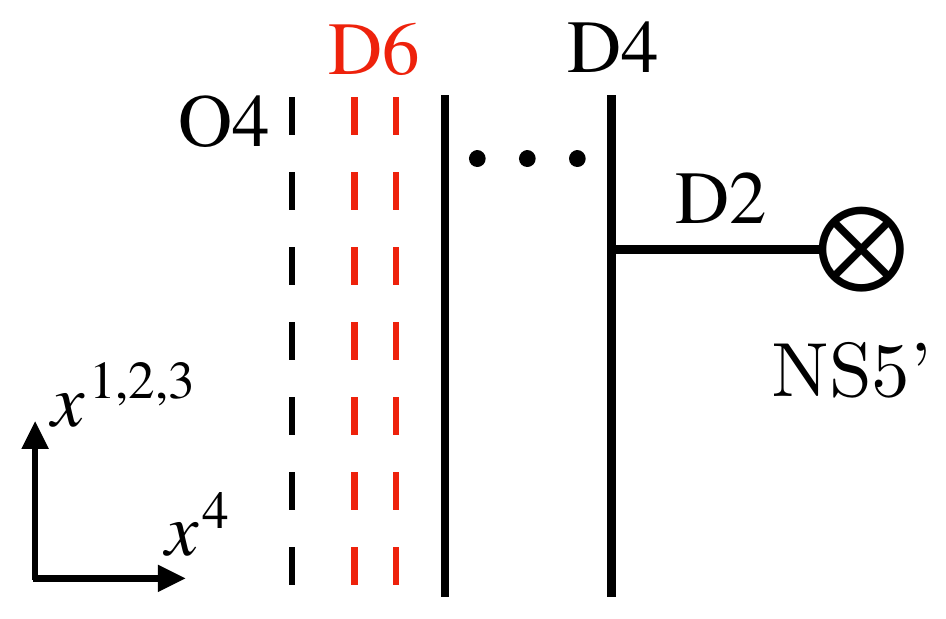}}}
\hspace{1cm}
\subfigure[]{\label{subfig:O4-tHooft-wedge2V}
\raisebox{.0cm}{\includegraphics[scale=.25]{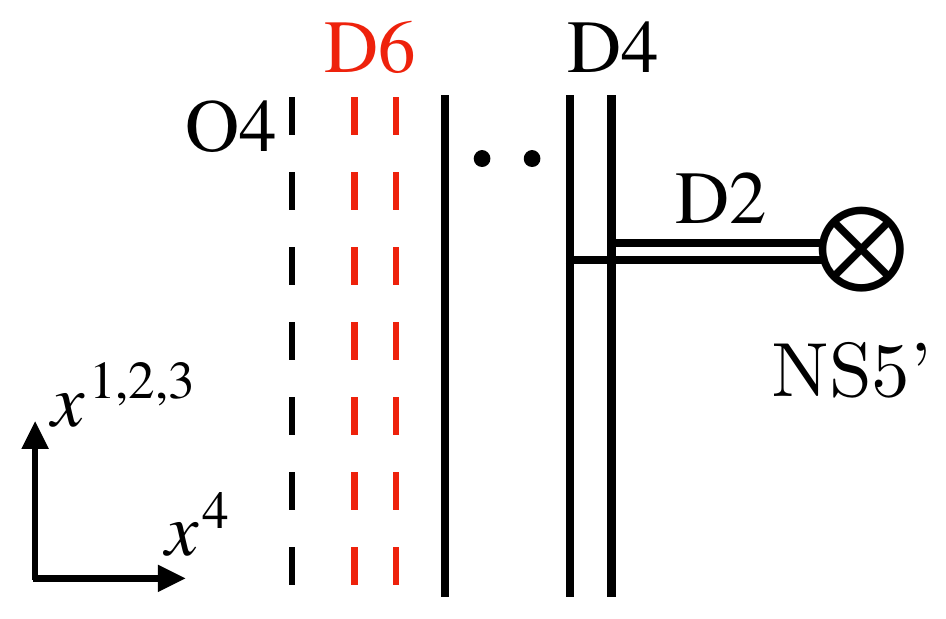}}}
\caption{(a): A realization of the minimal 't~Hooft operator corresponding to $V$ and $\bm{B}=\bm{e}_{n}$.
(b): A realization of the 't~Hooft operator corresponding to $\wedge^2 V$ and $\bm{B}=\bm{e}_{n-1}+\bm{e}_n$.
}
\label{fig:O4-thooft}
\end{figure}

As with the construction of the minimal 't~Hooft operator in the 4d $\mathcal{N}=2$ $U(N)$ SQCD, a minimal 't~Hooft line operator, corresponding to the fundamental representation $V$ of the Langlands dual of the gauge group, should be realized by a D2-brane stretched between a 
D4-brane and an NS5'-brane, as indicated in Figures~\ref{subfig:O4-4d} and~\ref{subfig:O4-tHooft-V}.
More generally we propose that for $k\leq n$, 
\begin{center}
\begin{minipage}{15cm}
$k$ D2-branes, stretched between the stack of $n$ D4-branes and a single NS5'-brane, realizes an 't~Hooft operator whose magnetic charge $\bm{B}\in\Lambda_\text{cochar}$ corresponds to~$\wedge^k V$, the rank-$k$ anti-symmetric representation of the Langlands dual~$G^\vee$ of the gauge group~$G$.
\end{minipage}
\end{center}
As an example, the brane configuration for~$\wedge^2V$  is shown in Figure~\ref{subfig:O4-tHooft-wedge2V}.

\subsection{$SO/USp$ SQCD and $\mathcal{N}=2^*$ theory on D4-branes and an O6-plane}
\label{sec:SQCD-2star-O6}

In Section \ref{sec:SQCD-O4}, we realized an $SO$ or $USp$ gauge group using an O4-plane. 
We can also use an O6-plane to engineer 
$SO(N)$ or $USp(N)$ gauge theories~\cite{Landsteiner:1997ei}. An O6-plane is extended in the same directions as those of D6-branes, namely in the $(x^0, x^1, x^2, x^3, x^7, x^8, x^9)$-directions. When an O6-plane is placed at $x^4 = x^5 = x^6 = 0$ the orientifold action on the spacetime coordinates is 
\begin{equation}\label{O6act}
(x^{0,1,2,3},x^{4,5,6},x^{7,8,9})
 \mapsto 
 (x^{0,1,2,3},-x^{4,5,6},x^{7,8,9}) \,.
\end{equation}
The brane configuration needs to be compatible with the orientifold action. 
\begin{figure}[t]
\centering
\subfigure[]{\label{subfig:O6}
\includegraphics[scale=.25]{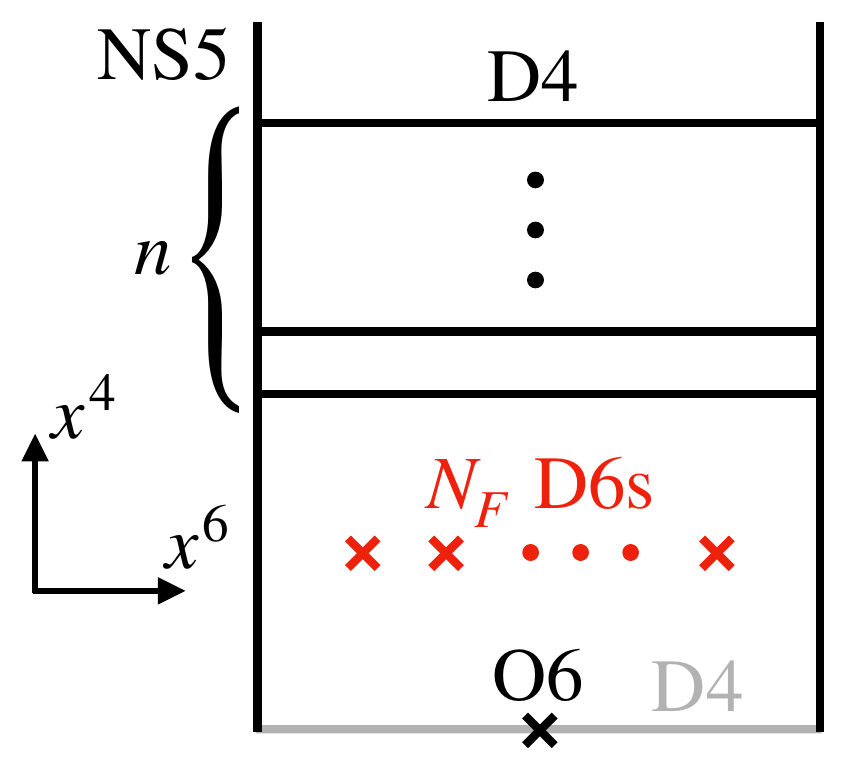}}
\hspace{1cm}
\subfigure[]{\label{subfig:O6-3d}
\raisebox{.0cm}{\includegraphics[scale=.25]{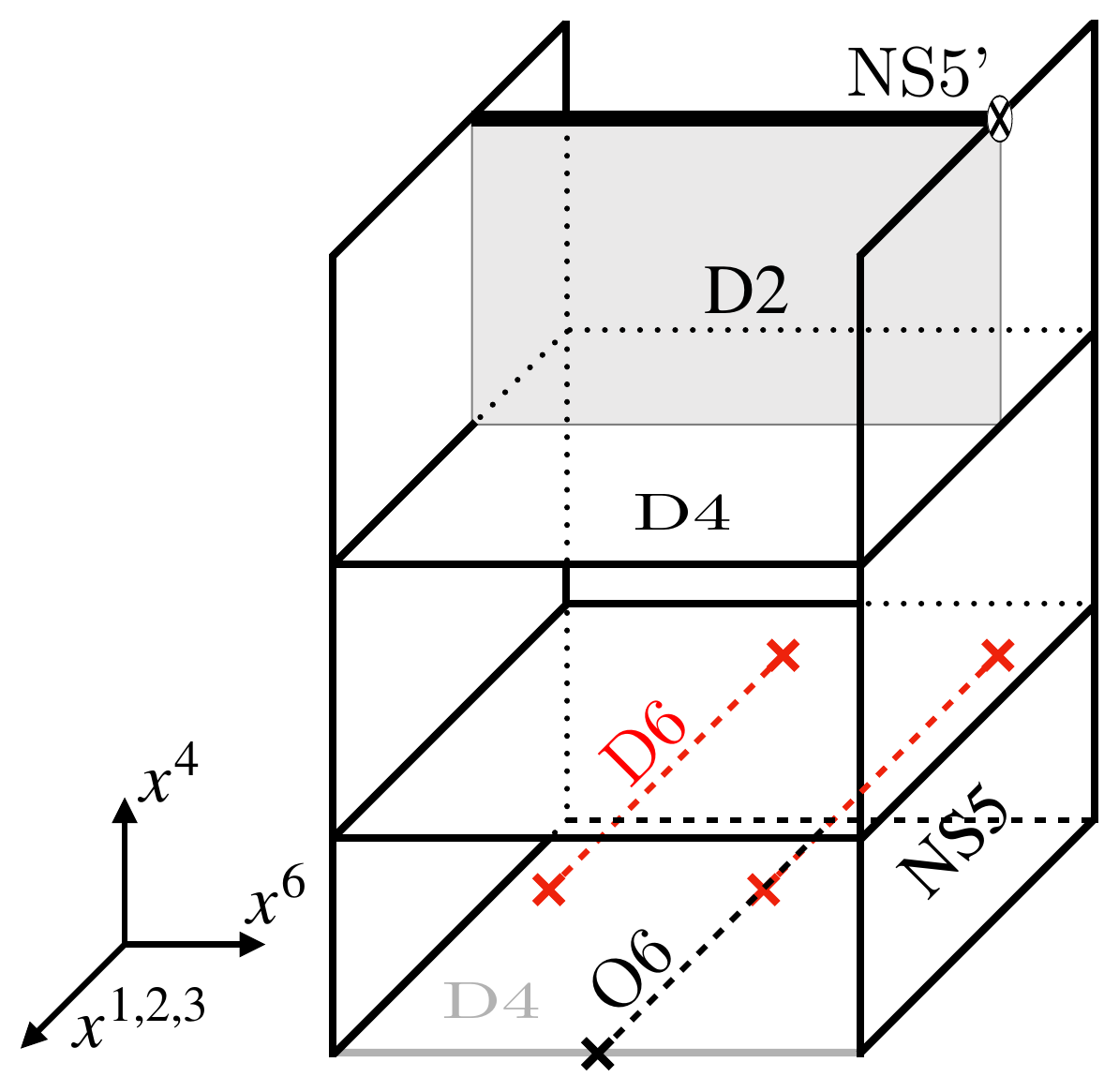}}}
\caption{(a): A brane construction of the $SO$/$USp$ SQCD with an O6-plane.
The (half) D4-brane on top of the O6-plane is present only when the gauge group is $SO(2n+1)$.
(b): A realization of the 't~Hooft operator corresponding to the fundamental representation~$V$ and $\bm{B}=\bm{e}_n$.
}
\label{fig:O6}
\end{figure}

As in the case with an O4-plane, different types of an O6-plane give different gauge groups. 
When the O6-plane is the O6$^+$-plane the $SO(N)$ gauge group is realized. $N$ is the number of D4-branes including the mirror images. When $N$ is even $(N=2n)$, the gauge group is $SO(2N)$. It is also possible to consider $N$ is odd $(N=2n+1)$ where a half D4-brane is stuck at the O6$^+$-plane in addition to $n$ physical D4-branes. Such a configuration gives rise to the $SO(2n+1)$ gauge group. When the O6-plane is the O6$^-$-plane we cannot have a D4-brane stuck at the orientifold. Then the number of the D4-branes including the mirror image is always even and $n$ physical D4-branes with the O6$^-$-plane yield the $USp(2n)$ gauge group. In each case, $N_F$ D6-branes between NS5-branes again introduce hypermultiplets in the fundamental representation of the realized gauge group. We summarize the symmetries of the systems in the following table.
\begin{center}
\begin{tabular}{c|cccc}
& O6$^-$ &  ${\rm O6}{}^+$ 
\\
\hline
D4 
 & $USp(N)$ & $SO(N)$ 
\\
D6
& $SO(2N_F)$ & $USp(2N_F)$ 
\\
\end{tabular}
\end{center}
The brane configuration with the O6-plane is depicted in Figure \ref{subfig:O6}.

\begin{figure}[t]
\centering
\subfigure[]{\label{subfig:O6-2star}
\includegraphics[scale=.25]{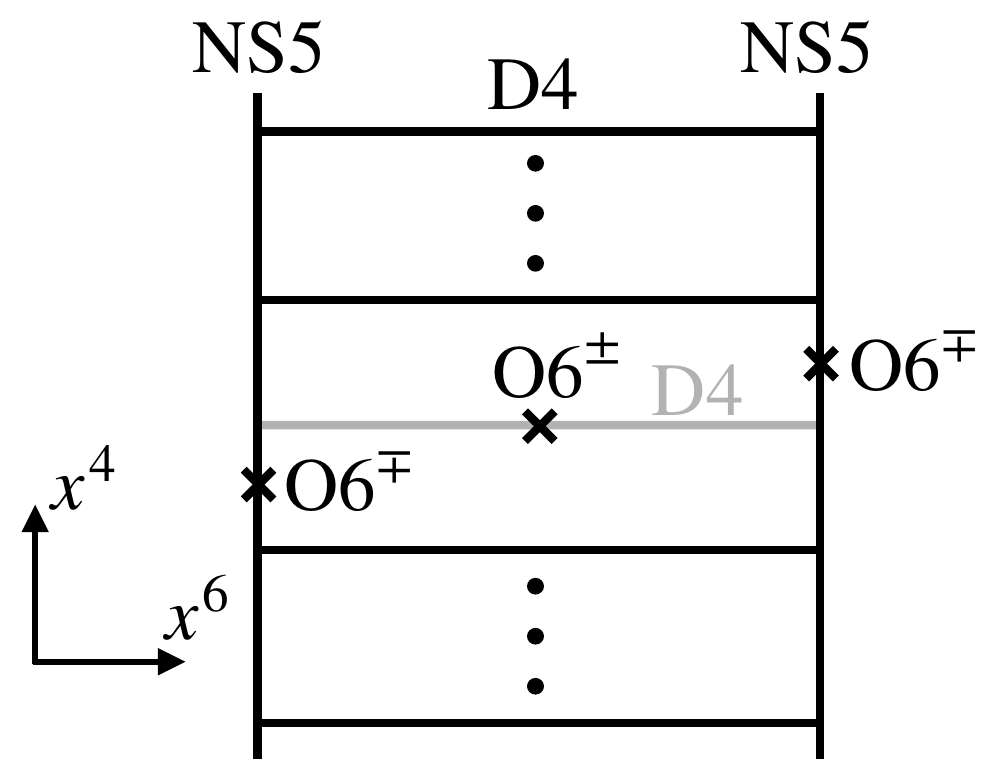}}
\hspace{1cm}
\subfigure[]{\label{subfig:O6-2star-4d}
\includegraphics[scale=.25]{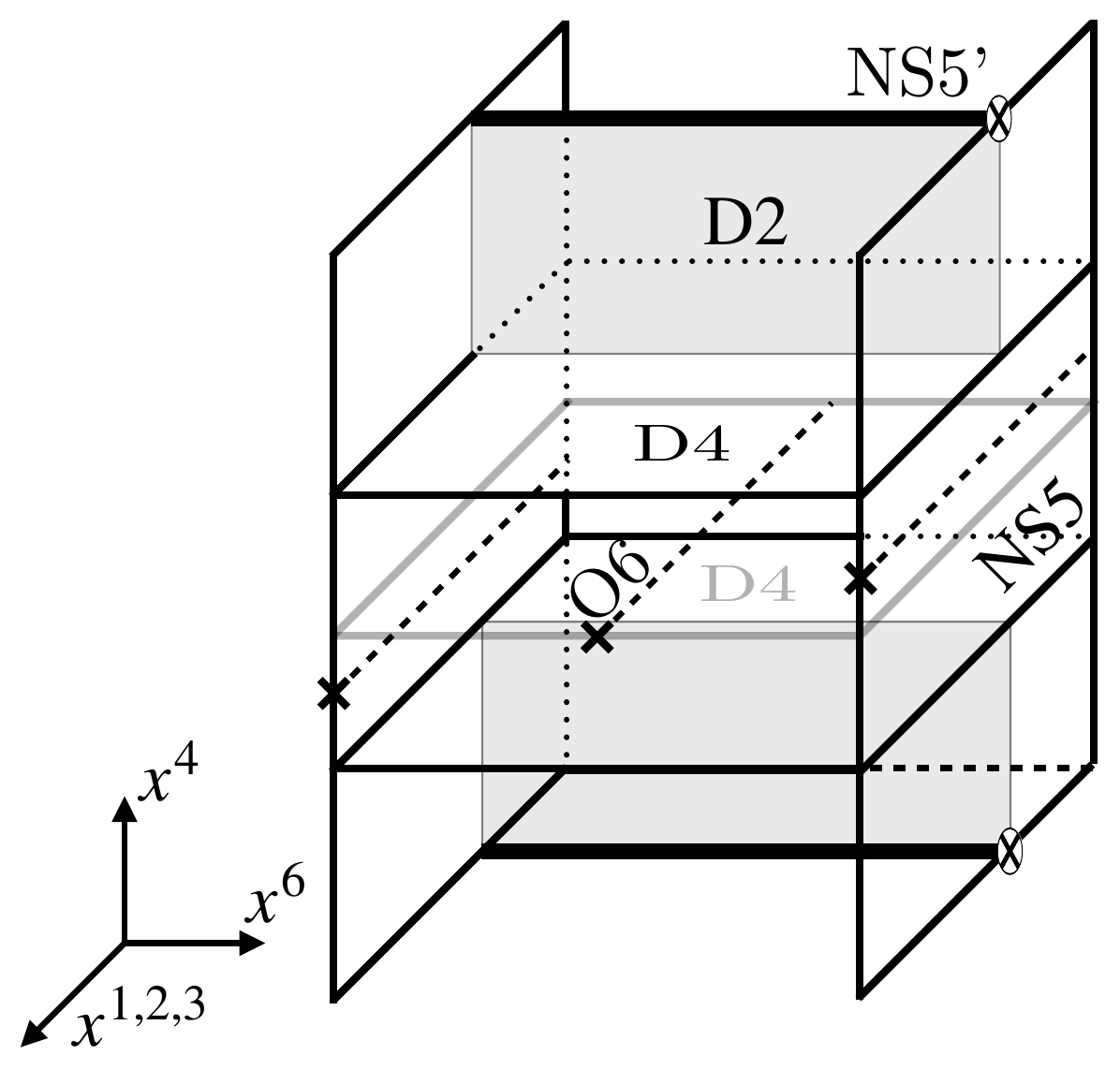}}
\caption{(a): A brane construction of the $SO$/$USp$ $\mathcal{N}=2^*$ theory with an O6-plane.
The (half) D4-brane on top of the O6-plane is present only when the gauge group is $SO(2n+1)$.
(b): A realization of the 't~Hooft operator corresponding to the fundamental representation~$V$ and $\bm{B}=\bm{e}_n$.
}
\label{fig:O6-2star}
\end{figure}

O6-planes can also be used to construct the $\mathcal{N}=2^{\ast}$ $SO$ or $USp$ gauge theory~\cite{Uranga:1998uj}. 
We compactify the $x^6$-direction on a circle with a shift as in~\eqref{shift-for-two-star}.
We introduce two O6-planes, one at~$(x^4,x^5,x^6)=(0,0,0)$ and the other at $(x^4,x^5,x^6)=\left(\frac12 {\rm Re}\, m, \frac12 {\rm Im}\,m,  \pi L\right)$.
The first O6-plane identifies spacetime points related by~\eqref{O6act}.
The identification due to the second O6-plane automatically follows from~\eqref{shift-for-two-star} and~\eqref{O6act}.
As in the $U(N)$ case we introduce a single NS5-brane at $x^6 = \pi L$  and let $N$ D4-branes, including the mirror images, end on it. 
The type of the O6-plane at $(x^4,x^5,x^6)=(0,0,0)$ determines the gauge group according to the table above.
The O6$^+$-plane yields the $SO(N)$ gauge group, and the O6$^-$-plane gives rise to the~$USp(N)$ gauge group with $N$ even.
The type of the other O6-plane determines the representation of the hypermultiplet arising from the fundamental strings that cross the NS5-brane and end on the D4-branes.
An O6$^-$-plane gives the rank-2 anti-symmetric representation, while an O6$^+$-plane gives the rank-2 symmetric representation.
Since the net RR charge of the O6-planes localized on a compact space has to vanish, we get an anti-symmetric representation for $SO(N)$ and a symmetric representation for $USp(N)$, corresponding to the $\mathcal{N}=2^*$ theory.
We depict the brane configuration in Figure~\ref{fig:O6-2star}.

\begin{figure}[t]
\centering
\subfigure[]{\label{subfig:O6-tHooft}
\raisebox{.5cm}{\includegraphics[scale=.25]{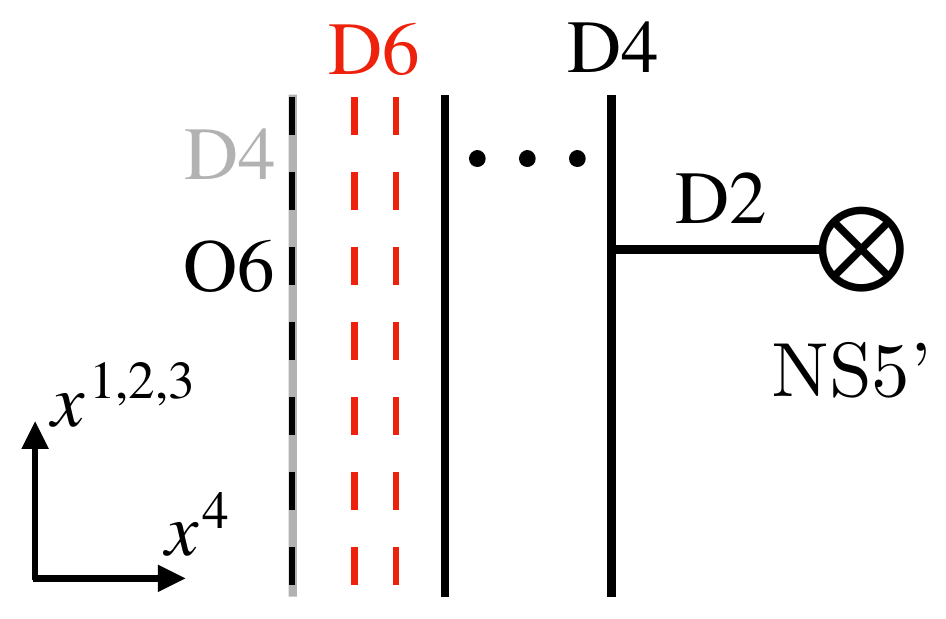}}}
\hspace{1cm}
\subfigure[]{\label{subfig:O6-tHooft-wedge2V}
\raisebox{.5cm}{\includegraphics[scale=.25]{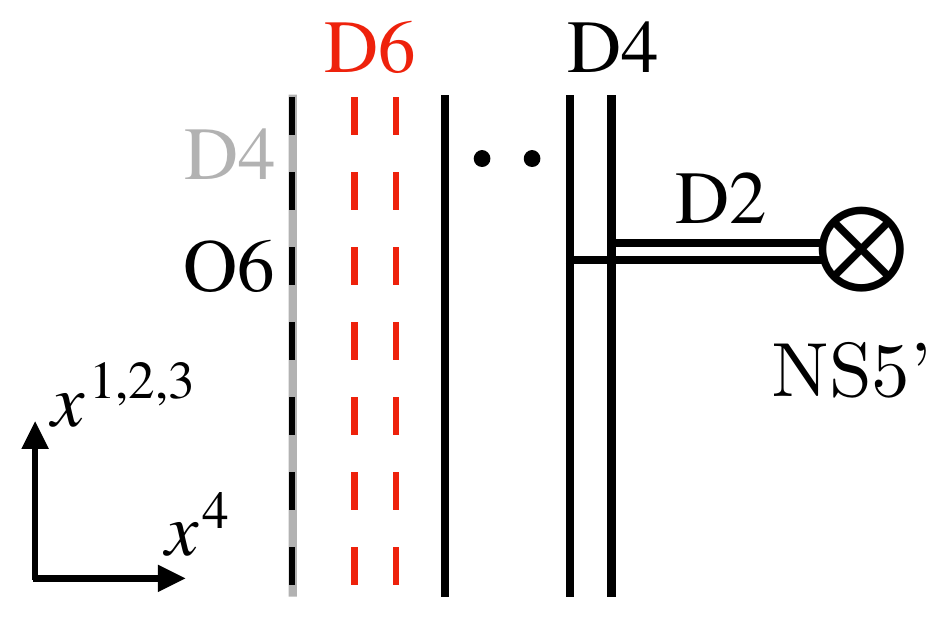}}}
\caption{(a):  A realization of the 't~Hooft operator corresponding to the minimal representation~$V$ and $\bm{B}=\bm{e}_n$ 
in the SQCD or the $\mathcal{N}=2^*$ theory.
In the $\mathcal{N}=2^*$ case, the D6-branes should be omitted, and the O6-plane that intersects an NS5-brane is present in the system but suppressed in the figure.
The D4-brane on top of the O6-plane, indicted by the gray line, is present only for the gauge group~$SO(2n+1)$.
(b): A realization of the 't~Hooft operator corresponding to $\wedge^2 V$ and $\bm{B}=\bm{e}_{n-1}+\bm{e}_n$.
}
\label{fig:O6-thooft}
\end{figure}

As in the $U(N)$ and the O4-plane cases, D2-branes ending on the stack of D4-branes and an NS5'-brane should realize 't~Hooft operators in the $SO/USp$ SQCD or $\mathcal{N} = 2^{\ast}$ theory constructed with O6-planes. 
We propose that for $k\leq N/2$, $k$ D2-branes stretched between the stack of D4-branes and a single NS5'-brane realize an 't~Hooft operator whose magnetic charge~$\bm{B} \in\Lambda_\text{cochar}$ corresponds to $\wedge^k V$, the rank-$k$ anti-symmetric representation. Examples of such 't~Hooft operators are shown in Figures \ref{subfig:O6-2star-4d} and~\ref{fig:O6-thooft}.
In Sections~\ref{sec:N2starSO} and~\ref{sec:N2starUSp} we will make use of these brane configurations with O6-planes to obtain the SQMs that describe monopole screening
 in the expectation values of 't~Hooft operators 
 in the $\mathcal{N}=2^{\ast}$ $SO/USp$ gauge theories. We will provide quantitive evidence for our proposal by comparing the supersymmetric indices of SQMs with the Moyal products (to be defined in Section \ref{sec:wall-crossing}) of the two minimal 't~Hooft operators. 

\subsection{SQMs on 't~Hooft operators}\label{sec:SQM-tHooft}

So far we have constructed 't~Hooft operators in the 4d $\mathcal{N}=2$ $U/SO/USp$ SQCD or $\mathcal{N}=2^{\ast}$ $U/SO/USp$ gauge theories using branes and orientifolds in type IIA string theory. A physical quantity associated with the 't~Hooft operators is the expectation value of their product. The expectation value may or may not contain non-perturbative contributions, namely monopole screening contributions. The brane construction is useful to understand the monopole screening contributions of the 't~Hooft operators~\cite{Brennan:2018yuj,Brennan:2018moe,Brennan:2018rcn}. 

\begin{figure}[t]
\centering
\subfigure[]{\label{subfig:screening1}
\includegraphics[scale=.25]{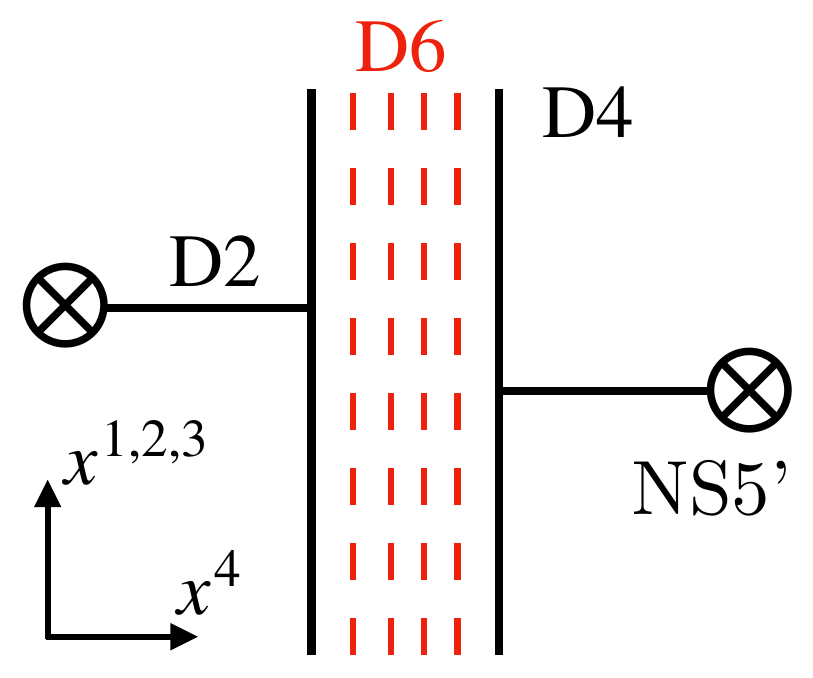}}
\hspace{0cm}
\subfigure[]{\label{subfig:screening2}
\includegraphics[scale=.25]{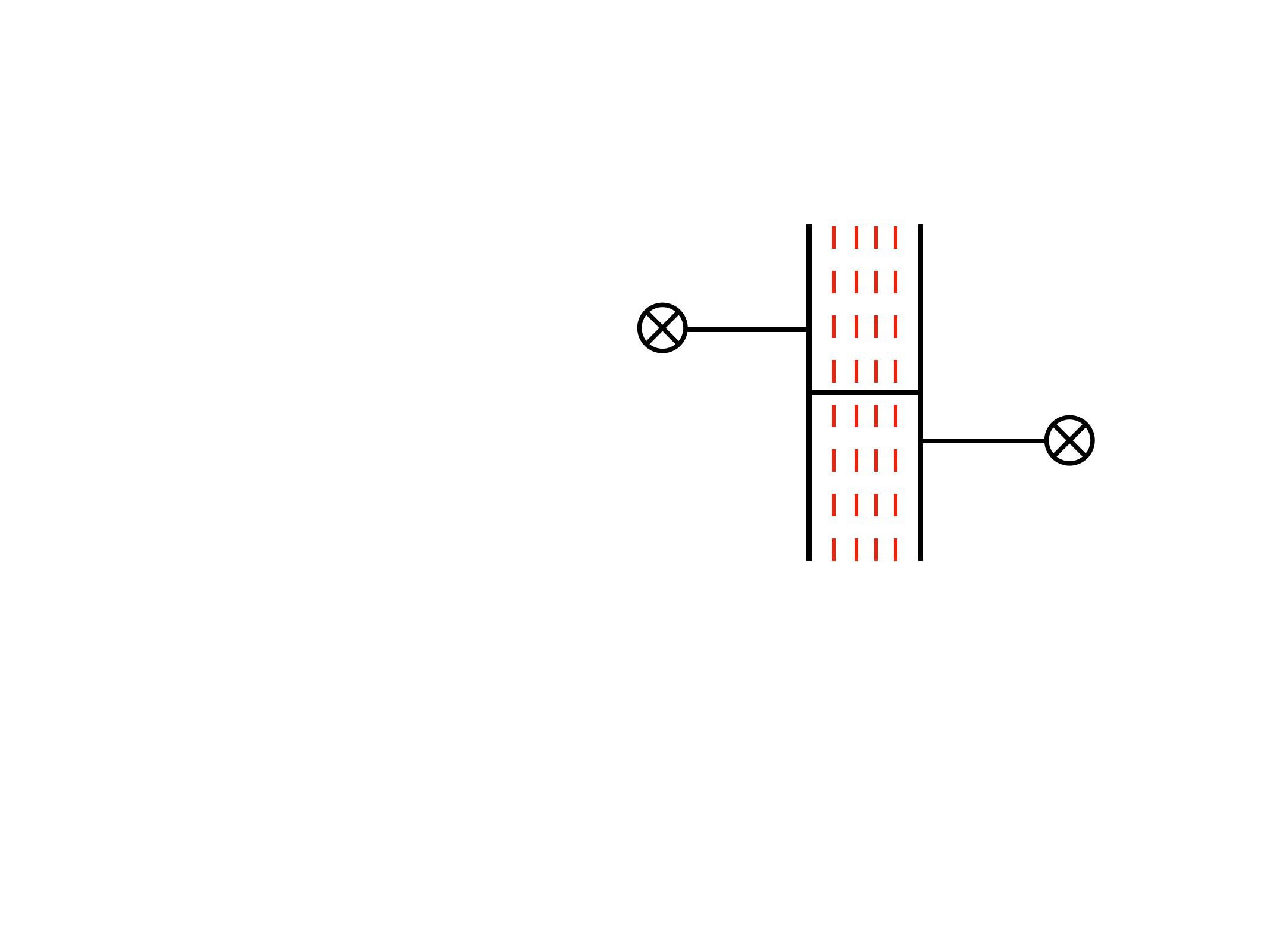}}
\hspace{0cm}
\subfigure[]{\label{subfig:screening3}
\includegraphics[scale=.25]{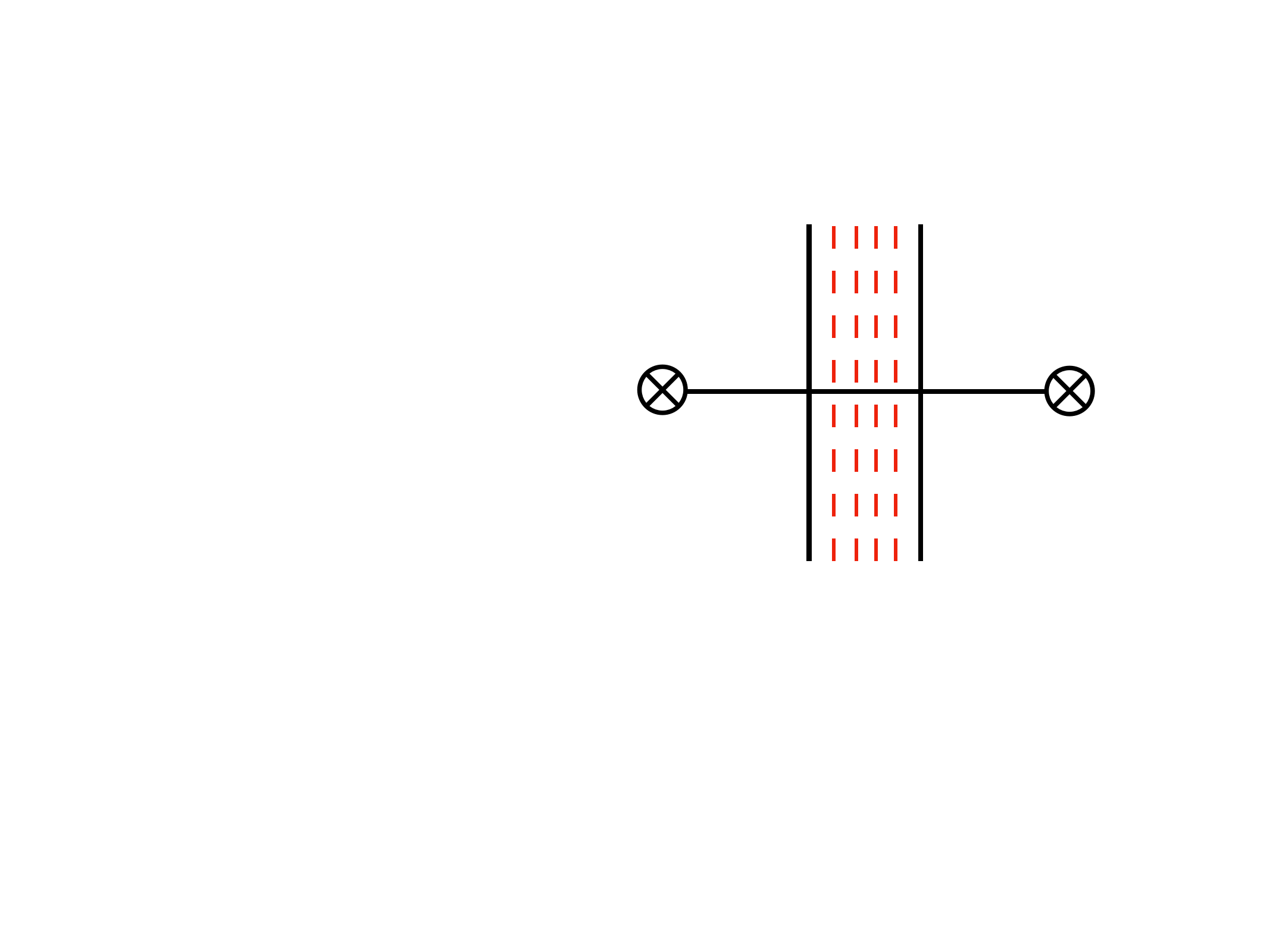}}
\hspace{1cm}
\subfigure[]{\label{subfig:screening4}
\includegraphics[width=.6cm]{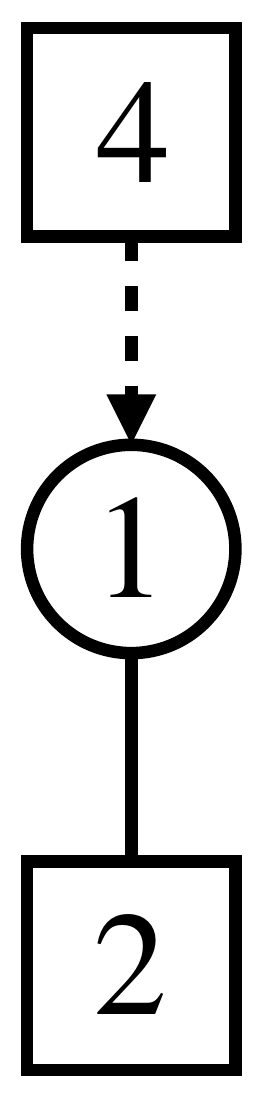}
\hspace{.1mm}
}
\caption{(a): The NS5'-branes and the D2-branes insert a product 't Hooft operator with total charge $\bm{B}=-\bm{e}_1 + \bm{e}_2$.
(b): We add a D2-brane between the two D4-branes; it represents a smooth monopole.
(c): The new D2-brane reconnects with the other D2-branes to form a single D2-brane that only end on NS5'-branes.
The system realizes the sector with $\bm{v}=0$, where the magnetic charge is screened completely.
(d): The $\mathcal{N}=(0,4)$ quiver diagram for the SQM that lives on the D2-brane in (c).
}
\label{fig:SQCD-screening}
\end{figure}

Let us look at an example of the 4d $\mathcal{N}=2$ $U(2)$ SQCD with $4$ flavors.  The brane configuration where we insert two minimal 't~Hooft operators with the total magnetic charge ${\bm B} = -\be_1 + \be_2$ is depicted in Figure \ref{subfig:screening1}. 
The charge may be screened by a smooth monopole, with the opposite charge, corresponding to a D2-brane stretched between the two D4-branes as in Figure~\ref{subfig:screening2}. 
By tuning the positions we can align and combine the three D2-branes as in Figure~\ref{subfig:screening3} to obtain a single D2-brane between the two NS5'-branes.
This configuration corresponds to the monopole screening sector with $\bv = {\bm 0}$. 
Since the D2-brane is also bounded by NS5-branes and NS5'-branes, the theory that lives on the D2-brane realizes a 1d theory, {\it i.e.}, an SQM, at low energies. 

The SQM is the dimensional reduction of a 2d $\mathcal{N}= (0, 4)$ supersymmetric gauge theory \cite{Hwang:2014uwa,Tong:2014yna,Brennan:2018moe,Hanany:2018hlz}. The matter content of the SQM can be read off from fundamental strings connecting D-branes in the configuration. 
The matter content is summarized as follows.
\begin{center}
\begin{tabular}{c|c|c|c}
& $\mathcal{N} = (0,4)$ multiplet &  representation & number 
\\
\hline
$n$ D2 - $n$ D2 & vector multiplet & adjoint of $U(n)$ & 1 
\\
$n$ D2 - $n'$ D2 & hypermultiplet & bifundamental of $U(n) \times U(n')$ & 1 
\\ 
$n$ D2 - $N_c$ D4 & hypermultiplet & fundamental of $U(n)$ & $N_c$
\\
$n$ D2 - $N_F$ D6 & short Fermi multiplet & fundamental of $U(n)$ & $N_F$ 
\end{tabular}
\end{center}
The leftmost column lists the branes connected by fundamental strings. 
Each number on the rightmost column is the number of supermultiplets in the representation indicated on the third column. For later use we include a row for the fundamental strings between a stack of D2-branes and another stack of D2-branes that are adjacently separated by an NS5'-brane. 
While a short Fermi multiplet has $\mathcal{N} = (0,4)$ on-shell supersymmetry~\cite{Tong:2014yna}, an~$\mathcal{N} = (0,2)$ subalgebra can be completed off-shell~\cite{Brennan:2018rcn}.
This is sufficient for SUSY localization.

Let us consider an example in Figure~\ref{fig:SQCD-screening}.
The SQM living on the D2-brane in Figure~\ref{subfig:screening3} can be represented by a quiver diagram in Figure~\ref{subfig:screening4}. 
A circular node with an integer~$n$ indicates a $U(n)$ gauge group\footnote{If the gauge group is different from $U$-type we will explicitly write the gauge group in a circular node.} and the corresponding $\mathcal{N}=(0, 4)$ vector multiplet. A solid line between two circular nodes would be an $\mathcal{N}=(0, 4)$ hypermultiplet in the bifundamental representation. 
On the other hand a solid line between one circular node and a square node, as appearing in Figure~\ref{subfig:screening4}, represents as many hypermultiplets in the fundamental representation as the number in the square. 
A dashed arrow between a circular node and a square node represents short Fermi multiplets in the fundamental representation, with their number shown in the square. 
The supersymmetric index of the SQM yields the monopole screening contribution~\cite{Brennan:2018yuj,Brennan:2018moe,Brennan:2018rcn,Assel:2019iae,Hayashi:2019rpw}.
We will use this technique to compute 't~Hooft operator correlators in $U(N)$ gauge theories in Section~\ref{sec:UNhigher}.

The same strategy works for the $SO/USp$ SQCD constructed with an O4-plane.
The SQM gauge group on the D2-branes is determined by the type of the O4-plane they intersect as follows~\cite{Dorey:2002ik}. 
 \begin{center}
\begin{tabular}{c|c|c|c}
&O4$^-$& $\widetilde{\text{O4}}{}^-$&  O4$^+$
\\
\hline
D2 & $USp(2n)$& $USp(2n)$ & $O(n)$
\end{tabular}
\end{center}
An O4$^-$- or $\widetilde{\text{O4}}{}^-$-plane requires the number of D2-branes, including the mirror images, to be even ($2n$).
An O4$^+$-plane allows it to be an arbitrary positive integer ($n$).
We emphasize that the gauge group is the $O$-type not the $SO$-type.
This will be important when we compare supersymmetric indices with Moyal products. 

The SQMs describing monopole screening contributions for 't~Hooft operators in $\mathcal{N}=2^{\ast}$ gauge theories may be obtained similarly. In those cases we do not have D6-branes and the SQMs are the dimensional reduction of 2d $\mathcal{N} = (4, 4)$ supersymmetric gauge theories \cite{Brennan:2018yuj}. For SQMs from 't~Hooft operators in the 4d $\mathcal{N} = 2^{\ast}$ $U(N)$ gauge theory, the fundamental strings between D-branes give supermultiplets as in the following table. 
\begin{center}
\begin{tabular}{c|c|c|c}
& $\mathcal{N} = (4, 4)$ multiplet &  representation & number 
\\
\hline
$n$ D2 - $n$ D2 & vector multiplet & adjoint of $U(n)$ & 1
\\
$n$ D2 - $n'$ D2 & hypermultiplet & bifundamental of $U(n) \times U(n')$ & 1
\\ 
$n$ D2 - $N_c$ D4 & hypermultiplet & fundamental of $U(n)$ & $N_c$
\end{tabular}
\end{center}
An $\mathcal{N}=(4,4)$ vector multiplet consists of an $\mathcal{N} = (0, 4)$ vector multiplet and an $\mathcal{N} = (0, 4)$ twisted hypermultiplet. An $\mathcal{N} = (4, 4)$ hypermultiplet is made from an $\mathcal{N} = (0, 4)$ hypermultiplet and an $\mathcal{N} = (0, 4)$ (long) Fermi multiplet.

As discussed in Section \ref{sec:SQCD-2star-O6}, we may change the 4d gauge group into $SO$ or $USp$ by introducing O6-planes. 
The 1d gauge group on the D2-branes which intersect with the O6-plane at $x^6 = 0$ changes due to the effect of the orientifold. The correspondence between the type of the O6-plane and the 1d gauge group is given as follows. 
\begin{center}
\begin{tabular}{c|c|c}
&O6$^-$ &   O6$^+$
\\
\hline
D2 & $O(n)$ &  $USp(2n)$
\end{tabular}
\end{center}
See for example~\cite{Hwang:2014uwa}.
The representations for the open strings that connect various D-branes are modified by the presence of an orientifold appropriately.

Let us also summarize the graphical notations we use for SQM quiver diagrams.
Throughout this paper we use the notation appropriate for the $\mathcal{N}=(0,4)$ supersymmetry.%
\footnote{%
In our previous paper~\cite{Hayashi:2019rpw} we mostly used the $\mathcal{N}=(0,2)$ notation for SQM quiver diagrams.
}
\begin{center}
\begin{tabular}{c|c}
 $\mathcal{N} = (0,4)$ multiplet & symbol
\\
\hline
 vector multiplet & circle
\\
hypermultiplet & straight solid line
\\
twisted hypermultiplet & wavy solid line
\\ 
short Fermi multiplet & dashed line 
\\ 
long Fermi multiplet & dash-dotted line 
\end{tabular}
\end{center}
For a short Fermi multiplet in the fundamental representation of a factor $U(n)$ in the gauge group, we use a dashed line with an arrow pointing toward the $U(n)$ node.
See Figures~\ref{subfig:screening4} and~\ref{subfig:UNscreening02} for representative examples.

\subsection{Extra term in the supersymmetric index}\label{sec:extra-term}
In Sections~\ref{sec:SO} and~\ref{sec:USp} we will make use of the brane construction and determine the SQMs which describe monopole screening in 4d $SO/USp$ gauge theories. The monopole screening contributions are given by certain supersymmetric indices of the SQMs. 
In fact there is a subtlety in the supersymmetric index computations.
What we here call the ``supersymmetric index'' is either a quantity obtained by the JK residue prescription applied to the poles away from infinities, or such a quantity modified by the ``extra term'' prescription below. We will use this nomenclature throughout this paper. 
We refer to Appendix~2.2 of~\cite{Hayashi:2019rpw} for a summary of the JK residue prescription.

Supersymmetric indices of SQMs are not only used for monopole screening contributions in expectation values of 't~Hooft operators but also used for instanton partition functions of 5d gauge theories that are ultraviolet (UV) complete. In those cases, the SQMs are the ADHM gauged quantum mechanics. Then the sum of the supersymmetric indices of the SQMs weighted by instanton fugacities basically yields the 5d instanton partition function.  Sometimes, however, typically when the number of flavors in the 5d theory is large, it has turned out that one needs to factor out an extra factor from the instanton partition function \cite{Bergman:2013ala, Bao:2013pwa, Hayashi:2013qwa, Bergman:2013aca, Hwang:2014uwa}. Namely in general we have
\begin{equation}\label{sqminstextra}
1 + \sum_{k=1}^{\infty}u^k Z_{\text{SQM}, k} = Z_{\text{5d inst}}Z_{\text{extra}},
\end{equation}
where the lefthand side is the sum of the supersymmetric indices of the AHDM quantum mechanics with the instanton fugacity $u$, $Z_{\text{5d inst}}$ is the 5d instanton partition function and $Z_{\text{extra}}$ is the extra factor. The extra factor contribution is decoupled from the 5d dynamics and may be quantitatively characterized as a factor which is independent of Coulomb branch moduli. 
The two factors $Z_{\text{5d inst}}$ and $Z_{\text{extra}}$ themselves can be expanded by the instanton fugacity $u$ as 
\begin{align}
Z_{\text{5d inst}} &= 1 + \sum_{k=1}^{\infty}Z_{\text{5d inst},k}u^k,\\
Z_{\text{extra}} &= 1 + \sum_{k=1}^{\infty}Z_{\text{extra},k}u^k.
\end{align}
Then \eqref{sqminstextra} becomes 
\begin{equation}\label{sqm.inst.exp}
1 + uZ_{\text{SQM}, 1} + \cdots = 1 + u\left(Z_{\text{5d inst}, 1} + Z_{\text{extra}, 1}\right) + \cdots,
\end{equation}
and in particular we have
\begin{equation}\label{1inst}
Z_{\text{5d inst}, 1} = Z_{\text{SQM}, 1} - Z_{\text{extra}, 1}.
\end{equation}
The one-instanton part of $Z_{\text{5d inst}}$ is obtained by subtracting the Coulomb branch moduli independent part of $Z_{\text{SQM},1}$. 

In fact the instanton partition function is not completely unrelated to a monopole screening contribution in expectation values of 't~Hooft operators. For example, in \cite{Ito:2011ea, Brennan:2018yuj}, a monopole screening contribution of  't~Hooft operators in the 4d $\mathcal{N}=2^{\ast}$ $U(N)$ or $SU(N)$ gauge theory was obtained by using a part of the 5d instanton partition function of the 5d $\mathcal{N}=1^{\ast}$ $U(N)$ or $SU(N)$ gauge theory.%
\footnote{%
See also~\cite{Mekareeya:2013ija} for a work that demonstrates the relation between monopole screening contributions in 4d and 5d instanton partition functions on ALE spaces.
} 
The relation is indeed expected from the Kronheimer correspondence~\cite{Kronheimer:MTh}.
Since we remove an extra factor from the 5d instanton partition function it is natural also to remove the corresponding part from the supersymmetric indices of SMQs describing monopole screening contributions.

Based on the observation we propose that we need to remove a part which is independent of the 4d Coulomb branch moduli from the supersymmetric indices of SQMs describing monopole screening for irreducible 't Hooft operators\footnote{For $\mathcal{N}=2^*$ theories, irreducible 't Hooft operators are those which are S-dual to Wilson operators in irreducible representations.} 
which arises from a single D2-brane in asymptotically free or superconformal theories\footnote{We will not apply this prescription to monopole screening contributions of $U(N)$ gauge theories in Section~\ref{sec:UNhigher} as they are not UV complete.}. 
In this case we subtract, rather than factor out, a Coulomb branch moduli%
\footnote{%
What we call the Coulomb branch moduli are the coefficients~$a_i$ in the sum~$\sum_i a_i\bm{e}_i$, which is a complex combination of the vev an adjoint scalar and the gauge holonomy along $S^1$.
The conjugate parameters~$b_i$ are the coefficients in the sum~$\sum_i b_i\bm{e}_i$, which is a complex combination of the vev of another adjoint scalar and the chemical potential for the magnetic charge~$\bm{v}$.
The 't~Hooft operator vevs depend on $a_i$ and $b_i$, as well as the $\Omega$-deformation parameter~$\epsilon_+$ and the mass parameters ($m_f$ for SQCD and $m$ for the $\mathcal{N}=2^*$ theory).
See~\cite{Ito:2011ea} for the precise definitions.
In this paper we use the convention of our previous paper~\cite{Hayashi:2019rpw}.
}
 independent term due to the relation \eqref{1inst} and we will call it an ``extra term". 
 Since \eqref{1inst} is valid for the one-instanton part the prescription is restricted to 't Hooft operators from a single D2-brane\footnote{\label{fn:extra} When $Z_{\text{extra, 1}} = 0$, then \eqref{sqm.inst.exp} gives
\begin{equation}
Z_{\text{5d inst}, 2} = Z_{\text{SQM}, 2} - Z_{\text{extra}, 2}.
\end{equation}
Hence we can also simply subtract the Coulomb branch independent term in this case also. We will see this type of examples in Section \ref{sec:USp-2star-wedge2v} and in Appendix \ref{sec:USp.adj}.}.
To extract the extra term we move deep inside a Weyl chamber of Cartan scalars in the 4d $\mathcal{N}$=2 vector multiplets where the exponentiated Coulomb branch moduli become good expansion parameters. Then we expand a monopole screening contribution by the Coulomb branch moduli and the extra term is given by the zeroth order of the expansion parameters. Considering a different Weyl chamber will give a gauge equivalent result. %
For $\mathcal{N}=2^{\ast}$ theories we do not subtract the whole Coulomb branch independent term but leave the number of zero weights for $\bv = {\bm 0}$ sectors so that the $\mathcal{N}=4$ limit reproduces the character of the representation under consideration.

In fact the extra term we determine by our prescription captures the Coulomb branch contribution to the ground state index of the SQM considered for the
4d $\mathcal{N} = 2$ $SU(2)$ gauge theory with four flavors in~\cite{Brennan:2018rcn}. The monopole screening contribution in the product of the two minimal 't~Hooft operators of the $U(2)$ gauge theory with four flavors is given by the supersymmetric index of the SQM in Figure \ref{subfig:screening4}. The result becomes \cite{Ito:2011ea,Brennan:2018rcn,Assel:2019iae,Hayashi:2019rpw}
\begin{equation}
Z_{\text{SQM}}(a_1, a_2) = -\frac{\prod_{f=1}^42\sinh\frac{a_1 - m_f - \epsilon_+}{2}}{2\sinh\frac{a_1 - a_2}{2}2\sinh\frac{a_1 - a_2 - 2\epsilon_+}{2}} - \frac{\prod_{f=1}^42\sinh\frac{a_2 - m_f - \epsilon_+}{2}}{2\sinh\frac{a_1 - a_2}{2}2\sinh\frac{a_1 - a_2 + 2\epsilon_+}{2}},
\end{equation}
when the Fayet-Iliopoulos (FI) parameter associated to the $U(1)$ gauge group is positive. 
To apply the extra term prescription to $SU(2)$ gauge group we set $a := a_1 = -a_2$. 
The Coulomb branch moduli independent part is 
\begin{equation}
Z_{\text{extra}} = -2\cosh\left(\sum_{f=1}^4m_f + \epsilon_+\right).
\end{equation}
Then subtracting the extra term yields
\begin{equation}
\begin{split}
&Z_{\text{SQM}}(a,-a) -  Z_{\text{extra}} \cr
\hspace{1cm}&= -\frac{\prod_{f=1}^42\sinh\frac{a - m_f - \epsilon_+}{2}}{2\sinh(a)2\sinh(a - \epsilon_+)} - \frac{\prod_{f=1}^42\sinh\frac{-a - m_f - \epsilon_+}{2}}{2\sinh(a)2\sinh(a + \epsilon_+)} + 2\cosh\left(\sum_{f=1}^4m_f + \epsilon_+\right),
\end{split}
\end{equation}
which reproduces the result in \cite{Brennan:2018rcn,Assel:2019iae}, and also is consistent with the CFT result using the AGT correspondence \cite{Ito:2011ea}. This analysis gives support for the extra term prescription. 

In Section \ref{sec:SO} and Section \ref{sec:USp}, we will apply this prescription for SQMs which arise in $SO$ and $USp$ gauge theories. In fact the monopole screening contribution in the expectation value of the 't~Hooft operator in the rank-2 anti-symmetric representation of the Langlands dual group in 4d $\mathcal{N}=2$ $SO/USp$ SQCD has another subtlety. We will comment on the subtlety in Appendix~\ref{sec:subtle} and circumvent the issue by choosing the number of flavors to be small enough.

\subsection{Wall-crossing and operator ordering}
\label{sec:wall-crossing}

In~\cite{Hayashi:2019rpw} we studied, for the $U(N)$ SQCD with $N_{\rm F} = 2N$ flavors, the expectation value of the product of several 't~Hooft operators with minimal charges 
\begin{equation}\label{prod.ops}
 T_{\bm B} =  T_1(s_1) \cdot T_2(s_2) \cdot \ldots \cdot T_{\ell}(s_\ell) 
 \quad \text{ in path integral}.
\end{equation}
Minimal charges refer to those which correspond to the fundamental or anti-fundamental representation of the Langlands dual of the gauge group, which is again $U(N)$.
Let $(x^1, x^2, x^3)$ be the Cartesian coordinates of~$\mathbb{R}^3$.
An $\Omega$-deformation in the $(x^1, x^2)$-space (with parameter~$\epsilon_+$) requires $T_a$'s to be on the 3-axis ($x^1=x^2=0$).
The parameter~$s_a$ in~(\ref{prod.ops}) is the $x^3$-value ($x^3=s_a$) of the $a$-th operator, and we assume that $s_a$'s are distinct.
The expectation value of~\eqref{prod.ops} depends only on the ordering of $s_a$'s.
If we regard $x^3$ as the Euclidean time, the ordering of $s_a$'s specified by a permutation $\sigma\in S_\ell$ as
\begin{equation} \label{ordering-position}
 s_{\sigma(1)} > s_{\sigma(2)} >\ldots > s_{\sigma(\ell)} 
\end{equation}
 is equivalent to the time ordering of $\widehat T_a$'s in canonical quantization
 \begin{equation}\label{prod.ops.ordered}
 \widehat T_{\sigma(1)} \cdot \widehat T_{\sigma(2)} \cdot \ldots  \cdot \widehat T_{\sigma(\ell)}.
\end{equation}

The SQMs that describe monopole screening in the product of $\ell$ operators have unitary gauge groups and the corresponding 
FI parameters~$\bm{\zeta}=(\zeta_a)_{a=1}^{\ell-1}$ related to the positions as  $\zeta_a=s_{a + 1} - s_a$ ($a=1,\ldots,\ell-1$) ~\cite{Brennan:2018yuj,Brennan:2018rcn}.
As we vary $\bm{\zeta}$ some SQMs exhibit wall-crossing, {\it i.e.}, a discrete change occurs in the supersymmetric index.
As shown in~\cite{Ito:2011ea} the expectation value of the product of (general) 't~Hooft operators is equal to the Moyal product of the expectation values of individual operators:
\begin{align}
\Braket{T_1 \cdot T_2 \cdot \ldots  \cdot T_\ell  } 
=
\Braket{T_{\sigma(1)}} * \Braket{T_{\sigma(2)}} * \ldots * \Braket{T_{\sigma(\ell)}} .
\end{align}
The possible dependence on the ordering of $s_a$'s is realized by the non-commutativity of the Moyal product, which is associative and is defined as%
\footnote{%
The definition~(\ref{moyal}) assumes that $a_i$ and $b_i$ are the coefficients in the expansions $\bm{a}=\sum_i a_i \bm{e}_i$,  $\bm{b}=\sum_i b_i \bm{e}_i$ with respect to an orthonormal basis $\{\bm{e}_i\}$ of the Cartan subalgebra.
}
\begin{align}\label{moyal}
(f\ast g)({\bm a}, {\bm b}) =\exp\left[-\epsilon_+ \sum_{k=1}^{\text{rank}\,G}\left(\partial_{b_k}\partial_{a'_k} - \partial_{a_k}\partial_{b'_k}\right)\right]f({\bm a}, {\bm b})g({\bm a}', {\bm b}')\Big|_{{\bm a}' = {\bm a}, {\bm b}' = {\bm b}}. 
\end{align}

Non-commutativity of the Moyal product as well as the brane construction of 't~Hooft operators suggest that wall-crossing is associated with a change in the ordering of operators.
In~\cite{Hayashi:2019rpw} we made the following conjectures for minimal operators in $U(N)$ theories:
\begin{center}
\begin{tabular}{c p{14cm}}
(i) & The supersymmetric indices of the SQMs coincide with the $Z_\text{mono}$'s read off from the Moyal products.
\\
\vspace{-3mm}
\\
(ii) & Wall-crossing can occur in the SQMs only across the walls in the space of FI parameters where the ordering of  distinct operators changes.
\end{tabular}
\end{center}
By explicit calculations, we confirmed the conjectures for the correlators involving up to three minimal operators.%
\footnote{%
The relation between wall-crossing and operator ordering was extended to monopole operators in three-dimensional $\mathcal{N}=4$ quiver gauge theories involving unitary gauge groups~\cite{Okuda:2019emk,Assel:2019yzd}.
}

Based on earlier discussions of this section we make modified versions of conjecture~(i).
As we saw in Section~\ref{sec:SQM-tHooft}, the brane constructions of some types of 't~Hooft operators in~$\mathcal{N}=2$~$U/SO/USp$ SQCD or~$\mathcal{N}=2^{\ast}$~$U/SO/USp$ gauge theories naturally give rise to SQMs that capture monopole screening effects.
In Section~\ref{sec:extra-term} we proposed, for $SO/USp$ theories, a prescription for computing the monopole screening contributions~$Z_\text{mono}$ using the SQMs, up to the subtleties to be discussed in Appendix~\ref{sec:subtle}.
We make the following conjectures.
\begin{center}
\begin{tabular}{c p{14cm}}
(i)' & For the SQCD and the $\mathcal{N}=2^*$ theory with gauge group $U(N)$,  and for the product of the operators with minuscule magnetic charges corresponding to~$\wedge^k V$ or~$\wedge^k\overline{V}$, the $Z_\text{mono}$'s computed as the supersymmetric indices by the JK prescriptions coincide with those read off from the Moyal products.
\\
\vspace{-3mm}
\\
(i)'' & For the SQCD and the $\mathcal{N}=2^*$ theory with gauge group $SO(N)$ or $USp(N)$, the correct $Z_\text{mono}$'s can be computed from the SQMs according to the extra term prescription formulated in Section~\ref{sec:extra-term}, up to the subtleties discussed in Appendix~\ref{sec:subtle}.
\end{tabular}
\end{center}
We will provide evidence for (i)' in Section~\ref{sec:UNhigher}, and for (i)'' in Sections~\ref{sec:SO} and~\ref{sec:USp}.

We also make conjecture (ii)', which says that in the $U(N)$ SQCD the statement of~(ii) holds for the operators $T_a$'s in (i)'.
We will explicitly check this for some examples in Section~\ref{sec:UNSQCD}.
We will discuss wall-crossing in the $SO/USp$ gauge groups in Section~\ref{sec:conclusion} and Appendix~\ref{sec:subtle}.

\section{'t~Hooft operators with higher charges in $U(N)$ gauge theories}
\label{sec:UNhigher}

In this section we study the 't~Hooft operators with non-minimal charges in the $U(N)$ SQCD and the $U(N)$ $\mathcal{N}=2^*$ theory.
In Section~\ref{sec:UNSQCD} we focus on the SQCD and compute some of the correlation functions involving non-minimal 't~Hooft operators, confirming the conjectures (i)' and (ii)' in Section~\ref{sec:wall-crossing} in these examples.
In Section~\ref{sec:UN-2star} we repeat the analysis for the $\mathcal{N}=2^*$ theory.

\subsection{$\mathcal{N}=2$ SQCD}
\label{sec:UNSQCD}
\begin{figure}[thb]
\centering
\subfigure[]{\label{subfig:UNHooft2nd1}
\includegraphics[scale=.5]{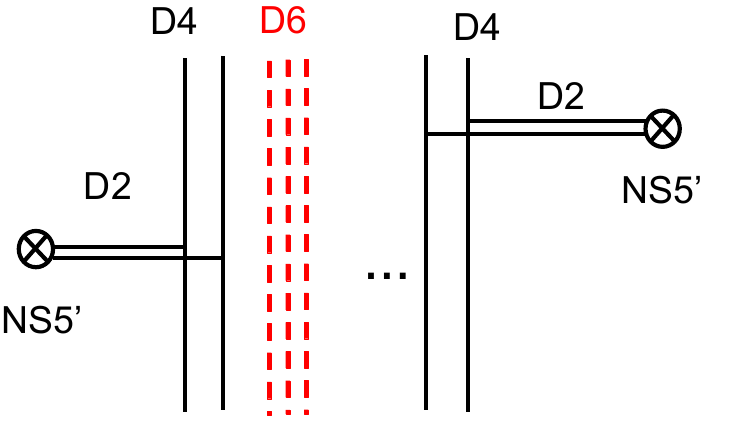}}
\hspace{-.5cm}
\subfigure[]{\label{subfig:UNHooft2nd2}
\includegraphics[scale=.5]{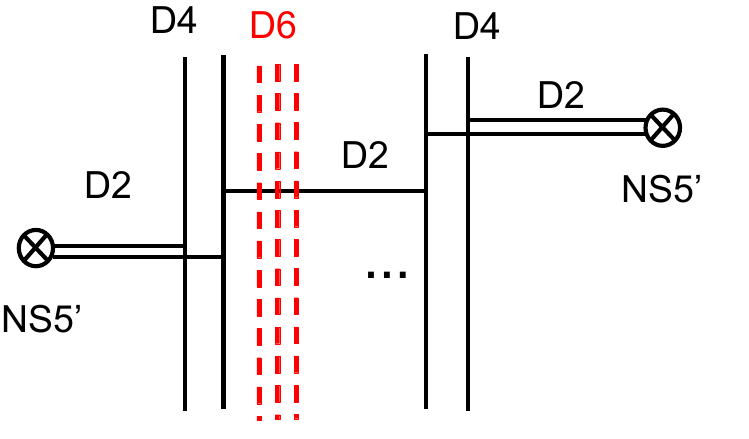}}
\hspace{0cm}
\subfigure[]{\label{subfig:UNHooft2nd3}
\includegraphics[scale=.5]{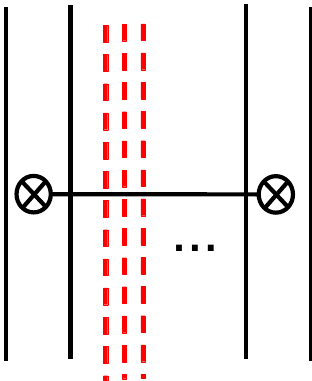}}
\hspace{.1mm}
\caption{(a) The brane configuration for an 't~Hooft loop with ${\bm B}={\bm e}_{N}+{\bm e}_{N-1}-{\bm e}_{1}-{\bm e}_{2}$. 
(b) A  D2-brane suspended  between two D4-branes is introduced to describe  't~Hooft-Polyakov monopole which screen the magnetic charge. 
 (c) After D2-brane moves with Hanany--Witten effects, we obtain  a brane configuration for the  SQM with ${\bm v}={\bm e}_{N}-{\bm e}_{1}$.}
\label{fig:UN2ndanti}
\end{figure}

We consider the 4d $U(N)$ gauge theory with $2N$ hypermultiplets in the fundamental representation.
  The minimal 't~Hooft line operator is either in the fundamental representation~$V$, or in the anti-fundamental representation~$\overline{V}$, both of the Langlands dual of the gauge group~$U(N)$, which is again $U(N)$.
The Moyal products of the minimal 't~Hooft operators were studied in \cite{Hayashi:2019rpw}. 
Here we consider products of non-minimal 't~Hooft operators.

\subsubsection{$\wedge^2 V$ and $ \wedge^2\overline{V}$}
\label{sec:UN-wedge2V-and-wedge2barV}
An example of a non-minimal operator is the 't~Hooft operator~$T_{\wedge^2 V}:=T_{\bm{B}={\bm e}_{N-1}+{\bm e}_N}$ in the rank-2 anti-symmetric representation~$\wedge^2 V$. (See Appendix~\ref{sec:thooft-QFT} for useful facts about Lie algebras.)
Its expectation value on~$S^1\times\mathbb{R}^3$ is determined by the localization formula~(\ref{TB.general}) and is given in terms of one-loop determinants as 
\begin{align}
\Braket{T_{\wedge^2 V}} = \sum_{1 \leq i < j \leq N}e^{b_i + b_j}Z_{ij}(\ba),
\end{align}
where%
\footnote{%
Throughout the paper, we use the short-hand notation $2\sinh\frac{a\pm b}{2} \equiv\prod_{s=\pm 1} (2\sinh\frac{a+ s b}{2})$, $2\sinh\frac{a\pm b \pm c}{2} \equiv \prod_{s,s'=\pm 1} (2\sinh\frac{a+ s b + s' c}{2})$, etc.
}
\begin{align}\label{UN-Zij}
Z_{ij}(\ba) = \left(\frac{\prod_{f=1}^{2N}2\sinh\frac{a_i - m_f}{2}2\sinh\frac{a_j - m_f}{2}}{\prod_{1 \leq k\neq i, j \leq N}2\sinh\frac{\pm a_i - a_k + \epsilon_+}{2}2\sinh\frac{\pm a_j - a_k + \epsilon_+}{2}}\right)^{\frac{1}{2}}.
\end{align}
The expectation value of the 't~Hooft line operator~$T_{\wedge^2\overline{V}}$ in the representation~$\wedge^2\overline{V}$ is 
\begin{align}
\Braket{T_{\wedge^2\overline{V}}} = \sum_{1 \leq i < j \leq N}e^{-b_i - b_j}Z_{ij}(\ba).
\end{align}

\subsubsection{$\wedge^2 V \times \wedge^2\overline{V}$}
\label{sec:UN-wedge2V-wedge2barV}

Let us consider the product of~$T_{\wedge^2 V}$ and $T_{\wedge^2\overline{V}}$. 
The Moyal product of the vevs may depend on the order because the two operators are distinct.
The product in one order is given by 
\begin{equation}\label{TL2L2}
\begin{split}
\Braket{T_{\wedge^2V}}\ast \Braket{T_{\wedge^2\overline{V}}} =&\sum_{1 \le h < i \le N, 1 \le j < k \le N \atop \{ h, i \} \cap  \{j, k \} =\emptyset }e^{b_h + b_i - b_j - b_k}Z_{hi}(\ba - \epsilon_+(\be_j + \be_k))Z_{jk}({\bm a} -\epsilon_+(\be_h + \be_i))\\
&+\sum_{1 \le i \neq j \le N} \sum_{1 \leq k \neq i, j \leq N} e^{b_i - b_j }Z_{ik}(\ba - \epsilon_+(\be_j + \be_k))Z_{jk}({\bm a} -\epsilon_+(\be_i + \be_k))\\
&+\sum_{1 \leq i < j \leq N}Z_{ij}(\ba - \epsilon_+(\be_i + \be_j))^2,
\end{split}
\end{equation}
and the product in the other order by
\begin{equation}\label{TL2bL2}
\begin{split}
\Braket{T_{\wedge^2\overline{V}}}\ast \Braket{T_{\wedge^2V}} =&\sum_{1 \le h < i \le N, 1 \le j < k \le N \atop \{ h, i \} \cap  \{j, k \} = \emptyset }e^{-b_h - b_i + b_j + b_k}Z_{hi}(\ba + \epsilon_+(\be_j + \be_k))Z_{jk}({\bm a} +\epsilon_+(\be_h + \be_i))\\
&+\sum_{1 \le i \neq j \le N} \sum_{1 \leq k \neq i, j \leq N} e^{- b_i + b_j }Z_{ik}(\ba + \epsilon_+(\be_j + \be_k))Z_{jk}({\bm a} + \epsilon_+(\be_i + \be_k))\\
&+\sum_{1 \leq  i <  j \leq N}Z_{ij}(\ba + \epsilon_+(\be_i + \be_j))^2.
\end{split}
\end{equation}
In each of these expressions there are two types of monopole screening contribution for the unscreened charge $\bm{B}={\bm e}_{N-1} +{\bm e}_N -{\bm e}_1 -{\bm e}_2 $: one characterized by the screened charge $\bv = \be_i - \be_j$ and the other by $\bv = {\bm 0}$.

\paragraph{$\bv = \be_i - \be_j$ ($i\neq j$).}

We first consider the monopole screening contributions in the sector~$\bv = \be_i - \be_j$. 
In~\eqref{TL2L2}, such a contribution can be read off by writing
\begin{equation}
\begin{split}
&e^{b_i - b_j }\sum_{1 \leq k \neq i, j \leq N}Z_{ik}(\ba - \epsilon_+(\be_j + \be_k))Z_{jk}({\bm a} -\epsilon_+(\be_i + \be_k))\\
&\hspace{4cm} = e^{b_i -b_j}Z_{\text{1-loop}}(\bv = \be_i - \be_j)Z^{\wedge^2V \times \wedge^2\overline{V}}_{\text{mono}}(\bv = \be_i - \be_j),
\end{split}
\end{equation}
where the one-loop part is
\begin{equation}\label{1loopL2L2}
\begin{split}
&Z_{\text{1-loop}}(\bv = \be_i - \be_j) \\
&\hspace{.5cm}= \left(\frac{\prod_{f=1}^{2N}2\sinh\frac{a_i - m_f}{2}2\sinh\frac{a_j - m_f}{2}}{2\sinh\frac{\pm(a_i - a_j)}{2}2\sinh\frac{\pm(a_i - a_j) + 2\epsilon_+}{2}\prod_{1\leq k\neq i, j \leq N}2\sinh\frac{\pm(a_i - a_k) + \epsilon_+}{2}2\sinh\frac{\pm(a_j - a_k) + \epsilon_+}{2}}\right)^{\frac{1}{2}}.
\end{split}
\end{equation}
We can read off the monopole screening contribution, which is given as
\begin{align}\label{Zmono-wedge2Vwedge2Vbar-eiej}
Z_{\text{mono}}^{\wedge^2V \times \wedge^2 \overline{V}}(\bv = \be_i - \be_j) = \sum_{1\leq k \neq i, j \leq N}\frac{\prod_{f=1}^{2N}2\sinh\frac{a_k - m_f - \epsilon_+}{2}}{\prod_{1\leq h \neq i, j, k \leq N}2\sinh\frac{a_k - a_h}{2}2\sinh\frac{-a_k + a_h + 2\epsilon_+}{2}}.
\end{align}
Similarly, for $\Braket{T_{\wedge^2\overline{V}}}\ast \Braket{T_{\wedge^2V}}$, we rewrite a part of the second line in \eqref{TL2bL2} as
\begin{equation}
\begin{split}
&e^{b_i - b_j }\sum_{1 \leq k \neq i, j \leq N} Z_{ik}(\ba + \epsilon_+(\be_j + \be_k))Z_{jk}({\bm a} + \epsilon_+(\be_i + \be_k))\\
&\hspace{4cm} =e^{b_i -b_j}Z_{\text{1-loop}}(\bv = \be_i - \be_j)Z^{\wedge^2 \overline{V} \times \wedge^2V}_{\text{mono}}(\bv = \be_i - \be_j),
\end{split}
\end{equation}
where the one-loop part is again ~\eqref{1loopL2L2}.
The monopole screening contribution in this case is
\begin{align}\label{Zmono-wedge2Vbarwedge2V-eiej}
Z_{\text{mono}}^{\wedge^2\overline{V} \times \wedge^2V}(\bv = \be_i - \be_j) = \sum_{1 \leq k \neq i, j \leq N}\frac{\prod_{f=1}^{2N}2\sinh\frac{a_k - m_f + \epsilon_+}{2}}{\prod_{1\leq h \neq i, j, k \leq N}2\sinh\frac{- a_k + a_h}{2}2\sinh\frac{a_k - a_h + 2\epsilon_+}{2}}.
\end{align}
The two expressions~\eqref{Zmono-wedge2Vwedge2Vbar-eiej} and~\eqref{Zmono-wedge2Vbarwedge2V-eiej} are similar but not obviously equal to each other.

\begin{figure}[t]
\centering
\subfigure[]{\label{fig:quiverantisym2}
\hspace{-4mm}
\includegraphics[scale=.3]{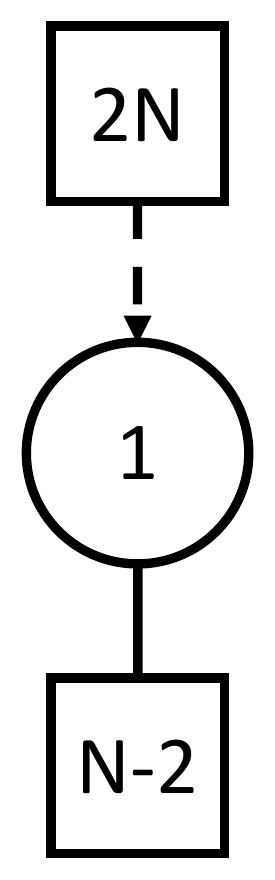}}
\hspace{2cm}
\subfigure[]{\label{fig:quiverantisym}
\hspace{-4mm}
\includegraphics[scale=.3]{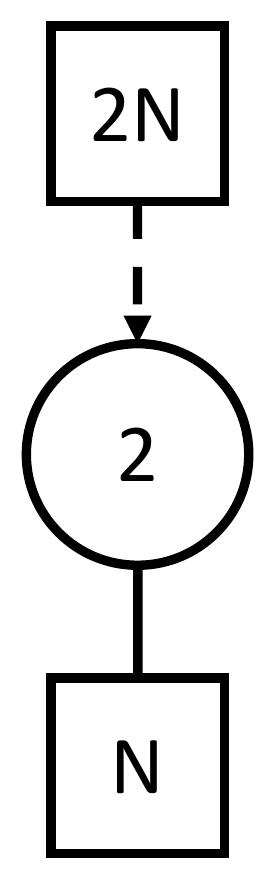}}
\hspace{.1mm}
\caption{
(a) The $\mathcal{N}=(0,4)$ quiver diagram describing the monopole screetning contribution to the $\bv = \be_i - \be_j$ sector in $\Braket{T_{\wedge^2 V}}\ast \Braket{T_{\wedge^2\overline{V}}}$ or $\Braket{T_{\wedge^2 \overline{V}}}\ast \Braket{T_{\wedge^2 V}}$.
(b) The $\mathcal{N}=(0,4)$ quiver diagram describing the monopole screetning contribution to the $\bv = {\bm 0}$ sector in $\Braket{T_{\wedge^2 V}} \ast \braket{ T_{\wedge^2 \overline{V}}}$ or $\Braket{T_{\wedge^2 \overline{V}}}\ast \Braket{T_{\wedge^2 V}}$.
}
\end{figure}

We now compare the expressions with the supersymmetric index of the SQM for the $\bm{v}=\bm{e}_i-\bm{e}_j$ sector of~$T_{{\bm e}_{N-1} +{\bm e}_N -{\bm e}_1 -{\bm e}_2 }$.
The SQM specialized to $i=N, j=1$ can be read off from 
the D-brane configuration
shown in Figure \ref{subfig:UNHooft2nd3}.
For general $i\neq j$ the SQM corresponds to
the quiver diagram 
in Figure~\ref{fig:quiverantisym2}. 
Its supersymmetric index is
\begin{equation}\label{WI.Uantisym2}
\begin{split}
Z(\bv = \be_i - \be_j, \zeta) = \oint_{JK(\zeta)}\frac{d\phi}{2\pi i} \frac{2\sinh(\epsilon_+)\prod_{f=1}^{2N}2\sinh\frac{\phi - m_f}{2}}{\prod_{1 \leq k \neq i, j \leq N}2\sinh\frac{\phi - a_k + \epsilon_+}{2}2\sinh\frac{-\phi + a_k + \epsilon_+}{2}}.
\end{split}
\end{equation}
For $\zeta >0$  the JK residue prescription, summarized in Appendix~2.2 of~\cite{Hayashi:2019rpw}, instructs us to evaluate the residues at the poles $\phi=a_k - \epsilon_+$ with $k \ne i, j$. For  $\zeta <0$, on the other hand, the residues are to be evaluated at $\phi=a_k + \epsilon_+$ with $k \ne i, j$.
This gives
\begin{align}\label{WIMoyal.Uantisym2}
Z_{\text{mono}}^{\wedge^2V \times \wedge^2\overline{V}}(\bv = \be_i - \be_j) &= Z(\bv = \be_i - \be_j, \zeta > 0), \nonumber \\
Z_{\text{mono}}^{\wedge^2\overline{V} \times \wedge^2V}(\bv = \be_i - \be_j) &= Z(\bv = \be_i - \be_j, \zeta < 0).
\end{align}
It can be checked that there occurs wall-crossing.%
\footnote{%
The SQM and the contour integral~(\ref{WI.Uantisym2}) are in fact identical to those which appear for the product of $T_V$ and $T_{\overline{V}}$ in the $U(N-2)$ SQCD with $2N$ flavors.
See~(3.17) of~\cite{Hayashi:2019rpw}.
It follows, from the discussion in Section~5 there, that for the $U(N)$ SQCD with $N_F$ flavors, $T_V$ and $T_{\overline{V}}$ exhibit wall-crossing for $N_F\geq 2N-2$.
}

\paragraph{$\bv = {\bm 0}$.}
\begin{figure}[t]
\centering
\subfigure[]{\label{subfig:UNHooft2nd4}
\includegraphics[scale=.5]{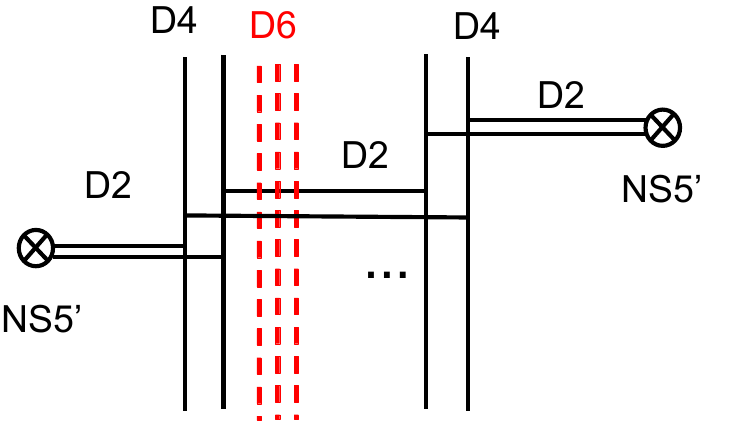}}
\hspace{0cm}
\subfigure[]{\label{subfig:UNHooft2nd5}
\includegraphics[scale=.5]{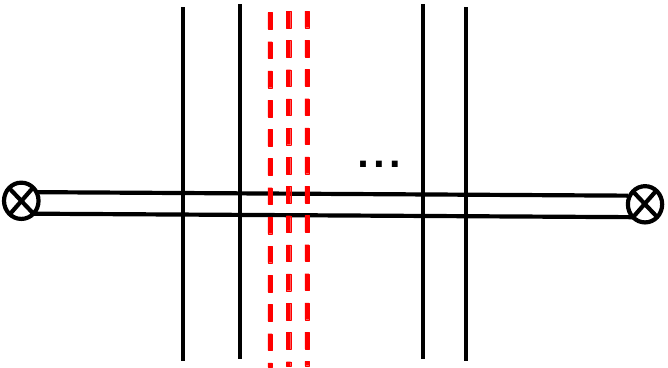}}
\hspace{.1mm}
\caption{
(b) two  D2-brane suspended  between D4-branes are introduced to describe  't~Hooft-Polyakov monopole which screen the magnetic charge. 
 (c) After Hanany--Witten effects, we obtain  a brane configuration for the  SQM with ${\bm v}={\bm 0}$.}
\label{fig:UN2nd2}
\end{figure}

We next consider the monopole screening contributions from the sector $\bv = {\bm 0}$.  
Such contributions in the  Moyal products $\Braket{T_{\wedge^2V}}\ast \Braket{T_{\wedge^2\overline{V}}}$ and $\Braket{T_{\wedge^2\overline{V}}}\ast \Braket{T_{\wedge^2 {V}}}$ are given by the last lines of \eqref{TL2L2} and \eqref{TL2bL2},
\begin{align}
Z_{\text{mono}}^{\wedge^2V \times \wedge^2\overline{V}}(\bv = 0)&=
\sum_{1 \le i < j \le N} \prod_{l=i, j} \frac{\prod_{f=1}^{2N}2\sinh\frac{a_l - m_f- \epsilon_+}{2}}{\prod_{k=1 \atop k \neq i,j}^N 2\sinh\frac{a_l - a_k }{2}2\sinh\frac{-a_l + a_k + 2\epsilon_+}{2}} ,\\
Z_{\text{mono}}^{\wedge^2\overline{V} \times \wedge^2V }(\bv = 0)&=\sum_{1 \le i < j \le N} \prod_{l=i, j} \frac{\prod_{f=1}^{2N}2\sinh\frac{a_l - m_f+ \epsilon_+}{2}}{\prod_{k=1 \atop k \neq i,j}^N 2\sinh\frac{a_l - a_k }{2}2\sinh\frac{-a_l + a_k -2\epsilon_+}{2}} .
\end{align}

 The D-brane configuration for ${\bm v}={\bm 0}$ is depicted in Figure~\ref{fig:UN2nd2}, and the quiver diagram for the SQM is in~Figure \ref{fig:quiverantisym}. 
 The supersymmetric index of the quiver theory is 
\begin{equation}\label{WI.Uantisym1}
\begin{split}
Z({\bm v}=0,\zeta) = &\frac{1}{2}\oint_{JK(\zeta)}\frac{d\phi_1}{2\pi i}\frac{d\phi_2}{2\pi i}\prod_{1 \leq i \neq j\leq 2}2\sinh\frac{\phi_i - \phi_j}{2}\\&\frac{\left(2\sinh\epsilon_+\right)^2\left(\prod_{1\leq i \neq j \leq 2}2\sinh\frac{\phi_i - \phi_j + 2\epsilon_+}{2}\right)\left(\prod_{i=1}^2\prod_{f=1}^{2N}2\sinh\frac{\phi_i - m_f}{2}\right)}{\prod_{i=1}^2\prod_{j=1}^N2\sinh\frac{\phi_i - a_j + \epsilon_+}{2}2\sinh\frac{-\phi_i + a_j + \epsilon_+}{2}}.
\end{split}
\end{equation}
For $\zeta>0$ the JK  residues are evaluated at
\begin{align}
&\{ \phi_{1}-a_i+\epsilon=0 \} \cap \{ \phi_{2}-a_j+\epsilon=0 \} \text{ for }  i, j =  1,2, \cdots, N,
\end{align}
and for $\zeta<0$ at
\begin{align}
&\{- \phi_{1}+a_i+\epsilon=0 \} \cap \{ -\phi_{2}+a_j+\epsilon=0 \} \text{ for }  i, j =  1,2, \cdots, N,
\end{align}
giving the relations
\begin{align}\label{WIMoyal.Uantisym1}
Z_{\text{mono}}^{\wedge^2V \times \wedge^2\overline{V}}(\bv = 0) = Z({\bm v}=0,\zeta > 0), \qquad Z_{\text{mono}}^{\wedge^2\overline{V} \times \wedge^2 {V}}(\bv = 0) = Z({\bm v}=0, \zeta < 0).
\end{align}
For small values of $N$ we checked that wall-crossing occurs between the positive and negative values of $\zeta$.
We obtain the relations
\begin{align}
\Braket{T_{{\bm e}_{N-1} +{\bm e}_N -{\bm e}_1 -{\bm e}_2 } }^{(\zeta>0)}=\Braket{T_{\wedge^2 V}}\ast\Braket{T_{\wedge^2 \overline{ V}}}, \nonumber \\
\Braket{T_{{\bm e}_{N-1} +{\bm e}_N -{\bm e}_1 -{\bm e}_2 } }^{(\zeta<0)}=\Braket{T_{\wedge^2 \overline{ V}}}\ast\Braket{T_{\wedge^2 V}}.
\end{align}
Here $\Braket{ \bullet }^{(\zeta>0)}$ (resp. $\Braket{ \bullet }^{(\zeta<0)}$) expresses the vev of the 't~Hooft operator  with the supersymmetric indices evaluated in the region $\zeta>0$ (resp. $\zeta <0$).

\subsection{$\mathcal{N}=2^{\ast}$ theory}
\label{sec:UN-2star}

It is also possible to consider monopole screening contributions in the expectation values of 't Hooft operators of the 4d $\mathcal{N}=2^{\ast}$ $U(N)$ gauge theory by utilizing the brane construction reviewed in Section \ref{sec:braneUNSQCD}. Then we can repeat the analysis which we have done in Section \ref{sec:UNSQCD}.

\subsubsection{$V \times \overline{V}$}\label{sec:UN-V-Vbar}
\begin{figure}[tb]
\centering
\subfigure[]{\label{subfig:UNscreening0}
\includegraphics[scale=.5]{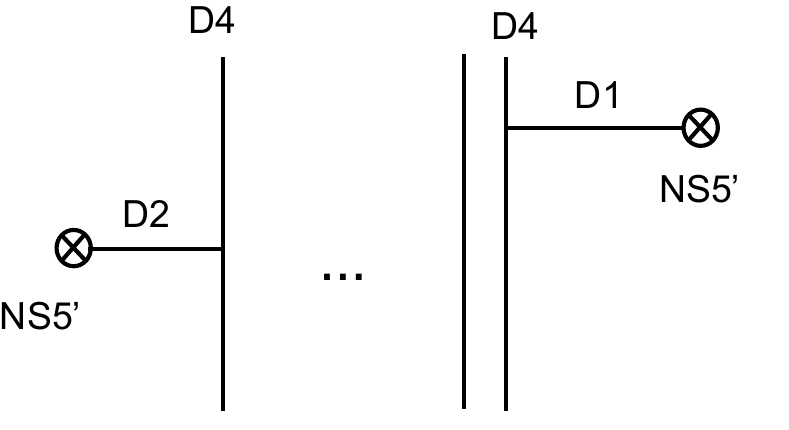}}
\hspace{.2cm}
\subfigure[]{\label{subfig:UNscreening01}
\includegraphics[scale=.5]{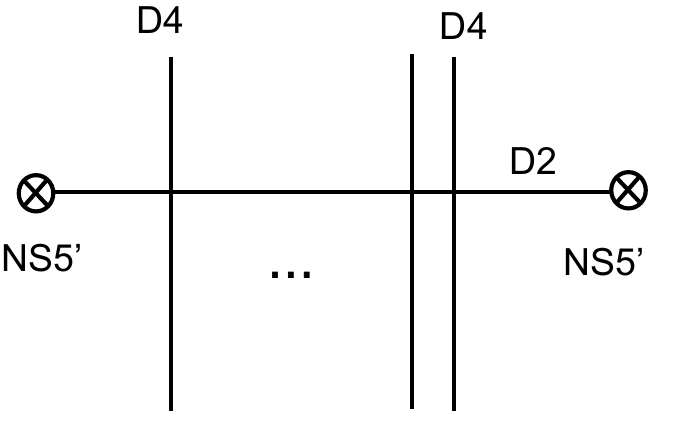}}
\hspace{.5cm}
\subfigure[]{\label{subfig:UNscreening02}
\hspace{-4mm}
\includegraphics[scale=.4]{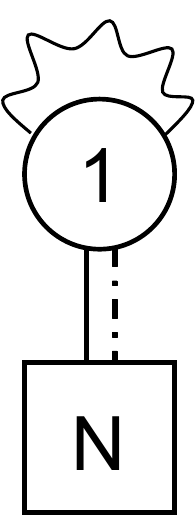}}
\hspace{.1mm}
\caption{(a) The brane configuration for an 't~Hooft loop with magnetic charge ${\bm B}={\bm e}_{N}-{\bm e}_{1}$.  
(b) A  D2-brane suspended  between two D4-branes is introduced to describe  't~Hooft-Polyakov monopole with magnetic charge
 ${\bm v}=-{\bm e}_{N}+{\bm e}_{1}$. When the vertical position of three segments of D2-branes coincide, 
the D2-branes combined into  signle D2-brane suspended between two NS5'-branes. Then the magnetic charge are completely screened. (c) The quiver diagram for SQM for the D2-brane world volume theory of (b). The solid line denotes 1d hypermultiplets. The dash-dotted line denotes long Fermi multiplets. The wavy line denotes a free twisted hypermultiplet.
}
\end{figure}

For the $\mathcal{N}=2^{\ast}$ theory we begin with the product of the minimal 't~Hooft operators, repeating the analysis we did for the SQCD in~\cite{Hayashi:2019rpw}.

The vevs $\langle T_V \rangle$ and $\langle T_{\overline{V}} \rangle$ are determined by the one-loop determinants  \eqref{1loopvm} and \eqref{1loophm} as
\begin{align}
\label{eq:lthantif}
\langle T_{ V }\rangle= \sum_{i=1 }^N e^{  b_{i}  } Z_{i}({\bm a}), \quad 
\langle T_{ \overline{V} }\rangle= \sum_{i=1 }^N e^{-  b_{i}  } Z_{i}({\bm a}),
\end{align}
with
\begin{align}
Z_{i}({\bm a})&=
\left(\prod_{k =1 \atop k \neq i  }^N  
 \frac{  \sinh  \frac{ a_{i}-a_k-m}{2}  \sinh \frac{a_k- a_{i}-m}{2} } 
{\sinh \frac{ a_i -a_k+\epsilon_+ }{2} \sinh  \frac{a_k -a_i+\epsilon_+}{2} }  \right) ^{\frac{1}{2}}.
\end{align}
Using the definition~(\ref{moyal}) of the Moyal product we get 
\begin{align}
\langle T_{ V }\rangle *
\langle T_{ \overline{V} }\rangle
&=\sum_{1 \le i  \neq j \le N}e^{ b_{i} -b_{j}}   Z_{\text{1-loop}}({\bm a}, \bv = \be_i - \be_j) + Z^{ V \times \overline{V}}_{\text{mono}}({\bm a}, \bv = {\bm 0}; \epsilon_+)  \, , \label{2star-VVbar-Moyal1} \\
\langle T_{ \overline{V} }\rangle *
\langle T_{ V }\rangle
&=\sum_{1 \le i  \neq j \le N}e^{ b_{i} -b_{j}}   Z_{\text{1-loop}}({\bm a}, \bv = \be_i - \be_j) + Z^{ V \times \overline{V} }_{\text{mono}}({\bm a}, \bv = {\bm 0}; -\epsilon_+)  \, ,\label{2star-VVbar-Moyal2}
\end{align}
with $Z^{ V \times  \overline{V}}_{\text{mono}}$ given as
\begin{align}
 Z^{ V \times  \overline{V}}_{\text{mono}}({\bm a}, \bv = {\bm 0}; \epsilon_+)
&=\sum_{i=1}^{N} \prod_{k =1 \atop k \neq i }^N \frac{2 \sinh \frac{a_i -a_k-m -\epsilon_+}{2} 2 \sinh \frac{a_k -a_i-m +\epsilon_+}{2} }{2 \sinh \frac{a_i -a_k}{2} 2 \sinh \frac{a_k -a_i+2 \epsilon_+ }{2} }.
\label{Zmonomoyal1}
\end{align}

Next we evaluate the monopole screening contribution in $\langle T_{\bm{B}={\bm e}_{N}- {\bm e}_{1}} \rangle$ using an SQM.
The D-brane configuration for the reduced magnetic charge  ${\bm v}={\bm 0}$ is shown in Figure~\ref{subfig:UNscreening01}. 
The corresponding $\mathcal{N}=(0,4)$ quiver diagram is depicted  in Figure~\ref{subfig:UNscreening02}.
 The localization formula for the supersymmetric index gives
\begin{align}
Z(\bv = {\bm 0},  \zeta)=-\oint_{JK(\zeta)} \frac{d \phi}{2 \pi i} 
 \frac{2 \sinh \epsilon_+ }{2 \sinh \frac{ \pm m - \epsilon_+}{2} } 
 \prod_{i =1}^{N} \frac{2 \sinh \frac{ \pm( \phi-a_i)-m }{2}  }
{ 2 \sinh \frac{ \pm( \phi-a_i)+\epsilon_+ }{2}} \, .
\label{eq:Wittenind1}
\end{align}
Here we chose the overall sign by hand, to obtain a match with the Moyal product below. 
The JK residues are evaluated at the poles $\phi= a_i  -\epsilon$ for $\zeta >0$,  and at the poles $\phi=a_i  +\epsilon$ for $\zeta <0$, both with $i=1, \cdots, N$.
This gives
\begin{align}
Z(\bv = {\bm 0},  \zeta>0)&=\sum_{i=1 }^{N} \mathop{\prod_{ k =1 \atop k  \neq   i  }}^{N} \frac{  2 \sinh \frac{a_i -a_k-m -\epsilon_+}{2} \, 2 \sinh \frac{a_k -a_i-m +\epsilon_+}{2}  }{   
 2 \sinh \frac{a_i -a_k}{2} \, 2 \sinh \frac{a_k -a_i+2\epsilon_+}{2} } \, ,
\label{eq:zmonoN2star11}
\\
Z(\bv = {\bm 0},  \zeta<0)&=\sum_{i=1  }^{N} \mathop{\prod_{ k =1 \atop k  \neq  i }}^{N} \frac{  2 \sinh \frac{a_i-a_k-m +\epsilon_+}{2} \, 2 \sinh \frac{a_k -a_i-m -\epsilon_+}{2}  }{   
 2 \sinh \frac{a_i -a_k+2\epsilon_+}{2} \, 2 \sinh \frac{a_k -a_i}{2} } \, .
\label{eq:zmonoN2star12}
\end{align}
We see that the supersymmetric indices \eqref{eq:zmonoN2star11} and \eqref{eq:zmonoN2star12} agree with the monopole screening contributions read off from the Moyal products \eqref{2star-VVbar-Moyal1} and~\eqref{2star-VVbar-Moyal2}.
 The residues at $\mathrm{Re} \, \phi=\pm \infty$ in \eqref{eq:Wittenind1} cancel each other and we have $Z(\zeta>0)=Z(\zeta<0)$: there is no wall-crossing.
We have the relations
\begin{align}
\langle T_{ V }\rangle * \langle T_{ \overline{V} }\rangle=\langle T_{ \overline{V} }\rangle * \langle T_{ V }\rangle
=\langle T_{{\bm e}_{N}- {\bm e}_{1}} \rangle .
\end{align}

\subsubsection{$\wedge^2 V\times \wedge^2 \overline{V} $}

Next we study monopole screening for  $ T_{{\bm e}_{N-1}+{\bm e}_{N}- {\bm e}_{1}- {\bm e}_{2}} $. 
The vevs of the 't~Hooft operators in the rank-2 anti-symmetric representations are given as
\begin{align}
\label{eq:lthantia}
\langle T_{\wedge^{2} {V} }\rangle
= \sum_{ 1 \le i, j  \le N } e^{b_{i}+b_j  } Z_{i j}({\bm a}),
 \quad \langle T_{\wedge^{2} \overline{V} }\rangle
= \sum_{ 1 \le i, j  \le N } e^{- b_{i}-b_j  } Z_{i j}({\bm a}) 
\end{align}
with
\begin{align}
Z_{i j}({\bm a})&= \prod_{l= i,j}^N \prod_{k=1 \atop k \neq i,j}^N
\left( \frac{  \sinh  \frac{ a_{l}-a_k-m}{2}  \sinh \frac{a_k- a_{i}-m}{2} } 
{\sinh \frac{ a_l -a_k+\epsilon_+ }{2} \sinh  \frac{a_k -a_l+\epsilon_+}{2} }  \right) ^{\frac{1}{2}}.
\end{align}
Their Moyal products are
\begin{align}
\langle T_{\wedge^{2} V }\rangle
\ast \langle T_{\wedge^{2} \overline{V} }\rangle
&=\sum_{1 \le  i <j \le N,  1 \le k <l \le N \atop \{ i,j \} \cap \{ k, l \} = \emptyset }
e^{ b_{i} +b_{j} -b_{k} -b_{l} }   Z_{\text{1-loop}}({\bm a}, \bv = \be_i + \be_j -\be_k - \be_l) \nonumber \\
&+\sum_{1 \le i  \neq j \le N}e^{ b_{i} -b_{j}}   Z_{\text{1-loop}}({\bm a}, \bv = \be_i - \be_j) Z^{\wedge^{2} V \times \wedge^{2} \overline{V}}_{\text{mono}}({\bm a}, \bv = \be_i - \be_j; \epsilon_+) \nonumber \\
&+ Z^{\wedge^{2} V \times \wedge^{2} \overline{V}}_{\text{mono}}({\bm a}, \bv = {\bm 0}; \epsilon_+)  \, ,
\\
 \langle T_{\wedge^{2} \overline{V} }\rangle \ast
\langle T_{\wedge^{2} V }\rangle
&=\sum_{1 \le  i <j \le N,  1 \le k <l \le N \atop \{i,j \} \cap \{k, l \} = \emptyset }
e^{ b_{i} +b_{j} -b_{k} -b_{l} }   Z_{\text{1-loop}}({\bm a}, \bv = \be_i + \be_j -\be_k - \be_l) \nonumber \\
&+\sum_{1 \le i  \neq j \le N}e^{ b_{i} -b_{j}}   Z_{\text{1-loop}}({\bm a}, \bv = \be_i - \be_j) Z^{\wedge^{2} V \times \wedge^{2} \overline{V}}_{\text{mono}}({\bm a}, \bv = \be_i - \be_j; -\epsilon_+) \nonumber \\
&+ Z^{\wedge^{2} V \times \wedge^{2} \overline{V}}_{\text{mono}}({\bm a}, \bv = {\bm 0}; -\epsilon_+)  \, .
\end{align}
Here $Z^{\wedge^{2} V \times \wedge^{2} \overline{V}}_{\text{mono}} ({\bm a}, {\bm v}; \epsilon_+)$ are given by
\begin{equation}\label{eq:ZmonoN2star1} 
Z^{\wedge^{2} V \times \wedge^{2} \overline{V}}_{\text{mono}}({\bm a}, \bv = \be_i - \be_j; \epsilon_+)
=\sum_{l=1 \atop l \neq i,j }^{N} \prod_{ k =1 \atop k  \neq  i, j , l  }^N \frac{  2 \sinh \frac{a_l -a_k-m -\epsilon_+}{2} \, 2 \sinh \frac{a_k -a_l-m -\epsilon_+}{2}  }{   
 2 \sinh \frac{a_l -a_k}{2} \, 2 \sinh \frac{a_k -a_l+2\epsilon_+}{2} }   
\end{equation}
with $i\neq j$, and 
\begin{equation}\label{eq:ZmonoN2star2} 
Z^{\wedge^{2} V \times \wedge^{2} \overline{V}}_{\text{mono}}({\bm a}, \bv = {\bm 0}; \epsilon_+) 
=\sum_{ 1 \le  i < j \le N}  \prod_{l=i,j} \prod_{k=1 \atop k \neq i,j}^N \frac{2  \sinh  \frac{a_l -a_k-m -\epsilon_+}{2} \, 2\sinh \frac{a_k -a_l-m+\epsilon_+}{2} }   
{2 \sinh \frac{a_l -a_k}{2}  \, 2 \sinh \frac{a_k -a_l+2\epsilon_+}{2} }  \, .
\end{equation}
We now compare these expressions with the supersymmetric indices of the SQMs.

\begin{figure}[tb]
\centering
\subfigure[]{\label{fig:quiverN2star1}
\hspace{-4mm}
\includegraphics[scale=.45]{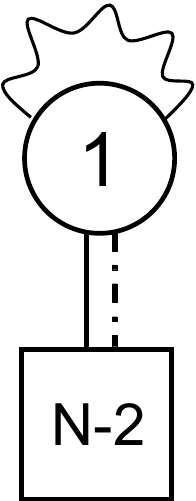}}
\hspace{3cm}
\subfigure[]{\label{fig:quiverN2star2}
\hspace{-4mm}
\includegraphics[scale=.45]{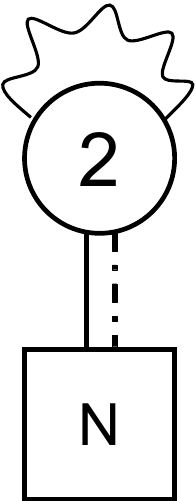}}
\hspace{.1mm}
\caption{
(a) The quiver diagram describing the monopole screening contribution to the $\bv = \be_i - \be_j$ sector in $\Braket{T_{\wedge^2 V}}\ast \Braket{T_{\wedge^2\overline{V}}}$ or $\Braket{T_{\wedge^2 \overline{V}}}\ast \Braket{T_{\wedge^2 V}}$.
(b) The quiver diagram describing the monopole screening contribution to the $\bv = {\bm 0}$ sector in $\Braket{T_{\wedge^2 V}} \ast \braket{ T_{\wedge^2 \overline{V}}}$ or $\Braket{T_{\wedge^2 \overline{V}}}\ast \Braket{T_{\wedge^2 V}}$.}
\end{figure}

\paragraph{$\bv = \be_i - \be_j$.}
The D-brane configuration for the monopole screening sector~$\bv = \be_i - \be_j$ would be represented by the same figure as Figure~\ref{subfig:UNHooft2nd3} except that the dashed lines for the D6-branes should be omitted.
The resulting SQM on the D2-branes has the $\mathcal{N}=(4,4)$ supersymmetry.
Its quiver diagram in the $\mathcal{N}=(0,4)$ notation is given
in Figure~\ref{fig:quiverN2star1}. The localization formula for the supersymmetric index gives
\begin{align}
Z({\bm v}={\bm e}_{N-1}- {\bm e}_{2}, \zeta)=-\oint_{JK(\zeta)} \frac{d \phi}{2 \pi i} 
 \frac{2 \sinh \epsilon_+ }{2 \sinh \frac{ \pm m + \epsilon_+}{2} } 
 \prod_{i =2}^{N-1} \frac{2 \sinh \frac{ \pm( \phi-a_i)-m }{2}  }
{ 2 \sinh \frac{ \pm( \phi-a_i)+\epsilon_+ }{2}} \, .
\label{eq:UN2starscreen1}
\end{align}
Again the overall sign is fixed by hand.
The JK residues are evaluated at $\phi= a_i  -\epsilon$ for $\zeta >0$,
and at $\phi= a_i +\epsilon$ for $\zeta <0$, both with $i=2, \cdots, N-1$, giving
\begin{align}
Z({\bm v}={\bm e}_{N-1}- {\bm e}_{2},\zeta>0)&=\sum_{l=2  }^{N-1} \prod_{ k =2 \atop k  \neq   l  }^{N-1} \frac{  2 \sinh \frac{a_l -a_k-m -\epsilon_+}{2} \, 2 \sinh \frac{a_k -a_l-m +\epsilon_+}{2}  }{   
 2 \sinh \frac{a_l -a_k}{2} \, 2 \sinh \frac{a_k -a_l+2\epsilon_+}{2} } \, ,
\label{eq:zmonoN2star23}
\\
Z({\bm v}={\bm e}_{N-1}- {\bm e}_{2}, \zeta<0)&=\sum_{l=2  }^{N-1} \prod_{ k =2 \atop k  \neq   l  }^{N-1} \frac{  2 \sinh \frac{a_l -a_k-m +\epsilon_+}{2} \, 2 \sinh \frac{a_k -a_l-m -\epsilon_+}{2}  }{   
 2 \sinh \frac{a_l -a_k}{2} \, 2 \sinh \frac{a_k -a_l+2\epsilon_+}{2} } \, .
\label{eq:zmonoN2star33}
\end{align}
The expressions~\eqref{eq:zmonoN2star23} and \eqref{eq:zmonoN2star33} match~\eqref{eq:ZmonoN2star1} and~\eqref{eq:ZmonoN2star2} specialized to $(i,j)=(2,N-1)$, respectively.
The residues at $\mathrm{Re} \, \phi=\pm \infty$  cancel,  $Z(\zeta>0)=Z(\zeta<0)$, and there is no wall-crossing.

\paragraph{$\bv = {\bm 0}$.}
The figure for the brane configuration would be the same as~Figure \ref{subfig:UNHooft2nd5} with the D6-branes omitted.
The SQM quiver is shown in Figure~\ref{fig:quiverN2star2}.
The supersymmetric index is given by the contour integral
\begin{align}
Z({\bm v}={\bm 0}, \zeta)&=\frac{1}{2}\oint_{JK(\zeta)} \frac{d \phi_1}{2 \pi i} \frac{d \phi_2}{2 \pi i}
\left(  \frac{2 \sinh \epsilon_+ }{2 \sinh \frac{ \pm m - \epsilon_+}{2} } \right)^2 
\nonumber \\
& \times
\prod_{1 \le \alpha \neq \beta \le 2} \frac{2 \sinh \frac{\phi_{\alpha}-\phi_{\beta} }{2} 2 \sinh \frac{\phi_{\alpha}-\phi_{\beta}+2 \epsilon_{+} }{2} }
{ 2 \sinh \frac{\phi_{\alpha}-\phi_{\beta} \pm m - \epsilon_{+} }{2}}
\prod_{\alpha =1}^2 \prod_{i =1}^N \frac{2 \sinh \frac{ \pm( \phi_{\alpha}-a_i)-m }{2}  }
{ 2 \sinh \frac{ \pm( \phi_{\alpha}-a_i)+\epsilon_+ }{2}} \, .
\label{eq:UN2starWiten}
\end{align}
The positive FI-parameter~$\zeta>0$ corresponds to JK parameter  $\bm{\eta} =(1,1)$.%
\footnote{%
See Appendix~2.2 of~\cite{Hayashi:2019rpw} for our convention regarding the JK parameter~$\bm\eta$.
}
In this region, the following sets of singular hyperplane arrangements  contribute according to the JK residue prescription:
\begin{align}
&\{ \phi_{1}-a_i+\epsilon=0 \} \cap \{ \phi_{2}-a_j+\epsilon=0 \} \text{ for }  i, j =  1,2, \cdots, N,
\label{eq:hyp1} \\
&\{ \phi_{1}-a_i+\epsilon=0 \} \cap \{ \phi_{2}-\phi_{1}\pm m+\epsilon=0 \} \text{ for } i=  1,2, \cdots, N, 
\label{eq:hyp2}\\
&\{ \phi_{2}-a_i+\epsilon=0 \} \cap \{ \phi_{1}-\phi_{2} \pm m+\epsilon=0 \} \text{ for }  i=  1,2, \cdots, N . 
\label{eq:hyp4} 
\end{align}
The JK residues for  \eqref{eq:hyp2} and \eqref{eq:hyp4}  turn out to vanish, and the supersymmetric index is given by the sum of  the residues associated with the singular hyperplane arrangements \eqref{eq:hyp1}. 
The supersymmetric index for $\zeta<0$ can be computed similarly.
We get
\begin{align}
Z({\bm v}={\bm 0},\zeta>0)&=
\sum_{ 1 \le  i < j \le N}  \prod_{l=i,j} \prod_{k=1 \atop k \neq i,j}^N \frac{ \sinh  \frac{a_l -a_k-m -\epsilon_+}{2}  \sinh \frac{a_k -a_l-m+\epsilon_+}{2} }   
{\sinh \frac{a_l -a_k}{2} \sinh \frac{a_k -a_l+2\epsilon_+}{2} } 
\label{eq:Witten2ndasy1} \, , \\
Z({\bm v}={\bm 0}, \zeta<0)&=
\sum_{ 1 \le  i < j \le N}  \prod_{l=i,j} \prod_{k=1 \atop k \neq i,j}^N \frac{ \sinh  \frac{a_l -a_k-m +\epsilon_+}{2}  \sinh \frac{a_k -a_l-m-\epsilon_+}{2} }   
{\sinh \frac{a_l -a_k-2\epsilon_+}{2} \sinh \frac{a_k -a_l}{2} } \, .
\label{eq:Witten2ndasy2}
\end{align}
The  supersymmetric indices \eqref{eq:Witten2ndasy1} and  \eqref{eq:Witten2ndasy2} agree with $Z^{\wedge^2 V \times \wedge^2 \overline{V}}_{\text{mono}} ({\bm v}=0)$ obtained from the Moyal product.   
For several small values of $N$ we checked that $Z(\zeta >0)=Z(\zeta <0)$, {\it i.e.}, there is no wall-crossing, so that
\begin{align}
\langle T_{{\bm e}_{N-1}+{\bm e}_{N}- {\bm e}_{1}- {\bm e}_{2}} \rangle
=\langle T_{\wedge^{2} V }\rangle
\ast \langle T_{\wedge^{2} \overline{V} }\rangle
= \langle T_{\wedge^{2} \overline{V} }\rangle
\ast \langle T_{\wedge^{2} V }\rangle \, .
\end{align}

\section{'t~Hooft operators in $SO(N)$ gauge theories}
\label{sec:SO}

It is possible to compute the expectation values of 't~Hooft operators of theories with a different gauge group. In this section we consider 't~Hooft operators in $B$ or $D$-type gauge theories, namely $SO(N)$ gauge theories. In particular we focus on 4d $\mathcal{N}=2$ $SO(N)$ gauge theory with $N-2$ hypermultiplets in the vector representation and the 4d $\mathcal{N}=2^{\ast}$ $SO(N)$ gauge theory. 

\subsection{$\mathcal{N}=2$ SQCD}
\label{sec:SOSQCD}
We start from 't~Hooft operators in the 4d $SO(N)$ gauge theory with $N-2$ hypermultiplets in the vector representation. The Langlands dual is different whether $N$ is even or odd. When $N = 2n$ the Langlands dual is $SO(2n)$, namely it is self-dual. When $N=2n+1$ the Langlands dual is $USp(2n)$. Then the minimal 't~Hooft line operator is in the fundamental representation of $SO(2n)$ or $USp(2n)$. 
In both cases the expectation value on~$S^1\times\mathbb{R}^3$ takes the form
\begin{align}\label{SO-TV-vev-form}
\Braket{T_V} = \sum_{i=1}^n\left(e^{b_i}Z_i(\ba) + e^{-b_i}Z_i(\ba)\right),
\end{align}
where 
\begin{align}
Z_i(\ba) = \left(\frac{\prod_{f=1}^{N-2}2\sinh\frac{\pm a_i-m_f}{2}
}{\prod_{1 \leq j \neq i \leq n}2\sinh\frac{\pm a_i \pm a_j + \epsilon_+}{2}}
\right)^{\frac{1}{2}}
\end{align}
for the $SO(2n)$ gauge theory, and 
\begin{align}\label{Zi-V-SOodd-SQCD}
Z_i(\ba) = \left(\frac{\prod_{f=1}^{N-2}2\sinh\frac{\pm a_i-m_f}{2}
}{2\sinh\frac{\pm a_i + \epsilon_+}{2}\prod_{1 \leq j \neq i \leq n}2\sinh\frac{\pm a_i \pm a_j + \epsilon_+}{2}}
\right)^{\frac{1}{2}}
\end{align}
for the $SO(2n+1)$ gauge theory.

\subsubsection{$V \times V$}
The simplest example of the Moyal product is the Moyal product between the expectation value of the minimal 't~Hooft operators. The explicit computation gives
\begin{equation}\label{SOVxV}
\begin{split}
& \Braket{T_V}\ast\Braket{T_V} \\
= &\sum_{1\leq i, j \leq n}\left(e^{b_i + b_j}Z_i(\ba + \epsilon_+\be_j)Z_j(\ba - \epsilon_+\be_i) + e^{-b_i - b_j}Z_i(\ba - \epsilon_+\be_j)Z_j(\ba + \epsilon_+\be_i)\right)\\
&+\sum_{1\leq i \neq  j \leq n}\left(e^{b_i - b_j}Z_i(\ba - \epsilon_+\be_j)Z_j(\ba - \epsilon_+\be_i) + e^{-b_i + b_j}Z_i(\ba + \epsilon_+\be_j)Z_j(\ba + \epsilon_+\be_i)\right)\\
&+\sum_{i=1}^n\left(Z_i(\ba - \epsilon_+\be_i)^2 + Z_i(\ba + \epsilon_+\be_i)^2\right)
\\
= &\sum_{1\leq i \leq n} (e^{2 b_i} + e^{-2b_i}) Z_i(\ba + \epsilon_+\be_i)Z_i(\ba - \epsilon_+\be_i)
\\
&+ \sum_{1\leq i < j\leq n} \sum_{s,t=\pm 1}e^{s b_i + t b_j}\Big(Z_i(\ba + t \epsilon_+\be_j)Z_j(\ba - s \epsilon_+\be_i)+ Z_i(\ba - t \epsilon_+\be_j)Z_j(\ba + s \epsilon_+\be_i)\Big)\\
&+\sum_{i=1}^n\left(Z_i(\ba - \epsilon_+\be_i)^2 + Z_i(\ba + \epsilon_+\be_i)^2\right).
\end{split}
\end{equation}
We have two types of the screening sector in \eqref{SOVxV}. One is characterized by $\bm{v}=\pm \bm{e}_i \pm \bm{e}_j$ ($1\leq i<j \leq n$, signs uncorrelated)%
\footnote{%
For $N=2n$ these correspond to all the coroots of~$SO(2n)$.
For $N=2n+1$ they correspond to the short coroots of~$SO(2n+1)$.
}
in the line second from the last in \eqref{SOVxV} and the other is the sector ${\bm v} = {\bm 0}$, which is given by the last line in \eqref{SOVxV}.
We now compare the monopole screening contributions in the Moyal product~\eqref{SOVxV} with the supersymmetric indices of the SQMs that describe monopole screening.

\paragraph{$\bv = \pm \be_i \pm \be_j\; (i \neq j)$.}

\begin{figure}[t]
\centering
\subfigure[]{\label{subfig:O4-tHooft-VxV-v=eNpluseN-1}
\raisebox{0mm}{\includegraphics[scale=.3]{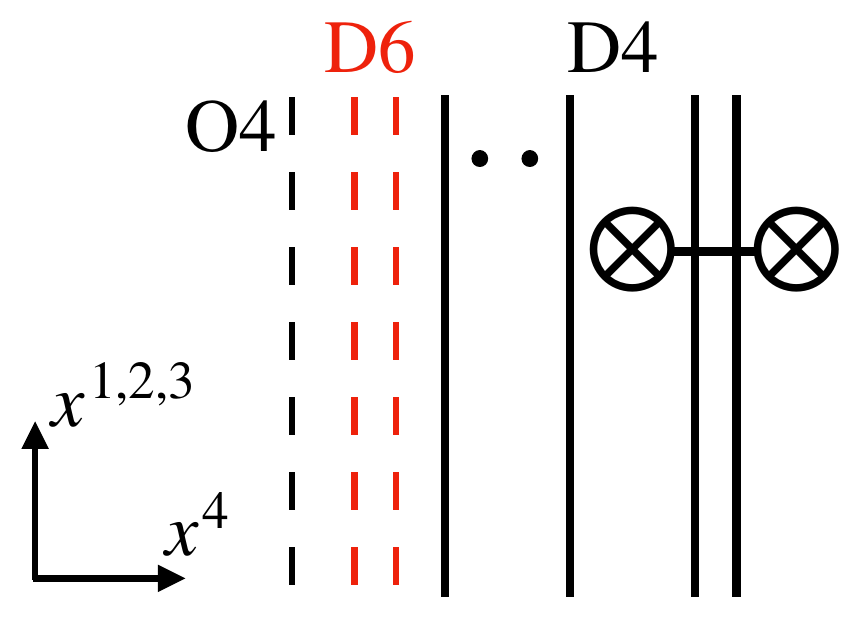}}}
\hspace{2cm}
\subfigure[]{\label{subfig:quiverSO-VxV-v=eNpluseN-1}
\hspace{-4mm}
\includegraphics[scale=.3]{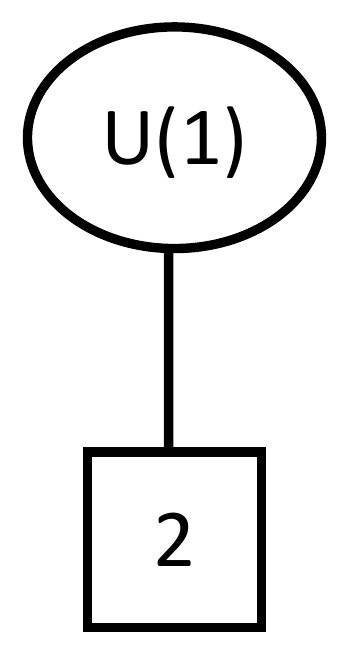}}
\caption{(a):  
The brane configuration for the monopole screening contribution to the $\bv = \bm{e}_{n-1}+\bm{e}_n$ sector in $T_{V}\cdot T_{V}$.
(b): The corresponding quiver diagram.
}
\label{fig:SO-VxV-v=eNpluseN-1}
\end{figure}
We focus on the sector $\bv = \be_{n-1} + \be_n$ since the other cases are related by Weyl reflections. For brane construction we focus on the realization of the gauge theory and 't~Hooft operators using an ${\rm O4}^-$-plane discussed in Section~\ref{sec:SQCD-O4}, although we could alternatively consider the realization using ${\rm O6}$-planes discussed in Section~\ref{sec:SQCD-2star-O6}. 
The brane configuration for the sector ${\bm v} = \bm{e}_{n-1}+\bm{e}_n$ is shown in~Figure~\ref{subfig:O4-tHooft-VxV-v=eNpluseN-1}.
The corresponding quiver diagram is depicted in Figure~\ref{subfig:quiverSO-VxV-v=eNpluseN-1}. Since the brane configuration leading to the SQM involves branes away from the orientifold, we have the same SQM for both $N=2n$ and $N=2n+1$. 
The supersymmetric index of the quiver theory is 
\begin{equation}\label{VxVuniversal}
\begin{split}
Z(\bv = \be_{n-1} + \be_n, \zeta) &= \oint_{JK(\zeta)} \frac{d\phi}{2\pi i}\frac{2\sinh\epsilon_+}{\prod_{i=n-1}^{n}2\sinh\frac{\phi - a_i + \epsilon_+}{2}2\sinh\frac{-\phi + a_i + \epsilon_+}{2}}\cr
&=\frac{1}{2\sinh\frac{a_{n-1} - a_n}{2}2\sinh\frac{-a_{n-1} + a_n + 2\epsilon_+}{2}} + \frac{1}{2\sinh\frac{a_{n} - a_{n-1}}{2}2\sinh\frac{-a_{n} + a_{n-1} + 2\epsilon_+}{2}}.
\end{split}
\end{equation}
for the both signs of the FI parameter $\zeta$. 

For the second line from the last of \eqref{SOVxV}, the contribution~$Z_{\text{mono}}(\bv = \be_{n-1} + \be_n; \ba)$ from the sector $\bv = \be_{n-1} + \be_n$ is determined by 
\begin{equation}\label{moyalSOTVVvee}
\begin{split}
&Z_{n-1}(\ba + \epsilon_+\be_n)Z_n(\ba - \epsilon_+\be_{n-1})+ Z_{n-1}(\ba -  \epsilon_+\be_n)Z_n(\ba + \epsilon_+\be{n-1})\cr
& = Z_{(n-1)n}(\ba)Z_{\text{mono}}(\bv = \be_{n-1} + \be_n; \ba),
\end{split}
\end{equation}
where $Z_{(n-1)n}(\ba)$ is the specialization to $i = n-1, j= n$ of
\begin{equation}\label{ZijSOeven}
Z_{ij}({\bm a}) = \left(\frac{\prod_{f=1}^{N-2}2\sinh\frac{\pm a_i - m_f}{2}2\sinh\frac{\pm a_j - m_f}{2}}{2\sinh\frac{\pm (a_i + a_j)}{2}2\sinh\frac{\pm (a_i + a_j) + 2\epsilon_+}{2}
\prod_{k\neq i, j}
2\sinh\frac{\pm a_i \pm a_k + \epsilon_+}{2}2\sinh\frac{\pm a_{ j}\pm a_k + \epsilon_+}{2}}\right)^{\frac{1}{2}}
\end{equation}
 for $N = 2n$, or
\begin{equation}\label{ZijSOodd}
\begin{split}
Z_{ij}({\bm a}) =& \left(\frac{\prod_{f=1}^{N-2}2\sinh\frac{\pm a_i - m_f}{2}2\sinh\frac{\pm a_j - m_f}{2}}{2\sinh\frac{\pm a_i + \epsilon_+}{2}2\sinh\frac{\pm a_j + \epsilon_+}{2}2\sinh\frac{\pm (a_i + a_j)}{2}2\sinh\frac{\pm (a_i + a_j) + 2\epsilon_+}{2}}\right.\\
&\hspace{4cm}\left.\times\frac{1}{\prod_{1 \leq k\neq i, k\neq j \leq n}2\sinh\frac{\pm a_i \pm a_k + \epsilon_+}{2}2\sinh\frac{\pm a_{ j} \pm a_k + \epsilon_+}{2}}\right)^{\frac{1}{2}}
\end{split}
\end{equation}
for $N = 2n+1$. 
Then from \eqref{moyalSOTVVvee} we have
\begin{equation}
\begin{aligned}
&Z_{\text{mono}}(\bv = \be_{n-1} + \be_n; \ba)\cr
&\qquad = \frac{1}{2\sinh\frac{a_{n-1} - a_n}{2}2\sinh\frac{-a_{n-1} + a_n + 2\epsilon_+}{2}} + \frac{1}{2\sinh\frac{a_{n} - a_{n-1}}{2}2\sinh\frac{-a_{n} + a_{n-1} + 2\epsilon_+}{2}},
\end{aligned}
\end{equation}
which reproduces \eqref{VxVuniversal}.

\paragraph{$\bv = {\bm 0}$.}
\begin{figure}[t]
\centering
\hspace{10mm}
\subfigure[]{\label{subfig:O4-tHooft-VxV-v=0}
\raisebox{4mm}{\includegraphics[scale=.3]{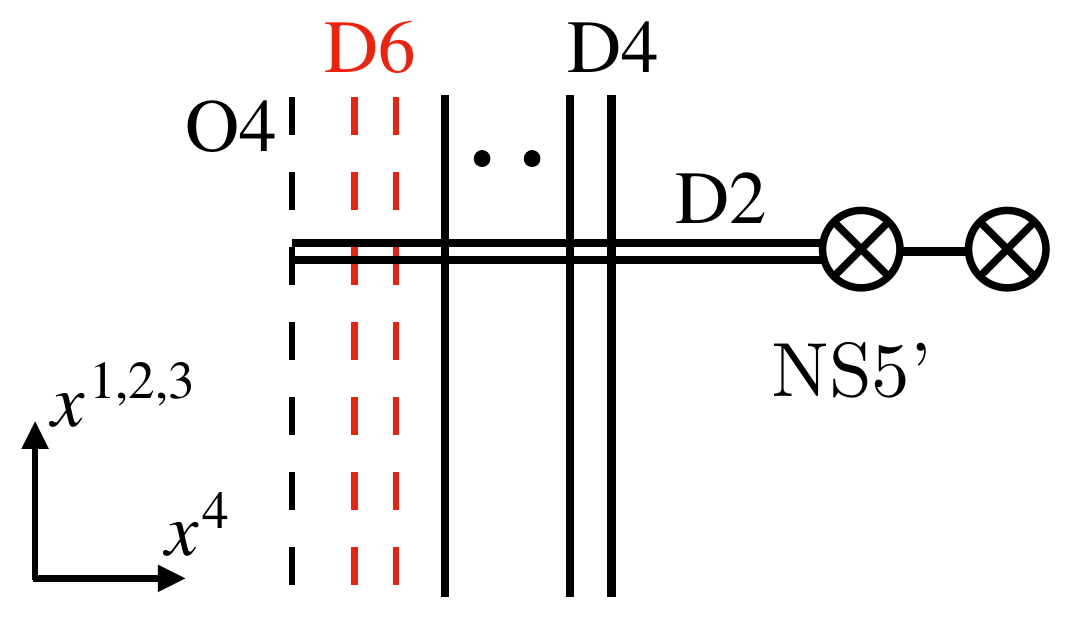}}}
\subfigure[]{\label{subfig:quiverSOv2}
\hspace{3mm}
\includegraphics[width=8cm]{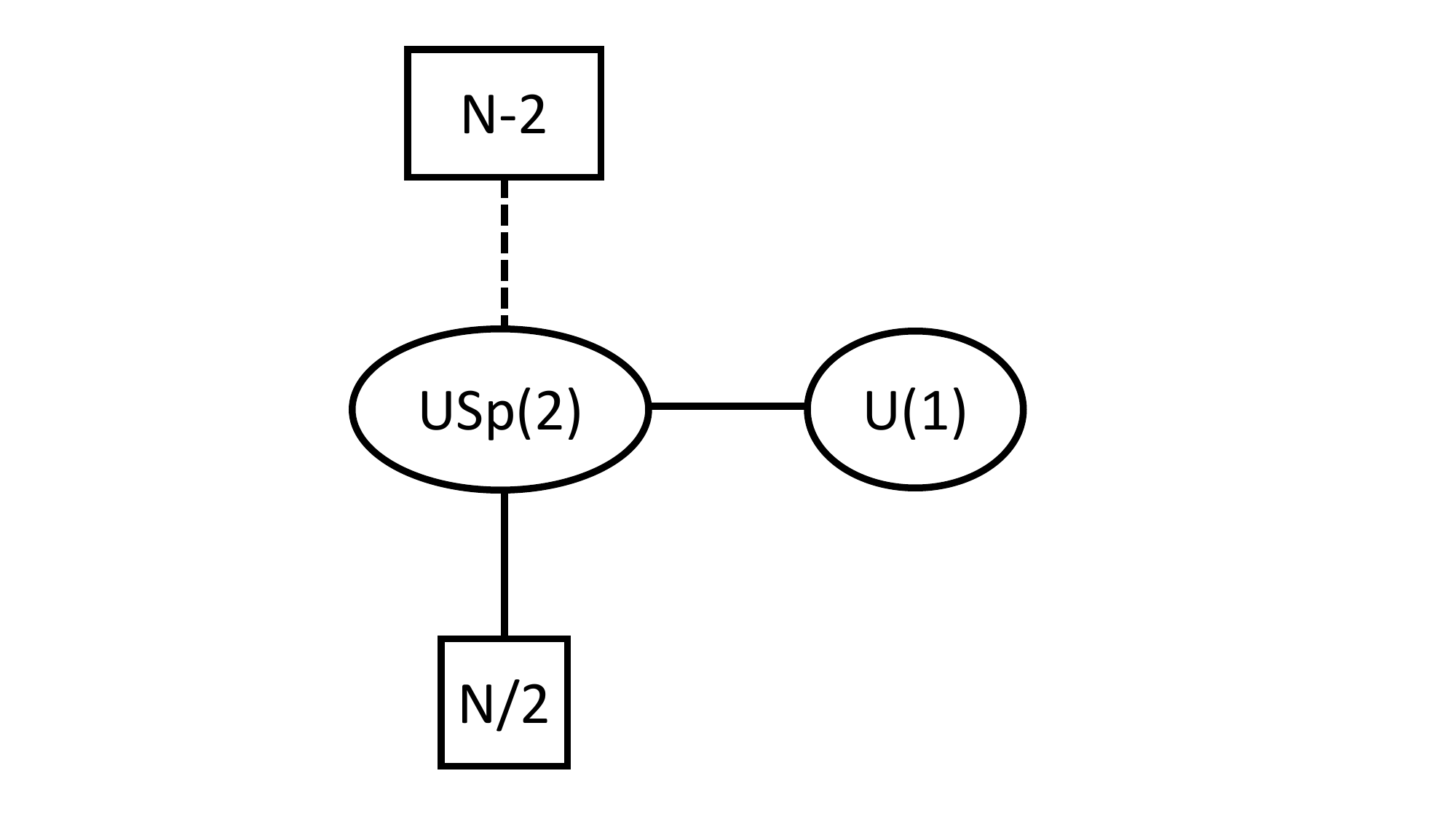}}
\caption{(a): The brane configuration for the monopole screening contribution to the $\bv = {\bm 0}$ sector in $T_{V}\cdot T_{V}$.
(b): The corresponding quiver diagram. 
}
\label{fig:SOv2}
\end{figure}
The brane configuration we use to read off the SQM for the sector ${\bm v} = {\bm 0}$ is shown in Figure~\ref{subfig:O4-tHooft-VxV-v=0}.
The corresponding quiver diagram is depicted in Figure \ref{subfig:quiverSOv2}. The supersymmetric index of the quiver theory is 
\begin{equation}\label{ZSO2nv1}
Z(\bv = {\bm 0}, \zeta)  = \frac{1}{2}\oint_{JK({ \zeta})}\frac{d\phi_1}{2\pi i}\frac{d\phi_2}{2\pi i}2\sinh(\pm \phi_1)\frac{\left(2\sinh\epsilon_+\right)^22\sinh(\pm \phi_1 + \epsilon_+)\left(\prod_{f=1}^{N-2}2\sinh\frac{\pm \phi_1 - m_f}{2}\right)}{\left(\prod_{i=1}^n2\sinh\frac{\pm \phi_1 \pm  a_i + \epsilon_+}{2}\right)2\sinh\frac{\pm \phi_1 \pm  \phi_2 + \epsilon_+}{2}}.
\end{equation}
for $N=2n$ and 
\begin{equation}\label{ZSO2np1v1}
Z(\bv = {\bm 0}, \zeta) = \frac{1}{2}\oint_{JK(\zeta)}\frac{d\phi_1}{2\pi i}\frac{d\phi_2}{2\pi i}2\sinh(\pm \phi_1)\frac{\left(2\sinh\epsilon_+\right)^22\sinh(\pm \phi_1 + \epsilon_+)\left(\prod_{f=1}^{N-2}2\sinh\frac{\pm \phi_1 - m_f}{2}\right)}{2\sinh\frac{\pm \phi_1 + \epsilon_+}{2}\left(\prod_{i=1}^n2\sinh\frac{\pm \phi_1 \pm  a_i + \epsilon_+}{2}\right)2\sinh\frac{\pm \phi_1 \pm  \phi_2 + \epsilon_+}{2}}.
\end{equation}
for $N=2n+1$. 
The FI parameter $\zeta$ is associated to the $U(1)$ gauge node in Figure \ref{subfig:quiverSOv2}. Namely for evaluating the integral \eqref{ZSO2nv1} and \eqref{ZSO2np1v1} with the JK residue prescription, we choose the reference vector $\bm\eta = (0, \zeta)$. In order to use the constructive definition of the JK residue, we deform the reference vector as $\bm\eta = (\delta, \zeta)$ with $|\delta| \ll  |\zeta|$. Then we find the relation
\begin{align}
\Braket{T_V}\ast\Braket{T_V}\Big|_{\bv = {\bm 0}} = Z(\bv={\bm 0}, \zeta),
\end{align}
for general $N$, irrespective of the sign of $\zeta$.
This relation is indeed expected because the product of two identical operators does not depend on the ordering.

\subsection{$\mathcal{N}=2^{\ast}$ theory}
\label{sec:N2starSO}

We now consider the $\mathcal{N}=2^{\ast}$ theory with gauge group $SO(N)$, {\it i.e.}, the 4d $SO(N)$ gauge theory with a hypermultiplet of mass $m$ in the adjoint representation. As in the SQCD case, the minimal 't~Hooft line operator corresponds to the fundamental representation of $SO(2n)$ for $N=2n$, and of $USp(2n)$ for $N=2n+1$. 
In both cases the expectation value of the minimal 't~Hooft operator~$T_V$ takes the same form as (\ref{SO-TV-vev-form}), namely 
\begin{align}\label{N2starSO-TV-vev-form}
\Braket{T_V} = \sum_{i=1}^n\left(e^{b_i}Z_i(\ba) + e^{-b_i}Z_i(\ba)\right),
\end{align}
with
\begin{equation}\label{Zi-SO-even-2star}
Z_i(\ba) = \left(\frac{
\prod_{1 \leq j \neq i \leq n}2\sinh\frac{\pm a_i \pm a_j -m}{2}
}{
\prod_{1 \leq j \neq i \leq n}2\sinh\frac{\pm a_i \pm a_j + \epsilon_+}{2}
}
\right)^{\frac{1}{2}}
\end{equation}
for the $SO(2n)$ gauge theory and 
\begin{equation}\label{Zi-SO-odd-2star}
Z_i(\ba) = \left(\frac{
2\sinh\frac{\pm a_i -m}{2}\prod_{1 \leq j \neq i \leq n}2\sinh\frac{\pm a_i \pm a_j -m}{2}
}{
2\sinh\frac{\pm a_i + \epsilon_+}{2}\prod_{1 \leq j \neq i \leq n}2\sinh\frac{\pm a_i \pm a_j + \epsilon_+}{2}
}
\right)^{\frac{1}{2}}
\end{equation}
for the $SO(2n+1)$ gauge theory. 

For gauge group $SO(3)$ the $\mathcal{N}=2^*$ theory trivially coincides with the SQCD with one flavor.
For $n=1$ indeed, (\ref{Zi-SO-odd-2star}) equals (\ref{Zi-V-SOodd-SQCD}) with the identification $m=m_{f=1}$.

For the 4d $\mathcal{N}=2^{\ast}$ $SO(N)$ gauge theory we also consider the 't~Hooft operator with magnetic charge~$\bm{B}=\bm{e}_{n-1}+\bm{e}_n$.
This corresponds to the rank-2 anti-symmetric representation~$\wedge^2 V$ of the Langlands dual group, which is $SO(2n)$ for gauge group $SO(2n)$ and $USp(2n)$ for gauge group $SO(2n+1)$.
The brane realization of the operator is similar to the one in Figure~\ref{subfig:O4-tHooft-wedge2V}; we remove D6-branes and replace O4 by ${\rm O6}^+$.
The expectation value of this operator, denoted as~$T_{\wedge^2V}$, takes the form
\begin{equation}\label{SOTwedgeV}
\begin{aligned}
\Braket{T_{\wedge^2V}} &= \sum_{1\leq i < j \leq n}\left(e^{b_i + b_j} + e^{-b_i - b_j}\right)Z_{ij}({\bm a}) 
\\
&\qquad\qquad
+ \sum_{1\leq i < j \leq n}\left(e^{b_i - b_j} + e^{-b_i + b_j}\right)Z'_{ij}({\bm a}) + Z_{\text{mono}}({\bm a}).
\end{aligned}
\end{equation}
The one-loop determinants are given as
\begin{equation}\label{N2starSOevenZij}
Z_{ij}({\bm a}) = \left(\frac{2\sinh\frac{\pm (a_i + a_j)-m\pm \epsilon_+}{2}
\prod_{k\neq i,j}
2\sinh\frac{\pm a_i \pm a_k -m}{2}2\sinh\frac{\pm a_{ j} \pm a_k -m}{2}}{2\sinh\frac{\pm (a_i + a_j)}{2}2\sinh\frac{\pm (a_i + a_j) + 2\epsilon_+}{2}
\prod_{k\neq i,j}
2\sinh\frac{\pm a_i \pm a_k + \epsilon_+}{2}2\sinh\frac{\pm a_{ j} \pm a_k + \epsilon_+}{2}}\right)^{\frac{1}{2}} ,
\end{equation}
\begin{equation}
Z'_{ij}({\bm a}) = \left(\frac{2\sinh\frac{\pm (a_i - a_j)-m\pm\epsilon_+}{2}
\prod_{k\neq i,j}
2\sinh\frac{\pm a_i \pm a_k -m}{2}2\sinh\frac{\pm a_{ j} \pm a_k -m}{2}}{2\sinh\frac{\pm (a_i - a_j)}{2}2\sinh\frac{\pm (a_i - a_j) + 2\epsilon_+}{2}
\prod_{k\neq i,j}
2\sinh\frac{\pm a_i \pm a_k + \epsilon_+}{2}2\sinh\frac{\pm a_{ j} \pm a_k + \epsilon_+}{2}}\right)^{\frac{1}{2}}
\end{equation}
for $N=2n$, and
\begin{equation}\label{N2starSOoddZij}
\begin{aligned}
Z_{ij}({\bm a}) &= \left(\frac{2\sinh\frac{\pm (a_i + a_j)-m\pm \epsilon_+}{2}
\prod_{k\neq i,j}
2\sinh\frac{\pm a_i \pm a_k -m}{2}2\sinh\frac{\pm a_{ j} \pm a_k -m}{2}}{2\sinh\frac{\pm (a_i + a_j)}{2}2\sinh\frac{\pm (a_i + a_j) + 2\epsilon_+}{2}
\prod_{k\neq i,j}
2\sinh\frac{\pm a_i \pm a_k + \epsilon_+}{2}2\sinh\frac{\pm a_{ j} \pm a_k + \epsilon_+}{2}}
\right.
\\
&\qquad\qquad\qquad
\times
\left. 
\frac{2\sinh\frac{\pm a_i-m}{2}2\sinh\frac{\pm a_j-m}{2}}{2\sinh\frac{\pm a_i +\epsilon_+}{2}2\sinh\frac{\pm a_j + \epsilon_+}{2}} \right)^{\frac{1}{2}},
\end{aligned}
\end{equation}
\begin{equation}\label{N2starSOoddZijp}
\begin{aligned}
Z'_{ij}({\bm a}) &= \left(\frac{2\sinh\frac{\pm (a_i - a_j)-m\pm \epsilon_+}{2}
\prod_{k\neq i,j}
2\sinh\frac{\pm a_i \pm a_k -m}{2}2\sinh\frac{\pm a_{ j} \pm a_k -m}{2}}{2\sinh\frac{\pm (a_i - a_j)}{2}2\sinh\frac{\pm (a_i - a_j) + 2\epsilon_+}{2}
\prod_{k\neq i,j}
2\sinh\frac{\pm a_i \pm a_k + \epsilon_+}{2}2\sinh\frac{\pm a_{ j} \pm a_k + \epsilon_+}{2}}
\right.
\\
&\qquad\qquad\qquad
\times
\left. 
\frac{2\sinh\frac{\pm a_i-m}{2}2\sinh\frac{\pm a_j-m}{2}}{2\sinh\frac{\pm a_i +\epsilon_+}{2}2\sinh\frac{\pm a_j + \epsilon_+}{2}} \right)^{\frac{1}{2}},
\end{aligned}
\end{equation}
for $N=2n+1$.
The last term~$Z_{\text{mono}}({\bm a})$ in~(\ref{SOTwedgeV}) is the monopole screening contribution corresponding to the zero weights in the rank-2 anti-symmetric representation\footnote{The rank-2 anti-symmetric representation of $SO(2n)$ is the adjoint representation of $SO(2n)$. The rank-2 anti-symmetric representation of $USp(2n)$ for $n\geq 2$ is reducible.  After subtracting a singlet, the remaining part has dimension $2n^2-n-1$.  Both the adjoint representation of $SO(2n)$ and the irreducible representation of $USp(2n)$ with dimension $2n^2-n-1$ are {\it quadi-minuscule}; all non-zero weights $\{\pm \bm{e}_i \pm \bm{e}_j|i<j\}$ lie in the same orbit under the Weyl group action.}. Unlike~$T_V$, the expectation value of $T_{\wedge^2 V}$ contains a monopole screening contribution which will be determined by the supersymmetric index of the corresponding SQM.

\subsubsection{$\wedge^2 V$}

\begin{figure}[t]
\centering
\subfigure[]{\label{subfig:O6-tHooft-wedge2V-v=0}
\includegraphics[scale=.3]{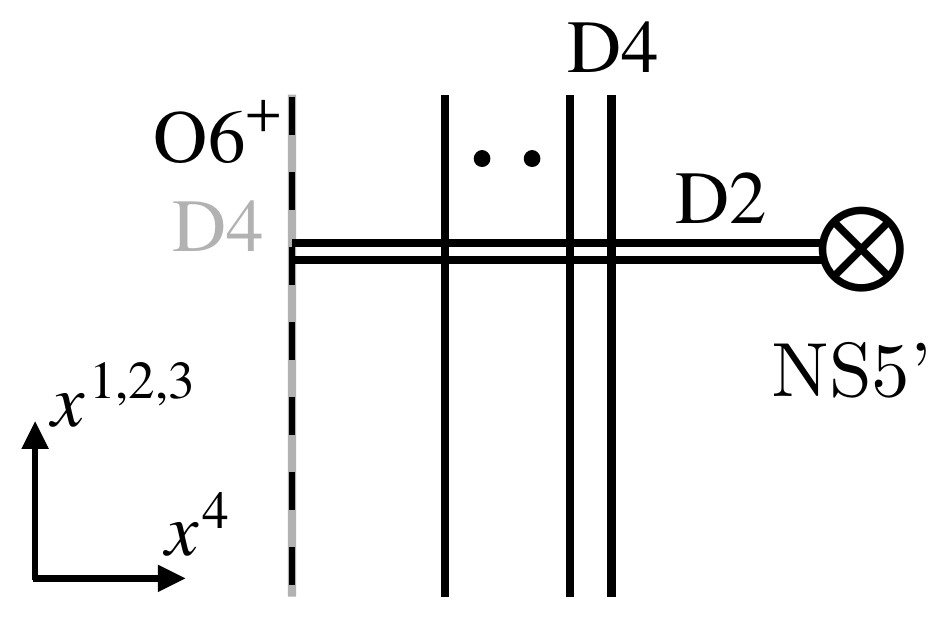}}
\hspace{1cm}
\subfigure[]{\label{subfig:quiverSO-2star-wedge2V}
\includegraphics[scale=.25]{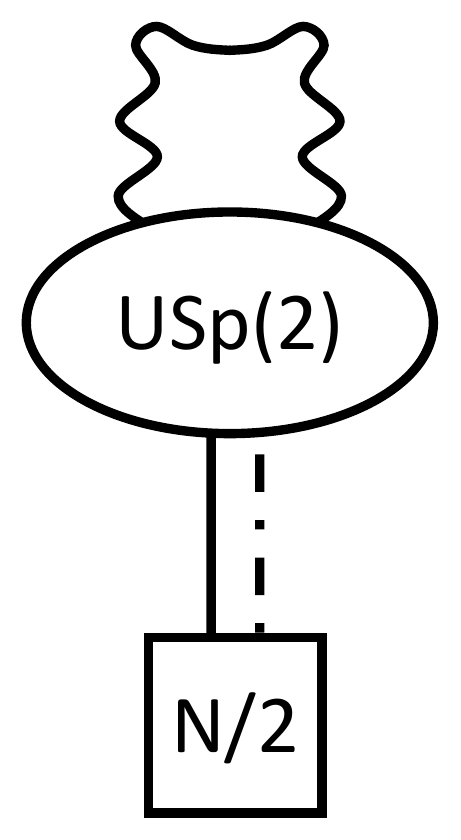}}
\caption{\label{fig:SO-2star-wedge2V}
(a): The brane configuration for the monopole screening of the 't~Hooft operator in the 4d $SO(N)$ $\mathcal{N}=2^*$ theory, corresponding to $\wedge^2 V$, $\bm{B}=\bm{e}_{n-1}+\bm{e}_n$, and $\bm{v}=0$.
The D4-brane on top of the O6-plane, indicted by the gray line, is present only for $N=2n+1$ and not for $N=2n$.
(b): The corresponding SQM quiver.
The box with the symbol $N/2$ represents $N$ half-hypermultiplets in the fundamental representation of $USp(2)$, which has $SO(N)$ as a 1d flavor symmetry group.
}
\end{figure}

We wish to determine the $\bm{v}=0$ term~$Z_{\text{mono}}({\bm a})$ in~(\ref{SOTwedgeV}), for the $\mathcal{N}=2^*$ theory using brane construction and an SQM. 
The brane configuration for the $\bm{v}=0$ sector is shown in Figure~\ref{subfig:O6-tHooft-wedge2V-v=0}, and the corresponding SQM quiver in~\ref{subfig:quiverSO-2star-wedge2V}.
The supersymmetric index is given by the contour integral
\begin{equation}\label{ZSO2nwedge2V-2star}
\begin{aligned}
Z_{\wedge^2 V}(\bv = {\bm 0}) =& 
 -
\frac{1}{2}
\oint_{JK(\eta)}\frac{d\phi}{2\pi i}
\frac{
2\sinh(\pm \phi)
\left(2\sinh\epsilon_+\right) 2\sinh(\pm \phi + \epsilon_+)
}{
2\sinh(\pm \phi+\frac{  \pm m  - \epsilon_+}{2} )
2\sinh( \frac{ \pm m -\epsilon_+}{2})
}
 \\
&
\qquad\qquad\qquad\qquad\qquad\qquad\qquad\qquad
\times
\frac{
\left(\prod_{i=1}^n
2\sinh\frac{\pm \phi \pm  a_i + m}{2}
\right)
}{
\left(\prod_{i=1}^n
2\sinh\frac{\pm \phi \pm  a_i + \epsilon_+}{2}
\right)
}
\end{aligned}
\end{equation}
for $N=2n$ and 
\begin{equation}\label{ZSO2np1wedge2V-2star}
\begin{aligned}
Z_{\wedge^2V}(\bv = {\bm 0})=&
 -
\frac{1}{2}
\oint_{JK(\eta)}\frac{d\phi}{2\pi i}
\frac{
2\sinh(\pm \phi)
\left(2\sinh\epsilon_+\right)2\sinh(\pm \phi + \epsilon_+)
}{
2\sinh(\pm \phi+\frac{  \pm m  - \epsilon_+}{2} )
2\sinh( \frac{ \pm  m  - \epsilon_+}{2})
}
 \\
&\qquad\qquad\qquad\qquad\times
\frac{
2\sinh\frac{\pm\phi+m}{2}
\left(\prod_{i=1}^n
2\sinh\frac{\pm \phi \pm  a_i + m}{2}
\right)
}{
2\sinh\frac{\pm\phi+\epsilon_+}{2}
\left(\prod_{i=1}^n
2\sinh\frac{\pm \phi \pm  a_i + \epsilon_+}{2}
\right)
}
\end{aligned}
\end{equation}
for $N=2n+1$.

Then $Z_{\wedge^2V}(\bv = {\bm 0})$ is given by the sum of the residues of the appropriate poles. 
For example when the JK parameter is positive, $\eta > 0$, 
the relevant poles are at $\phi= - \frac{\pm m -\epsilon_+}{2}, - \frac{\pm m-\epsilon_+}{2}+\pi i, -(\pm a_i+\epsilon_+)$ for $N=2n$.
For $N=2n+1$, in addition to the same poles, we should also include the pole at $\phi=-\epsilon_+$.
We propose that we need to include the overall minus signs in~(\ref{ZSO2nwedge2V-2star}) and (\ref{ZSO2np1wedge2V-2star}) by hand. 
To 
obtain the monopole screening contribution 
from $Z_{\wedge^2V}(\bv = {\bm 0})$ we need to remove the extra term from $Z_{\wedge^2V}(\bv = {\bm 0})$ as discussed in Section \ref{sec:extra-term}. For $N=2n$ we believe that the extra term is given by
\begin{equation}\label{extraSO2n}
Z_{\text{extra} }= \frac{\cosh\epsilon_+}{2\cosh\frac{m\pm \epsilon_+}{2}},
\end{equation}
which we checked explicitly for $n=2, 3, 4, 5$. For $N=2n+1$ we believe that the extra term is given by
\begin{equation}\label{extraSOodd}
Z_{\text{extra}} = -\frac{\cosh\epsilon_+}{2\cosh\frac{m\pm \epsilon_+}{2}},
\end{equation}
which we checked explicitly for $n=2, 3, 4, 5$. Accoding to the extra term pescription in Section~\ref{sec:extra-term}, $Z_{\text{mono}}(\ba)$ in \eqref{SOTwedgeV} is given by
\begin{equation}\label{Zwedge2VSOmono}
Z_{\text{mono}}(\ba) = Z_{\wedge^2 V}(\bv = {\bm 0}) - Z_{\text{extra}}.
\end{equation}
We will check the necessity of the signs as well as the extra term prescription by comparing the vevs $T_{\wedge^2V}$ for $N=4$ (below) $N=5$ (in Section~\ref{sec:USp}), and $N=6$ (below) with the vevs of the 't~Hooft operators corresponding to the adjoint representation for gauge groups $U(2)\times U(2)$, $USp(4)$, and $U(4)$, respectively. 

\paragraph{$SO(4)$ versus  $SU(2)\times SU(2)$.} 
For $N=4$, we note that $so(4)=su(2)\oplus su(2)$, and that $\wedge^2 V$ of $so(4)$ is the adjoint representation.
This motivates us to compare $\langle T_{\wedge^2 V}\rangle \equiv \langle T_{\wedge^2 V}^{SO(4)}\rangle$ for $SO(4)$ with (two copies of) the vev $\langle T^{\, SU(2)}_\text{adj}\rangle $ of the 't~Hooft operator which is S-dual to the adjoint Wilson operator. 
The result of Seciotion~\ref{sec:UN-V-Vbar} implies\footnote{
The vev of the adjoint 't Hooft operator in $\mathcal{N}=2^{\ast}$ $U(N)$ and $\mathcal{N}=2^{\ast}$ $SU(N)$ should be related as
\begin{align}\label{sunadjun}
\Braket{T^{SU(N)}_{\text{adj}}} = \Braket{T^{U(N)}_{\text{adj}}} -1,
\end{align}
where the traceless condition is imposed for $SU(N)$. We subtract one because of the difference between the number of zero weights of the adjoint representation of $SU(N)$ and that of the adjoint representation of $U(N)$. In particular for $N=2$, the relation is 
\begin{equation}
\Braket{T^{SU(2)}_{\text{adj}}} = \Braket{T^{U(2)}_{\text{adj}}}\Big|_{\substack{a_1 - a_2 \to a\\b_1 - b_2 \to b}} -1.
\end{equation}
}
\begin{equation}
\langle T^{\, SU(2)}_\text{adj} \rangle 
=
    (e^{b}+ e^{ - b})
    \left(
    \frac{2\sinh \frac{\pm a \pm \epsilon_+ - m}{2}}
    {
    2\sinh\frac{\pm a+2\epsilon_+}{2}
    2\sinh\frac{\pm a}{2}
    }
    \right)^{1/2}
    +Z_\text{mono}^{SU(2)}
\end{equation}
with
\begin{equation}\label{su2adjmono}
Z_\text{mono}^{SU(2)}= 
  \frac{
  2\sinh\frac{a-\epsilon_+ \pm m}{2}
  }
  {
  2\sinh\frac{a}{2}
  2\sinh\frac{a -2 \epsilon_+}{2}
  }
    + 
  \frac{
  2\sinh\frac{-a-\epsilon_+ \pm m}{2}
  }
  {
  2\sinh\frac{-a}{2}
  2\sinh\frac{-a -2 \epsilon_+}{2}
  } -1
  .
\end{equation}
It can be checked that
\begin{equation}
\begin{aligned}
Z_{\wedge^2V}^{SO(4)}(\bv ={\bm 0}) 
-
Z_\text{mono}^{SU(2)}|_{
\substack{
a\rightarrow a_1+a_2 \\
b\rightarrow b_1+b_2
}
}
-Z_\text{mono}^{SU(2)}|_{
\substack{
a\rightarrow a_1-a_2 \\
b\rightarrow b_1-b_2
}
}
&
=
\frac{\cosh \epsilon_+}{2\cosh\frac{m\pm \epsilon_+}{2}},\label{SO4vsSU2SU2v1}
\end{aligned}
\end{equation}
where $a_{1,2}$ and $b_{1,2}$ are the parameters of the $SO(4)$ theory.
Indeed we can see that \eqref{SO4vsSU2SU2v1} is precisely equal to \eqref{extraSO2n} and hence we have the equality
\begin{align}
Z_{\text{mono}}^{SO(4)}(\ba) =  Z_\text{mono}^{SU(2)}|_{
\substack{
a\rightarrow a_1+a_2 \\
b\rightarrow b_1+b_2
}
}
+Z_\text{mono}^{SU(2)}|_{
\substack{
a\rightarrow a_1-a_2 \\
b\rightarrow b_1-b_2
}
},
\end{align}
where $Z^{SO(4)}_{\text{mono}}(\ba)$ is given by \eqref{Zwedge2VSOmono} with $N=4$.
This further implies
\begin{equation}
\Braket{T_{\wedge^2V}^{SO(4)}} = \Braket{T_{\text{adj}}^{SU(2)}}\Big|_{
\substack{
a\rightarrow a_1-a_2 \\
b\rightarrow b_1-b_2
}
} + \Braket{T_{\text{adj}}^{SU(2)}}\Big|_{
\substack{
a\rightarrow a_1-a_2 \\
b\rightarrow b_1-b_2
}
}.
\end{equation}
We interpret the three brackets as the vevs of the adjoint 't~Hooft operator in the gauge theories with gauge groups $SO(4)$, $SU(2)$, and $SU(2)$ from left to right.

\paragraph{$SO(6)$ versus $SU(4)$.}
For $N=6$, the relation $so(6)=su(4)$ motivates us to compare $\langle T_{\wedge^2 V}\rangle \equiv \langle T_{\wedge^2 V}^{SO(6)}\rangle$ for $SO(6)$ with the vev of the 't~Hooft operator $T^{\, SU(4)}_\text{adj}$ corresponding to the adjoint representation of $SU(4)$.
Since $ \langle T_\text{adj}^{\, U(4)} \rangle = \langle T_V^{\, U(4)} \rangle  *\langle T_{\overline{V}}^{\, U(4)} \rangle$ this implies\footnote{Note that $\Braket{T_\text{adj}^{\, U(4)}}$ is related to $\Braket{T_\text{adj}^{\, SU(4)}}$ by \eqref{sunadjun}.} 
\begin{equation}
 \langle T_\text{adj}^{\, SU(4)} \rangle 
= 
\sum_{1\leq i<j\leq 4}  (e^{b_i - b_j } +e^{b_j - b_i })Z_i(\bm{a}- \epsilon_+ \bm{e}_j)  Z_j(\bm{a} - \epsilon_+ \bm{e}_i)
     +Z_\text{mono}^{SU(4)} ,
\end{equation}
\begin{equation}
Z_\text{mono}^{SU(4)} = \sum_{i=1}^4 Z_i(\bm{a}- \epsilon_+ \bm{e}_i)^2 -1,
\end{equation}
where
\begin{align}
Z_i({\bm a}) = \left(\prod_{j\neq i}\frac{2\sinh\frac{ a_i - a_j - m}{2}2\sinh\frac{-a_i + a_j -m}{2}}{2\sinh\frac{ a_i - a_j + \epsilon_+}{2}2\sinh\frac{-a_i + a_j + \epsilon_+}{2}}\right)^{\frac{1}{2}}. \label{Zi}
\end{align}
The vev $\langle T_\text{adj}^{\, SU(4)} \rangle$ depends only on the differences $a_i-a_j$ and $b_i-b_j$.
Note that the Dynkin diagram for $SO(6)$ is idential to that for $SU(4)$. 
From the identifications of the nodes, which correspond to simple roots, we identify the parameters $(a_1-a_2,a_2-a_3,a_3-a_4)$ of $SU(4)$ with $(a_2-a_3,a_1-a_2,a_2+a_3)$ of $SO(6)$.
We make similar identifications for $b_i$.
We find that
\begin{equation}
Z_{\wedge^2V}^{SO(6)}(\bv=0) - Z_{\text{mono}}^{SU(4)}|_{\substack{
 (a_1-a_2,a_2-a_3,a_3-a_4)\rightarrow (a_2-a_3,a_1-a_2,a_2+a_3) 
 \\
\hspace{-4mm} (b_1-b_2,b_2-b_3,b_3-b_4)\rightarrow (b_2-b_3,b_1-b_2,b_2+b_3) 
}}
 =
\frac{\cosh \epsilon_+}{2\cos\frac{m\pm \epsilon_+}{2}} , \label{SO6vsSU4v1}
\end{equation}
where the right hand side is independent of $a_i$ and $b_i$. Again \eqref{SO6vsSU4v1} precisely agrees with \eqref{extraSO2n}
and hence we have the equality
\begin{equation}
Z_{\text{mono}}^{SO(6)}(\ba) = Z_{\text{mono}}^{SU(4)}|_{\substack{
 (a_1-a_2,a_2-a_3,a_3-a_4)\rightarrow (a_2-a_3,a_1-a_2,a_2+a_3) 
 \\
\hspace{-4mm} (b_1-b_2,b_2-b_3,b_3-b_4)\rightarrow (b_2-b_3,b_1-b_2,b_2+b_3) 
}},
\end{equation}
where $Z^{SO(6)}_{\text{mono}}(\ba)$ is given by \eqref{Zwedge2VSOmono} with $N=6$. Namley, we have
\begin{equation} \label{SO6-U4-extra}
\begin{aligned}
\langle T_{\wedge^2 V}^{SO(6)}\rangle 
 = 
 \left.
 \langle T_\text{adj}^{\, SU(4)} \rangle \right|_{\substack{
 (a_1-a_2,a_2-a_3,a_3-a_4)\rightarrow (a_2-a_3,a_1-a_2,a_2+a_3) 
 \\
\hspace{-4mm} (b_1-b_2,b_2-b_3,b_3-b_4)\rightarrow (b_2-b_3,b_1-b_2,b_2+b_3) 
 }}.
 \end{aligned}
\end{equation}
The equality is in accord with the discussion in Section~\ref{sec:extra-term}.

\subsubsection{$V\times V$}

The Moyal product between the expectation value of the minimal 't~Hooft operators is again given by~(\ref{SOVxV}) with $Z_i$ given by~(\ref{Zi-SO-even-2star}) or~(\ref{Zi-SO-odd-2star}).
We have two types of the screening sector.

\paragraph{$\bv = \pm \be_i \pm \be_j\; (i \neq j)$.}

\begin{figure}[t]
\centering
\subfigure[]{\label{subfig:O6-tHooft-VxV-v=eN+eN-1}
\raisebox{0mm}{\includegraphics[scale=.3]{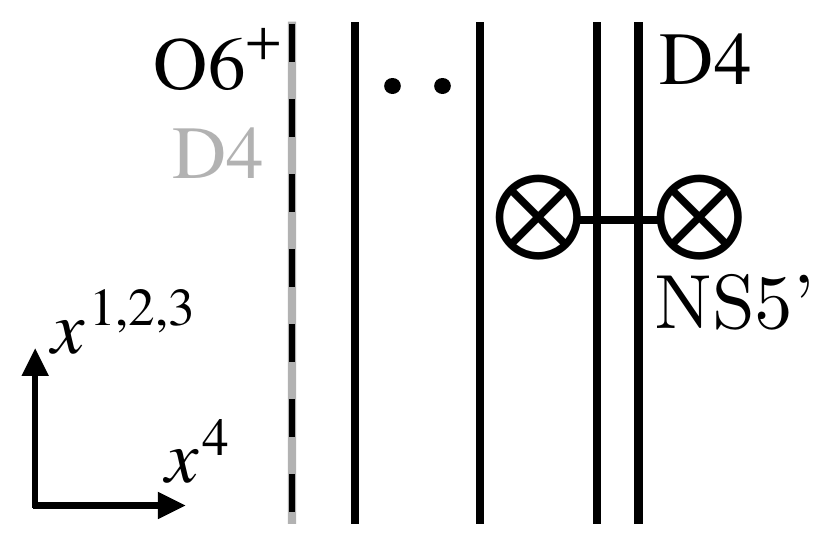}}}
\hspace{2cm}
\subfigure[]{\label{subfig:quiverSO-2star_VxV-v=eN+eN-1}
\hspace{-4mm}
\includegraphics[scale=.3]{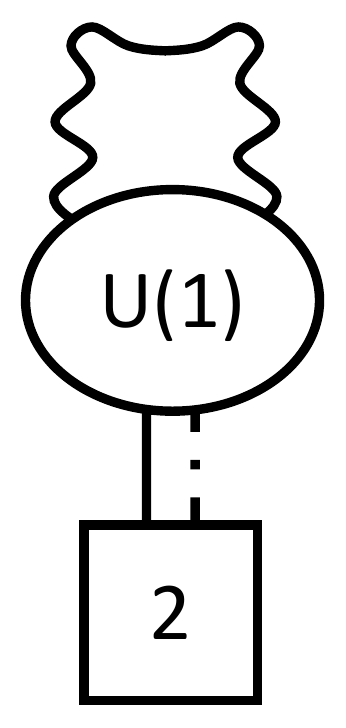}}
\caption{(a):  The brane configuration for the monopole screening contribution to the $\bv = \bm{e}_{n-1}+\bm{e}_n$ sector in $T_{V}\cdot T_{V}$.
(b): The corresponding quiver diagram.
}
\label{fig:SO-2star-VxV-v=eNpluseN-1}
\end{figure}
The terms in the line second from the last in \eqref{SOVxV} with the one-loop determinant \eqref{Zi-SO-even-2star} and \eqref{Zi-SO-odd-2star} corresponds to the monopole screening sectors~$\bv = \pm \be_i \pm \be_j\; (i \neq j)$. We only focus on $\bv = \be_{n-1} + \be_n$ as the other choices are related by Weyl reflections. The monopole screening contribution $Z_{\text{mono}}\left(\bv =\be_{n-1} + \be_n;\ba \right)$ is given by
\begin{equation}\label{moyalN2starSOTVVvee}
\begin{split}
&Z_{n-1}(\ba + \epsilon_+\be_n)Z_n(\ba - \epsilon_+\be_{n-1})+ Z_{n-1}(\ba -  \epsilon_+\be_n)Z_n(\ba + \epsilon_+\be_{n-1})\cr
& = Z_{(n-1)n}(\ba)Z_{\text{mono}}(\bv = \be_{n-1} + \be_n; \ba),
\end{split}
\end{equation}
where $Z_{(n-1)n}(\ba)$ is the specialization to~$i = n-1, j= n$ of either \eqref{N2starSOevenZij} for $N = 2n$, or \eqref{N2starSOoddZij} for $N = 2n+1$. 
We determine~$Z_{\text{mono}}(\bv = \be_{n-1} + \be_n; \ba)$ from~(\ref{moyalN2starSOTVVvee}) to be
\begin{equation}\label{zmono-from-moyal-SO-VxV-v=nn-1}
\begin{split}
&Z_{\text{mono}}(\bv = \be_{n-1} + \be_n; \ba)=\frac{2\sinh\frac{a_{n-1} - a_n - m + \epsilon_+}{2}2\sinh\frac{-a_{n-1} + a_n - m - \epsilon_+}{2}}{2\sinh\frac{a_{n-1} - a_n + 2\epsilon_+}{2}2\sinh\frac{-a_{n-1} + a_n}{2}}  \cr
&\qquad\qquad\qquad\qquad\qquad\qquad
+ \frac{2\sinh\frac{a_{n} - a_{n-1} - m + \epsilon_+}{2}2\sinh\frac{-a_{n} + a_{n-1} - m - \epsilon_+}{2}}{2\sinh\frac{a_{n} - a_{n-1} + 2\epsilon_+}{2}2\sinh\frac{-a_{n} + a_{n-1}}{2}}.
\end{split}
\end{equation}

The brane configuration for this sector is shown in Figure~\ref{subfig:O6-tHooft-VxV-v=eN+eN-1}, and the SQM quiver in Figure~\ref{subfig:quiverSO-2star_VxV-v=eN+eN-1}. 
This is the special case $N=2$ of Figure~\ref{subfig:UNscreening02}. 
The supersymmetric index can be obtained from the result \eqref{eq:zmonoN2star11} or \eqref{eq:zmonoN2star12}, and it is identical to the right hand side of~(\ref{zmono-from-moyal-SO-VxV-v=nn-1}), as expected.

\paragraph{$\bv =\bm{0}$.}

\begin{figure}[t]
\centering
\subfigure[]{\label{subfig:O6-tHooft-VxV-v=0}
\raisebox{2mm}{\includegraphics[scale=.3]{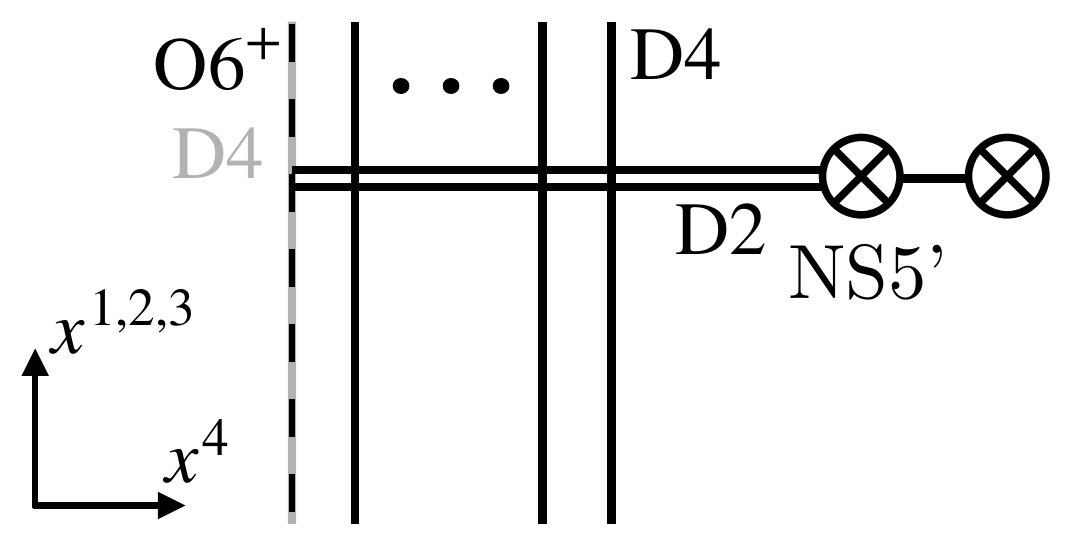}}}
\hspace{2cm}
\subfigure[]{\label{subfig:quiverSO-2star}
\includegraphics[width=4cm]{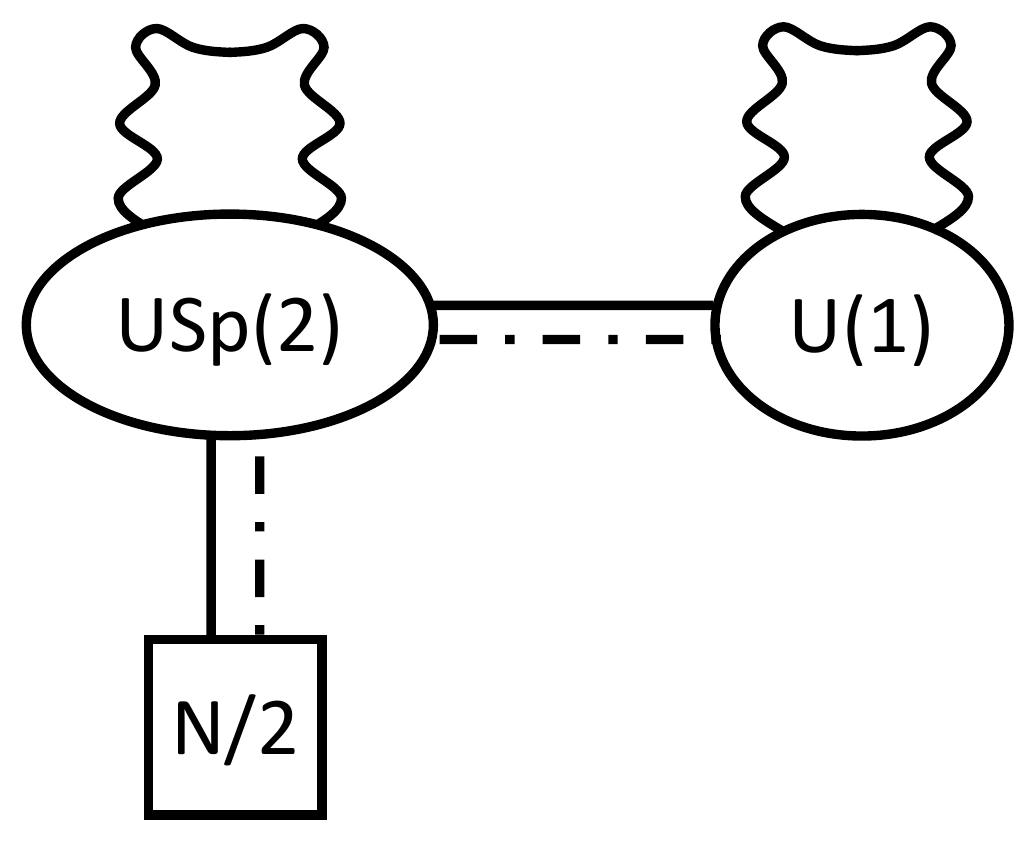}}
\caption{%
(a): The brane configuration for the monopole screening sector $\bv = {\bm 0}$ of $T_{V}\cdot T_{V}$ in the $SO(N)$ $\mathcal{N}=2^*$ theory.
(b): The corresponding quiver diagram. The wavy lines represent $\mathcal{N}=(0,4)$ twisted hypermultiplets while the dash-dotted lines represent  $\mathcal{N}=(0,4)$ long Fermi multiplets.
}
\label{fig:quiverSO-2star}
\end{figure}

The last line of~(\ref{SOVxV}) is the contribution from the monopole screening sector ${\bm v} = {\bm 0}$.
The brane configuration for this sector is shown in Figure~\ref{subfig:O6-tHooft-VxV-v=0}, and the SQM quiver in Figure~\ref{subfig:quiverSO-2star}.%
The supersymmetric index is given by the contour integral
\begin{equation}\label{ZSO2nv1-2star}
\begin{aligned}
Z(\bv = {\bm 0}, \zeta) =&
\frac{1}{2}
\oint_{JK(\zeta)}\frac{d\phi_1}{2\pi i}\frac{d\phi_2}{2\pi i}
\frac{
2\sinh(\pm \phi_1)
\left(2\sinh\epsilon_+\right)^22\sinh(\pm \phi_1 + \epsilon_+)
}{
2\sinh(\pm \phi_1+\frac{ \pm m  - \epsilon_+}{2} )
(2\sinh(\frac{m\pm\epsilon_+}{2}))^2
}
 \\
&
\times
\frac{
\left(\prod_{i=1}^n
2\sinh\frac{\pm \phi_1 \pm  a_i + m}{2}
\right)
2\sinh\frac{\pm \phi_1 \pm  \phi_2 + m}{2}
}{
\left(\prod_{i=1}^n
2\sinh\frac{\pm \phi_1 \pm  a_i + \epsilon_+}{2}
\right)
2\sinh\frac{\pm \phi_1 \pm  \phi_2 + \epsilon_+}{2}
}.
\end{aligned}
\end{equation}
for $N=2n$ and 
\begin{equation}\label{ZSO2np1v1-2star}
\begin{aligned}
Z(\bv = {\bm 0}, \zeta)=&
\frac{1}{2}
\oint_{JK(\zeta)}\frac{d\phi_1}{2\pi i}\frac{d\phi_2}{2\pi i}
\frac{
2\sinh(\pm \phi_1)
\left(2\sinh\epsilon_+\right)^22\sinh(\pm \phi_1 + \epsilon_+)
}{
2\sinh(\pm \phi_1+\frac{ \pm m -\epsilon_+}{2} )
(2\sinh(\frac{m\pm\epsilon_+}{2}))^2
}
 \\
&\qquad\times
\frac{
2\sinh\frac{\pm\phi_1+m}{2}
\left(\prod_{i=1}^n
2\sinh\frac{\pm \phi_1 \pm  a_i + m}{2}
\right)
2\sinh\frac{\pm \phi_1 \pm  \phi_2 + m}{2}
}{
2\sinh\frac{\pm\phi_1+\epsilon_+}{2}
\left(\prod_{i=1}^n
2\sinh\frac{\pm \phi_1 \pm  a_i + \epsilon_+}{2}
\right)
2\sinh\frac{\pm \phi_1 \pm  \phi_2 + \epsilon_+}{2}
}
\end{aligned}
\end{equation}
for $N=2n+1$. 
The FI parameter $\zeta$ is associated to the $U(1)$ gauge node in Figure \ref{fig:quiverSO-2star}, and  we evaluate the integral \eqref{ZSO2nv1-2star} and \eqref{ZSO2np1v1-2star} with the JK residue prescription using the JK parameter~$\bm\eta = (0, \zeta)$.
More precisely,  to use the constructive definition of the JK residue, we deform the JK parameter to $\bm\eta = (\delta, \zeta)$ with $|\delta| \ll |\zeta|$. 
We checked that $\Braket{T_V}\ast\Braket{T_V}\Big|_{\bv = {\bm 0}} = Z(\zeta)$ for $N=4,5,...,11$ and $\zeta>0$, as expected.

\section{'t~Hooft operators in $USp(2n)$ gauge theories}
\label{sec:USp}

In this section we consider the expectation values of 't~Hooft operators in $USp(2n)$ gauge theories. In particular we focus on the 4d $\mathcal{N}=2$ $USp(2n)$ gauge theory with $2n+2$ flavors and also the 4d $\mathcal{N}=2^{\ast}$ $USp(2n)$ gauge theory. 
 
\subsection{$\mathcal{N}=2$ SQCD}
\label{sec:USpSQCD}
In this subsection we consider the $USp(2n)$ gauge theory with $2n+2$ hypermultiplets in the fundamental representation. 
The Langlands dual of $USp(2n)$ is $SO(2n+1)$.
The magnetic charge of the minimal 't~Hooft operator~$T_V$ corresponds to the vector (fundamental) representation~$V$ of $SO(2n+1)$.
Unlike in $U(N)$ and $SO(N)$ gauge theories, in a $USp(2n)$ theory even the minimal 't~Hooft operator exhibits monopole screening.
Its expectation value on~$S^1\times\mathbb{R}^3$ takes the form
\begin{align}
\Braket{T_V} = \sum_{i=1}^n\left(e^{b_i} + e^{-b_i}\right)Z_i(\ba) + Z_{\text{mono}}(\ba), \label{TF.USp}
\end{align}
where 
\begin{align}\label{USpTZi}
Z_i(\ba) = \left(\frac{\prod_{f=1}^{2n+2}2\sinh\frac{\pm a_i - m_f}{2}}{2\sinh(\pm a_i)2\sinh(\pm a_i + \epsilon_+)\prod_{1 \leq j \neq i \leq n}2\sinh\frac{\pm a_i \pm  a_j + \epsilon_+}{2}}\right)^{\frac{1}{2}},
\end{align}
and $Z_{\text{mono}}(\ba)$ is the contribution from the monopole screening sector specified by the zero weight~$\bm{v}=0$ in~$V$.

We will determine $Z_{\text{mono}}(\ba)$ using an SQM in Section~\ref{sec:USp.fund}.
We will study the product of two copies of $T_V$ in Section~\ref{sec:USpVxV} using the Moyal product and SQMs.

\subsubsection{$V$}
\label{sec:USp.fund}

\begin{figure}[t]
\centering
\subfigure[]{\label{subfig:O4-USp-tHooft-V-v=0}
\raisebox{4mm}{\includegraphics[scale=.3]{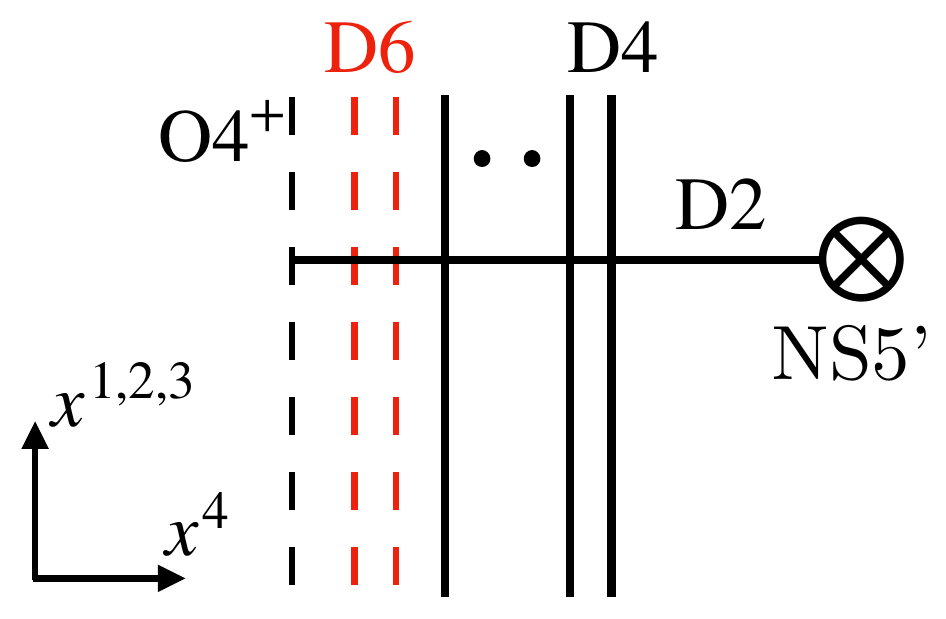}}}
\subfigure[]{\label{subfig:quiverUSp1}
\hspace{-4mm}
\includegraphics[width=8cm]{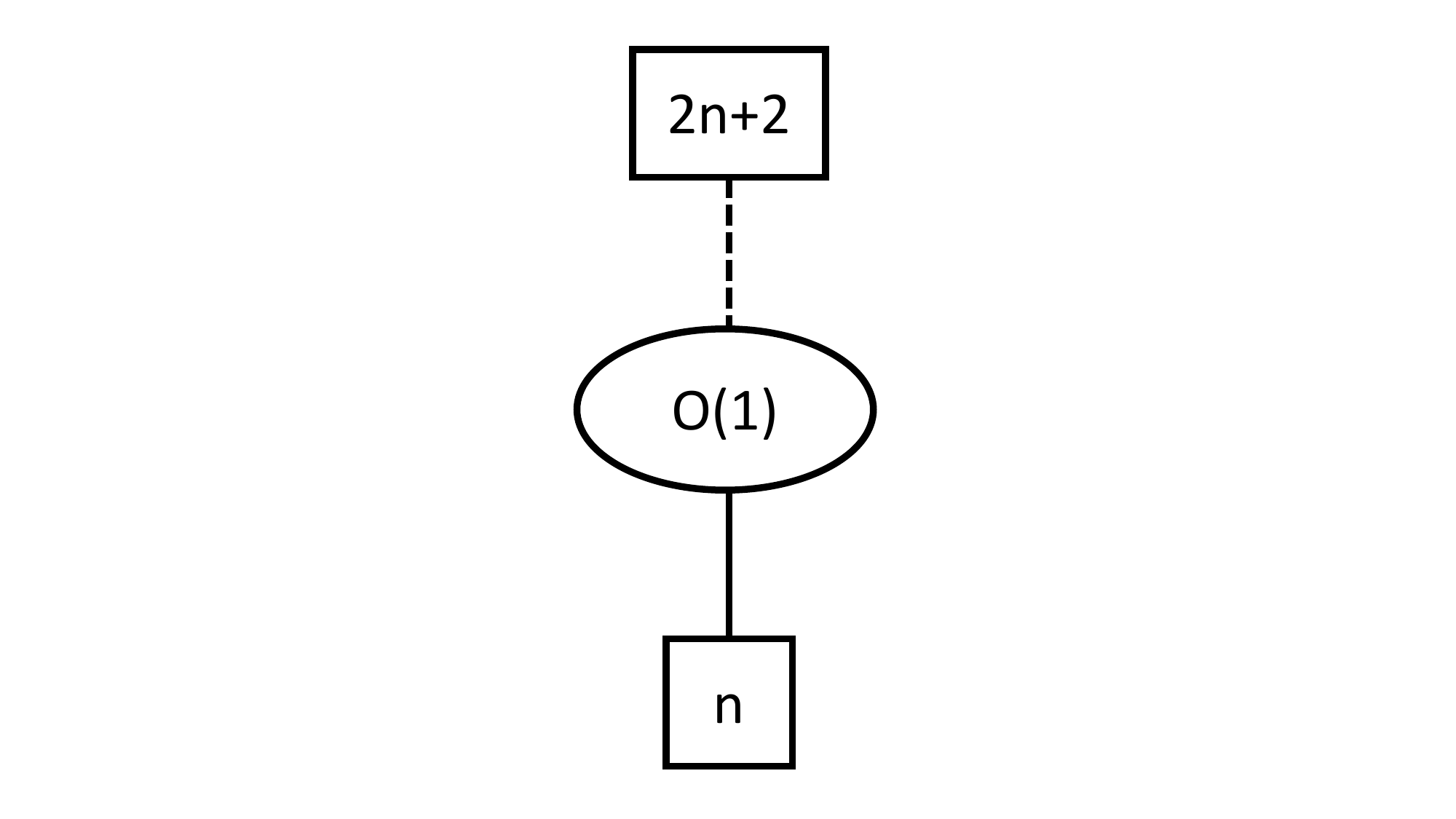}}
\caption{(a): The brane configuration for the monopole screening contribution to the $\bv = {\bm 0}$ sector in $T_V$.
(b): The corresponding quiver diagram.   
}
\label{fig:quiverUSp2_V_v=0}
\end{figure}

Here we compute $Z_{\text{mono}}(\ba)$ in \eqref{TF.USp}. 
The brane configuration for monopole screening  is shown in Figure~\ref{subfig:O4-USp-tHooft-V-v=0}, and the corresponding SQM quiver in Figure~\ref{subfig:quiverUSp1}. 
The non-connected group~$O(1)\simeq\mathbb{Z}_2$ consists of two connected components, $O(1)_+=\{1\}$ and $O(1)_-=\{-1\}$. 
Since these are discrete the supersymmetric index involves no integral.
By averaging the contribution
\begin{align}
Z_{V+}(\bv ={\bm 0})= \frac{\prod_{f=1}^{2n+2}2\sinh\frac{m_f}{2}}{\prod_{i=1}^n2\sinh\frac{\pm a_i + \epsilon_+}{2}}.
\end{align}
 from $O(1)_+$ and the contribution
\begin{align}
Z_{V-}(\bv = {\bm 0})= \frac{\prod_{f=1}^{2n+2}2\cosh\frac{m_f}{2}}{\prod_{i=1}^n2\cosh\frac{\pm a_i + \epsilon_+}{2}}
\end{align}
from $O(1)_-$, we obtain the 
supersymmetric index of the SQM in Figure \ref{subfig:quiverUSp1}
\begin{equation}\label{ZmonoUSp}
\begin{split}
Z_V(\bv = {\bm 0})&= 
\frac{1}{2}(Z_ {V+}(\bv = {\bm 0})  + Z_{V-}(\bv ={\bm 0}))\\
&=\frac{1}{2}\left(\frac{\prod_{f=1}^{2n+2}2\sinh\frac{m_f}{2}}{\prod_{i=1}^n2\sinh\frac{\pm a_i + \epsilon_+}{2}} + \frac{\prod_{f=1}^{2n+2}2\cosh\frac{m_f}{2}}{\prod_{i=1}^n2\cosh\frac{\pm a_i + \epsilon_+}{2}}\right).
\end{split}
\end{equation}
In this case there is no extra term, 
and hence 
$Z_{\text{mono}}(\ba)$ in \eqref{TF.USp} is given by \eqref{ZmonoUSp}.

When $n=1$ the gauge group $USp(2)$ is isomorphic to $SU(2)$. 
The vector representation~$V$ is the same as the adjoint representation of the Langlands dual group $SO(3)$. 
The monopole screening contribution to~$\langle T_V\rangle$ in the $SU(2)$ SQCD was computed in \cite{Ito:2011ea, Brennan:2018rcn, Assel:2019iae}, using the AGT correspondence, the Born-Oppenheimer approximation for the Coulomb branch contribution, and an improvement of the SQM by a completed brane configuration.
Our formula \eqref{ZmonoUSp} with $n=1$ has an expression different from those which appear in~\cite{Ito:2011ea, Brennan:2018rcn, Assel:2019iae}, but is in fact equal to them.
The specialization to $\epsilon_+=0$ of~\eqref{ZmonoUSp} with $n=1$, involving two terms each containing $\sinh$ only or $\cosh$ only, was obtained in Appendix~E of~\cite{Ito:2011ea}, where the correspondence between line operators and the $SL(2,\mathbb{C})$ holonomies on the four-punctured sphere was studied.

\subsubsection{$V \times V$}\label{sec:USpVxV}
Next we turn to the product of two copies of~$T_V$.
The Moyal product is given by
\begin{equation}\label{USp.TFTF}
\begin{split}
\Braket{T_V}\ast\Braket{T_V} = &\sum_{1\leq i, j \leq n}\left(e^{b_i + b_j}Z_i(\ba + \epsilon_+\be_j)Z_j(\ba - \epsilon_+\be_i) + e^{-b_i - b_j}Z_i(\ba - \epsilon_+\be_j)Z_j(\ba + \epsilon_+\be_i)\right)\\
&+\sum_{1\leq i \neq  j \leq n}
\hspace{-2mm}
\left(e^{b_i - b_j}Z_i(\ba - \epsilon_+\be_j)Z_j(\ba - \epsilon_+\be_i) + e^{-b_i + b_j}Z_i(\ba + \epsilon_+\be_j)Z_j(\ba + \epsilon_+\be_i)\right)\\
&+\sum_{i=1}^n\left(e^{b_i} + e^{-b_i}\right)Z_i(\ba)\left(Z_{\text{mono}}(\ba - \epsilon_+\be_i) + Z_{\text{mono}}(\ba + \epsilon_+\be_i) \right)\\
&+\sum_{i=1}^n\left(Z_i(\ba - \epsilon_+\be_i)^2 + Z_i(\ba + \epsilon_+\be_i)^2\right) + Z_{\text{mono}}(\ba)^2.
\end{split}
\end{equation}
By the Weyl group action reviewed in Appendices~\ref{sec:so-odd} and~\ref{app:usp}, the monopole screening contributions are classified into three types:
$\bv = \pm \be_i \pm \be_j$ for $1 \leq i < j \leq n$, $\bv = \pm \be_i$ for $i=1, \cdots, n$ and $\bv = {\bm 0}$. 
We will study them one by one.

\paragraph{$\bv =\pm \be_i \pm \be_j\; (i \neq j)$.}
We focus on the sector $\bv = \be_{n-1} + \be_n$ since the other cases are related by Weyl reflections. 
The brane configuration is the one given in Figure~\ref{subfig:O4-tHooft-VxV-v=eNpluseN-1}, where the O4-plane is taken to be an ${\rm O4}^+$-plane.
The SQM is given by the quiver in Figure~\ref{subfig:quiverSO-VxV-v=eNpluseN-1}.
The supersymmetric index giving the monopole screening contribution is~\eqref{VxVuniversal}. 

We can also read off~$Z_{\text{mono}}(\bv = \be_{n-1} + \be_n; \ba)$~from the Moyal product~\eqref{USp.TFTF} by writing
\begin{equation}\label{moyalUSpTVVvee}
\begin{split}
&\quad \  Z_{n-1}(\ba + \epsilon_+\be_n)Z_n(\ba - \epsilon_+\be_{n-1})  + Z_{n}(\ba + \epsilon_+\be_{n-1})Z_{n-1}(\ba - \epsilon_+\be_n)\cr
&= Z_{(n-1)n}(\ba)Z_{\text{mono}}(\bv = \be_{n-1} + \be_n; \ba),
\end{split}
\end{equation}
where $Z_{(n-1)n}(\ba)$ is the specialization of
\begin{equation}\label{ZijUSp}
\begin{aligned}
Z_{ij}({\bm a}) &= \Bigg(\frac{\prod_{f=1}^{N_F}2\sinh\frac{\pm a_i - m_f}{2}2\sinh\frac{\pm a_j - m_f}{2}}{2\sinh\frac{\pm (a_i + a_j)}{2}2\sinh\frac{\pm (a_i + a_j) + 2\epsilon_+}{2}
{\displaystyle \mathop{\prod_{1 \leq k\leq n}}_{k\neq i, j}}
2\sinh\frac{\pm a_i \pm a_k + \epsilon_+}{2}2\sinh\frac{\pm a_{ j} \pm a_k + \epsilon_+}{2}}\\
&\hspace{2.5cm}\times\frac{1}{2\sinh(\pm a_i) 2\sinh(\pm a_j) 2\sinh(\pm a_i + \epsilon_+) 2\sinh(\pm a_j + \epsilon_+) }\Bigg)^{\frac{1}{2}}
\end{aligned}
\end{equation}
to $(i,j)=(n-1,n)$.
From \eqref{moyalUSpTVVvee} we obtain
\begin{equation}
\begin{aligned}
&\quad Z_{\text{mono}}(\be = \be_{n-1} + \be_n; \ba)  \\
&= \frac{1}{2\sinh\frac{a_{n-1} - a_n}{2}2\sinh\frac{-a_{n-1} + a_n + 2\epsilon_+}{2}} + \frac{1}{2\sinh\frac{a_{n} - a_{n-1}}{2}2\sinh\frac{-a_{n} + a_{n-1} + 2\epsilon_+}{2}}.
\end{aligned}
\end{equation}
The right hand side is the supersymmetric index~(\ref{VxVuniversal}) of the SQM specified by the quiver in Figure~\ref{subfig:quiverSO-VxV-v=eNpluseN-1}, as expected.

\begin{figure}[t]
\centering
\subfigure[]{\label{subfig:O4-USp-tHooft-VxV-v=eN}
\raisebox{7mm}{\includegraphics[width=5cm]{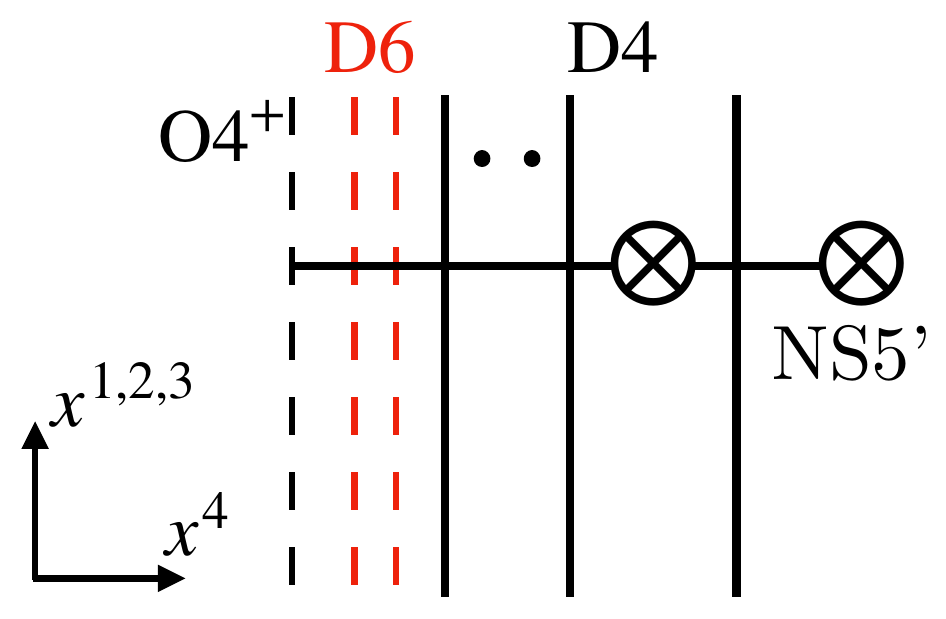}}}
\hspace{2cm}
\subfigure[]{\label{subfig:quiverUSp3}
\raisebox{5mm}{\includegraphics[width=4cm]{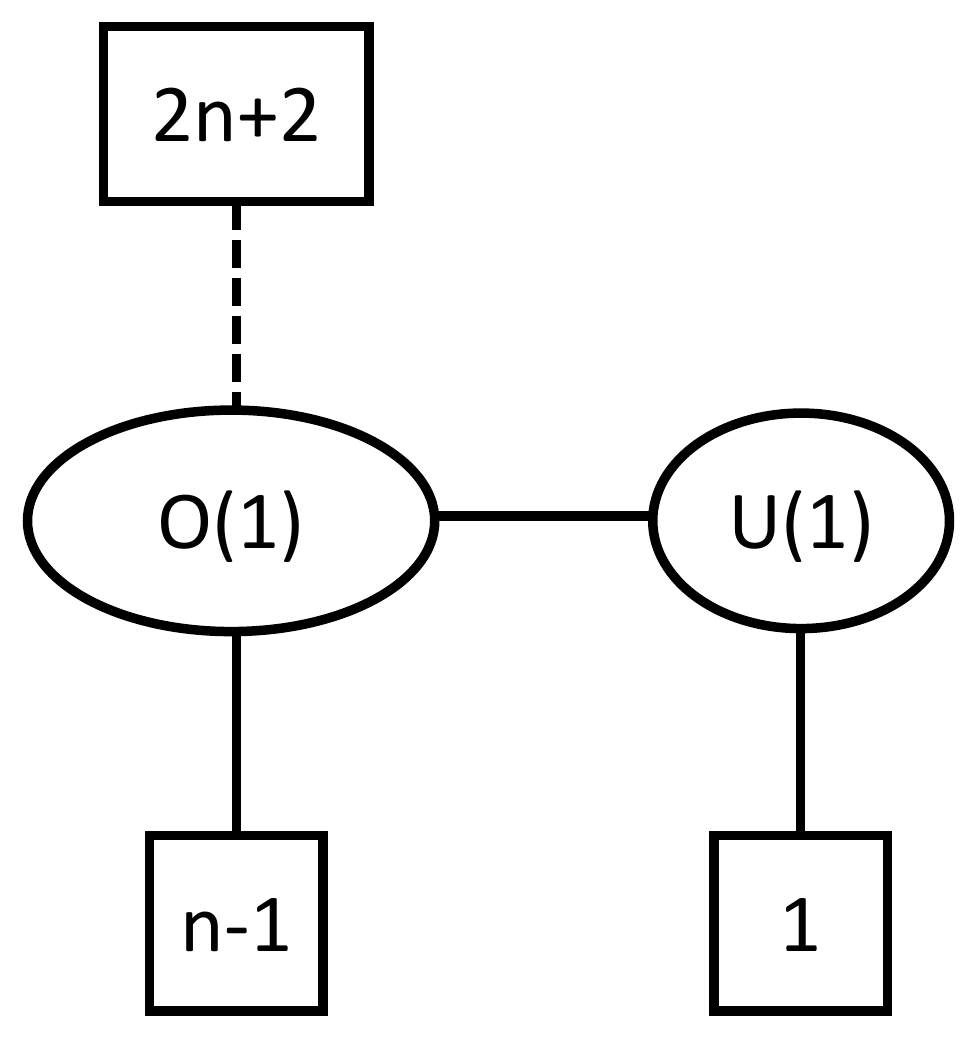}}}
\caption{%
(a): The brane configuration for the monopole screening sector~$\bm{v}=\bm{e}_N$ of $T_V\cdot T_V$.
(b): The corresponding SQM quiver diagram.
}
\label{fig:quiverUSp3}
\end{figure}

\paragraph{$\bv = \pm \be_i$.}
The brane configuration and the quiver diagram for the SQM which describe  monopole screening for the $\bv = \pm \be_i$ sector are depicted in Figure~\ref{fig:quiverUSp3}. 
We again use the $\mathcal{N}=(0, 4)$ notation. 
The supersymmetric index consists of contributions from two sectors, $O(1)_+-U(1)$ and $O(1)_--U(1)$. The contribution from the $O(1)_+-U(1)$ sector is given by
\begin{equation}
\begin{split}
Z_{+}(\bv = \pm \be_i, \zeta) = \oint_{JK(\zeta)}\frac{d\phi}{2\pi i}\frac{2\sinh(\epsilon_+)\prod_{f=1}^{2n+2}2\sinh\frac{m_f}{2}}{2\sinh\frac{\pm \phi + \epsilon_+}{2}2\sinh\frac{\pm (\phi - a_i) + \epsilon_+}{2}\prod_{1 \leq j \neq i \leq n}2\sinh\frac{\pm a_j + \epsilon_+}{2}}.
\end{split}
\end{equation}
On the other hand, the contribution from the $O(1)_--U(1)$ sector is given by
\begin{equation}\label{WI-.USp1}
\begin{split}
Z_{-}(\bv = \pm \be_i, \zeta) = \oint_{JK(\zeta)}\frac{d\phi}{2\pi i}\frac{2\sinh(\epsilon_+)\prod_{f=1}^{2n+2}2\cosh\frac{m_f}{2}}{2\cosh\frac{\pm \phi + \epsilon_+}{2}2\sinh\frac{\pm (\phi - a_i) + \epsilon_+}{2}\prod_{1 \leq j \neq i \leq n}2\cosh\frac{\pm a_j + \epsilon_+}{2}}.
\end{split}
\end{equation}
Note that for the evaluation of the integral \eqref{WI-.USp1}, one needs to include the pole $\phi = -\epsilon_+ + i\pi$ for $\zeta > 0$ and $\phi = \epsilon_+ + i\pi$ for $\zeta < 0$. 
The supersymmetric index 
for the $\bv = \pm \be_i$ sector is the average of the two contributions
\begin{align}\label{WI.USp1}
Z(\bv = \pm \be_i, \zeta) = \frac{1}{2}(Z_{+}(\bv = \pm \be_i, \zeta) + Z_{-}(\bv = \pm \be_i, \zeta)).
\end{align}

The evaluation of the JK residue for the supersymmetric index gives the relation
\begin{align}\label{WIMoyalUSp1}
Z(\bv = \pm \be_i, \zeta) = Z_{\text{mono}}(\ba - \epsilon_+\be_i) + Z_{\text{mono}}(\ba + \epsilon_+\be_i)
\end{align}
for $\zeta > 0$ and $\zeta < 0$.
The right hand side of~(\ref{WIMoyalUSp1}) appears in the third line of~(\ref{USp.TFTF}).

\paragraph{$\bv = {\bm 0}$.}

\begin{figure}[t]
\centering
\hspace{10mm}
\subfigure[]{\label{subfig:O4-USp-tHooft-VxV-v=0}
\raisebox{4mm}{\includegraphics[scale=.3]{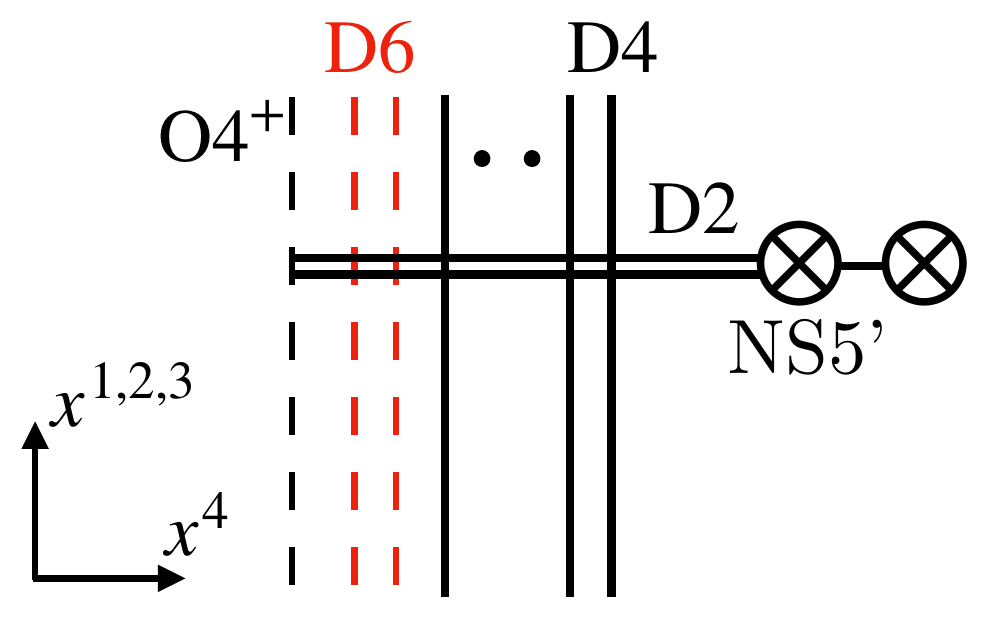}}}
\subfigure[]{\label{subfig:quiverUSp2_VxV_v=0}
\hspace{3mm}
\includegraphics[width=8cm]{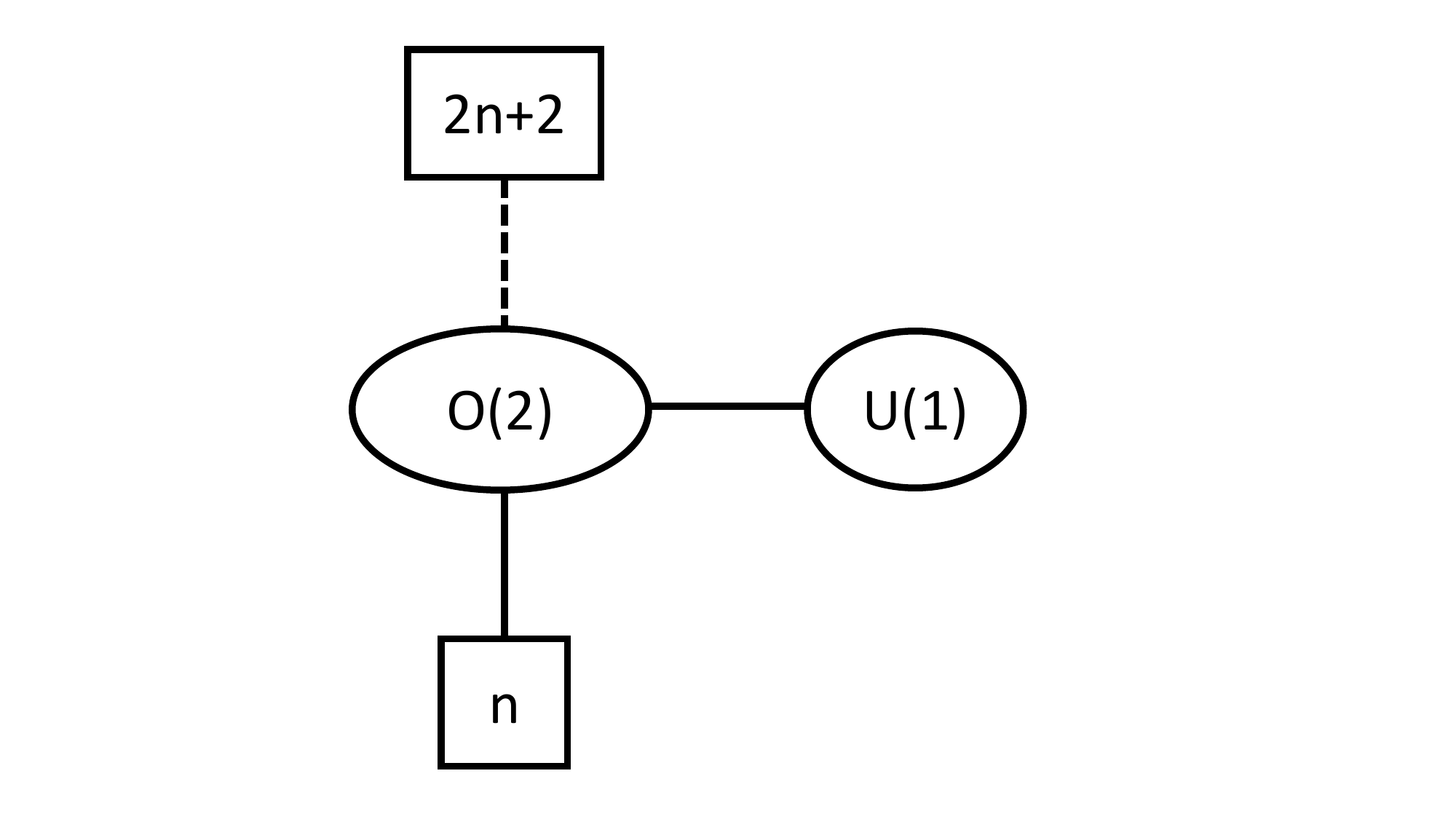}}
\caption{(a): The brane configuration for the monopole screening contribution to the $\bv = {\bm 0}$ sector in $T_V \cdot T_V$.
(b): The corresponding quiver diagram. 
}
\label{fig:quiverUSp2_VxV_v=0}
\end{figure}

The brane configuration for the $\bv = {\bm 0}$ sector and the corresponding quiver diagram of the SQM are depicted in Figures~\ref{subfig:O4-USp-tHooft-VxV-v=0} and~\ref{subfig:quiverUSp2_VxV_v=0}, respectively.  
The supersymmetric index again consists of two contributions, for~$O(2)_+-U(1)$ and $O(2)_--U(1)$. The contribution from the $O(2)_+-U(1)$ sector is given by
\begin{equation}
\begin{split}
Z_+(\bv = 0, \zeta) = \oint\frac{d\phi_1}{2\pi i}\frac{d\phi_2}{2\pi i}\frac{\left(2\sinh(\epsilon_+)\right)^2\prod_{f=1}^{2n+2}2\sinh\frac{\pm\phi_1 - m_f}{2}}{2\sinh\frac{\pm\phi_1 \pm \phi_2 + \epsilon_+}{2}\prod_{i=1}^n2\sinh\frac{\pm\phi_1 \pm a_i + \epsilon_+}{2}}.
\end{split}
\end{equation}
The contribution from the $O(2)_--U(1)$ sector is 
\begin{equation}
\begin{split}
Z_-(\bv = 0, \zeta) = \oint\frac{d\phi_2}{2\pi i}\frac{2\cos(\epsilon_+)2\sinh(\epsilon_+)\prod_{f=1}^{2n+2}2\sinh(m_f)}{2\sinh(\pm \phi_2 + \epsilon_+)\prod_{i=1}^n2\sinh(\pm a_i + \epsilon_+)}.
\end{split}
\end{equation}
The supersymmetric index is then given by
\begin{align}\label{WI.USp2}
Z(\bv = 0, \zeta) = \frac{1}{2}(Z_+(\bv = 0, \zeta) + Z_-(\bv = 0, \zeta)).
\end{align}

Computing the integral \eqref{WI.USp2} using the JK residue prescription, and comparing it with~(\ref{USpTZi})-(\ref{ZmonoUSp}), we obtain the relation,  for $n=1, 2$,
\begin{align}\label{WIMoyalUSp2}
Z(\bv = {\bm 0}, \zeta) = \sum_{i=1}^n\left(Z_i(\ba - \epsilon_+\be_i)^2 + Z_i(\ba + \epsilon_+\be_i)^2\right) + Z_{\text{mono}}(\ba)^2
\end{align}
for the both signs of $\zeta$.
In particular $\frac{1}{4}Z_{\text{mono}+}^2$ and $\frac{1}{4}Z_{\text{mono}-}^2$ come from the residues in $\frac{1}{2}Z_+(\bv = {\bm 0}, \zeta)$ and $\frac{1}{2}Z_{\text{mono}+} Z_{\text{mono}-}$ is equal to $\frac{1}{2}Z_-(\bv = {\bm 0}, \zeta)$.  
We conjecture that~(\ref{WIMoyalUSp2}) holds for general $n$, for the both signs of $\zeta$.

\subsection{$\mathcal{N}=2^{\ast}$ theory}
\label{sec:N2starUSp}

We now switch to the $\mathcal{N}=2^{\ast}$ $USp(2n)$ gauge theory, which is the $\mathcal{N}=2$ $USp(2n)$ gauge theory with a hypermultiplet in the adjoint representation. 
The expectation value of the minimal 't~Hooft operator in the $\mathcal{N}=2^{\ast}$ theory takes the same form as \eqref{TF.USp}, namely
\begin{equation}
\Braket{T_V} = \sum_{i=1}^n\left(e^{b_i} + e^{-b_i}\right)Z_i(\ba) + Z_{\text{mono}}(\ba), \label{TF.N2starUSp}
\end{equation}
but the functions $Z_i({\bm a})$ and $Z_{\text{mono}}({\bm a})$ are different. 
The contribution $Z_i({\bm a})$ is determined by the general formulas~(\ref{1loopvm}) and~(\ref{1loophm}) for the one-loop determinants as
\begin{align}\label{starUSpTZi}
Z_i(\ba) = \left(\frac{2\sinh\frac{\pm 2a_i - m \pm \epsilon_+}{2}\prod_{1\leq j\neq i \leq n}2\sinh\frac{\pm a_i \pm a_j - m}{2}}{2\sinh(\pm a_i)2\sinh(\pm a_i + \epsilon_+)\prod_{1 \leq j \neq i \leq n}2\sinh\frac{\pm a_i \pm  a_j + \epsilon_+}{2}}\right)^{\frac{1}{2}}.
\end{align}
We will compute the monopole screening contribution~$Z_{\text{mono}}({\bm a})$ from the supersymmetric index of an SQM in Section~\ref{sec:USp-Nstar-V}.
For $m = \epsilon_+ = 0$ we expect from S-duality that~$\Braket{T_V}$ becomes equal to the vev of the Wilson operator in representation~$V$ of the $\mathcal{N}=4$ $SO(2n+1)$ gauge theory.
This vev is simply the character of the representation~\cite{Ito:2011ea}.
Thus we expect the equation
\begin{align}
\Braket{T_V} = \sum_{i=1}^n\left(e^{b_i} + e^{-b_i}\right) + 1 \label{N4TVUSp}
\end{align}
to hold.
In Section~\ref{sec:USp-Nstar-V} we will choose by hand the overall sign of the contour integral for~$Z_{\text{mono}}({\bm a})$ so that~(\ref{N4TVUSp}) holds.

We also consider the vev of the 't~Hooft operator $T_{\wedge^2 V}$, which takes the form
\begin{equation}\label{Twedge2V.N2starUSp}
\begin{split}
\Braket{T_{\wedge^2V}} &= \sum_{1\leq i < j \leq n}\left(e^{b_i + b_j} + e^{-b_i - b_j}\right)Z_{ij}({\bm a}) + \sum_{1\leq i < j \leq n}\left(e^{b_i - b_j} + e^{-b_i + b_j}\right)Z'_{ij}({\bm a}) \cr
&\hspace{0.5cm} + \sum_{i=1}^n\left(e^{b_i} + e^{-b_i}\right)Z_i({\bm a})Z'_{\text{mono}, i}({\bm a}) + Z''_{\text{mono}}({\bm a}).
\end{split}
\end{equation}
The one-loop determinants are given by the general formulas~(\ref{1loopvm}) and~(\ref{1loophm}) as
\begin{equation}\label{Zij.N2starUSp}
\begin{split}
Z_{ij}(\ba)=& \left(\frac{2\sinh\frac{\pm(a_i + a_j) - m \pm \epsilon_+}{2}2\sinh\frac{\pm 2a_i - m \pm \epsilon_+}{2}2\sinh\frac{\pm 2a_j - m \pm \epsilon_+}{2}}{2\sinh(\pm a_i)2\sinh(\pm a_j)2\sinh(\pm a_i + \epsilon_+)2\sinh(\pm a_j + \epsilon_+)}\right.\cr
&
\times\left.\frac{\prod_{1\leq k\neq i,j \leq n}2\sinh\frac{\pm a_i \pm a_k - m}{2}2\sinh\frac{\pm a_j \pm a_k - m}{2}}{2\sinh\frac{\pm(a_i + a_j)}{2}2\sinh\frac{\pm (a_i + a_j) + 2\epsilon_+}{2}\prod_{1 \leq k \neq i, j \leq n}2\sinh\frac{\pm a_i \pm  a_k + \epsilon_+}{2}2\sinh\frac{\pm a_j \pm  a_k + \epsilon_+}{2}}\right)^{\frac{1}{2}},
\end{split}
\end{equation}
\begin{equation}\label{Zpij.N2starUSp}
\begin{split}
Z'_{ij}(\ba)=& \left(\frac{2\sinh\frac{\pm(a_i - a_j) - m \pm \epsilon_+}{2}2\sinh\frac{\pm 2a_i - m \pm \epsilon_+}{2}2\sinh\frac{\pm 2a_j - m \pm \epsilon_+}{2}}{2\sinh(\pm a_i)2\sinh(\pm a_j)2\sinh(\pm a_i + \epsilon_+)2\sinh(\pm a_j + \epsilon_+)}\right.\cr
&\times\left.\frac{\prod_{1\leq k\neq i,j \leq n}2\sinh\frac{\pm a_i \pm a_k - m}{2}2\sinh\frac{\pm a_j \pm a_k - m}{2}}{2\sinh\frac{\pm(a_i  - a_j)}{2}2\sinh\frac{\pm (a_i - a_j) + 2\epsilon_+}{2}\prod_{1 \leq k \neq i, j \leq n}2\sinh\frac{\pm a_i \pm  a_k + \epsilon_+}{2}2\sinh\frac{\pm a_j \pm  a_k + \epsilon_+}{2}}\right)^{\frac{1}{2}},
\end{split}
\end{equation}
and \eqref{starUSpTZi}.
The quantities~$Z'_{\text{mono}, i}({\bm a})$ and~$Z''_{\text{mono}, i}({\bm a})$ are the monopole screening contributions, which we will compute using SQMs in Section~\ref{sec:USp-2star-wedge2v}.
For $m = \epsilon_+ = 0$, from S-duality, we expect the vev~(\ref{sec:USp-2star-wedge2v}) to reduce to the character of the adjoint representation of $SO(2n+1)$.
Namely we expect the relation
\begin{align}
\Braket{T_{\wedge^2V}} = \sum_{1 \leq i <  j \leq n }\left(e^{b_i + b_j} + e^{b_i - b_j} + e^{-b_i + b_j} + e^{-b_i-b_i}\right) + \sum_{i=1}^n\left(e^{b_i} + e^{-b_i}\right) + n. \label{N4Twedge2VUSp}
\end{align} 
to hold.

\subsubsection{$V$}\label{sec:USp-Nstar-V}

\begin{figure}[t]
\centering
\subfigure[]{\label{subfig:O6-USp-tHooft-V-v=0}
\raisebox{3mm}{\includegraphics[scale=.3]{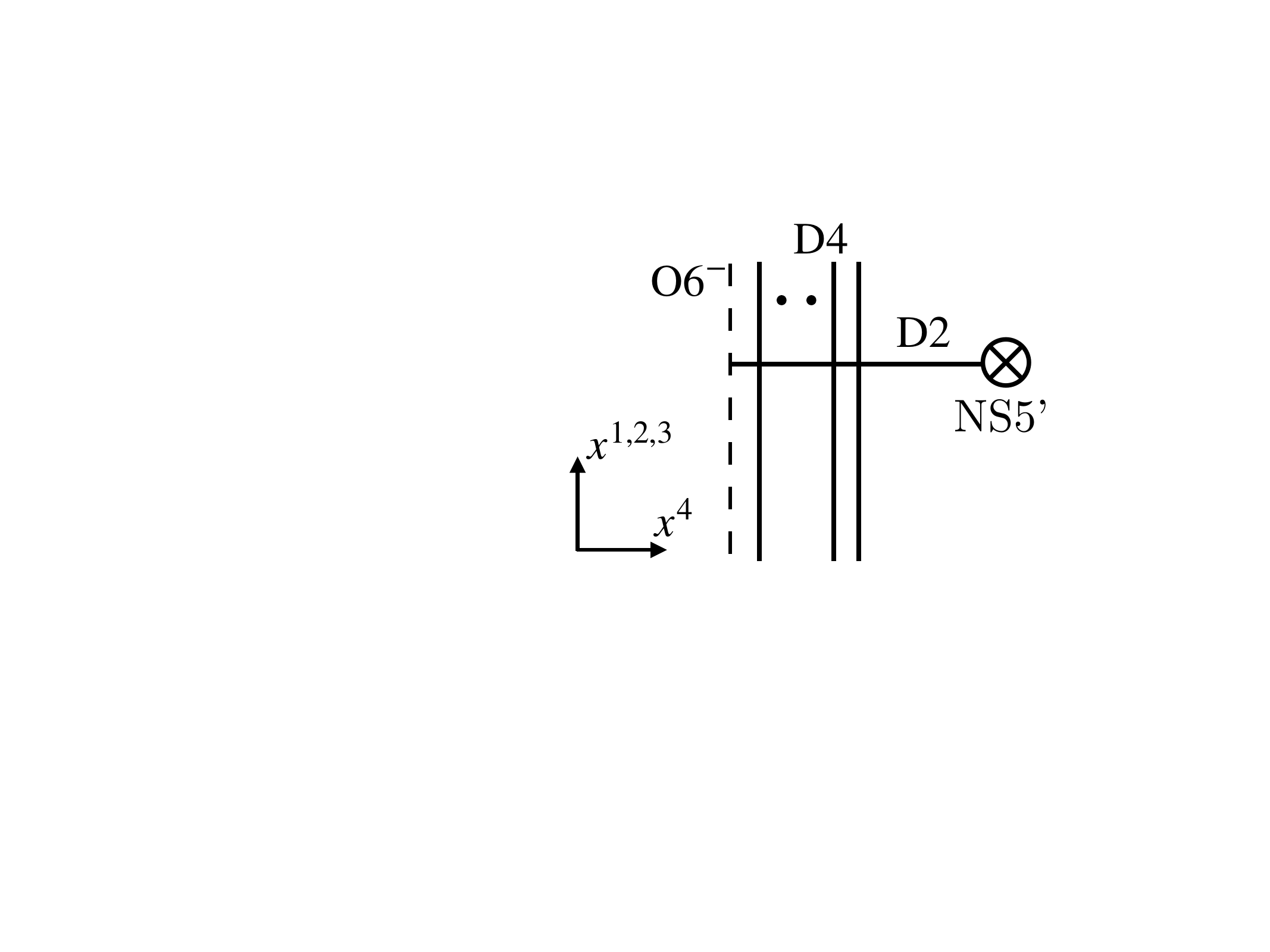}}}
\hspace{2cm}
\subfigure[]{\label{subfig:quiverUSp9}
\hspace{-4.5mm}
\includegraphics[width=2cm]{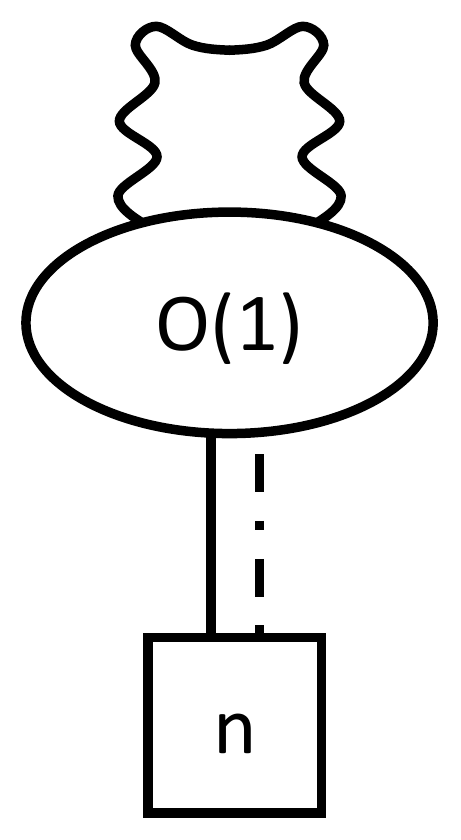}}
\caption{(a): The brane configuration for the monopole screening sector $\bm{v}=0$ of~$T_V$ in the $\mathcal{N}=2^{\ast}$ $USp(2n)$ gauge theory.
(b): The corresponding quiver diagram.
}
\label{fig:quiverUSp9}
\end{figure}

We first compute the contribution $Z_{\text{mono}}({\bm a})$ in~(\ref{TF.N2starUSp}). 
The brane configuration and the SQM quiver for the $\bm{v}=\bm{0}$ sector are shown in Figures~\ref{subfig:O6-USp-tHooft-V-v=0} and~\ref{subfig:quiverUSp9}, respectively.
The supersymmetric index consists of two contributions, corresponding to the two components $O(1)_+$ and $O(1)_-$ of $O(1)=\{\pm 1\}$. 
These contributions are
\begin{align}
Z_{V+}(\bv ={\bm 0})
 = \prod_{i=1}^n\frac{2\sinh\frac{\pm a_i + m}{2}}{2\sinh\frac{\pm a_i + \epsilon_+}{2}},
\qquad
Z_{V-}(\bv = {\bm 0})
 = \prod_{i=1}^n\frac{2\cosh\frac{\pm a_i + m}{2}}{2\cosh\frac{\pm a_i + \epsilon_+}{2}}.
\end{align}
The supersymmetric index is then 
\begin{align}\label{ZmonostarUSpV}
Z_V(\bv ={\bm 0}) &= \frac{1}{2}(Z_{V+}(\bv ={\bm 0}) + Z_{V-}(\bv = {\bm 0}))\cr
&=\frac{1}{2}\left(\prod_{i=1}^n\frac{2\sinh\frac{\pm a_i + m}{2}}{2\sinh\frac{\pm a_i + \epsilon_+}{2}} + \prod_{i=1}^n\frac{2\cosh\frac{\pm a_i + m}{2}}{2\cosh\frac{\pm a_i + \epsilon_+}{2}}\right).
\end{align}
We chose the overall sign by hand so that~(\ref{N4TVUSp}) holds when we set $m = \epsilon_+ = 0$. In this case also there is no extra term,
and hence 
$Z_{\text{mono}}(\ba)$ in \eqref{TF.N2starUSp} is given by \eqref{ZmonostarUSpV}.

When $n=1$, the monopole screening contribution should be equal to the one which arises in $\Braket{T_\text{adj}}$ of the $\mathcal{N}=2^{\ast}$ $SU(2)$ gauge theory. The monopole screening contribution $Z_{\text{mono}}^{SU(2)}$ is given in \eqref{su2adjmono}, and \eqref{su2adjmono} as a function equals  \eqref{ZmonostarUSpV} with $2a_1 = a$ specialized to~$n=1$ due to trigonometric identities, even though they have different expressions.

\paragraph{$SO(5)$\text{ versus }$USp(4)$.} We now compare $\Braket{T_{\wedge^2 V}^{SO(5)}}$ of the $\mathcal{N}=2^{\ast}$ $SO(5)$ gauge theory and $\Braket{T_V^{USp(4)}}$ of the $\mathcal{N}=2^{\ast}$ $USp(4)$ gauge theory. 
We expect that they are non-trivially related because of the isomorphism of the Lie algebras $so(5) \simeq usp(4)$.

The 1-loop determinants in $\Braket{T_{\wedge^2 V}^{SO(5)}}$ are the specializations of~\eqref{N2starSOoddZij} and~\eqref{N2starSOoddZijp} to $n=2$:
\begin{align}
Z_{12}^{SO(5)}(a_1, a_2) &= \left(\frac{2\sinh\frac{\pm(a_1 + a_2) - m \pm \epsilon_+}{2}2\sinh\frac{\pm a_1 + \epsilon_+}{2}2\sinh\frac{\pm a_2 - m}{2}}{2\sinh\frac{\pm(a_1 + a_2)}{2}2\sinh\frac{\pm(a_1 + a_2) + 2\epsilon_+}{2}2\sinh\frac{\pm a_1 + \epsilon_+}{2}2\sinh\frac{\pm a_2 + \epsilon_+}{2}}\right)^{\frac{1}{2}},\\
Z'^{SO(5)}_{12}(a_1, a_2) &= \left(\frac{2\sinh\frac{\pm(a_1  - a_2) - m \pm \epsilon_+}{2}2\sinh\frac{\pm a_1 + \epsilon_+}{2}2\sinh\frac{\pm a_2 - m}{2}}{2\sinh\frac{\pm(a_1 - a_2)}{2}2\sinh\frac{\pm(a_1 -  a_2) + 2\epsilon_+}{2}2\sinh\frac{\pm a_1 + \epsilon_+}{2}2\sinh\frac{\pm a_2 + \epsilon_+}{2}}\right)^{\frac{1}{2}}.
\end{align}
On the other hand the 1-loop determinants in $\Braket{T_{V}^{USp(4)}}$ can be obtained from \eqref{starUSpTZi} as
\begin{align}
Z_1^{USp(4)}(\tilde{a}_1, \tilde{a}_2) &= \left(\frac{2\sinh\frac{\pm 2\tilde{a}_1 - m \pm \epsilon_+}{2}2\sinh\frac{\pm \tilde{a}_1 \pm \tilde{a}_2 - m}{2}}{2\sinh(\pm\tilde{a}_1)2\sinh(\pm \tilde{a}_1 + \epsilon_+)2\sinh\frac{\pm \tilde{a}_1 \pm  \tilde{a}_2 + \epsilon_+}{2}}\right)^{\frac{1}{2}},
\\
Z_2^{USp(4)}(\tilde{a}_1, \tilde{a}_2) &= \left(\frac{2\sinh\frac{\pm 2\tilde{a}_2 - m \pm \epsilon_+}{2}2\sinh\frac{\pm \tilde{a}_1 \pm  \tilde{a}_2 - m}{2}}{2\sinh(\pm \tilde{a}_1)2\sinh(\pm  \tilde{a}_2 + \epsilon_+)2\sinh\frac{\pm \tilde{a}_1 \pm  \tilde{a}_2 + \epsilon_+}{2}}\right)^{\frac{1}{2}},
\end{align}
where $\tilde{a}_1, \tilde{a}_2$ are the Coulomb branch moduli of $USp(4)$.  We find the relations among the one-loop determinants
\begin{align}\label{SO5USp41loop}
Z_{12}^{SO(5)}(\tilde{a}_1 + \tilde{a}_2, \tilde{a}_1 - \tilde{a}_2) = Z_1^{USp(4)}(\tilde{a}_1, \tilde{a}_2), \quad Z'^{SO(5)}_{12}(\tilde{a}_1 + \tilde{a}_2, \tilde{a}_1 - \tilde{a}_2) = Z_2^{USp(4)}(\tilde{a}_1, \tilde{a}_2).
\end{align}

Let us now compare the monopole screening contributions to $\Braket{T_{\wedge^2V}^{SO(5)}}$ and $\Braket{T_V^{USp(4)}}$. 
The monopole screening contribution to $\Braket{T_{\wedge^2V}^{SO(5)}}$ is given by the contour integral~\eqref{ZSO2np1wedge2V-2star} specialized to~$n=2$. 
The reducible representation $\wedge^2V$ of $usp(4)$ contains a non-trivial irreducible representation whose Dynkin label is $[0,1]$.
The monopole screening contribution to the vev of the 't~Hooft operator~$\Braket{T_{[0,1]}^{SO(5)}}$ that corresponds to the irreducible representation
is obtained by subtracting an extra term
from $ Z_{\wedge^2V}^{SO(5)}(\bv = {\bm 0}) =
$\eqref{ZSO2np1wedge2V-2star}. 
Since we consider the irreducible representation corresponding to the Dynkin label [0, 1] instead of the reducible representation $\wedge^2 V$, we add $1$ to \eqref{extraSOodd} and consider the extra term 
\begin{align}
Z_{\text{extra}}^{\Braket{T^{SO(5)}_{[0,1]}}}&=\eqref{extraSOodd} + 1  = \frac{2\cosh(m) + \cosh(\epsilon_+)}{2\cosh\frac{m \pm \epsilon_+}{2}}.\label{extraSO5}
\end{align}
Then the monopole contribution in the expectation value of $\Braket{T_{[0,1]}^{SO(5)}}$ is given by
\begin{equation}
Z_{\text{mono}}^{\Braket{T^{SO(5)}_{[0,1]}}} = 
Z_{\wedge^2V}^{SO(5)}(\bv = {\bm 0}) - Z_{\text{extra}}^{\Braket{T^{SO(5)}_{[0,1]}}}.
\end{equation}
On the other hand, the monopole screening contribution to $\Braket{T_V^{USp(4)}}$ is \eqref{ZmonostarUSpV} specialized to~$n=2$.  
Indeed we have the following relation for monopole screening contributions
\begin{equation}\label{SO5USp4mono}
Z_{\text{mono}}^{\Braket{T_{[0,1]}^{SO(5)}}}\Big|_{\substack{
a_1\rightarrow \tilde{a}_1+\tilde{a}_2 \\
a_2\rightarrow \tilde{a}_1 - \tilde{a}_2}} 
= Z_{\text{mono}}^{\Braket{T_V^{USp(4)}}}(\tilde{a}_1, \tilde{a}_2).
\end{equation}

The results \eqref{SO5USp41loop} and \eqref{SO5USp4mono} give the relation between the vevs
\begin{equation}
\Braket{T_{[0,1]}^{SO(5)}}\Big|_{\substack{(a_1, a_2) \to (\tilde{a}_1 + \tilde{a}_2, \tilde{a}_1 - \tilde{a}_2)\\
(b_1+b_2, b_1 - b_2) \to (\tilde{b}_1, \tilde{b}_2)}} = \Braket{T_V^{USp(4)}}.
\end{equation} 
The relations between the $SO(5)$ and $USp(4)$ parameters are the opposite for the electric ($a_i$ and $\tilde{a}_i$) and magnetic ($b_i$ and $\tilde{b}_i$) parameters because the simple roots and coroots of $SO(5)$ are the simple coroots and roots of $USp(4)$, respectively.

\subsubsection{$\wedge^2 V$}\label{sec:USp-2star-wedge2v}

\begin{figure}[t]
\centering
\subfigure[]{\label{subfig:O6-USp-tHooft-wedge2V-v=eN}
\includegraphics[width=5cm]{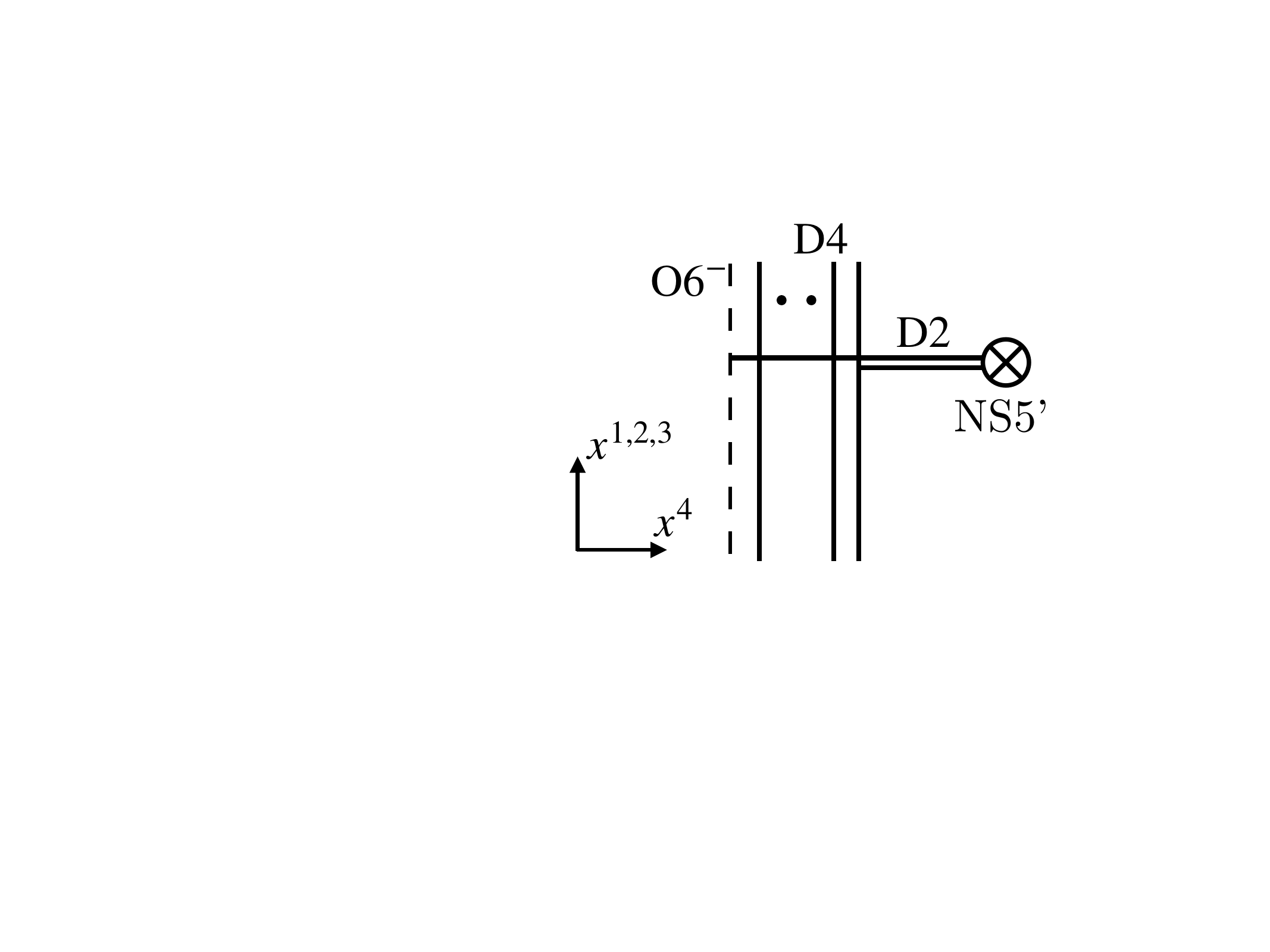}}
\hspace{2.5cm}
\subfigure[]{
\hspace{-2.7mm}
\includegraphics[width=2cm]{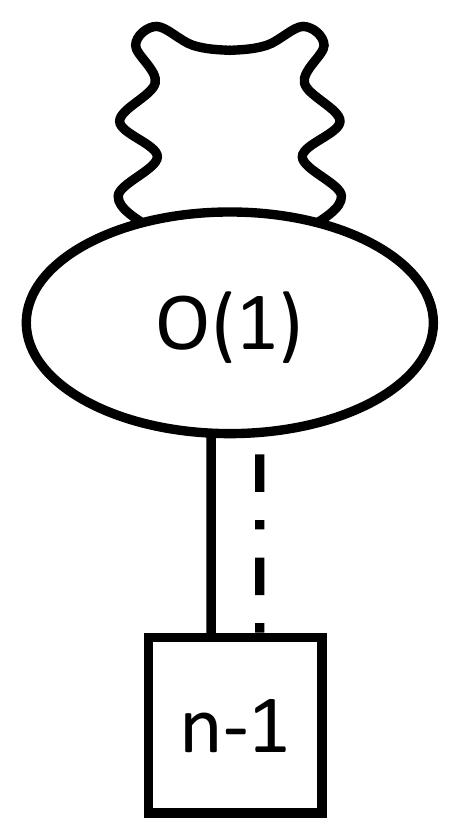}}
\caption{\label{fig:O6-USp-tHooft-wedge2V-v=eN}
(a). The brane configuration that describes the monopole screening sector $\bm{v}=\be_N$ and the contribution $Z'_{\text{mono},N}({\bm a})$ in $\Braket{T_{\wedge^2V}}$. (b). The corresponding quiver diagram.
}
\label{fig:quiverUSp_2star_v2}
\end{figure}

The expectation value $\Braket{T_{\wedge^2 V}}$ of \eqref{Twedge2V.N2starUSp} contains two types of  monopole screening contribution, $Z'_{\text{mono}, i}(\ba)$ and $Z''_{\text{mono}}(\ba)$. 

We first consider $Z'_{\text{mono}, i}(\ba)$. The brane configuration and the SQM quiver for $Z'_{\text{mono}, i}(\ba)$ are given in Figure \ref{fig:O6-USp-tHooft-wedge2V-v=eN}. The SQM quiver is the same as  the one in Figure \ref{subfig:quiverUSp9} except that the number of the hypermultiplets is $n-1$ in this case. Utilizing the result \eqref{ZmonostarUSpV} the supersymmetric index of the SQM becomes 
\begin{align}\label{ZmonostarUSp2V1}
Z'_{\wedge^2V, i}(\bv ={\bm e}_i)= \frac{1}{2}\left(\prod_{1\leq j(\neq i) \leq n}\frac{2\sinh\frac{\pm a_j + m}{2}}{2\sinh\frac{\pm a_j + \epsilon_+}{2}} + \prod_{1\leq j(\neq i) \leq n}\frac{2\cosh\frac{\pm a_j + m}{2}}{2\cosh\frac{\pm a_j + \epsilon_+}{2}}\right).
\end{align}
In this case there is no extra term, 
and 
$Z'_{\text{mono},i}(\ba)$ in \eqref{Twedge2V.N2starUSp} is given by \eqref{ZmonostarUSp2V1}.

\begin{figure}[t]
\centering
\subfigure[]{\label{subfig:O6-USp-tHooft-wedge2V-v=0}
\includegraphics[width=5cm]{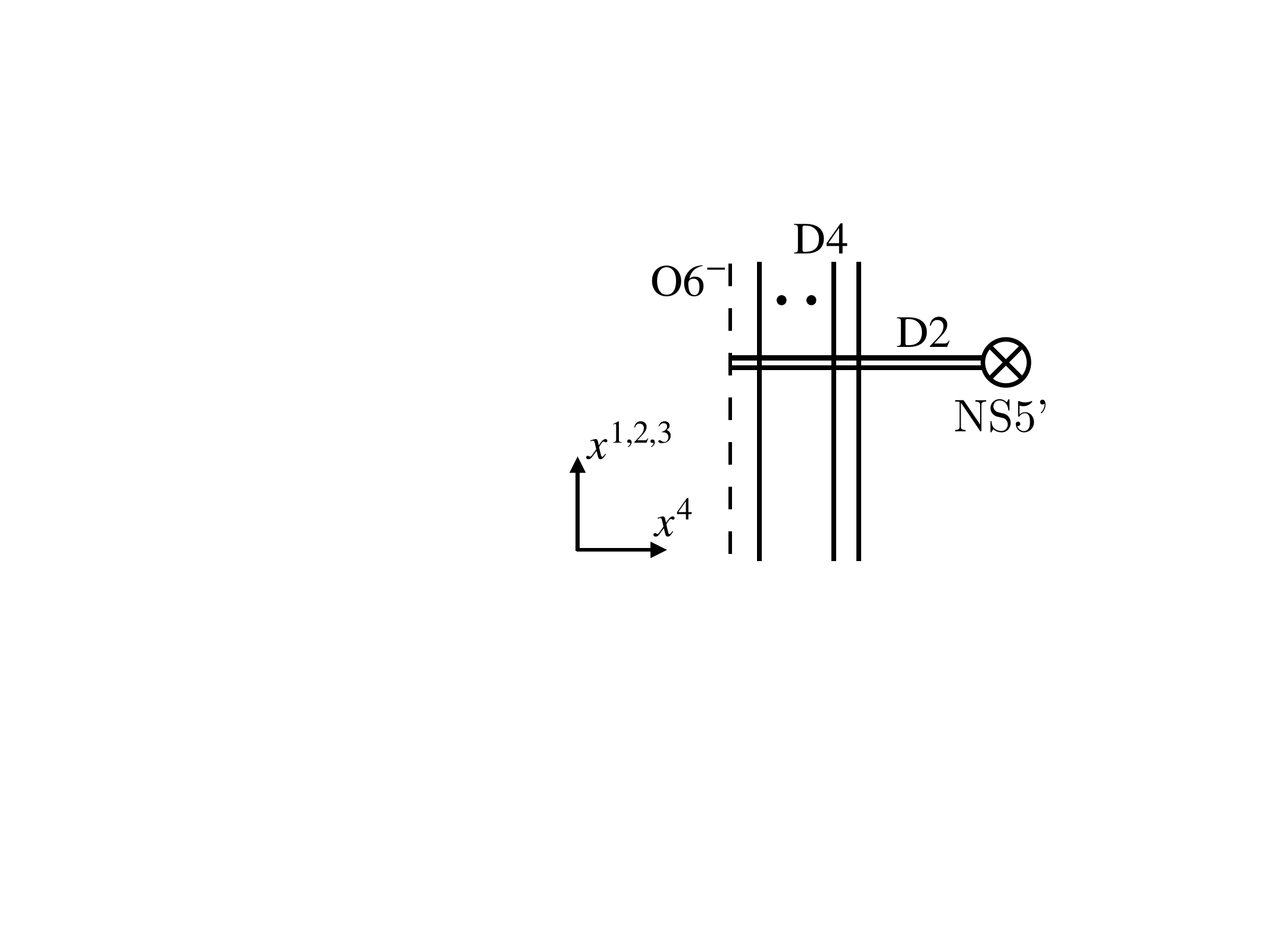}}
\hspace{2.5cm}
\subfigure[]{\label{subfig:quiverUSp-2star-O2}
\hspace{-2.7mm}
\includegraphics[width=2cm]{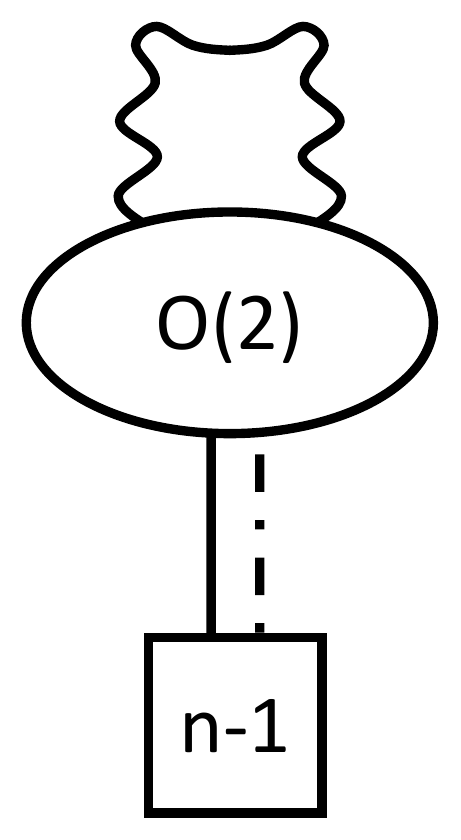}}
\caption{(a): The brane configuration for the monopole screening sector~$\bm{v}=\bm{0}$ of the 't~Hooft operator~$T_{\wedge^2V}$.
(b): The corresponding SQM quiver, which computes~$Z''_{\text{mono}}$.
}
\label{fig:quiverUSp2V2}
\end{figure}
For the other contribution $Z''_{\text{mono}}(\ba)$, the brane configuration and the SQM quiver are depicted in Figure \ref{fig:quiverUSp2V2}. The monopole screening contribution $Z''_{\text{mono}}(\ba)$ is given by the supersymmetric index of the SQM up to an extra term. 
The gauge group of the SQM is the non-connected group~$O(2)$ which has two connected components.
The first component $O(2)_+$ consists of elements with unit determinant, while the elements of the other component $O(2)_-$ have determinants equal to~$-1$.
The supersymmetric index receives contributions from the two components.
The contribution from $O(2)_+$  is 
\begin{equation}
Z''_{\wedge^2 V+}(\bv = {\bm 0})
 = \oint_{JK(\eta)}\frac{d\phi}{2\pi i} \frac{2\sinh(\epsilon_+)}{2\sinh\frac{\pm m - \epsilon_+}{2}}\frac{\prod_{i=1}^n 2\sinh\frac{\pm \phi \pm a_i + m}{2}}{\prod_{i=1}^n2\sinh\frac{\pm \phi \pm a_i + \epsilon_+}{2}},
\end{equation}
and that from $O(2)_-$ is
\begin{equation}
Z''_{\wedge^2 V-}(\bv = {\bm 0})= \frac{2\cosh(\epsilon_+)}{2\cosh\frac{\pm m - \epsilon_+}{2}}\frac{\prod_{i=1}^n 2\sinh\left(\pm a_i + m\right)}{\prod_{i=1}^n2\sinh\left(\pm a_i + \epsilon_+\right)}.
\end{equation}
The contribution from the $O(2)_+$ sector is given by a contour integral and the JK residue computation yields
\begin{equation}
\begin{split}
Z''_{\wedge^2 V+}(\bv = {\bm 0})
= &\sum_{i=1}^n\left(\frac{2\sinh\frac{\pm(2a_i-\epsilon_+)+m}{2}}{2\sinh(a_i)2\sinh(-a_i + \epsilon_+)}\prod_{j \neq i}\frac{2\sinh\frac{\pm(a_i + a_j - \epsilon_+) + m}{2}2\sinh\frac{\pm (a_i - a_j - \epsilon_+) + m}{2}}{2\sinh\frac{a_i \pm a_j}{2}2\sinh\frac{-a_i \pm a_j + 2\epsilon_+}{2}}\right.\cr
&\left.+\frac{2\sinh\frac{\pm(2a_i+\epsilon_+)+m}{2}}{2\sinh(-a_i)2\sinh(a_i + \epsilon_+)}\prod_{j \neq i}\frac{2\sinh\frac{\pm(a_i + a_j + \epsilon_+) + m}{2}2\sinh\frac{\pm (a_i - a_j + \epsilon_+) + m}{2}}{2\sinh\frac{-a_i \pm a_j}{2}2\sinh\frac{a_i \pm a_j + 2\epsilon_+}{2}}\right),
\end{split}
\end{equation}
which is independent of the JK parameter~$\eta$. 
Since the monopole screening contribution in $\Braket{T_V}$ does not have an extra term, we can apply the extra term prescription to the monopole screening contribution $Z''_{\text{momo}}(\ba)$ in $\Braket{T_{\wedge^2V}}$ \footnote{See footnote \ref{fn:extra}. }. 
The contribution from $O(2)_-$ has no integration but contains a non-trivial Coulomb branch moduli independent term, {\it i.e.}, an extra term defined in Section~\ref{sec:extra-term}, which we believe is given by
\begin{equation}\label{USp-wedge2V-extra}
Z_{\text{extra}-} = \frac{2\cosh(\epsilon_+)}{2\cosh\frac{m\pm \epsilon_+}{2}}.
\end{equation}
We explicitly checked this for $n=1,2, 3, 4, 5$. 
Hence the monopole screening contribution~$Z''_{\text{mono}}(\ba)$ in \eqref{Twedge2V.N2starUSp} is 
\begin{equation}\label{ZmonostarUSp2V2}
Z''_{\text{mono}}(\ba) = 
\frac{1}{2}\left(Z''_{\wedge^2V+}(\bv={\bm 0})+ Z''_{\wedge^2V-}(\bv ={\bm 0})\right) - Z_{\text{extra}},
\end{equation}
where $Z_{\text{extra}} = \frac{1}{2}Z_{\text{extra}-}$.

For $n=1$, 
we expect that $\Braket{T^{USp(2)}_{\wedge^2 V}}$ equals $\Braket{T^{SU(2)}_{\text{adj}}}$.
The monopole contribution to the former, given by the specialization of~\eqref{ZmonostarUSp2V2} to $n=1$, indeed coincides precisely with 
\eqref{su2adjmono} with $2a_1 = a$.
 This is evidence for the validity of~\eqref{ZmonostarUSp2V2} and for the extra term prescription we proposed in Section~\ref{sec:extra-term}.

\subsubsection{$V \times V$}

The Moyal product~$\Braket{T_V} \ast \Braket{T_V}$ takes the same form as~\eqref{USp.TFTF}, but $Z_i({\bm a})$ and $Z_{\text{mono}}({\bm a})$ are given by \eqref{starUSpTZi} and \eqref{ZmonostarUSpV} respectively. 
There are three types of monopole screening sectors: $\bv = \pm \be_i \pm \be_j$~($1 \leq i < j \leq n$), ${\bm v} = \pm {\bm e}_i$~($1\leq i\leq n$), and ${\bm v} = {\bm 0}$. We will compute their contributions from the supersymmetric indices of the corresponding SQMs and compare them with the Moyal product. 
We will determine the overall signs of the indices by requring that the index reduces,
after setting~$m = \epsilon_+ = 0$, to the square of a character
\begin{equation}\label{N4USpTVTV}
\begin{split}
\Braket{T_V} \ast \Braket{T_V}\Big|_{m = \epsilon_+ = 0}
&= \sum_{i=1}^n (e^{2b_i}+e^{-2b_i} +2e^{b_i} + 2e^{-b_i}) 
\cr
&\quad
+ \sum_{1 \leq i < j \leq n}\left(2e^{b_i + b_j} + 2e^{-b_i - b_j}+2e^{b_i - b_j} + 2e^{-b_i + b_j} \right) +2n+1.
\end{split}
\end{equation}

\paragraph{$\bv = \pm \be_i \pm \be_j$ for $1 \leq i < j \leq n$.} First we consider the sector $\bv = \pm \be_i \pm \be_j$ for $1 \leq i < j \leq n$.
The brane configuration for the sector $\bv = \be_{n-1} + \be_n$ and the quiver diagram for the monopole screening sector are the ones in Figure~\ref{fig:SO-VxV-v=eNpluseN-1}, except that ${\rm O6}^+$ should be replaced by ${\rm O6}^-$. Hence the monopole screening contribution is exactly the same as the right hand side of~\eqref{zmono-from-moyal-SO-VxV-v=nn-1}. 

On the other hand, the monopole screening contribution in this sector appears in the first or the second line of the Moyal product \eqref{USp.TFTF} with the one-loop determinant given in~\eqref{starUSpTZi}. 
We focus on the sector $\bv = \be_{n-1} + \be_n$ and the monopole screening contribution $Z_{\text{mono}}(\bv = \be_{n-1} + \be_n; \ba)$ in the Moyal product is computed by 
\begin{equation}\label{moyalN2starUSpTVVvee}
\begin{split}
&Z_{n-1}(\ba + \epsilon_+\be_n)Z_n(\ba - \epsilon_+\be_{n-1})+ Z_{n-1}(\ba -  \epsilon_+\be_n)Z_n(\ba + \epsilon_+\be_{n-1})\cr
& = Z_{(n-1)n}(\ba)Z_{\text{mono}}(\bv = \be_{n-1} + \be_n; \ba),
\end{split}
\end{equation}
where  $Z_{(n-1)n}(\ba)$ is \eqref{Zij.N2starUSp} with $i=n-1$ and $j=n$. 
We find that~$Z_{\text{mono}}(\bv = \be_{n-1} + \be_n; \ba)$ is again precisely the right hand side of~\eqref{zmono-from-moyal-SO-VxV-v=nn-1} as expected.

\begin{figure}[t]
\centering
\subfigure[]{\label{subfig:O6-USp-tHooft-VxV-v=eN}
\raisebox{3mm}{\includegraphics[scale=.3]{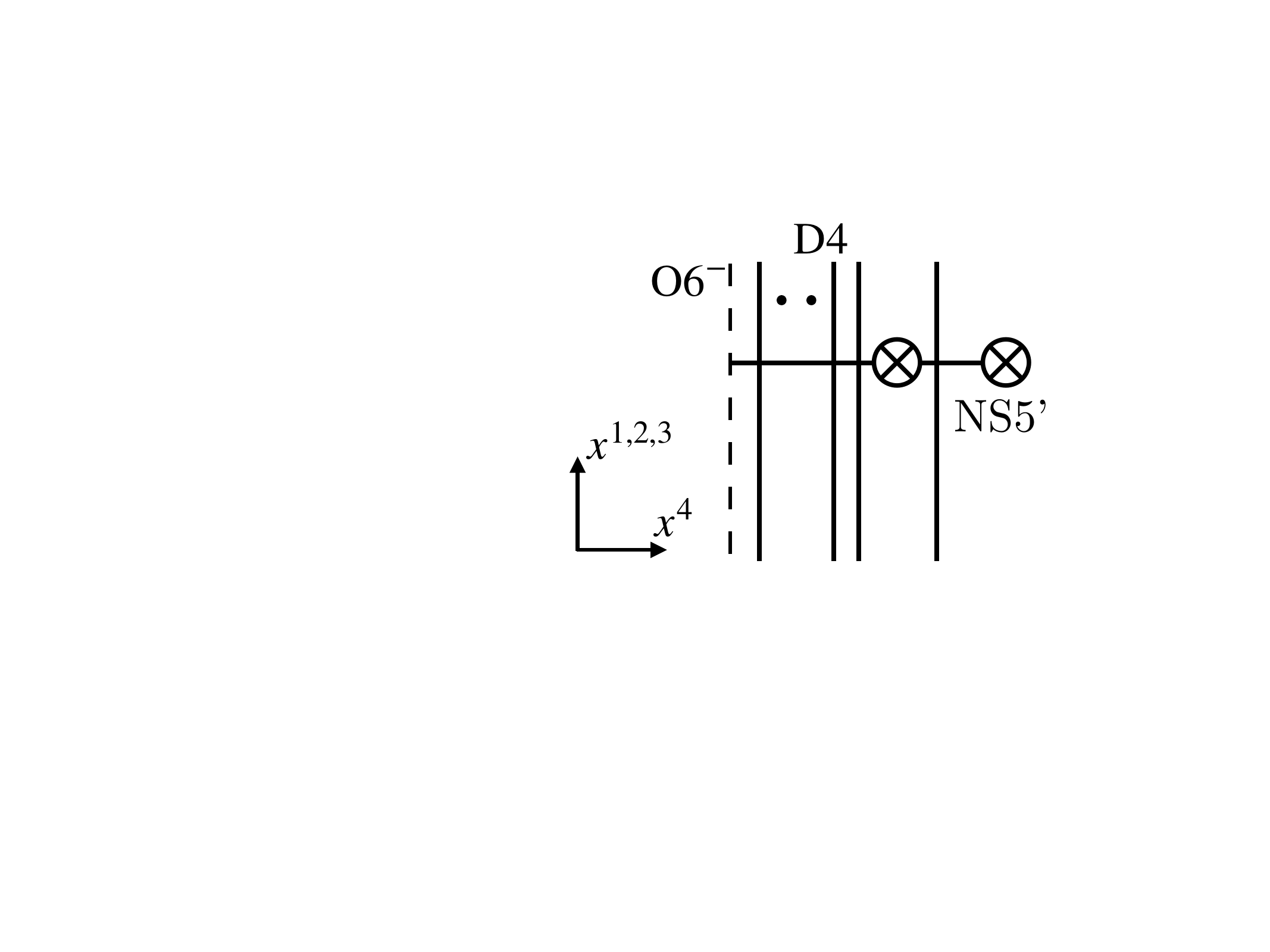}}}
\hspace{2cm}
\subfigure[]{\label{subfig:quiverUSp10}
\hspace{-4.5mm}
\includegraphics[width=4cm]{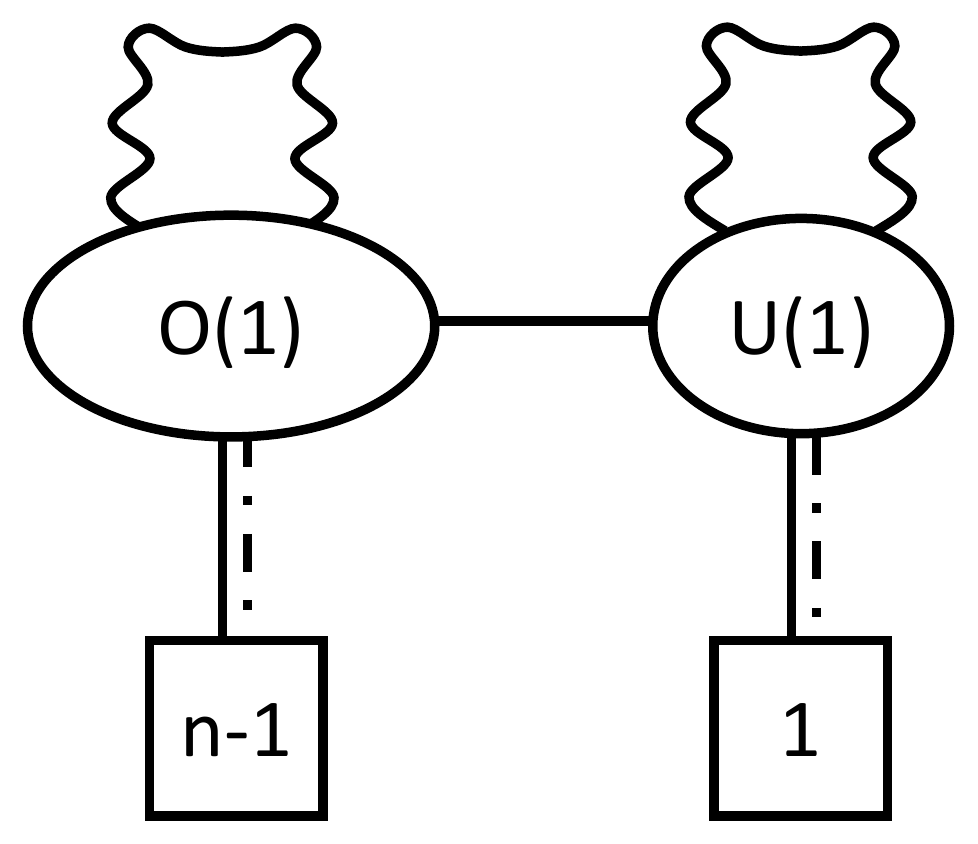}}
\caption{(a): The brane configuration for the monopole screening sector $\bv = \pm \be_i$ of~$T_V\cdot T_V$ in the $\mathcal{N}=2^{\ast}$ $USp(2n)$ gauge theory.
(b): The corresponding quiver diagram.
}
\label{fig:quiverUSp10}
\end{figure}

\paragraph{${\bm v} = \pm {\bm e}_i$.} Next we consider the sector ${\bm v} = \pm {\bm e}_i$. 
The brane configuration and the quiver diagram for the monopole screening sector are depicted in Figures~\ref{subfig:O6-USp-tHooft-VxV-v=eN} and~\ref{subfig:quiverUSp10}. 
The supersymmetric index  again consists of two contributions. The contribution from the $O(1)_+ - U(1)$ sector is given by
\begin{equation}
\begin{split}
&Z_{+}
(\bv = \pm \be_i, \zeta) \cr
&= -\oint_{JK(\zeta)}\frac{d\phi}{2\pi i}\frac{2\sinh(\epsilon_+)}{2\sinh\frac{\pm m - \epsilon_+}{2}}\frac{2\sinh\frac{\pm \phi + m}{2}}{2\sinh\frac{\pm \phi + \epsilon_+}{2}}\frac{2\sinh\frac{\pm (\phi - a_i) + m}{2}}{2\sinh\frac{\pm (\phi - a_i) + \epsilon_+}{2}}\frac{\prod_{1 \leq j \neq i \leq n}2\sinh\frac{\pm a_j + m}{2}}{\prod_{1 \leq j \neq i \leq n}2\sinh\frac{\pm a_j + \epsilon_+}{2}}.
\end{split}
\end{equation}
On the other hand, the $O(1)_- - U(1)$ contributes
\begin{equation}
\begin{split}
&Z_{-}
(\bv = \pm \be_i, \zeta) \cr
&= -\oint_{JK(\zeta)}\frac{d\phi}{2\pi i}\frac{2\sinh(\epsilon_+)}{2\sinh\frac{\pm m- \epsilon_+}{2}}\frac{2\cosh\frac{\pm \phi + m}{2}}{2\cosh\frac{\pm \phi + \epsilon_+}{2}}\frac{2\sinh\frac{\pm (\phi - a_i) + m}{2}}{2\sinh\frac{\pm (\phi - a_i) + \epsilon_+}{2}}\frac{\prod_{1 \leq j \neq i \leq n}2\cosh\frac{\pm a_j + m}{2}}{\prod_{1 \leq j \neq i \leq n}2\cosh\frac{\pm a_j + \epsilon_+}{2}}.
\end{split}
\end{equation}
The overall sign of each contribution was chosen so that we obtain
\begin{align}\label{eq:USp-2star-condition}
\left[\frac{1}{2}\left(Z_{+}
(\bv = \pm \be_i, \zeta) + Z_{-}
(\bv = \pm \be_i, \zeta)\right)\right]_{m =\epsilon_+ = 0} = 2,
\end{align}
which is the coefficient of $e^{\pm b_i}$ in~\eqref{N4USpTVTV}. 
The supersymmetric index is 
\begin{align}\label{WI.starUSpvbi}
Z({\bm v} = \pm {\bm e}_i, \zeta) = \frac{1}{2}\left(Z_{+}
(\bv = \pm \be_i, \zeta) + Z_{-}
(\bv = \pm \be_i, \zeta) \right).
\end{align}
By evaluating the JK residues we find the relation
\begin{equation}
Z({\bm v} = \pm {\bm e}_i, \zeta) = Z_{\text{mono}}({\bm a} - \epsilon_+{\bm e}_i) + Z_{\text{mono}}({\bm a} + \epsilon_+{\bm e}_i)
\end{equation}
for both $\zeta > 0$ and $\zeta < 0$, where the function~$Z_{\text{mono}}$ is given in~(\ref{ZmonostarUSpV}).
This is indeed the monopole screening contribution to the Moyal product~\eqref{USp.TFTF} from the sector~${\bm v} = \pm {\bm e}_i$.

\begin{figure}[t]
\centering
\subfigure[]{\label{subfig:O6-USp-tHooft-VxV-v=0}
\raisebox{3mm}{\includegraphics[scale=.3]{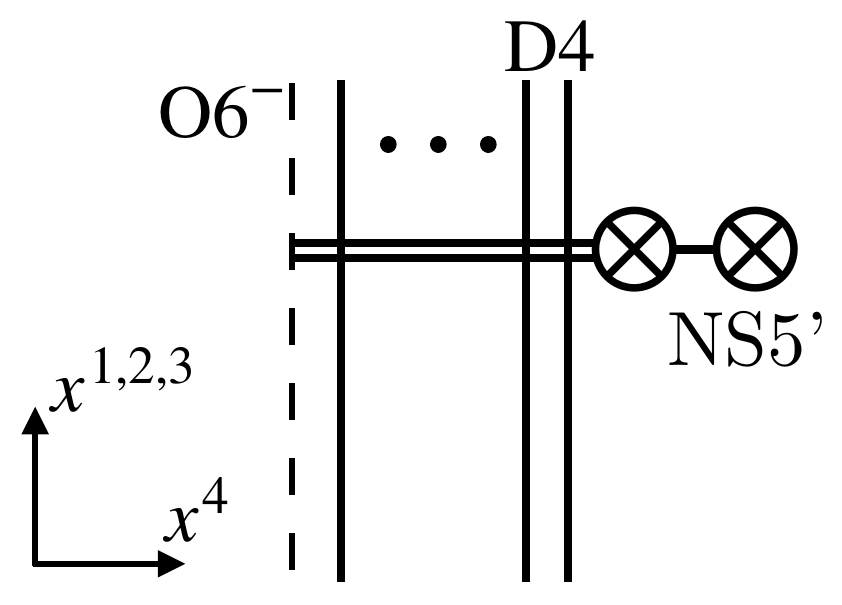}}}
\hspace{2cm}
\subfigure[]{\label{subfig:quiverUSp11}
\hspace{-4mm}
\includegraphics[width=4cm]{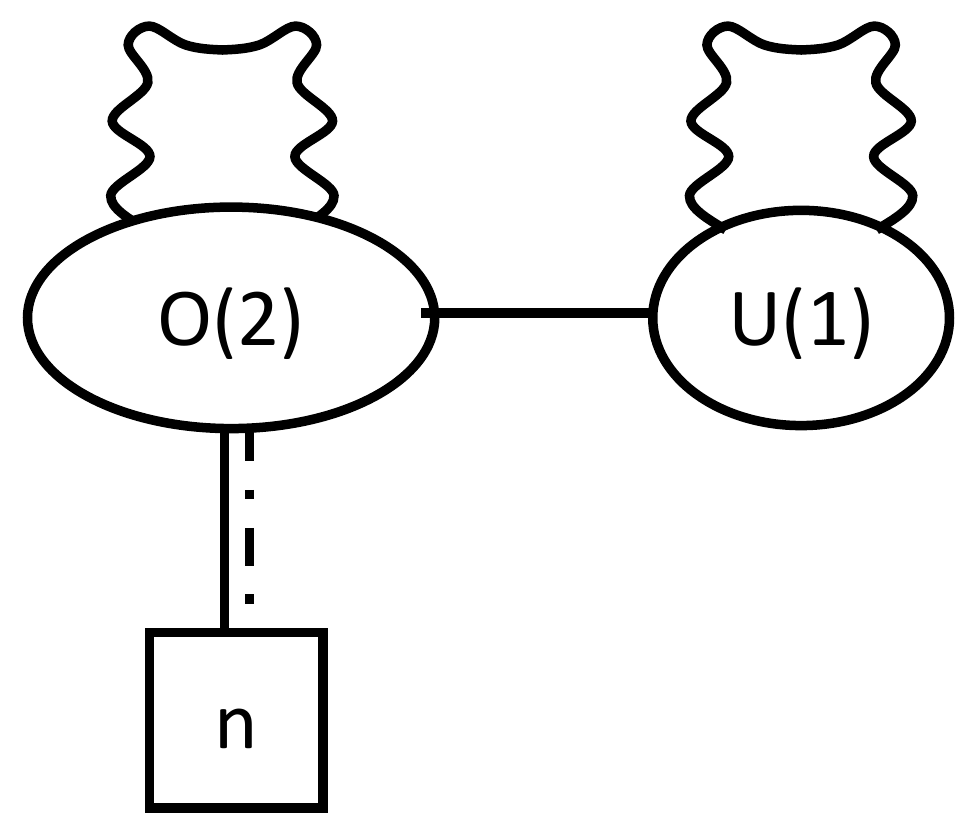}}
\caption{(a): The brane configuration for the monopole screening sector $\bv =\bm{0}$ of~$T_V\cdot T_V$ in the $\mathcal{N}=2^{\ast}$ $USp(2n)$ gauge theory.
(b): The corresponding quiver diagram.
}
\label{fig:quiverUSp11}
\end{figure}

\paragraph{${\bm v} = {\bm 0}$.} The other sector is ${\bm v} = {\bm 0}$. The brane configuration and the quiver diagram for the SQM describing the monopole screening contribution are depicted in Figures~\ref{subfig:O6-USp-tHooft-VxV-v=0} and~\ref{subfig:quiverUSp11}, respectively. 
Again the supersymmetric index consists of two contributions. 
The contribution from the $O(2)_+ - U(1)$ sector is 
\begin{equation}\label{N2starUSpZ+}
\begin{split}
Z_+
(\bv = 0, \zeta) = \oint_{JK(\zeta)}\frac{d\phi_1}{2\pi i}\frac{d\phi_2}{2\pi i}\frac{\left(2\sinh(\epsilon_+)\right)^2}{\left(2\sinh\frac{\pm m - \epsilon_+}{2}\right)^2}\frac{2\sinh\frac{\pm\phi_1 \pm \phi_2 + m}{2}\prod_{i=1}^n2\sinh\frac{\pm\phi_1 \pm a_i + m}{2}}{2\sinh\frac{\pm\phi_1 \pm \phi_2 + \epsilon_+}{2}\prod_{i=1}^n2\sinh\frac{\pm\phi_1 \pm a_i + \epsilon_+}{2}}.
\end{split}
\end{equation}
The contribution from the  $O(2)_- - U(1)$ sector is 
\begin{equation}\label{N2starUSpZ-}
\begin{aligned}
&\quad Z_-
(\bv = {\bm 0}, \zeta)
\\
& = -\oint_{JK(\zeta)}\frac{d\phi}{2\pi i}\frac{2\sinh(\epsilon_+)}{2\sinh\frac{\pm m - \epsilon_+}{2}}\frac{2\cosh(\epsilon_+)}{2\cosh\frac{\pm m - \epsilon_+}{2}}\frac{2\sinh(\pm \phi + m)\prod_{i=1}^n2\sinh\left(\pm a_i + m\right)}{2\sinh(\pm \phi + \epsilon_+)\prod_{i=1}^n2\sinh\left(\pm a_i + \epsilon_+\right)}.
\end{aligned}
\end{equation}
We chose the overall signs for \eqref{N2starUSpZ+} and \eqref{N2starUSpZ-} so that we reproduce the constant term in \eqref{N4USpTVTV} after setting $m = \epsilon_+ = 0$, namely
\begin{align}
\left[\frac{1}{2}\left(Z_+
(\bv = 0, \zeta) + Z_-
(\bv = {\bm 0}, \zeta)\right) \right]_{m =\epsilon_+ = 0}  = 2n+1.
\end{align}
The supersymmetric index is then given by
\begin{align}\label{WI.starUSpv0}
Z(\bv = {\bm 0}, \zeta) = \frac{1}{2}\left(Z_+
(\bv = 0, \zeta) + Z_-
(\bv = {\bm 0}, \zeta)\right).
\end{align}

The Moyal product of the ${\bm v} = {\bm 0}$ sector in \eqref{USp.TFTF} is the last line of \eqref{USp.TFTF}. By explicitly evaluating \eqref{WI.starUSpv0} we find the relation
\begin{equation}
Z(\bv = {\bm 0}, \zeta) = \sum_{i=1}^n\left(Z_i(\ba - \epsilon_+\be_i)^2 + Z_i(\ba + \epsilon_+\be_i)^2\right) + Z_{\text{mono}}(\ba)^2
\end{equation}
for both $\zeta > 0$ and $\zeta < 0$.

\section{Conclusion and discussion}
\label{sec:conclusion}

In this paper we calculated by supersymmetric localization the expectation values of  't~Hooft operators on $S^1\times\mathbb{R}^3$ in theories with gauge groups $U(N)$, $SO(N)$ and $USp(N)$.
Let us here specify the magnetic charge of an operator by a representation of the Langlands dual group.
For the SQCD and the $\mathcal{N}=2^*$ theory with gauge group $U(N)$, we computed the expectation values in the cases $\wedge^2 V $ and  $\wedge^2 \overline V $.
We also studied the product $\wedge^2 V \times \wedge^2 \overline V$, {\it i.e.}, we computed the correlator of an 't~Hooft operator with the charge corresponding to $\wedge^2 V$ and another operator with the charge corresponding to $\wedge^2 \overline V$.
For the $SO(N)$ SQCD we did the computation for $V\times V$.
For the $SO(N)$ $\mathcal{N}=2^*$ theory we computed the vev for $\wedge^2 V$ and $V\times V$.
For the $USp(N)$ SQCD we studied $V$ and $V\times V$.
For the $USp(N)$ $\mathcal{N}=2^*$ theory we computed the vevs for~$V$, $\wedge^2 V$, and $V\times V$.

As we mentioned in the introduction, we did not distinguish different global structures (gauge group topologies and discrete theta angles~\cite{Aharony:2013hda}) associated with a given gauge algebra.
The determination of the precise global structure associated with a brane system is an interesting and delicate problem.
The recent work~\cite{Gukov:2020btk} and its sequel may shed light on this.
The global structures that arise in the brane constructions we consider must admit at least  't~Hooft operators whose magnetic charges correspond to the anti-symmetric representations~$\wedge^k V$.

Let us comment on the global structures of ``$SO(N)$ gauge theories'' realized by branes and an orientifold.
In some of such realizations the global structure is actually that of~$O(N)$.%
\footnote{%
There are exceptions.
For example the gauge group of a theory with spinor matter~\cite{Zafrir:2015ftn} cannot be~$O(N)$.
}
For example it is believed that D3-branes on top of an orientifold 3-plane (${\rm O3}^-$ or $\widetilde{\rm O3}{}^-$) lead to the $O(N)$, rather than $SO(N)$, gauge group~\cite{Garcia-Etxebarria:2015wns}.
Also, the ADHM construction of $k$ $USp$ instantons requires $O(k)$, rather than $SO(k)$, as the ADHM gauge group, and this implies that the gauge group for $k$ D-instantons on top of an orientifold 3-plane  (${\rm O3}^+$) realizes $O(k)$~\cite{Dorey:2002ik}.
The following observation suggests that the gauge group arising from our brane realizations of ``$SO(2n)$ gauge theories'' is also $O(2n)$ (or its covering).
For $SO(2n)$ the representation $\wedge^n V$ is reducible, to the sum of (imaginary-)self-dual and (imaginary-)anti-self-dual parts.
But for $O(2n)$ it is irreducible.
We can construct the 't~Hooft operator corresponding to $\wedge^n V$ by $n$ D2-branes ending on a single NS5'-brane, as we explained in Section~\ref{sec:brane}.
But there appears to be no freedom to split the 5-brane into two, implying that the gauge group is more likely to be $O(2n)$ than $SO(2n)$.

In the main text we did not study wall-crossing in the $SO/USp$ theories, and this is related to the subtleties studied in Appendix~\ref{sec:subtle}.
The brane construction in~\ref{sec:SQCD-O4} makes it plausible that conjectures~(ii) and (ii)' in Section~\ref{sec:wall-crossing} extend to the $SO/USp$ SQCDs and that wall-crossing occurs.
But to confirm this we need to consider the product of different kinds of operators.
The simplest choice would be~$T_V\cdot T_{\wedge^2 V}$, but the computations involving $T_{\wedge^2 V}$ are subtle as discussed in Appendix~\ref{sec:subtle}. 
The subtleties may be avoided in cases of a small number of flavors and the monopole screening contributions in some examples are computed in Appendix \ref{sec:subtle}.  However, the monopole screening contributions do not exhibit wall-crossings in those cases with a small number of flavors, which was also the case for $U(N)$ SQCDs~\cite{Hayashi:2019rpw}.
It would be important to generalize the results by sorting out these subtleties and unambiguously determine if the $SO/USp$ SQCDs and more general conformal or asymptotically gauge theories exhibit wall-crossing.
This might involve a careful extension of the analysis of~\cite{Hori:2014tda} to the cases with non-abelian gauge groups without FI parameters and with higher-order poles at infinity.
A study of wall-crossing based on naive computations is summarized in Appendix~\ref{sec:naive-conformal}.
Also, it would be desirable to give a proper understanding of the extra term prescription in Section~\ref{sec:extra-term}.
According to the proposal in~\cite{Brennan:2018rcn}, corrections to the monopole screening contributions computed by the JK residue prescription are the contributions to the supersymmetric index from the Coulomb or the mixed Higgs-Coulomb branches of the SQM. In fact the application of the extra term prescription to the $SU(2)$ SQCD with $N_f=4$ flavors was able to reproduce the Coulomb branch contribution in \cite{Brennan:2018rcn} as we have seen in Section \ref{sec:extra-term}. We suspect that the extra terms computed in other examples may also correspond to the correction terms from ground states on a Coulomb branch. 
We regard it as an important open problem to establish a clear understanding of the relation between the corrections and the extra terms. Also the analogy with instanton partition functions implies that the extra term may not be always independent from Coulomb branch moduli, which has been also pointed out in \cite{Brennan:2018rcn}. So far we have seen that the extra terms for monopole screening contributions in 't Hooft operators in the rank-2 antisymmetric representaiton of Langlands dual groups in $\mathcal{N} = 2^{\ast}$  $SO(N)$ and $USp(2N)$ gauge thoeries are simply Coulomb branch independent parts of the supersymmetric indices of SQMs. The argument in Section \ref{sec:extra-term} suggests that a Coulomb branch dependent extra term may possibly arise when we consider an 't Hooft operator in a higher rank antisymmetric representation ({\it e.g.} rank-3 antisymmetric representation) in $\mathcal{N} = 2^{\ast}$  $SO(N)$ and $USp(2N)$ gauge theories and it would be interesting to check if this is indeed the case.

Before this work, localization for 't~Hooft operators, especially with monopole screening, had been done only for~$SU(N)$ and~$U(N)$ gauge groups; consequently applications such as~\cite{Xie:2013lca,Mekareeya:2013ija,Xie:2013vfa,Bullimore:2013xsa,Tachikawa:2015iba,Honda:2017cnz,Maruyoshi:2020cwy} and Section 8.2 of~\cite{Okuda:2019emk} had been limited to these groups.
Our results here should be useful when extending these applications to $SO(N)$ and $USp(N)$.
It would also be interesting to explore S-duality between Wilson and 't~Hooft operators on $\mathbb{S}^4_b$~\cite{Okuda:2014fja} and $S^1\times S^3$~\cite{Gang:2012yr} for various gauge groups and for $\mathcal{N}=2^*$ and SQCDs.
This paper only treated the hypermultiplets only in special representations of the gauge groups; it would be desirable to generalize to other representations along the line of~\cite{Shadchin:2005mx}.

Another natural extension of~\cite{Hayashi:2019rpw} and the present work is to consider exceptional gauge groups.
The set-ups for instanton counting with exceptional gauge groups in~\cite{Benvenuti:2010pq, Keller:2011ek, Hanany:2012dm, Keller:2012da, Cremonesi:2014xha, Haghighat:2014vxa, DelZotto:2016pvm, Hayashi:2017jze, Kim:2018gjo, Hayashi:2018bkd, DelZotto:2018tcj, Gu:2019dan, Kim:2019uqw, Gu:2020fem} may be useful. 
Also, the 't~Hooft operators in $SO/USp$ SQCD  with the magnetic charges corresponding to $\wedge^2 V$ and higher anti-symmetric representations, which exhibit some subtleties and of which we did a preliminary study in Appendix~\ref{sec:subtle}, deserve further investigation.
Finally, as explained in Section~\ref{sec:SQM-tHooft}, our SQMs arise from D2-branes bounded by NS5- and NS5'-branes.
It would be nice to extend the 2d $\mathcal{N}=(0,4)$ brane box models  of~\cite{Hanany:2018hlz} by including orientifolds and also by allowing D-branes to end on NS5(')-branes. 
Our SQMs would be the dimensional reduction of such 2d theories.

\section*{Acknowledgements}
We wouldl like to thank Joonho Kim for useful discussions.
The work of H.H. is supported in part by JSPS KAKENHI Grant Number JP18K13543, and that of T.O. by Grant Number JP16K05312.
The work of Y.Y. is supported in part by JSPS KAKENHI Grant Number JP16H06335 and also by World Premier International Research Center Initiative (WPI), MEXT Japan.

\appendix

\section{Useful facts about $U(N)$, $SO(N)$, and $USp(N)$}
\label{sec:thooft-QFT}

In this appendix we collect some relevant facts that we use in the main text.
For the basic notions such as various lattices, we recommend~\cite{MR1363490} and Appendix A of~\cite{Gukov:2006jk}.

\subsection{$U(N)$}

We denote by $\bm{e}_i$ ($i=1,\ldots ,N$) the orthonormal basis of the Cartan subalgebra of $U(N)$, identified with its dual.
We summarize the information about the Lie algebra of $ U(N)$ in the following table.
\begin{center}
\begin{tabular}{c|c}
simple (co)roots & $\bm{e}_i-\bm{e}_{i+1}$ ($ 1\leq i\leq N-1$)\\ 
(co)roots & $  \bm{e}_i - \bm{e}_j$ ($i\neq j$) \\
fundamental (co)weights & $\bm{e}_1+\ldots +\bm{e}_j$ ($1\leq j \leq N$)\\
Weyl group & $S_N$
\end{tabular}
\end{center}
The Weyl group acts by permuting $\bm{e}_i$ ($i=1,\ldots,N$).

For the 't~Hooft operator corresponding to the rank-$k$ anti-symmetric representation~$\wedge^k V$ of $U(N)$, we imitate the convention in~\cite{Brennan:2018yuj} and write $\bm{B}=\bm{e}_{N-k+1}+ \bm{e}_{N-i+2} +\ldots+  \bm{e}_{N}$ ($k=1,\ldots,N$) to specify their magnetic charges.
For the 't~Hooft operators corresponding to complex conjugate of the rank-$k$ anti-symmetric representation~$\wedge^k\overline{V}$, we write $\bm{B}=-\bm{e}_{1}- \bm{e}_{2} -\ldots -  \bm{e}_{k}$ ($k=1,\ldots,N$) to specify their magnetic charges.
The representations $\wedge^k V$ and~$\wedge^k\overline{V}$ are minuscule, in the sense that all the weights in each representation are in a single Weyl group orbit.
The adjoint representation~$V\otimes\overline{V}$ is quasi-minuscule~\cite{MR2388163} in the sense that all the non-zero weights are in a single Weyl group orbit.%
\footnote{%
For a general gauge group $G$, if the magnetic charge of an 't~Hooft operator corresponds to a minuscule representation of the Langlands dual~$G^\vee$, the vev is completely determined by the one-loop formulas~(\ref{1loopvm}) and~(\ref{1loophm}).
If the magnetic charge corresponds to a quasi-minuscule representation, the vev is determined by the one-loop formulas except a single $b_i$-independent term which is a monopole screening contribution.
}

The (co)character lattice of $U(N)$ is given by
\begin{equation}
\Lambda_\text{cochar}(U(N)) = 
 \Lambda_\text{char}(U(N)) 
=
\bigoplus_{i=1}^N \mathbb{Z} \bm{e}_i .
\end{equation}

\subsection{$SO(2n)$}

We denote by $\bm{e}_i$ ($i=1,\ldots,n$) the orthonormal basis of the Cartan subalgebra identified with its dual.
We summarize the information about the Lie algebra of $SO(2n)$ with $n\geq 3$ in the following table.
\begin{center}
\begin{tabular}{c|c}
simple (co)roots & $\bm{e}_i-\bm{e}_{i+1}$ ($i=1,\ldots,n-1$), $\bm{e}_{n-1} + \bm{e}_n$ \\
(co)roots &  $\pm \bm{e}_i \pm \bm{e}_j$ ($1\leq i<j\leq n$, signs uncorrelated) \\
fundamental (co)weights & $\bm{e}_1+\ldots +\bm{e}_j$ ($1\leq j \leq n-2$), $\displaystyle\frac{1}{2}\left(\sum_{i=1}^{n-1}\bm{e}_i\pm \bm{e}_n\right)$ \\
Weyl group & $S_n\ltimes (\mathbb{Z}_2)^{n-1}$
\end{tabular}
\end{center}
The Weyl group acts by permuting $\bm{e}_i$ ($i=1,\ldots,N$) as well as by flipping the signs of an even number of $\bm{e}_i$'s simultaneously.

For the 't~Hooft operators corresponding to the first $n-2$ fundamental coweights and anti-symmetric representations, we use the coweight $\bm{B}=\bm{e}_{n-i+1}+ \bm{e}_{n-i+2} +\ldots+  \bm{e}_{n}$ ($i=1,\ldots,n-2$), which is Weyl equivalent to the coweights, to specify their magnetic charges.
The vector representation~$V$ is minuscule.
The adjoint representation~$\wedge^2V$ is quasi-minuscule.

The compact real form $SO(2n)$ is neither simply connected nor of adjoint type.%
\footnote{%
A semisimple Lie group is said to be of adjoint type if its center is trivial.
}
The cocharacter lattice is given by
\begin{equation}
\Lambda_\text{cochar}(SO(2n)) = 
\Lambda_\text{char}(SO(2n)) = 
\bigoplus_{i=1}^N \mathbb{Z} \bm{e}_i .
\end{equation}
For $n\geq 2$, its universal cover $Spin(2n)$ is a double cover of $SO(2n)$, corresponding to the fact that $\Lambda_\text{coweight}(SO(2n))/\Lambda_\text{cochar}(SO(2n))=\pi_1(SO(2n))=\mathbb{Z}_2$.
The center of $SO(2n)$ is $\Lambda_\text{cochar}(SO(2n))/\Lambda_\text{coroot}(SO(2n))=\mathbb{Z}_2$.%
\footnote{%
The center of $Spin(2n)$ is $\mathbb{Z}_2\times \mathbb{Z}_2$ for $n$ even, and is $\mathbb{Z}_4$ for $n$ odd (and $\geq 3$).
}

\subsection{$SO(2n+1)$}
\label{sec:so-odd}

We denote by $\bm{e}_i$ ($i=1,\ldots,n$) the orthonormal basis of the Cartan subalgebra identified with its dual.
We summarize the information about the Lie algebra of $SO(2n+1)$:
\begin{center}
\begin{tabular}{c|c}
simple roots & $\bm{e}_i-\bm{e}_{i+1}$ ($i=1,\ldots,n-1$), $ \bm{e}_n$ \\
roots &  $\pm \bm{e}_i \pm \bm{e}_j$ ($1\leq i<j\leq n$), $\pm \bm{e}_i$ ($i=1,\ldots,n$) \\
simple coroots &$\bm{e}_i-\bm{e}_{i+1}$ ($i=1,\ldots,n-1$), $ \bm 2{e}_n$ \\
coroots &  $\pm \bm{e}_i \pm \bm{e}_j$ ($1\leq i<j\leq n$), $\pm 2 \bm{e}_i$ ($i=1,\ldots,n$) \\
fundamental weights & $\bm{e}_1+\ldots +\bm{e}_j$ ($1\leq j \leq n-1$), $\displaystyle\frac{1}{2}\left(\sum_{i=1}^{n-1}\bm{e}_i+ \bm{e}_n\right)$ \\
fundamental coweights & $\bm{e}_1+\ldots +\bm{e}_i$ ($1\leq i \leq n$) \\
Weyl group & $S_n\ltimes (\mathbb{Z}_2)^n$
\end{tabular}
\end{center}
The double signs in roots and coroots are uncorrelated.
The Weyl group acts by permuting $\bm{e}_i$ ($i=1,\ldots,N$) as well as by flipping the signs of an arbitrary number of $\bm{e}_i$'s simultaneously.

For the 't~Hooft operators corresponding to the fundamental coweights and the anti-symmetric representations $\wedge^k V$ of $USp(2n)$, we use 
\begin{equation}
\bm{B}=\bm{e}_{n-k+1}+ \bm{e}_{n-i+2} +\ldots+  \bm{e}_{n} \qquad  (k=1,\ldots,n) 
\end{equation}
to specify their magnetic charges.
The vector representation~$V$ of $USp(2n)$ is minuscule.
The adjoint representation of $USp(2n)$, which is obtained from $\wedge^2V$ by subtracting a singlet corresponding to the symplectic form, is quasi-minuscule for $n\geq 2$.

The compact real form $SO(2n+1)$ is of adjoint type, and its Langlands dual $USp(2n)$ is simply connected.
The cocharacter lattice is given by
\begin{equation}
\Lambda_\text{cochar}(SO(2n+1))
= \Lambda_\text{char}(USp(2n)) = \Lambda_\text{weight}(USp(2n))
 = \bigoplus_{i=1}^N \mathbb{Z} \bm{e}_i .
\end{equation}

\subsection{$USp(2n)$}\label{app:usp}

The relevant information about the Lie algebra of $USp(2n)$ is obtained from the table given above for the Langlands dual~$SO(2n+1)$ by exchanging roots and coroots, as well as by exchanging weights and coweights.

For the 't~Hooft operators corresponding to the fundamental coweights and the anti-symmetric representations of $SO(2n+1)$, we use the coweight
\begin{equation}
\bm{B}=\bm{e}_{n-i+1}+ \bm{e}_{n-i+2} +\ldots+  \bm{e}_{n} \quad  (i=1,\ldots,n-1)
\end{equation}
 to specify their magnetic charges.
The vector representation~$V$ of $SO(2n+1)$ is quasi-minuscule.

The compact real form $USp(2n)$ is simply connected, and its Langlands dual $SO(2n+1)$ is of adjoint type.
The cocharacter lattice is given by
\begin{equation}
 \Lambda_\text{cochar}(USp(2n))
=\Lambda_\text{char}(SO(2n+1))
=
\Lambda_\text{root}(SO(2n+1))
= \bigoplus_{i=1}^N \mathbb{Z} \bm{e}_i
\end{equation}

\section{Formulas for one-loop determinants}
\label{sec:formulas}

In this appendix we collect useful formulas which we use in the computations of the expectations values of 't Hooft operators.

\subsection{One-loop determinants for 't~Hooft operator vevs in 4d $\mathcal{N}=2$ theories}
\label{sec:one-loop-4d}

The expectation value of an 't~Hooft line operator with magnetic charge ${\bm B}$  takes the form~\cite{Ito:2011ea}
\begin{align}
\Braket{T_{\bm B} } = \mathop{\sum_{\bm{v} \in \bm{B}+ \Lambda_{\text{cort}}}}_{|\bm v|\leq |\bm B|} e^{\bm{v}\cdot \bm{b}} Z_\text{1-loop}(\bm{a}, {\bm m}, \epsilon_+;\bm{v}) Z_\text{mono}({\bm a}, {\bm m}, \epsilon_+;\bm B,\bm v), \label{TB.general} 
\end{align}
where ${\bm v}$ labels monopole screening sectors.
The total one-loop determinant is the product
\begin{align}
Z_\text{1-loop}(\bm{a}, {\bm m}, \epsilon_+ ;\bm{v})=Z_{\text{1-loop}}^{\text{vm}} ( {\bm a}, \epsilon_+; \bm{v} ) \prod_f Z_{\text{1-loop}}^{\text{hm},R_f}( \bm{a}, m_f, \epsilon_+ ;\bm{v}) .
\end{align}
of the contribution~$Z_{\text{1-loop}}^{\text{vm}}$ from the  vector multiplet for gauge group $G$ and the contributions $Z_{\text{1-loop}}^{\text{hm},R_f}$ from matter hypermultiplets in the representations $R_f$ of $G$.
The one-loop determinant of  the vector multiplet is
\begin{align}
Z_{\text{1-loop}}^{\text{vm}}  ({\bm a}, \epsilon_+ ;\bm{v} )
= \prod_{ {\bm \alpha} \in \text{root}, |{\bm \alpha}\cdot \bm{v}| \neq 0}\prod_{k=0}^{|{\bm \alpha}\cdot \bm{v}| - 1}\left[2\sinh\left(\frac{{\bm \alpha}(\bm{a}) + \left(|{\bm \alpha}(\bm{v})|- 2k\right)\epsilon_+}{2}\right)\right]^{-\frac{1}{2}} , \label{1loopvm}
\end{align}
where the symbol ``$\text{root}$'' denotes the set of roots.
The one-loop determinant of the hypermultiplet in a representation $R$ of $G$ with a mass parameter $m$ is 
\begin{align}
Z_{\text{1-loop}}^{\text{hm},R}  ({\bm a}, m, \epsilon_+ ;\bm{v} )= 
\mathop{\prod_{{\bm \rho} \in \Delta(R)}}_{ |{\bm \rho} \cdot \bm{v}|\neq 0}
\prod_{k=0}^{|{\bm \rho}\cdot \bm{v}| - 1}\left[2\sinh\left(\frac{{\bm \rho}(\bm{a}) - m + \left(|{\bm \rho}(\bm{v})|-1 - 2k\right)\epsilon_+}{2}\right)\right]^{\frac{1}{2}}, \label{1loophm}
\end{align}
where  $\Delta (R)$ is the set of the weights  in $R$. 

\subsection{One-loop determinants in SQMs}
\label{sec:one-loop-SQM}

In terms of one-loop determinants for $\mathcal{N}=(0,4)$ supermultiplets, the supersymmetric index of the SQM takes the form
\begin{equation}\label{Z-SQM-JK}
Z = \pm \oint_{JK} Z_\text{vec} Z_\text{hyp}  Z_{\widetilde{\text{hyp}}} Z_\text{fer} .
\end{equation}
We denote the gauge group and the flavor symmetry group by $H$ and $F$, respectively.
For the JK residue prescription indicated in~(\ref{Z-SQM-JK}), we refer the reader to the early references~\cite{Benini:2013xpa,Hwang:2014uwa,Cordova:2014oxa,Hori:2014tda} and our previous work~\cite{Hayashi:2019rpw}.
We choose the overall sign in~(\ref{Z-SQM-JK}) by hand in each example.
For the precise one-loop determinants, we use the formulas given in~\cite{Hwang:2014uwa,Hwang:2016gfw}.%
\footnote{%
The poles that contribute to the contour integral~(\ref{Z-SQM-JK}) via the residue prescription depend crucially on the precise charge assignments associated with the $\sinh$ factors in the denominator of the integrand.
For example, the expression $1/(2\sinh\frac{\phi-a+\epsilon_+}{2})$ corresponding to $U(1)$ charge $+1$ and another expression $-1/(2\sinh\frac{-\phi +a-\epsilon_+}{2})$ corresponding to charge $-1$ lead to different results.
}
We use the short-hand notation $2\sinh\frac{a\pm b}{2} \equiv (2\sinh\frac{a+ b}{2})(2\sinh\frac{a - b}{2})$.

For the $\mathcal{N}=(0,4)$ vector multiplet the one-loop determinant is given by
\begin{equation}
Z_\text{vec} = 
\frac{1}{|W|}
\prod_{i=1}^{r}\frac{d\phi_i}{2\pi i}
\prod_{\alpha\in\text{root}} 2\sinh\frac{\alpha(\phi)}{2}
\prod_{\alpha\in\Delta(\text{adj})}
 2\sinh\frac{\alpha(\phi)+2\epsilon_+}{2}
  .
\end{equation}
Here $|W|$ is the order of the Weyl group $W$, $r$ is the rank, ``root'' is the set of roots, and $\Delta(\text{adj})$ is the set of weights in the adjoint representation including zero weights with multiplicity $r$, all with respect to $H$.

For the $\mathcal{N}=(0,4)$ hypermultiplet in representation $R_\text{hyp}$ of $H\times F$, the one-loop determinant is
\begin{equation}
Z_\text{hyp}  = \frac{1}{\displaystyle \prod_{w\in\Delta(R_\text{hyp})} 2\sinh\frac{\pm w(\phi,m)+\epsilon_+}{2}} ,
\end{equation}
where $\Delta(R)$ denotes the set of  weights in a representation $R$, and $(\phi,m)$ is an element of the complexified Cartan subalgebra of $H\times F$.

For the $\mathcal{N}=(0,4)$ twisted hypermultiplet, we restrict to the case that it transforms in the adjoint representation of a simple Lie subgroup of $H$ and in the fundamental representation of $SU(2)\subseteq F $.
The one-loop determinant is
\begin{equation}
 Z_{\widetilde{\text{hyp}}} = \frac{1}{\displaystyle \prod_{w\in\Delta(\text{adj})} 2\sinh\frac{w(\phi)\pm m -\epsilon_+}{2}} 
\end{equation}

For the $\mathcal{N}=(0,4)$ short Fermi multiplet in representation $R_\text{fer}$ of $H\times F$, the one-loop determinant is given by%
\footnote{%
Gauge invariance gives some restriction.
For example
$Z_\text{fer}  = \prod_{f=1}^{N_F}  2\sinh\frac{\phi -m_f}{2}$ is invariant under $\phi\rightarrow\phi+2\pi$ (large gauge transformation) only for $N_F$ even.
For $N_F$ odd, one can restore invariance by including the contribution $e^{i k\phi}$ from a Chern-Simons action with a half-odd integer level $k$.
}
\begin{equation} \label{1d-1-loop-short-Fermi}
Z_\text{fer}  = \prod_{w\in\Delta(R_\text{fer})} 2\sinh\frac{w(\phi,m)}{2}.
\end{equation}
The  $\mathcal{N}=(0,4)$ long Fermi multiplet in representation $R_\text{fer}$ consists of one short Fermi multiplet in $R_\text{fer}$ and another in~$\bar{R}_\text{fer}$.\footnote{%
In our previous paper~\cite{Hayashi:2019rpw} we actually used~(\ref{1d-1-loop-short-Fermi}) here, not (A.10) of that paper, which contains a typo.
}

\section{On correlators involving $T_{\wedge^2 V}$ in $SO/USp$ SQCD}
\label{sec:subtle}

In Sections~\ref{sec:SOSQCD} and~\ref{sec:USpSQCD}, we considered the expectation values of the minimal 't~Hooft operator~$T_V$ in the $\mathcal{N}=2$ $SO/USp$ SQCD with the number~$N_F$ of flavors for which the theory is conformal.
This is $N_F = N-2$ for $SO(N)$ and $N_F = N+2$ for $USp(N)$.
Here we study the expectation value of the 't~Hooft operator~$T_{\wedge^2 V}$ in the rank-2 anti-symmetric representation of the Langlands dual of the gauge group.
We relegated this case to the present appendix because there is a subtlety in the computation of their monopole screening contributions. 
The subtlety disappears when we consider SQCDs with less flavors and hence we present calculations involving $\Braket{T_{\wedge^2 V}}$ using an SQCD with less flavors.

\subsection{Correlators involving $T_{\wedge^2 V}$ in $SO(N)$ SQCD}
\label{subsec:SO-SQCD-wedge2V}

We start with monopole screening contributions involving $T_{\wedge^2 V}$ in the $SO(N)$ SQCD. 

\subsubsection{$\wedge^2 V$}
\label{sec:SO-SQCD-wedge2V}

We first consider the expectation value of the 't~Hooft operator $T_{\wedge^2 V}$ in the 4d $\mathcal{N}=2$ $SO(N)$ gauge theory with $N_F$ hypermultiplets in the vector representation, assuming that the operator $T_{\wedge^2 V}$ exists in the theory. 
We require the number of the flavors to satisfy~$N_F \leq N-2$ so that the theory is asymptotically free or superconformal.
The magnetic charge of the operator $T_{\wedge^2 V}$ is $\bm{B}=\bm{e}_{n-1}+\bm{e}_n$, which is the highest weight (in our convention) of the rank-2 anti-symmetric representation of the Langlands dual of the gauge group.
(We recall that the Langlands dual is $SO(2n)$ for gauge group $SO(2n)$, and $USp(2n)$ for gauge group $SO(2n+1)$.)
The brane realization of the operator with the magnetic charge~$\bm{B}=\bm{e}_{n-1}+\bm{e}_n$ is shown in Figure~\ref{subfig:O4-tHooft-wedge2V}.

The expectation value of this operator 
$T_{\wedge^2V}$ takes the same form as \eqref{SOTwedgeV}, namely
\begin{equation}\label{SOSQCDTwedgeV}
\begin{aligned}
\Braket{T_{\wedge^2V}} &= \sum_{1\leq i < j \leq n}\left(e^{b_i + b_j} + e^{-b_i - b_j}\right)Z_{ij}({\bm a}) 
+ \sum_{1\leq i < j \leq n}\left(e^{b_i - b_j} + e^{-b_i + b_j}\right)Z'_{ij}({\bm a}) + Z_{\text{mono}}({\bm a}).
\end{aligned}
\end{equation}
The one-loop determinant parts $Z_{ij}(\ba)$, $Z'_{ij}(\ba)$ are different from those in 4d $\mathcal{N}=2^{\ast}$ $SO(N)$ gauge theory and 
in the case of the SQCD they are given by
\begin{equation}\label{ZijSOevenSQCD}
Z_{ij}({\bm a}) = \left(\frac{\prod_{f=1}^{N_F}2\sinh\frac{\pm a_i - m_f}{2}2\sinh\frac{\pm a_j - m_f}{2}}{2\sinh\frac{\pm (a_i + a_j)}{2}2\sinh\frac{\pm (a_i + a_j) + 2\epsilon_+}{2}
\prod_{k\neq i, j}
2\sinh\frac{\pm a_i \pm a_k + \epsilon_+}{2}2\sinh\frac{\pm a_{j}\pm a_k + \epsilon_+}{2}}\right)^{\frac{1}{2}}
\end{equation}
and 
\begin{equation}
Z'_{ij}({\bm a}) = \left(\frac{\prod_{f=1}^{N_F}2\sinh\frac{\pm a_i - m_f}{2}2\sinh\frac{\pm a_j - m_f}{2}}{2\sinh\frac{\pm (a_i - a_j)}{2}2\sinh\frac{\pm (a_i - a_j) + 2\epsilon_+}{2}
\prod_{k\neq i, j}
2\sinh\frac{\pm a_i \pm a_k + \epsilon_+}{2}2\sinh\frac{\pm a_{j}\pm a_k + \epsilon_+}{2}}\right)^{\frac{1}{2}}
\end{equation}
for the $SO(2n)$ gauge theory, and
\begin{equation}\label{ZijSOoddSQCD}
\begin{split}
Z_{ij}({\bm a}) =& \left(\frac{\prod_{f=1}^{N_F}2\sinh\frac{\pm a_i - m_f}{2}2\sinh\frac{\pm a_j - m_f}{2}}{2\sinh\frac{\pm a_i + \epsilon_+}{2}2\sinh\frac{\pm a_j + \epsilon_+}{2}2\sinh\frac{\pm (a_i + a_j)}{2}2\sinh\frac{\pm (a_i + a_j) + 2\epsilon_+}{2}}\right.\\
&\hspace{4cm}\left.\times\frac{1}{\prod_{1 \leq k\neq i, k\neq j \leq n}2\sinh\frac{\pm a_i \pm a_k + \epsilon_+}{2}2\sinh\frac{\pm a_{j} \pm a_k + \epsilon_+}{2}}\right)^{\frac{1}{2}}
\end{split}
\end{equation}
and
\begin{equation}
\begin{split}
Z'_{ij}({\bm a}) =& \left(\frac{\prod_{f=1}^{N_F}2\sinh\frac{\pm a_i - m_f}{2}2\sinh\frac{\pm a_j - m_f}{2}}{2\sinh\frac{\pm a_i + \epsilon_+}{2}2\sinh\frac{\pm a_j + \epsilon_+}{2}2\sinh\frac{\pm (a_i - a_j)}{2}2\sinh\frac{\pm (a_i - a_j) + 2\epsilon_+}{2}}\right.\\
&\hspace{4cm}\left.\times\frac{1}{\prod_{1 \leq k \neq i, k\neq j \leq n}2\sinh\frac{\pm a_i \pm a_k + \epsilon_+}{2}2\sinh\frac{\pm a_{j} \pm a_k + \epsilon_+}{2}}\right)^{\frac{1}{2}}
\end{split}
\end{equation}
for the $SO(2n+1)$ gauge theory. 
The last term~$Z_{\text{mono}}({\bm a})$ in 
\eqref{SOSQCDTwedgeV} is the monopole screening contribution corresponding to the zero weights in the rank-2 anti-symmetric representation.

\begin{figure}[t]
\centering
\hspace{1.5cm}
\subfigure[]{\label{subfig:O4-tHooft-wedge2V-v=0}
\raisebox{3mm}{\includegraphics[scale=.3]{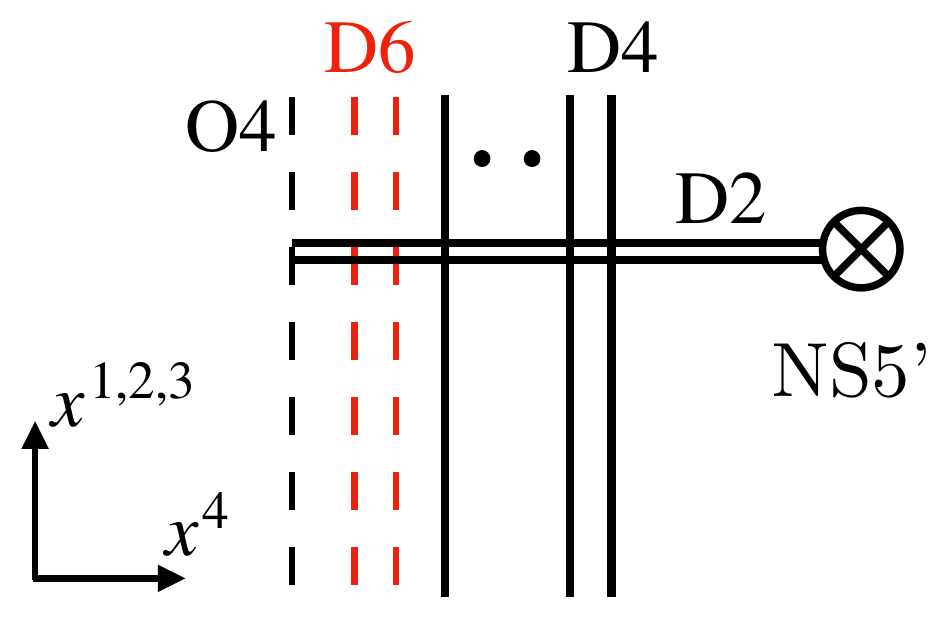}}}
\subfigure[]{\label{subfig:quiverSOwedge}
\hspace{-4mm}
\includegraphics[width=8cm]{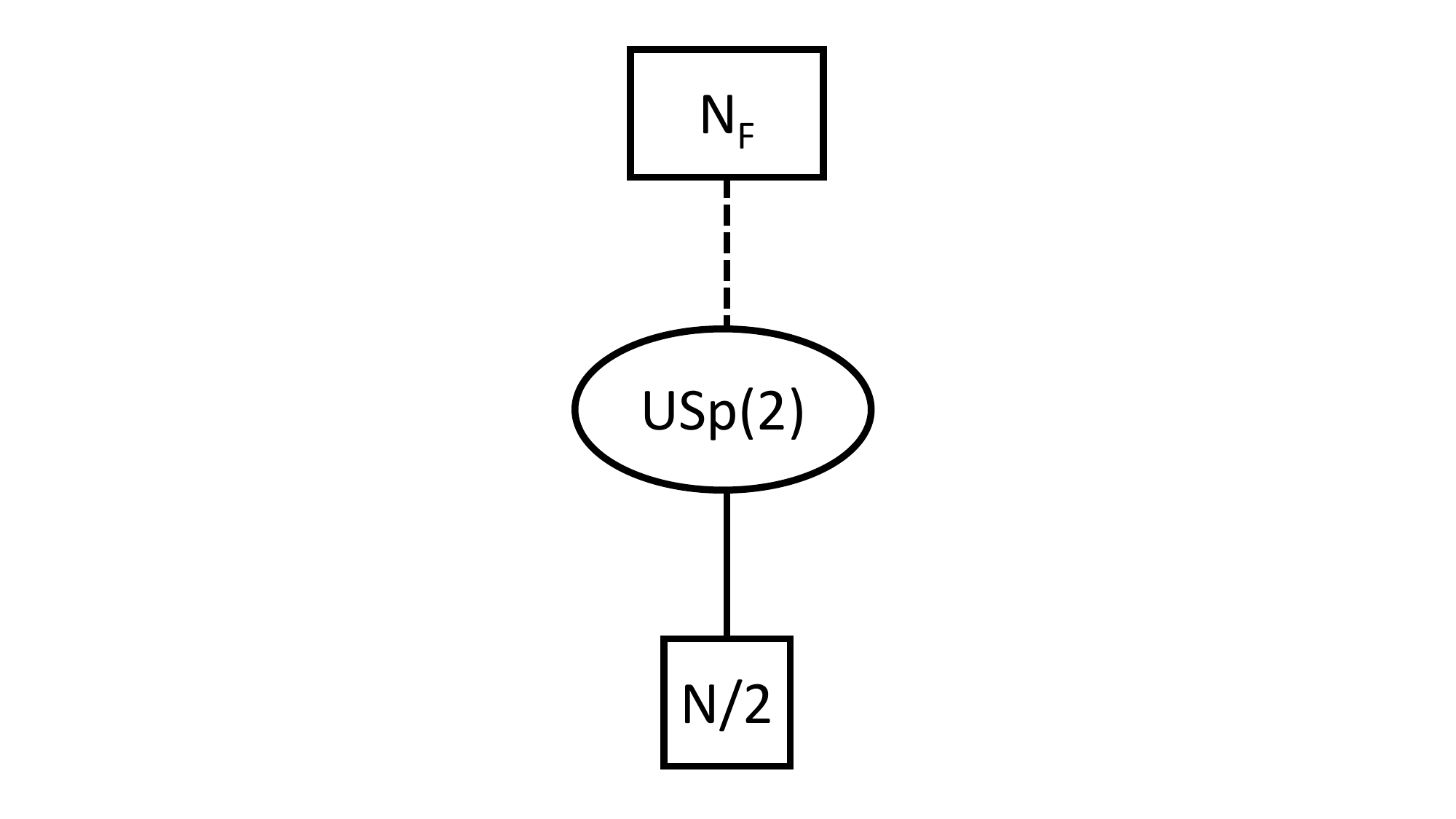}}
\caption{(a): The brane configuration for the monopole screening contribution to the $\bv = {\bm 0}$ sector in $\Braket{T_{\wedge^2V}}$.
(b): The corresponding quiver diagram.
}
\label{fig:SOwedge}
\end{figure}

Let us determine $Z_{\text{mono}}({\bm a})$. 
The brane configuration for monopole screening is shown in Figure~\ref{subfig:O4-tHooft-wedge2V-v=0}.
The vertical black dashed line represents an ${\rm O4}^-$-plane for $N=2n$, and an $\widetilde{\text{O4}}^-$-plane for $N=2n+1$.
In addition there are $N_F$ D6-branes, $n$ D4-branes, and two D2-branes ending on a single NS5'-brane.
The corresponding quiver diagram is given in Figure~\ref{subfig:quiverSOwedge}.
Here we use the $\mathcal{N}=(0, 4)$ notation. 
If the number of the hypermultiplets is $n + \frac{1}{2}$, it implies $n$ hypermultiplets in the fundamental representation and a half-hypermultiplet in the fundamental representation. 
The supersymmetric index of the quiver theory is 
\begin{equation}\label{ZSOewedgev1}
Z_{\wedge^2V}(\bv = {\bm 0}) = \frac{1}{2}\oint_{JK(\eta)}\frac{d\phi}{2\pi i}2\sinh(\pm \phi)\frac{2\sinh(\epsilon_+)2\sinh(\pm \phi + \epsilon_+)\left(\prod_{f=1}^{N_F}2\sinh\frac{\pm \phi - m_f}{2}\right)}{\prod_{i=1}^n2\sinh\frac{\pm \phi \pm  a_i + \epsilon_+}{2}}.
\end{equation}
for $N=2n$ and 
\begin{equation}\label{ZSOowedgev1}
Z_{\wedge^2 V}(\bv = {\bm 0}) = \frac{1}{2}\oint_{JK(\eta)}\frac{d\phi}{2\pi i}2\sinh(\pm \phi)\frac{2\sinh(\epsilon_+)2\sinh(\pm \phi + \epsilon_+)\left(\prod_{f=1}^{N_F}2\sinh\frac{\pm \phi - m_f}{2}\right)}{2\sinh\frac{\pm \phi + \epsilon_+}{2}\left(\prod_{i=1}^n2\sinh\frac{\pm \phi \pm  a_i + \epsilon_+}{2}\right)}.
\end{equation}
for $N=2n+1$. 

In the case of $N_F = N-2$ where the theory becomes superconformal, the integrand of \eqref{ZSOewedgev1} and \eqref{ZSOowedgev1} has a higher order pole at the infinites $\phi = \pm \infty$. A higher order pole in the supersymmetric index of an SQM also appears in the computation of the instanton partition function of a 5d gauge theory with a large number of flavors or a 5d $SU(N)$ gauge theory with a large Chern-Simons level . For example, an SQM for the instanton partition function of the $SU(3)$ gauge theory with $N_F$ flavors with the Chern-Simons level $\kappa = 5 - \frac{N_F}{2}$ has been considered in \cite{Gaiotto:2015una}. The integrand of the supersymmetric index of the corresponding ADHM quantum mechanics has a higher order pole and the computation of the supersymmetric index was carried out by adding pseudo hypermultiplets which make the higher order pole disappear. Since the integrand \eqref{ZSOewedgev1} and \eqref{ZSOowedgev1} has a higher order pole at the inifinites, we might need a similar technique to compute the integral. Here instead we will use less number of flavors such that \eqref{ZSOewedgev1} and \eqref{ZSOowedgev1} do not have a pole at the infinities. Namely we assume $N_F \leq N-5$.

In the cases where $N_F \leq N-5$, we can evaluate the integral \eqref{ZSOewedgev1} and \eqref{ZSOowedgev1} in a usual way.  
Namely we take the contributions from the poles $\phi = \pm a_i - \epsilon_+$ for $\eta > 0$ or $\phi = \pm a_i + \epsilon_+$ for $\eta < 0$. In this case the JK residue of the integral for the both sets of the poles yields the same result regardless of the sign of $\eta$, and they are given by
\begin{equation}\label{ZSOewedgev2}
\begin{split}
Z_{\wedge^2V}(\bv = {\bm 0})=\sum_{i=1}^n&\left(\frac{2\sinh(a_i-\epsilon_+)2\sinh(-a_i+2\epsilon_+)\prod_{f=1}^{N_F}2\sinh\frac{a_i - m_f  - \epsilon_+}{2}2\sinh\frac{-a_i - m_f  + \epsilon_+}{2}}{\prod_{1 \leq k\neq i\leq n}2\sinh\frac{a_i \pm  a_k}{2}2\sinh\frac{-a_i \pm  a_k + 2\epsilon_+}{2}}\right.\\
&\hspace{-.5cm}\left.+\frac{2\sinh(-a_i-\epsilon_+)2\sinh(a_i+2\epsilon_+)\prod_{f=1}^{N_F}2\sinh\frac{a_i - m_f + \epsilon_+}{2}2\sinh\frac{-a_i - m_f - \epsilon_+}{2}}{\prod_{1 \leq k\neq i\leq n}2\sinh\frac{- a_i \pm a_k}{2}2\sinh\frac{a_i \pm a_k + 2\epsilon_+}{2}} \right).
\end{split}
\end{equation}
for $N=2n$ and 
\begin{equation}\label{ZSOowedgev2}
\begin{split}
Z_{\wedge^2V}(\bv = {\bm 0}) =\sum_{i=1}^n&\left(\frac{2\sinh(a_i-\epsilon_+)2\sinh(-a_i+2\epsilon_+)\prod_{f=1}^{N_F}2\sinh\frac{a_i - m_f  - \epsilon_+}{2}2\sinh\frac{-a_i - m_f  + \epsilon_+}{2}}{2\sinh\frac{a_i}{2}2\sinh\frac{-a_i + 2\epsilon_+}{2}\prod_{1 \leq k\neq i\leq n}2\sinh\frac{a_i \pm  a_k}{2}2\sinh\frac{-a_i \pm  a_k + 2\epsilon_+}{2}}\right.\\
&\hspace{-0.5cm}\left.+\frac{2\sinh(-a_i-\epsilon_+)2\sinh(a_i+2\epsilon_+)\prod_{f=1}^{N_F}2\sinh\frac{a_i - m_f + \epsilon_+}{2}2\sinh\frac{-a_i - m_f - \epsilon_+}{2}}{2\sinh\frac{-a_i}{2}2\sinh\frac{a_i + 2\epsilon_+}{2}\prod_{1 \leq k\neq i\leq n}2\sinh\frac{- a_i \pm a_k}{2}2\sinh\frac{a_i \pm a_k + 2\epsilon_+}{2}} \right).
\end{split}
\end{equation}
for $N=2n+1$. 
In this case there is no extra term, which we checked for $N=5, 6, 7, 8, 9, 10, 11, 12$, and 
$Z_{\text{mono}}({\bm a})$ in \eqref{SOSQCDTwedgeV} is given by \eqref{ZSOewedgev2} or \eqref{ZSOowedgev2}  for $SO(2n)$ or for $SO(2n+1)$ repsecitvely when $N_F \leq N-5$.

\subsubsection{$\wedge^2 V \times V$}

Let us then consider the expectation value of a product involving $T_{\wedge^2 V}$. The simple case is the product  between $T_{\wedge^2 V}$ and $T_V$ which we considered in Section \ref{sec:SOSQCD}.  
Since we consider the Moyal product of different operators we may need to be careful of the order of the product. Here we do not consider all the screening sectors but focus on the most screened sector which is characterized by $\bv = \be_i$. 
The $\bv = \be_i$ sector in the Moyal product $\Braket{T_{\wedge^2 V}} \ast \Braket{T_V}$ is given by
\begin{equation}\label{SOwedge2VVv1}
\begin{split}
&\Braket{T_{\wedge^2 V}} \ast \Braket{T_V}\Big|_{\bv=\be_i} \cr
&=e^{b_i}\sum_{1 \leq j \neq i\leq n}\Big(Z_{ij}({\bm a} - \epsilon_+{\bm e}_j)Z_j({\bm a} - \epsilon_+({\bm e}_i + {\bm e}_j)) 
\\
&\qquad\qquad\qquad
+Z'_{ij}({\bm a} + \epsilon_+{\bm e}_j)Z_j({\bm a} - \epsilon_+({\bm e}_i - {\bm e}_j))\Big)
+e^{b_i}Z_{\text{mono}}({\bm a} + \epsilon_+{\bm e}_i)Z_i({\bm a})\cr
&=:e^{b_i}Z_{\text{1-loop}}({\bm v} = {\bm e}_i)Z_{\text{mono}}^{\wedge^2 V \times V}({\bm v} = {\bm e}_i).
\end{split}
\end{equation}
One the other hand the sector $\bv= \be_i$ in the Moyal product in the order  $\Braket{T_{V}} \ast \Braket{T_{\wedge^2 V}}$ is 
\begin{equation}\label{SOwedge2VVv2}
\begin{split}
&\Braket{T_V} \ast \Braket{T_{\wedge^2 V}}\Big|_{\bv = \be_i} \cr
&=e^{b_i}\sum_{1 \leq j \neq i\leq n} \Big(Z_j({\bm a} + \epsilon_+({\bm e}_i + {\bm e}_j))Z_{ij}({\bm a} + \epsilon_+{\bm e}_j)
\\
&\qquad\qquad\qquad
+Z_j({\bm a} + \epsilon_+ ({\bm e}_i - {\bm e}_j))Z'_{ij}({\bm a} - \epsilon_+{\bm e}_j)\Big)
+e^{b_i}Z_i({\bm a})Z_{\text{mono}}({\bm a} - \epsilon_+{\bm e}_i)\cr
&=:e^{b_i}Z_{\text{1-loop}}({\bm v} = {\bm e}_i)Z_{\text{mono}}^{V \times \wedge^2 V}({\bm v} = {\bm e}_i). 
\end{split}
\end{equation}

\begin{figure}[t]
\centering
\subfigure[]{\label{subfig:O4-tHooft-wedge2V-Va}
\raisebox{.0cm}{\includegraphics[scale=.25]{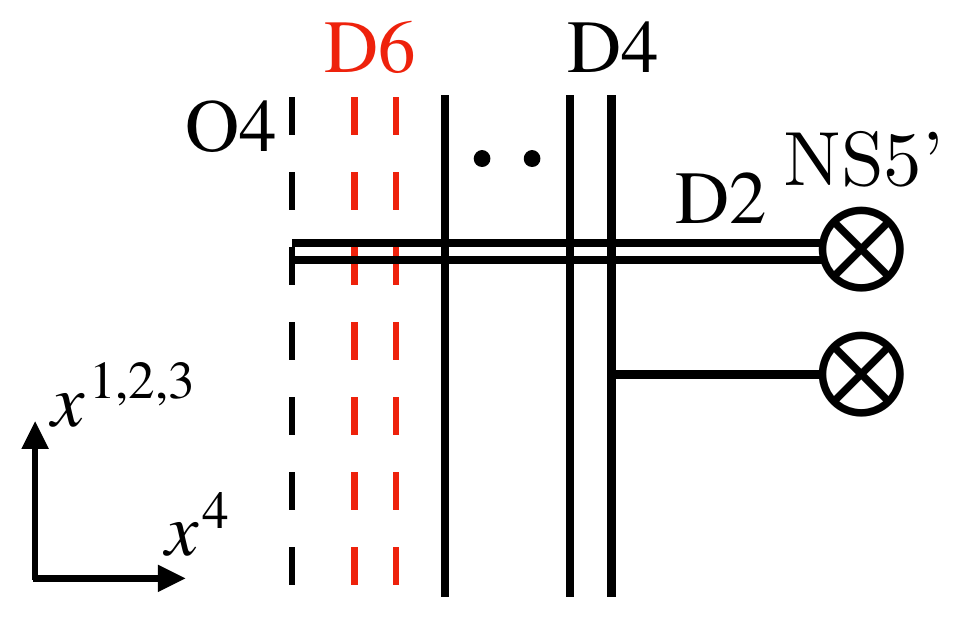}}}
\hspace{1cm}
\subfigure[]{\label{subfig:O4-tHooft-wedge2V-Vb}
\raisebox{.0cm}{\includegraphics[scale=.25]{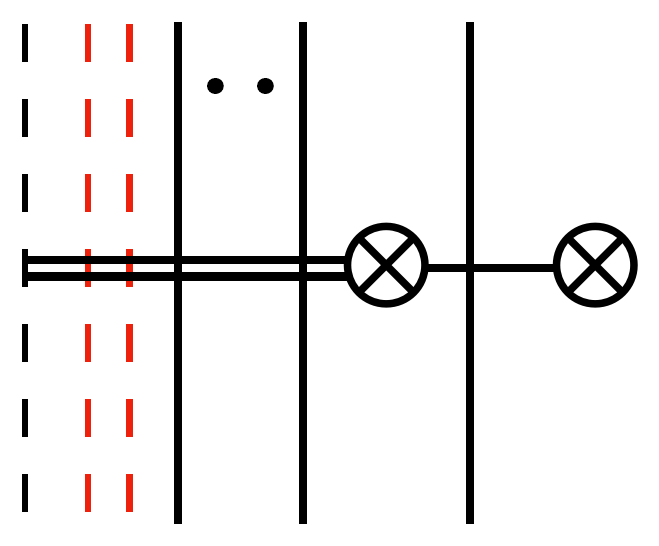}}}
\hspace{1cm}
\subfigure[]{\label{subfig:O4-tHooft-wedge2V-Vc}
\raisebox{.0cm}{\includegraphics[scale=.25]{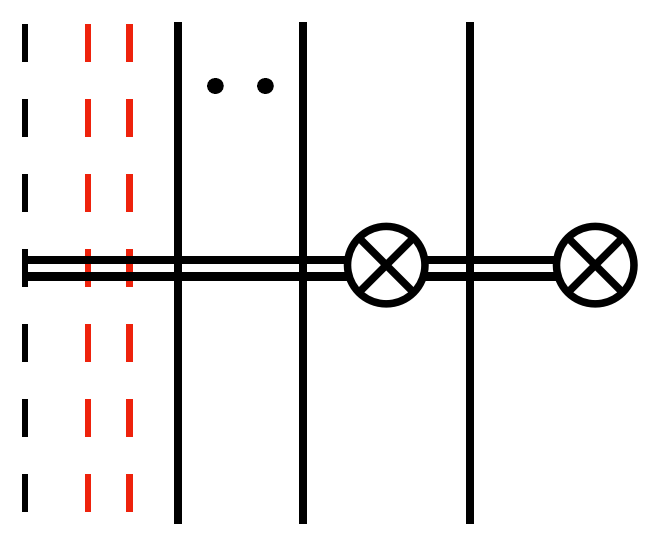}}}
\caption{(a): The brane configuration corresponding to the monopole screening sector~$\bv = + {\bm e}_N$ for the product 't~Hooft operator $T_{\wedge^2 V}\cdot T_{V}$.
(b): A brane configuration obtained by moving the top NS5'-brane to the left.
(c): A brane configuration obtained by moving the bottom NS5'-brane to the left.
}
\label{fig:O4-tHooft-wedge2V-V}
\end{figure}
\begin{figure}[t]
\centering
\subfigure[]{\label{fig:quiverSOv3}
\includegraphics[width=8cm]{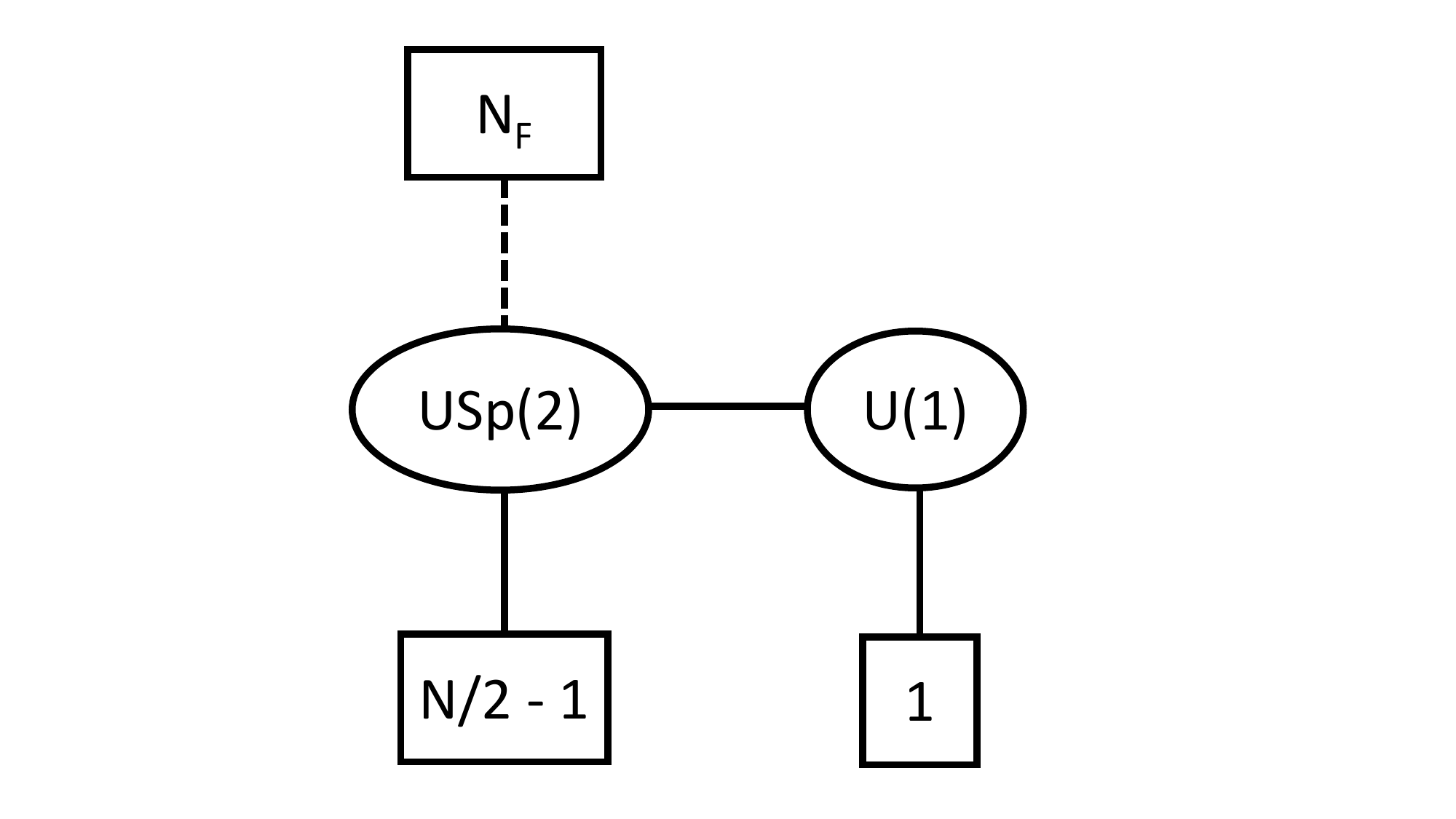}}
\subfigure[]{\label{fig:quiverSOv4}
\includegraphics[width=8cm]{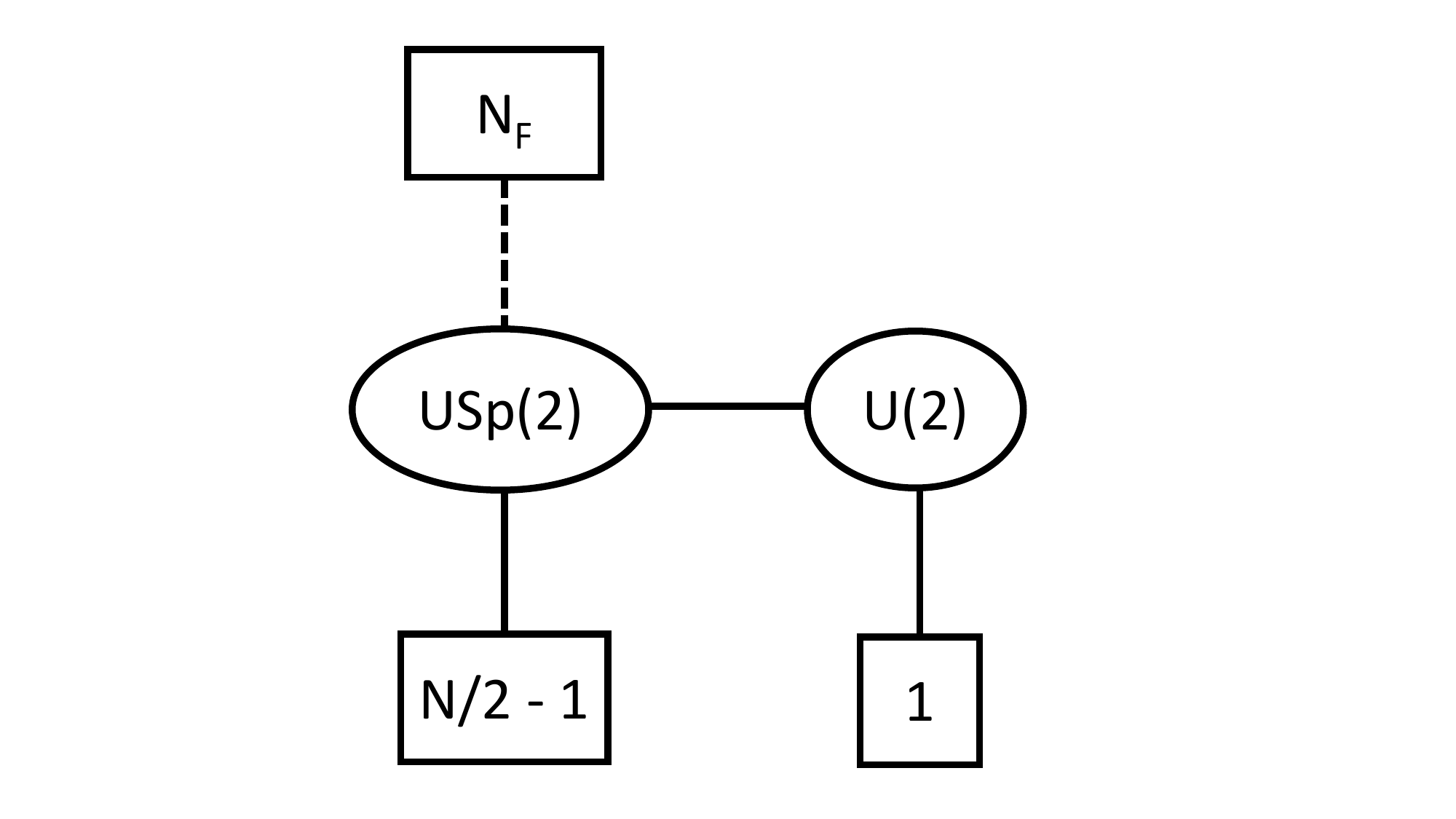}}
\caption{(a): The quiver diagram read off from Figure~\ref{subfig:O4-tHooft-wedge2V-Vb}.
(b): The quiver diagram read off from Figure~\ref{subfig:O4-tHooft-wedge2V-Vc}.
}
\label{fig:quiverSO-A}
\end{figure}

We wish to compare the quantities~$Z_{\text{mono}}^{\wedge^2 V \times V}({\bm v} = {\bm e}_i)$ and $Z_{\text{mono}}^{V \times \wedge^2 V}({\bm v} = {\bm e}_i)$ in~(\ref{SOwedge2VVv1}) and (\ref{SOwedge2VVv2}) with the supersymmetric indices of the corresponding SQMs.
To read off the SQMs which describe monopole screening we use brane configurations shown in Figure~\ref{fig:O4-tHooft-wedge2V-V}. 
Two ways of moving the NS5'-branes via the Hanany--Witten transition yield two different SQMs that compute monopole screening contributions.
The SQMs obtained from the configurations in Figures~\ref{subfig:O4-tHooft-wedge2V-Vb} and~\ref{subfig:O4-tHooft-wedge2V-Vb} are depicted in Figures~\ref{fig:quiverSOv3} and~\ref{fig:quiverSOv4}, respectively.
Since the configurations are related by Hanany--Witten transitions, the two SQMs are dual to each other and possess identical indices.
It is thus enough to consider one of the SQMs.
Let us focus on the SQM specified by the quiver in~Figure \ref{fig:quiverSOv3}. Its index is given by 
\begin{equation}\label{ZSO2nv2}
\begin{aligned}
Z(\bv = \be_i, \zeta) &= \frac{1}{2}\oint_{JK( \zeta)}\frac{d\phi_1}{2\pi i}\frac{d\phi_2}{2\pi i}2\sinh(\pm \phi_1)
\\
&\qquad\times
\frac{\left(2\sinh\epsilon_+\right)^22\sinh(\pm \phi_1 + \epsilon_+)\prod_{f=1}^{N_F}2\sinh\frac{\pm \phi_1 - m_f}{2}}{\left(\prod_{1 \leq j \neq i \leq n}2\sinh\frac{\pm \phi_1 \pm  a_j + \epsilon_+}{2}\right)2\sinh\frac{\pm \phi_1 \pm  \phi_2 + \epsilon_+}{2}2\sinh\frac{\pm(\phi_2 - a_i + \epsilon_+)}{2}}
\end{aligned}
\end{equation}
for $N=2n$ and 
\begin{equation}\label{ZSO2np1v2}
\begin{split}
Z(\bv = \be_i, \zeta) &= \frac{1}{2}\oint_{JK( \zeta)}\frac{d\phi_1}{2\pi i}\frac{d\phi_2}{2\pi i}2\sinh(\pm \phi_1)\frac{\left(2\sinh\epsilon_+\right)^22\sinh(\pm \phi_1 + \epsilon_+)\prod_{f=1}^{N_F}2\sinh\frac{\pm \phi_1 - m_f}{2}}{2\sinh\frac{\pm \phi_1 + \epsilon_+}{2}\prod_{1\leq j\neq i\leq n}2\sinh\frac{\pm \phi_1 \pm  a_j + \epsilon_+}{2}}\cr
&\hspace{6cm}\times\frac{1}{2\sinh\frac{\pm \phi_1 \pm  \phi_2 + \epsilon_+}{2}2\sinh\frac{\pm(\phi_2 - a_i + \epsilon_+)}{2}}
\end{split}
\end{equation}
for $N=2n+1$. We can explicitly perform the JK residue computations for \eqref{ZSO2nv2} and \eqref{ZSO2np1v2} with the JK parameter $\bm\eta = (\delta, \zeta)$ where $|\delta| \ll  |\zeta|$. 
Comparing the results with the monopole screening contributions \eqref{SOwedge2VVv1} and \eqref{SOwedge2VVv2} yields the relations
\begin{equation}
Z_{\text{mono}}^{\wedge^2 V \times V}({\bm v} = {\bm e}_i) = Z(\bv = \be_i, \zeta<0), \qquad  Z_{\text{mono}}^{V \times \wedge^2 V}({\bm v} = {\bm e}_i) = Z(\bv = \be_i, \zeta > 0).
\end{equation}
We also checked that $Z(\bv = \be_i, \zeta > 0) = Z(\bv = \be_i, \zeta < 0)$ for $(N, N_F)=(8, 3), (7, 2)$.


\subsection{Correlators involving $T_{\wedge^2 V}$ in $USp(N)$ SQCD}
\label{sec:USP-SQCD-wedge2V}

We next consider monopole screening contributions involving $T_{\wedge^2 V}$ in an $USp(N)$ SQCD. 

\subsubsection{$\wedge^2 V$}
\label{sec:USp.adj}

The next case is the expectation value of the 't~Hooft operator $T_{\wedge^2 V}$ in the 4d $\mathcal{N}=2$ $USp(2n)$ gauge theory with $N_F$ hypermultiplets in the vector representation, assuming the operator $T_{\wedge^2 V}$ exists in the theory. The number of the flavors satisfy $N_F \leq 2n+2$ for focusing on asymptotic free or superconformal field theories. 
$\wedge^2 V$ represents the rank-2 anti-symmetric representation of the Langlands dual group which is $SO(2n+1)$, which is the adjoint representation of $SO(2n+1)$. 
The expectation value of the 't~Hooft operator $T_{\wedge^2 V}$ takes the same form as \eqref{Twedge2V.N2starUSp}, namely
\begin{equation}\label{Twedge2V.USpSQCD}
\begin{split}
\Braket{T_{\wedge^2V}} &= \sum_{1\leq i < j \leq n}\left(e^{b_i + b_j} + e^{-b_i - b_j}\right)Z_{ij}({\bm a}) + \sum_{1\leq i < j \leq n}\left(e^{b_i - b_j} + e^{-b_i + b_j}\right)Z'_{ij}({\bm a}) \cr
&\hspace{0.5cm} + \sum_{i=1}^n\left(e^{b_i} + e^{-b_i}\right)Z_i({\bm a})Z'_{\text{mono}, i}({\bm a}) + Z''_{\text{mono}}({\bm a}),
\end{split}
\end{equation}
where $Z_i$ is defined in~\eqref{USpTZi}, $Z_{ij}$ in~(\ref{ZijUSp}), and $Z'_{ij}$ by
\begin{equation}
\begin{split}
Z'_{ij}({\bm a}) &= \Bigg(\frac{\prod_{f=1}^{N_F}2\sinh\frac{\pm a_i - m_f}{2}2\sinh\frac{\pm a_j - m_f}{2}}{2\sinh\frac{\pm (a_i - a_j)}{2}2\sinh\frac{\pm (a_i - a_j) + 2\epsilon_+}{2}
\hspace{-3mm}
{\displaystyle\prod_{1 \leq k\neq i, k\neq j \leq n}}
\hspace{-3mm}
2\sinh\frac{\pm a_i \pm a_k + \epsilon_+}{2}2\sinh\frac{\pm a_i \pm a_k + \epsilon_+}{2}}\cr
&\hspace{2cm}\times\frac{1}{2\sinh(\pm a_i) 2\sinh(\pm a_j) 2\sinh(\pm a_i + \epsilon_+) 2\sinh(\pm a_j + \epsilon_+) }\Bigg)^{\frac{1}{2}}.
\end{split}
\end{equation}

\begin{figure}[t]
\centering
\hspace{2cm}
\subfigure[]{\label{subfigO4-USp-tHooft-wedge2V-v=eN}
\raisebox{5mm}{\includegraphics[width=5cm]{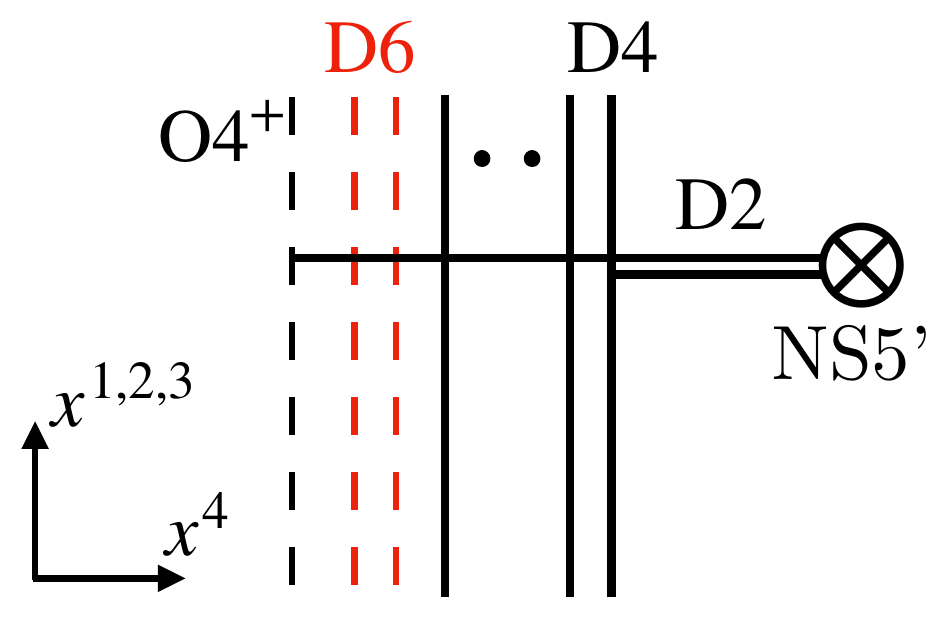}}}
\subfigure[]{\label{subfig:quiverUSp5}
\hspace{-4.5mm}
\includegraphics[width=8cm]{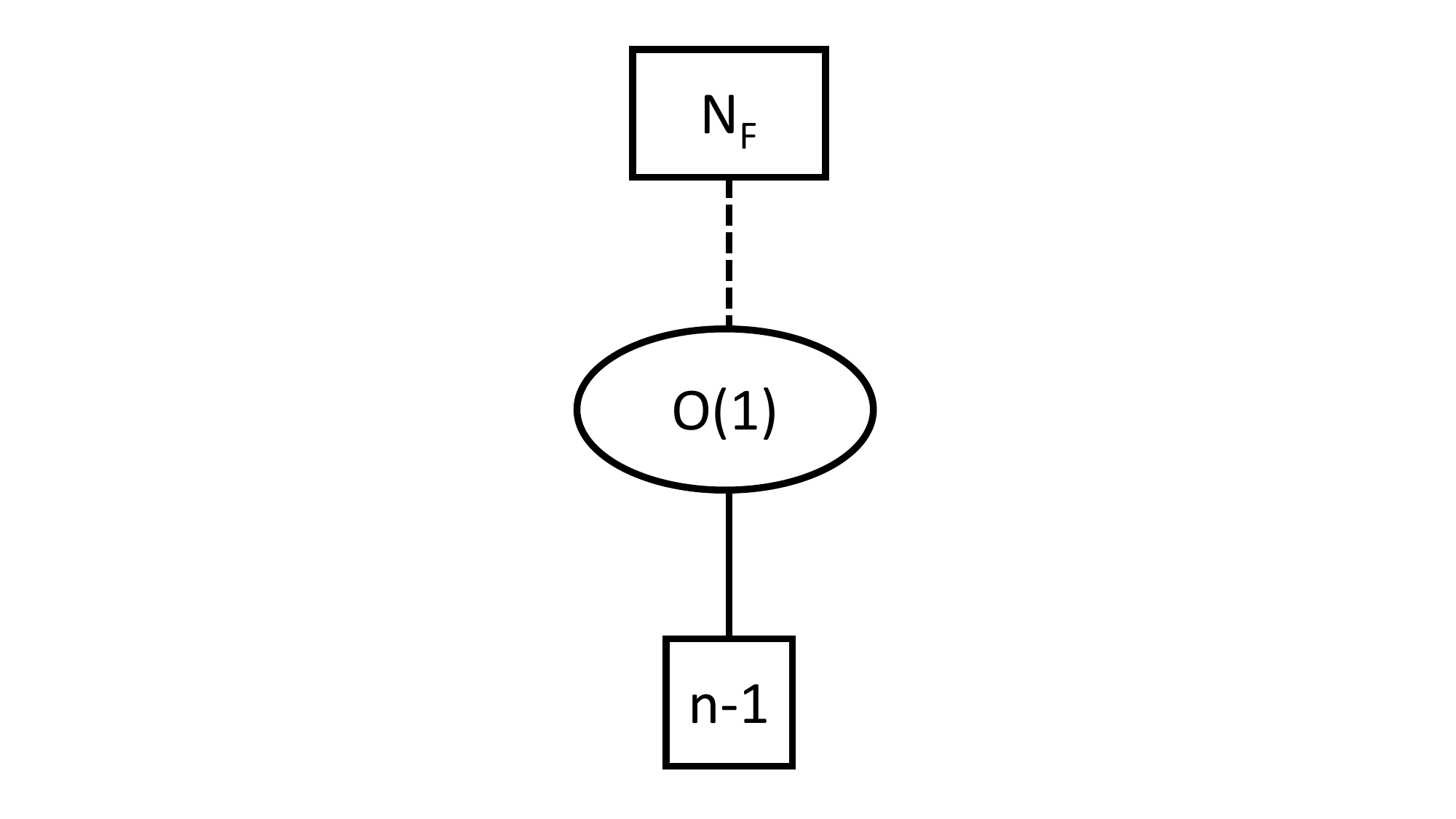}}
\caption{(a). The brane configuration that describes the monopole screening sector $\bm{v}=\be_N$ and the contribution $Z'_{\text{mono},N}({\bm a})$ in $\Braket{T_{\wedge^2V}}$. (b). The corresponding quiver diagram.
}
\label{fig:quiverUSpv2_O1}
\end{figure}

As can be seen from the explicit form of \eqref{Twedge2V.USpSQCD}, we have two types of the monopole screening contributions $Z'_{\text{mono}, i}(\ba)$ and $Z''_{\text{mono}}(\ba)$. 
The SQM which describes the contribution $Z'_{\text{mono},i}({\bm a})$ can be read off from the brane configuration in Figure~\ref{subfigO4-USp-tHooft-wedge2V-v=eN}, which leads to the quiver in Figure~\ref{subfig:quiverUSp5}. 
The computation of the supersymmetric index of the quiver in Figure \ref{subfig:quiverUSp5} is similar to the one for the quiver theory in Figure \ref{subfig:quiverUSp1}. Since the computation is parallel to \eqref{ZmonoUSp}, 
we here simply write the final result, which is 
\begin{equation}\label{ZpmonoUSp}
\begin{split}
Z'_{\wedge^2V, i}(\bv={\bm e}_i ) =\frac{1}{2}\left(\frac{\prod_{f=1}^{N_F}2\sinh\frac{m_f}{2}}{\prod_{1\leq j\neq i \leq n}2\sinh\frac{\pm a_j + \epsilon_+}{2}} + \frac{\prod_{f=1}^{N_F}2\cosh\frac{m_f}{2}}{\prod_{1\leq j \neq i \leq n}2\cosh\frac{\pm a_j + \epsilon_+}{2}}\right).
\end{split}
\end{equation}
In this case there is no extra term, 
 and 
$Z'_{\text{mono}, i}(\ba)$ in \eqref{Twedge2V.USpSQCD} is given by \eqref{ZpmonoUSp}.

\begin{figure}[t]
\centering
\hspace{2cm}
\subfigure[]{\label{subfig:O4-USp-tHooft-wedge2V-v=0}
\raisebox{5mm}{\includegraphics[width=5cm]{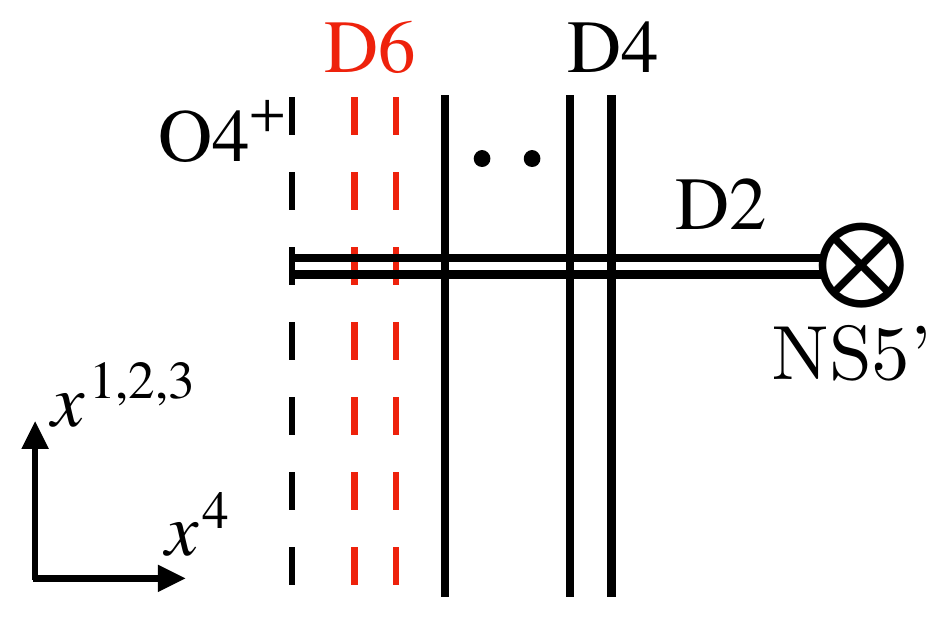}}}
\subfigure[]{\label{subfig:quiverUSp4}
\hspace{-4.5mm}
\includegraphics[width=8cm]{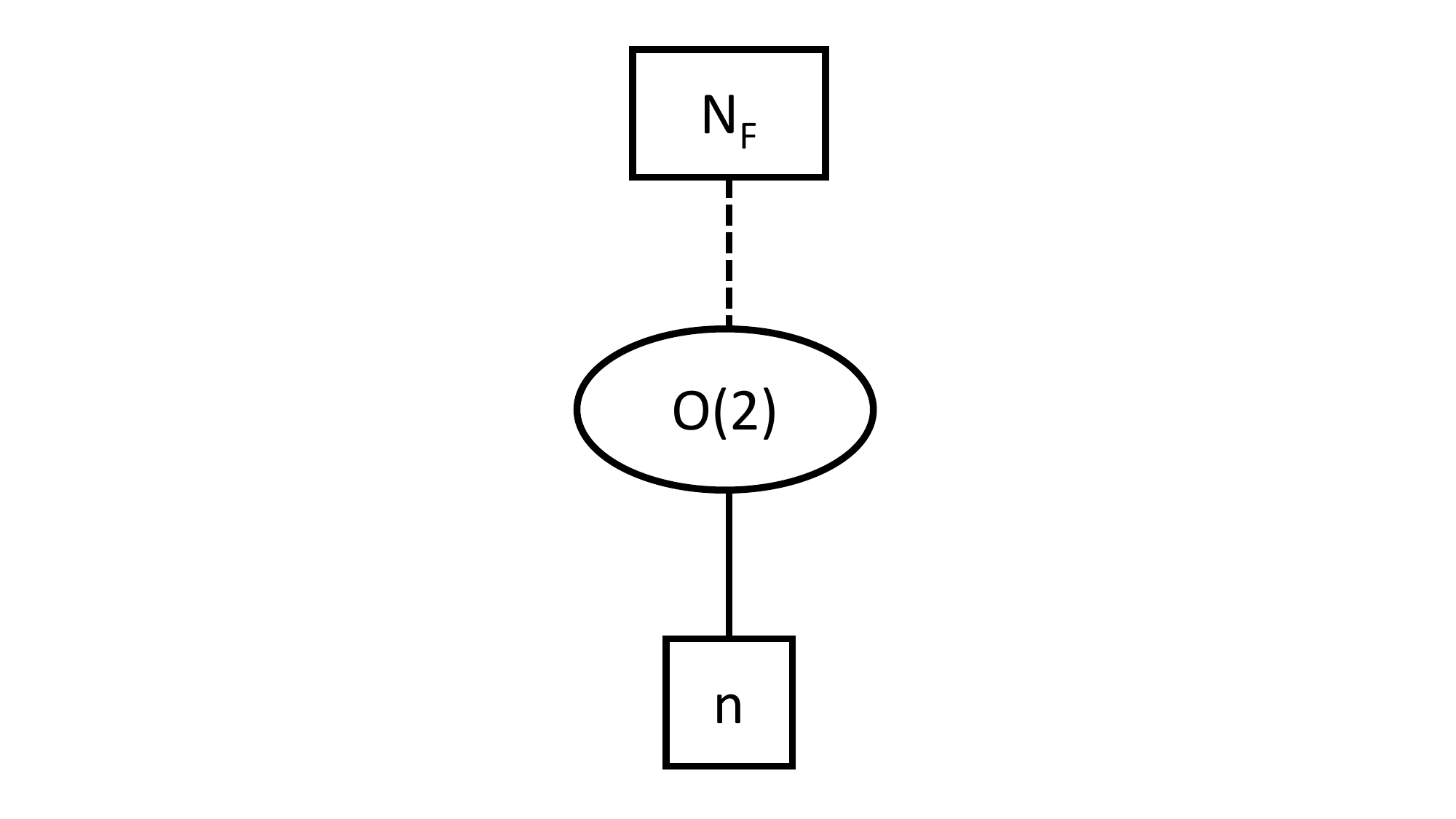}}
\caption{(a): The brane configuration for the monopole screening sector~$\bm{v}=\bm{0}$ of the 't~Hooft operator~$T_{\wedge^2V}$.
(b): The corresponding SQM quiver, which computes~$Z''_{\text{mono}}$.
}
\label{fig:quiverUSpv2_O2}
\end{figure}
The other monopole screening contribution $Z''_{\text{mono}}({\bm a})$ corresponds to the zero weights of the adjoint representation of $SO(2n+1)$ and the brane configuration in Figure~\ref{subfig:O4-USp-tHooft-wedge2V-v=0} for the sector gives rise to the quiver theory depicted in Figure~\ref{subfig:quiverUSp4}. Similar to the $O(1)$ case in Section \ref{sec:USp.fund}, the $O(2)$ integral can be split into two contributions $O(2)_+$ and $O(2)_-$. The contribution from $O(2)_+$ is given by
\begin{align}\label{Zmonowedge2VUSp}
Z''_{\wedge^2V+}(\bv ={\bm 0})&= \oint_{JK(\eta)}\frac{d\phi}{2\pi i}\frac{2\sinh(\epsilon_+)\prod_{f=1}^{N_F}2\sinh\frac{\pm\phi - m_f}{2}}{\prod_{i=1}^n2\sinh\frac{\pm\phi \pm a_i + \epsilon_+}{2}}.
\end{align}
The integrand in \eqref{Zmonowedge2VUSp} also has a higher order pole at the infinities $\phi = \pm \infty$ in the superconformal case with $N_F = 2n+2$. In order to avoid the poles at the infinites, we choose $N_F \leq 2n-1$. Then the integral of \eqref{Zmonowedge2VUSp} can be evaluated by taking the sum of the contributions from the poles $\phi = \pm a_i - \epsilon_+$ or $\phi = \pm a_i + \epsilon_+$ and the both give
\begin{equation}\label{Twedge2V.USp+}
\begin{aligned}
Z''_{\wedge^2V+}(\bv ={\bm 0})&=\sum_{i=1}^n\left(\frac{\prod_{f=1}^{N_F}2\sinh\frac{a_i - m_f - \epsilon_+}{2}2\sinh\frac{-a_i - m_f + \epsilon_+}{2}}{2\sinh(a_i)2\sinh(-a_i+\epsilon_+)\prod_{1\leq j\neq i\leq n}2\sinh\frac{a_i \pm a_j}{2}2\sinh\frac{-a_i \pm a_j + 2\epsilon_+}{2}}\right.\\
&\hspace{1cm}\left.+\frac{\prod_{f=1}^{N_F}2\sinh\frac{-a_i - m_f - \epsilon_+}{2}2\sinh\frac{a_i - m_f + \epsilon_+}{2}}{2\sinh(-a_i)2\sinh(a_i+\epsilon_+)\prod_{1\leq j\neq i\leq n}2\sinh\frac{- a_i \pm a_j}{2}2\sinh\frac{a_i  \pm a_j + 2\epsilon_+}{2}}\right).
\end{aligned}
\end{equation}
On the other hand, the contribution from $O(2)_-$ is 
\begin{equation}\label{Twedge2V.USp-}
\begin{split}
Z''_{\wedge^2V-}(\bv ={\bm 0})=\frac{2\cos(\epsilon_+)\prod_{f=1}^{N_F}2\sinh(m_f)}{\prod_{i=1}^n2\sinh(\pm a_i + \epsilon_+)}.
\end{split}
\end{equation}
Then the supersymmetric index of the quiver in Figure \ref{subfig:quiverUSp4} is given by the average of \eqref{Twedge2V.USp+} and \eqref{Twedge2V.USp-}, namely
\begin{equation}\label{Twedge2V.USpApp}
Z''_{\wedge^2V}(\bv ={\bm 0})= \frac{1}{2}(Z''_{\wedge^2V+}(\bv ={\bm 0})+ Z''_{\wedge^2V-}(\bv ={\bm 0})).
\end{equation}
Since the monopole screening contribution in $\Braket{T_V}$ of the theory does not have an extra term, we can apply the extra term prescription to the monopole screening contribution $Z''_{\text{momo}}(\ba)$ in $\Braket{T_{\wedge^2V}}$ \footnote{See footnote \ref{fn:extra}}. It turns out that in this case also 
there is no extra term, which we checked for $n=1, 2, 3, 4, 5$, and $Z''_{\text{mono}}({\bm a})$ in \eqref{Twedge2V.USpSQCD} is given by \eqref{Twedge2V.USpApp} for $N_F \leq 2n-1$.

\subsubsection{$\wedge^2 V \times V$}

It is also possible to consider the expectation value of the product of $T_{\wedge^2 V}$ and $T_V$. The expectation value may depend on the order of the operators generally. Here we also focus on the screening sector characterized by $\bv = \be_i$ and $\bv = {\bm 0}$. 

\paragraph{${\bm v} = {\bm e}_i$.} 
The monopole screening contribution of $\Braket{T_{\wedge^2V}}\ast\Braket{T_V}$ in the sector ${\bm v}=\be_i$ is given by 
\begin{equation}
\begin{split}
&\Braket{T_{\wedge^2V}}\ast\Braket{T_V}\Big|_{\bv=\be_i}\cr
=&e^{b_i}Z_i({\bm a})Z'_{\text{mono}, i}({\bm a}) Z_{\text{mono}}({\bm a} - \epsilon_+{\bm e}_i) + e^{b_i}Z''_{\text{mono}}({\bm a} + \epsilon_+{\bm e}_i)Z_i({\bm a})\cr
&+e^{b_i}\sum_{1 \leq j \neq i\leq n}\left(Z_{ij}({\bm a} + \epsilon_+(-{\bm e}_j))Z_j({\bm a} - \epsilon_+({\bm e}_i + {\bm e}_j)) +Z'_{ij}({\bm a} + \epsilon_+{\bm e}_j)Z_j({\bm a} - \epsilon_+({\bm e}_i - {\bm e}_j))\right)\cr
=:&e^{b_i}Z_{\text{1-loop}}(\bv = \be_i)Z_{\text{mono}}^{\wedge^2 V \times V}({\bm v}  =\be_i).\label{USpTwedge2VVvi}
\end{split}
\end{equation}
Similarly the monopole screening contribution of $\Braket{T_V}\ast\Braket{T_{\wedge^2 V}}$ in the sector ${\bm v}=\be_i$ is given by 
\begin{equation}
\begin{split}
&\Braket{T_V}\ast\Braket{T_{\wedge^2 V}}\Big|_{\bv = \be_i}\cr
=&e^{b_i}Z_{\text{mono}}({\bm a} + \epsilon_+ {\bm e}_i)Z_i({\bm a})Z'_{\text{mono}, i}({\bm a}) + e^{b_i}Z_i({\bm a})Z''_{\text{mono}}({\bm a} - \epsilon_+ {\bm e}_i)\cr
&+e^{b_i}\sum_{1 \leq j \neq i\leq n}\left(Z_j({\bm a} + \epsilon_+({\bm e}_i + {\bm e}_j))Z_{ij}({\bm a} - \epsilon_+(-{\bm e}_j))
+Z_j({\bm a} + \epsilon_+({\bm e}_i - {\bm e}_j))Z'_{ij}({\bm a} - \epsilon_+{\bm e}_j)\right)\cr
=:&e^{b_i}Z_{\text{1-loop}}(\bv = \be_i)Z_{\text{mono}}^{V \times \wedge^2 V}({\bm v}  =\be_i).\label{USpTVwedge2Vvi}
\end{split}
\end{equation}

\begin{figure}[t]
\centering
\subfigure[]{\label{fig:quiverUSp7v1}
\includegraphics[width=8cm]{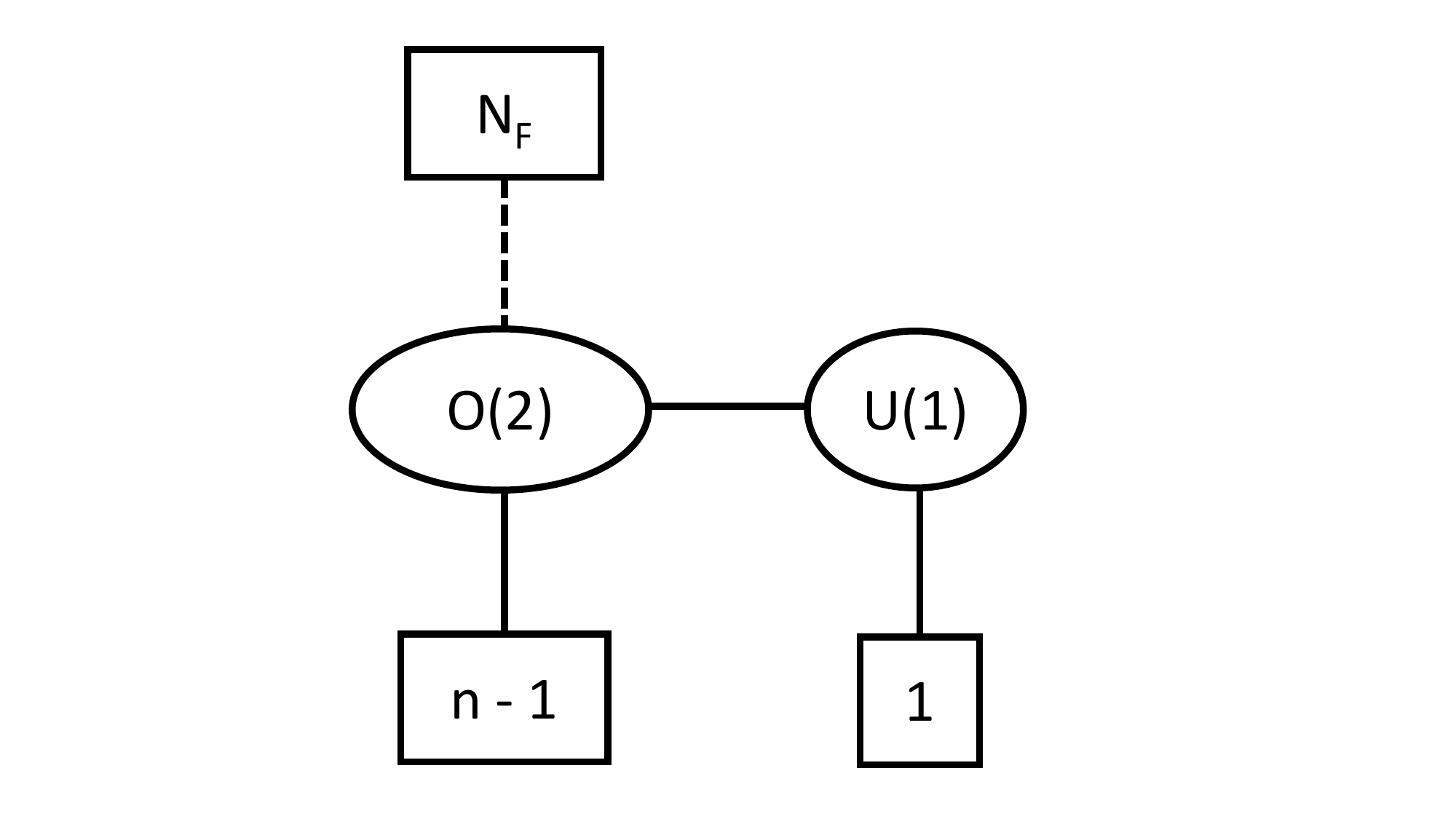}}
\subfigure[]{\label{fig:quiberUSp7v2}
\includegraphics[width=8cm]{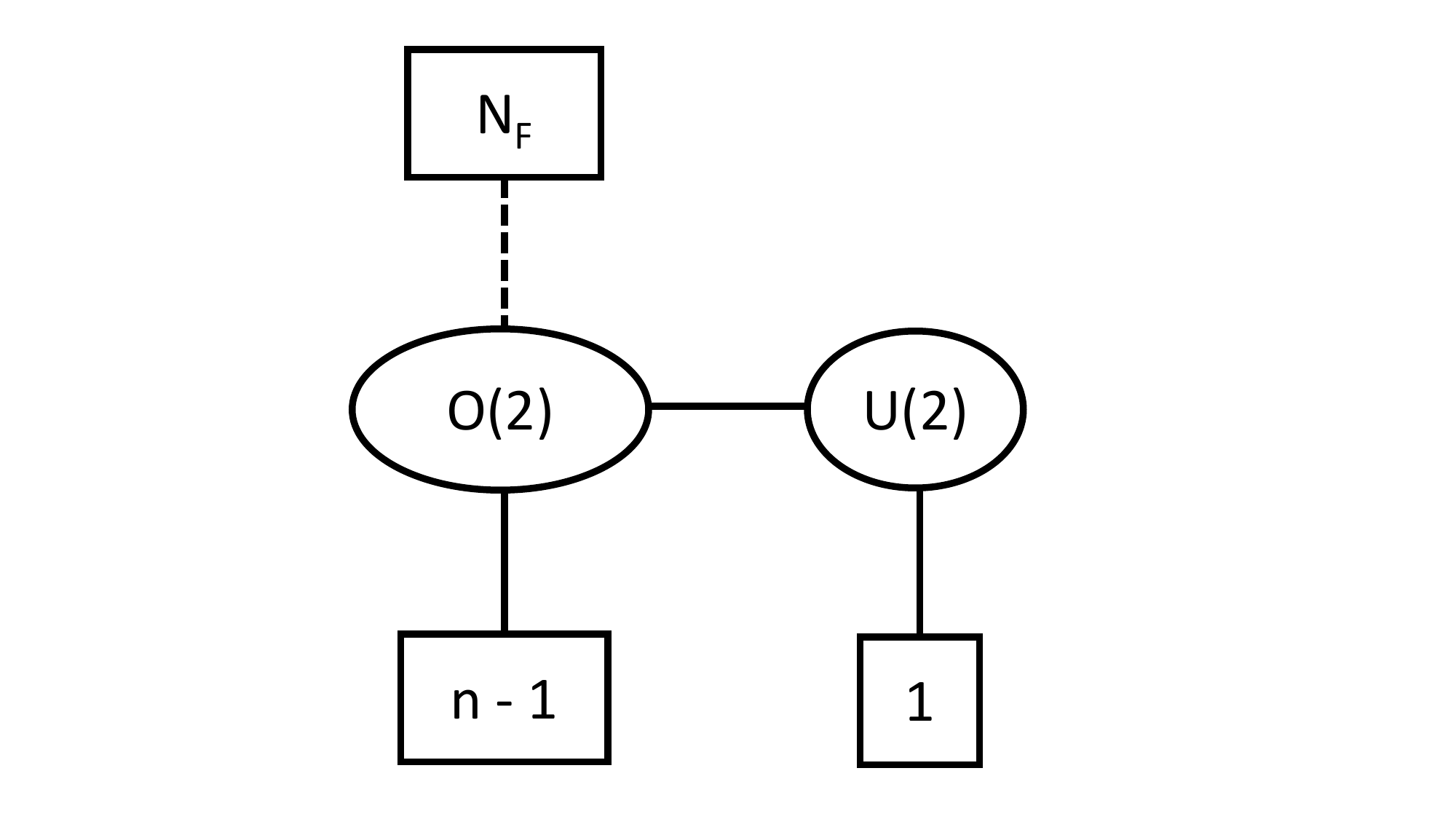}}
\caption{The two possible quiver diagrams describing the monopole screening contribution of the $\bv = {\bm e}_i$ sector in $\Braket{T_{\wedge^2 V}}\ast\Braket{T_V}$ or $\Braket{T_V}\ast\Braket{T_{\wedge^2 V}}$.
}
\label{fig:quiverUSp7}
\end{figure}

We compute the monopole screening contributions $Z_{\text{mono}}^{\wedge^2 V \times V}({\bm v}  =\be_i)$ and $Z_{\text{mono}}^{V \times \wedge^2 V}({\bm v}  =\be_i)$ from the supersymmetric index of the corresponding SQM. 
The brane configurations for these sectors with $i=N$ are those in Figures~\ref{subfig:O4-tHooft-wedge2V-Va} and~\ref{subfig:O4-tHooft-wedge2V-Vb}, where we take the O4-plane to be of the type ${\rm O4}^+$.
These configurations yield the two quiver theories depicted in Figure~\ref{fig:quiverUSp7}. 
However the brane configurations yielding the quiver theories in Figure \ref{fig:quiverUSp7} are related by moving D-branes and hence they are dual to each other. We will use the simpler quiver theory in Figure \ref{fig:quiverUSp7v1} to compute the supersymmetric index.

The supersymmetric index consists of contributions from two sectors, $O(2)_+-U(1)$ and $O(2)_--U(1)$. The contribution from the $O(2)_+-U(1)$ sector is given by
\begin{equation}\label{USpO2pU1}
\begin{split}
Z_+(\bv = \be_i, \zeta) = \oint_{JK(\zeta)}\frac{d\phi_1}{2\pi i}\frac{d\phi_2}{2\pi i}\frac{\left(2\sinh(\epsilon_+)\right)^2\prod_{f=1}^{2n+2}2\sinh\frac{\pm\phi_1 - m_f}{2}}{2\sinh\frac{\pm\phi_1 \pm \phi_2 + \epsilon_+}{2}\prod_{1\leq j\neq i \leq n}2\sinh\frac{\pm\phi_1 \pm a_j + \epsilon_+}{2}2\sinh\frac{\pm(\phi_2 - a_i) + \epsilon_+}{2}},
\end{split}
\end{equation}
where $\zeta$ is the FI parameter for the $U(1)$ gauge group. The contribution from the $O(2)_--U(1)$ sector is 
\begin{equation}\label{USpO2mU1}
\begin{split}
Z_-(\bv = \be_i, \zeta) = \oint_{JK(\zeta)}\frac{d\phi_2}{2\pi i}\frac{2\cos(\epsilon_+)2\sinh(\epsilon_+)\prod_{f=1}^{2n+2}2\sinh(m_f)}{2\sinh(\pm \phi_2 + \epsilon_+)\prod_{1\leq j\neq i\leq n}2\sinh(\pm a_j + \epsilon_+)2\sinh\frac{\pm(\phi_2 - a_i) + \epsilon_+}{2}}.
\end{split}
\end{equation}
The supersymmetric index is then given by
\begin{align}\label{ZUSpwedge2VVei}
Z(\bv = \be_i, \zeta) = \frac{1}{2}(Z_+(\bv = \be_i, \zeta) + Z_-(\bv = \be_i, \zeta)).
\end{align}

We compare \eqref{ZUSpwedge2VVei} with (\ref{USpTwedge2VVvi}, \ref{USpTVwedge2Vvi}) calculated from the Moyal product. 
We checked the following equality explicitly for $n=2$ and $N_F=3$, and believe that it holds generally:
\begin{align}
Z(\bv = \be_i, \zeta) = Z_{\text{mono}}^{\wedge^2 V \times V}({\bm v}  =\be_i) = Z_{\text{mono}}^{V\times \wedge^2 V}({\bm v}  =\be_i).
\end{align}
for both $\zeta > 0$ and $\zeta < 0$.

\paragraph{$\bv = {\bm 0}$.}
The $\bv = {\bm 0}$ sector in the Moyal product of the form $\Braket{T_{\wedge^2 V}}\ast\Braket{T_V}$ is given by
\begin{equation}
\begin{split}
\Braket{T_{\wedge^2 V}} \ast \Braket{T_V}\Big|_{\bv = {\bm 0}}
&=\sum_{i=1}^n\Big(Z_i({\bm a} - \epsilon_+{\bm e}_i)^2Z'_{\text{mono}, i}({\bm a} - \epsilon_+{\bm e}_i) 
\\
&\qquad\qquad
+ Z_i({\bm a} + \epsilon_+{\bm e}_i)^2Z'_{\text{mono}, i}({\bm a} + \epsilon_+{\bm e}_i)\bigg)
+Z_{\text{mono}}({\bm a})Z''_{\text{mono}}({\bm a})\cr
&:=Z_{\text{mono}}^{\wedge^2 V \times V}({\bm v} = {\bm 0}), \label{USpTwedge2VVv0}
\end{split}
\end{equation}
The $\bv = {\bm 0}$ sector in the Moyal product for the other order yields 
\begin{equation}
\begin{aligned}
\Braket{T_V} \ast \Braket{T_{\wedge^2 V}}\Big|_{\bv = {\bm 0}}
&= \sum_{i=1}^n\Big(Z_i({\bm a} - \epsilon_+{\bm e}_i)^2Z'_{\text{mono}, i}({\bm a} - \epsilon_+{\bm e}_i) 
\\
&\qquad\qquad
+ Z_i({\bm a} + \epsilon_+{\bm e}_i)^2Z'_{\text{mono}, i}({\bm a} + \epsilon_+{\bm e}_i)\Big)
+Z_{\text{mono}}({\bm a})Z''_{\text{mono}}({\bm a})\cr
&:=Z_{\text{mono}}^{V \times \wedge^2 V}({\bm v} = {\bm 0}). \label{USpTVwedge2Vv0}
\end{aligned}
\end{equation}
The sector arise from ${\bm v} = ({\bm e}_i) + (-{\bm e}_i), (-{\bm e}_i) + ({\bm e}_i), {\bm 0} + {\bm 0}$. Note that from the explicit form of \eqref{USpTwedge2VVv0} and \eqref{USpTVwedge2Vv0} the both contributions give the same result in this sector. 

\begin{figure}[t]
\centering
\subfigure[]{\label{subfig:O4-USp-tHooft-wedge2VxV-v=0}
\raisebox{4mm}{\includegraphics[width=5cm]{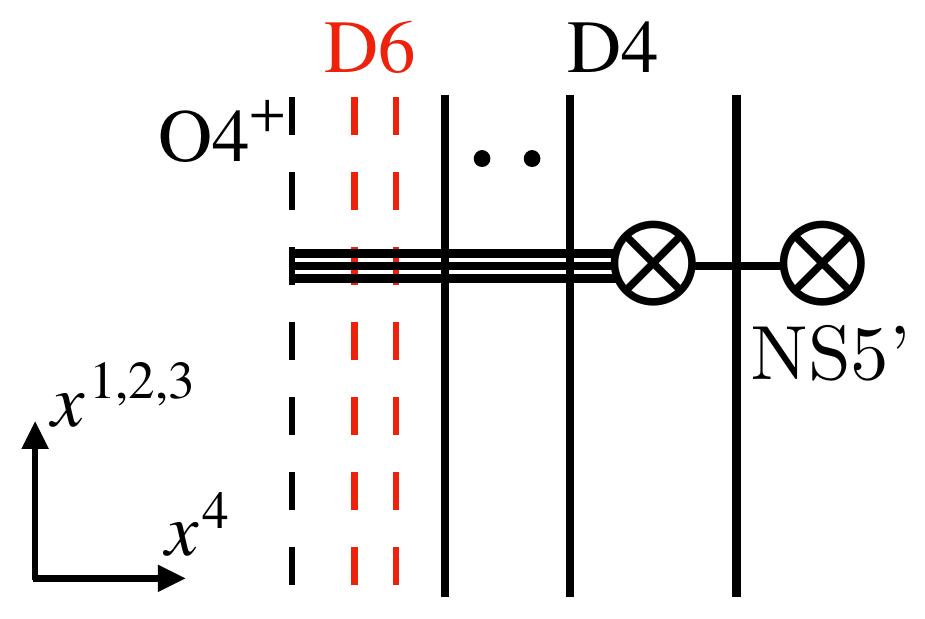}}}
\hspace{1cm}
\subfigure[]{\label{subfig:quiverUSp8}
\includegraphics[width=8cm]{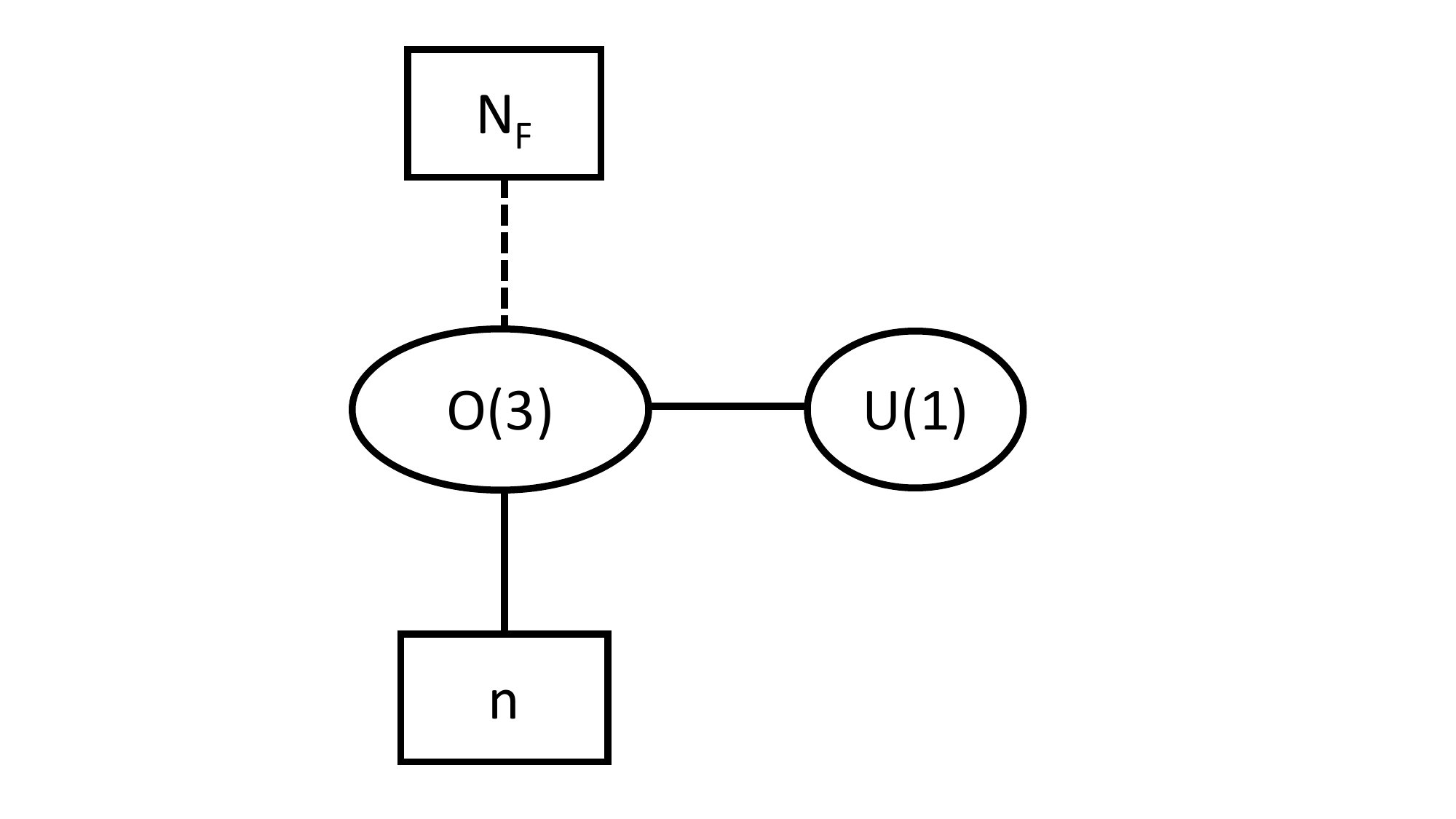}}
\caption{(a): The brane configuration for the monopole screening sector~$\bv = {\bm 0}$ of $T_{\wedge^2 V}\cdot T_{V}$.
(b): The corresponding quiver diagram.
}
\label{fig:quiverUSp8}
\end{figure}

Let us compare~$Z_{\text{mono}}^{V \times \wedge^2 V}({\bm v} = {\bm 0})$ in~(\ref{USpTVwedge2Vv0}) with the supersymmetric index of the corresponding SQM. The brane configuration shown in Figure~\ref{subfig:O4-USp-tHooft-wedge2VxV-v=0} gives the quiver depicted in Figure~\ref{subfig:quiverUSp8}. 
The supersymmetric index of the quiver theory can be computed in a similar manner. The contribution consists of contributions from two components, $O(3)_+-U(1)$ and $O(3)_--U(1)$. The contribution from $O(3)_+-U(1)$ is given by
\begin{equation}\label{USpO3pU1}
\begin{split}
Z_+(\bv = {\bm 0}, \zeta) &= \frac{1}{2}\oint_{JK(\zeta)}\frac{d\phi_1}{2\pi i}\frac{d\phi_2}{2\pi i}2\sinh\frac{\pm\phi_1}{2}2\sinh\frac{\pm\phi_1 + 2\epsilon_+}{2}\left(2\sinh\epsilon_+\right)^2\cr
&\hspace{-1cm}\times \frac{\left(\prod_{f=1}^{N_F}2\sinh\frac{\phi_1-m_f}{2}2\sinh\frac{m_f}{2}\right)}{\left(\prod_{i=1}^n2\sinh\frac{\pm\phi_1\pm a_i + \epsilon_+}{2}2\sinh\frac{\pm a_i + \epsilon_+}{2}\right)\left(2\sinh\frac{\pm\phi_1 \pm \phi_2 + \epsilon_+}{2}2\sinh\frac{\pm \phi_2 + \epsilon_+}{2}\right)},
\end{split}
\end{equation}
and the contribution from $O(3)_- - U(1)$ is given by
\begin{equation}\label{USpO3mU1}
\begin{split}
Z_-(\bv = {\bm 0}, \zeta) &= \oint_{JK(\zeta)}\frac{d\phi_1}{2\pi i}\frac{d\phi_2}{2\pi i }2\cosh\frac{\pm \phi_1}{2}2\cosh\frac{\pm\phi_1 + 2\epsilon_+}{2}\left(2\sinh\epsilon_+\right)^2\cr
&\hspace{-1cm}
\times \frac{\left(\prod_{f=1}^{N_F}2\sinh\frac{\pm \phi_1 + m_f}{2}2\cosh\frac{m_f}{2}\right)}{\left(\prod_{i=1}^n2\sinh\frac{\pm\phi_1\pm a_i + \epsilon_+}{2}2\cosh\frac{\pm a_i + \epsilon_+}{2}\right)\left(2\sinh\frac{\pm\phi_1 \pm \phi_2 + \epsilon_+}{2}2\sinh\frac{\pm \phi_2 + \epsilon_+}{2}\right)}.
\end{split}
\end{equation}
Then the  supersymmetric index is given by
\begin{equation}\label{ZUSpwedge2VV0}
Z({\bm v} = {\bm 0}, \zeta) = \frac{1}{2}\left(Z_+(\bv = {\bm 0}, \zeta)  + Z_-(\bv = {\bm 0}, \zeta) \right).
\end{equation}

We can then compare the supersymmetric index \eqref{ZUSpwedge2VV0} with the Moyal product \eqref{USpTwedge2VVv0} or \eqref{USpTVwedge2Vv0}. For the explicit evaluation of the supersymmetric index \eqref{ZUSpwedge2VV0}, we use the JK parameter~$\bm\eta = (\delta, \zeta)$ with $|\delta| \ll   |\zeta|$. Then we find 
\begin{equation}
Z({\bm v} = {\bm 0}, \zeta) =  Z_{\text{mono}}^{\wedge^2 V \times V}({\bm v}={\bm 0}) =  Z_{\text{mono}}^{V \times \wedge^2 V}({\bm v} = {\bm 0}),
\end{equation}
for $\zeta > 0$ and $\zeta < 0$, which we checked for $n=2$ and $N_F = 3$. 

\subsection{Naive extension to superconformal cases}
\label{sec:naive-conformal}

In superconformal cases for $SO(N)$ gauge theories and $USp(2n)$ gauge theories, we mentioned the subtlety of the presenece of higher order poles in \eqref{ZSOewedgev1}, \eqref{ZSOowedgev1} and \eqref{Zmonowedge2VUSp}. When we chose the cases of $N_F \leq N-5$ for the $SO(N)$ SQCD and of $N_f \leq 2n-1$ for the $USp(2n)$ SQCD, where there are no poles at infinities, the monopole screening contributions in $\Braket{T_{\wedge^2 V}}$ become \eqref{ZSOewedgev2}, \eqref{ZSOowedgev2} and \eqref{Twedge2V.USp+} respectively. Here we consider consequences when we simply set the $N_F$ to be the number of flavors when the theories become conformal in \eqref{ZSOewedgev2}, \eqref{ZSOowedgev2}, \eqref{ZpmonoUSp},  \eqref{Twedge2V.USp+} and \eqref{Twedge2V.USp-}.

For the $SO(N)$ gauge theory, the correlator $\Braket{T_{\wedge^2 V} \cdot T_V}$ and $\Braket{T_V \cdot T_{\wedge^2 V}}$ has a monopole screening contribution in the $\bv = \be_i$ sector and it is given by \eqref{ZSO2nv2} for $N=2n$ and \eqref{ZSO2np1v2} for $N=2n+1$ when $N_F \leq N-5$. Again let us set $N_F = N-2$ and take the same poles of the integral \eqref{ZSO2nv2} and \eqref{ZSO2np1v2} as those which we used in the less flavor cases, and then compare the sum of the residues of the poles with the result from the corresponding part of the Moyal product.

Let us denote the sets of poles by $S_1^{SO(8)}, S_{1'}^{SO(8)}, S_{2}^{SO(8)}, S_{2'}^{SO(8)}$ which are respectively related to the poles in \eqref{ZSO2nv2} from the JK parameter ${\bm \eta}$ with $\left(\delta > 0, \zeta < 0\right)$, $\left(\delta < 0, \zeta < 0\right)$, $\left(\delta > 0, \zeta > 0\right)$ and $\left(\delta < 0, \zeta > 0\right)$ in the case of $(N, N_F) = (8, 3)$. For example  in the sector $\bv = \be_4$, we have
\begin{equation}\label{so8pole1}
\begin{split}
S_{1}^{SO(8)}:\{(\phi_1, \phi_2)\} =& \{ (0 , \epsilon _+ ), 
 (a_4 , a_4+\epsilon _+ ), 
 (-a_4-2 \epsilon _+ , a_4+\epsilon _+ ),
 (-a_1-\epsilon _+ , a_4+\epsilon _+ ),\cr
 &(-a_1-\epsilon _+ , a_1+2 \epsilon _+ ),
 (a_1-\epsilon _+ , a_4+\epsilon _+ ),
 (a_1-\epsilon _+ , 2 \epsilon _+-a_1 ),\cr
& (-a_2-\epsilon _+ , a_4+\epsilon _+ ),
 (-a_2-\epsilon _+ , a_2+2 \epsilon _+ ),
 (a_2-\epsilon _+ , a_4+\epsilon _+ ),\cr
& (a_2-\epsilon _+ , 2 \epsilon _+-a_2 ),
 (-a_3-\epsilon _+ , a_4+\epsilon _+ ),
 (-a_3-\epsilon _+ , a_3+2 \epsilon _+ ),\cr
& (a_3-\epsilon _+ , a_4+\epsilon _+ ),
 (a_3-\epsilon _+ , 2 \epsilon _+-a_3 ),
 (\epsilon _+-a_1 , 2 \epsilon _+-a_1 ),\cr
& (a_1+\epsilon _+ , a_1+2 \epsilon _+ ),
 (\epsilon _+-a_2 , 2 \epsilon _+-a_2 ),
 (a_2+\epsilon _+ , a_2+2 \epsilon _+ ),\cr
& (\epsilon _+-a_3 , 2 \epsilon _+-a_3 ),
 (a_3+\epsilon _+ , a_3+2 \epsilon _+ )
\},
\end{split}
\end{equation}
When we evaluate the residues of \eqref{ZSO2nv2} in the case of $(N, N_F) = (8, 6)$ and $\bv = \be_4$ using the poles of $S_1^{SO(8)}$, $S_{1'}^{SO(8)}$, then their sum give the same result and both agree with $Z_{\text{mono}}^{\wedge^2 V \times V}({\bm v} = {\bm e}_4)$ in \eqref{SOwedge2VVv1} with $(N, N_F) =(8, 6)$. Similarly the sum of the residues from the poles of $S_{2}^{SO(8)}$, $S_{2'}^{SO(8)}$  for $(N, N_F) = (8, 6)$ gives the same result and it agrees with $Z_{\text{mono}}^{V \times \wedge^2 V}({\bm v} = {\bm e}_4)$ in \eqref{SOwedge2VVv2} with $(N, N_F) =(8, 6)$. In this case the result depends on the order, namely $Z_{\text{mono}}^{\wedge^2 V \times V}({\bm v} = {\bm e}_4) \neq Z_{\text{mono}}^{V \times \wedge^2 V}({\bm v} = {\bm e}_4)$.

Similarly we denote the sets of poles of \eqref{ZSO2np1v2} by $S_1^{SO(7)}, S_{1'}^{SO(7)}, S_{2}^{SO(7)}, S_{2'}^{SO(7)}$ which are respectively related to the poles from the JK parameter ${\bm \eta}$ with $\left(\delta > 0, \zeta < 0\right)$, $\left(\delta < 0, \zeta < 0\right)$, $\left(\delta > 0, \zeta > 0\right)$ and $\left(\delta < 0, \zeta > 0\right)$ in the case of $(N, N_F) = (7, 2)$. The sum of the residues of \eqref{ZSO2np1v2} from the poles of $S_{1}^{SO(7)}$ for $(N, N_F) = (7, 5)$ in the sector $\bv = \be_3$ is the same as the sum of the residues from the poles of $S_{1'}^{SO(7)}$ with the same $N$ and $N_F$ and the result agrees with $Z_{\text{mono}}^{\wedge^2 V \times V}({\bm v} = {\bm e}_3)$ in \eqref{SOwedge2VVv1} with $(N, N_F) =(7, 5)$. On the other hand, the sum of the residues from the poles of $S_2^{SO(7)}$ and that from $S_{2'}^{SO(7)}$ for $(N, N_F) = (7, 5)$ in the sector $\bv = \be_3$ give the same result which agrees with $Z_{\text{mono}}^{V \times \wedge^2 V}({\bm v} = {\bm e}_3)$ in \eqref{SOwedge2VVv2} with $(N, N_F) =(7, 5)$. In this case again we observe $Z_{\text{mono}}^{\wedge^2 V \times V}({\bm v} = {\bm e}_3)\neq Z_{\text{mono}}^{V \times \wedge^2 V}({\bm v} = {\bm e}_3)$.

We then move on to monopole screening contributions of the same type of the correlator in the $USp(2n)$ SQCD. The monopole screening contribution in the $\bv = \be_i$ sector  in the case of $N_F \leq 2n-1$ is given by \eqref{ZUSpwedge2VVei} based on \eqref{USpO2pU1} and \eqref{USpO2mU1} and the monopole screening contribution in the $\bv = {\bm 0}$ sector in the case of $N_F \leq 2n-1$ is \eqref{ZUSpwedge2VV0} based on \eqref{USpO3pU1} and \eqref{USpO3mU1}. Here we consider the sum of the residues using the same poles in the less flavor cases for the number of flavors $N_F = 2n+2$ and compare the result with that from the Moyal product. 

We start from the monopole screening contribution in the $\bv = \be_i$ sector. Let the sets of poles of \eqref{USpO2pU1} from the JK parameter ${\bm \eta}$ with $\left(\delta > 0, \zeta < 0\right)$, $\left(\delta < 0, \zeta < 0\right)$, $\left(\delta > 0, \zeta > 0\right)$ and $\left(\delta < 0, \zeta > 0\right)$ be $S_{1+}^{USp(4)}$, $S_{1'+}^{USp(4)}$, $S_{2+}^{USp(4)}$ and $S_{2'+}^{USp(4)}$ respectively for $(n, N_F)=(2, 3)$. Also we denote the sets of poles of \eqref{USpO2mU1} from the JK parameter $\zeta$ with $\zeta < 0$ and $\zeta > 0$ by $S_{1-}^{USp(4)}$ and $S_{2-}^{USp(4)}$ respectively for $(n, N_F) = (2, 3)$. Then we use these poles for computing the residues of \eqref{USpO2pU1} and \eqref{USpO2mU1} in the case of $(n, N_F) = (2, 6)$ and take the average like \eqref{ZUSpwedge2VVei}. The contribution from the sets $S_{1+}^{USp(4)}$ and $S_{1-}^{USp(4)}$ is the same as that from the pole sets $S_{1'+}^{USp(4)}$ and $S_{1-}^{USp(4)}$, and the contribution in the sector $\bv = \be_1$ is equal to $Z_{\text{mono}}^{\wedge^2 V \times V}({\bm v}  =\be_1)$ with $(n, N_F) = (2, 6)$. Similarly the contribution from the sets $S_{2+}^{USp(4)}$ and $S_{2-}^{USp(4)}$ is the same as that from the pole sets $S_{2'+}^{USp(4)}$ and $S_{2-}^{USp(4)}$, and the contribution in the sector $\bv = \be_1$ is equal to $Z_{\text{mono}}^{V\times \wedge^2 V}({\bm v}  =\be_1)$ with $(n, N_F) = (2, 6)$. In this case the Moyal product depends on the order and $Z_{\text{mono}}^{\wedge^2 V \times V}({\bm v}  =\be_1) \neq Z_{\text{mono}}^{V\times \wedge^2 V}({\bm v}  =\be_1)$. 

Finally we consider the monopole screening contribution in the $\bv = {\bm 0}$ sector. We denote the sets of poles of \eqref{USpO3pU1} from the JK parameter ${\bm \eta}$ with $\left(\delta > 0, \zeta < 0\right)$, $\left(\delta < 0, \zeta < 0\right)$, $\left(\delta > 0, \zeta > 0\right)$ and $\left(\delta < 0, \zeta > 0\right)$ by $\widetilde{S}_{1+}^{USp(4)}$, $\widetilde{S}_{1'+}^{USp(4)}$, $\widetilde{S}_{2+}^{USp(4)}$ and $\widetilde{S}_{2'+}^{USp(4)}$ respectively for $(n, N_F)=(2, 3)$. Also let the sets of poles of \eqref{USpO3mU1} from the JK parameter ${\bm \eta}$ with $\left(\delta > 0, \zeta < 0\right)$, $\left(\delta < 0, \zeta < 0\right)$, $\left(\delta > 0, \zeta > 0\right)$ and $\left(\delta < 0, \zeta > 0\right)$ be $\widetilde{S}_{1-}^{USp(4)}$, $\widetilde{S}_{1'-}^{USp(4)}$, $\widetilde{S}_{2-}^{USp(4)}$ and $\widetilde{S}_{2'-}^{USp(4)}$ respectively for $(n, N_F) = (2, 3)$. We then evaluate the integral of \eqref{USpO3pU1} and \eqref{USpO3mU1} with $(n, N_F) = (2, 6)$ using the poles and take the average like \eqref{ZUSpwedge2VV0}. In this case the contributions from the pairs of the sets $(\widetilde{S}_{1+}^{USp(4)}, \widetilde{S}_{1-}^{USp(4)})$, $(\widetilde{S}_{1'+}^{USp(4)}, \widetilde{S}_{1'-}^{USp(4)})$, $(\widetilde{S}_{2+}^{USp(4)}, \widetilde{S}_{2-}^{USp(4)})$, $(\widetilde{S}_{2'+}^{USp(4)}, \widetilde{S}_{2'-}^{USp(4)})$ all give the same result and it agrees with $Z_{\text{mono}}^{\wedge^2 V \times V}({\bm v}  ={\bm 0})  = Z_{\text{mono}}^{V\times \wedge^2 V}({\bm v}  ={\bm 0})$ for $(n, N_F)  = (2, 6)$.

\bibliography{refs}

\end{document}